# Postcranial Pneumaticity in Dinosaurs and the Origin of the Avian Lung

by

Mathew John Wedel

B.S. (University of Oklahoma) 1997

A dissertation submitted in partial satisfaction of the

requirements for the degree of

Doctor of Philosophy

in

Integrative Biology

in the

Graduate Division

of the

University of California, Berkeley

Committee in charge:

Professor Kevin Padian, Co-chair
Professor William Clemens, Co-chair
Professor Marvalee Wake
Professor David Wake
Professor John Gerhart

Spring 2007

The dissertation of Mathew John Wedel is approved:

| | | | |
|---|---|---|---|
| Co-chair | ______________________________ | Date | __________ |
| Co-chair | ______________________________ | Date | __________ |
| | ______________________________ | Date | __________ |
| | ______________________________ | Date | __________ |
| | ______________________________ | Date | __________ |

University of California, Berkeley

Spring 2007

Postcranial Pneumaticity in Dinosaurs and the Origin of the Avian Lung

by Mathew John Wedel

Abstract

Postcranial Pneumaticity in Dinosaurs and the Origin of the Avian Lung

by

Mathew John Wedel

Doctor of Philosophy in Integrative Biology

University of California, Berkeley

Professor Kevin Padian, Co-chair

Professor William Clemens, Co-chair

Among extant vertebrates, postcranial skeletal pneumaticity is present only in birds. In birds, diverticula of the lungs and air sacs pneumatize specific regions of the postcranial skeleton. The relationships among pulmonary components and the regions of the skeleton that they pneumatize form the basis for inferences about the pulmonary anatomy of non-avian dinosaurs. Fossae, foramina and chambers in the postcranial skeletons of pterosaurs and saurischian dinosaurs are diagnostic for pneumaticity.

In basal saurischians only the cervical skeleton is pneumatized. Pneumatization by cervical air sacs is the most consilient explanation for this pattern. In more derived sauropods and theropods pneumatization of the posterior dorsal, sacral, and caudal vertebrae indicates that abdominal air sacs were also present. The presence of abdominal air sacs in sauropods is also indicated by a pneumatic hiatus (a gap in the pneumatization of the vertebral column) in Haplocanthosaurus. Minimally, most sauropods and theropods had a dorsally attached diverticular lung plus air sacs both anterior and posterior to the lung, and thus had all of the pulmonary prerequisites for flow-through lung ventilation like that of birds.

Pneumaticity reduced the mass of the postcranial skeleton in sauropods and theropods. I propose the Air Space Proportion (ASP) as a measure of the proportional volume of air in pneumatic bones. The mean ASP of a sample of sauropod and theropod vertebrae is 0.61. This means that, on average, air occupied more than half of the volume of pneumatic saurischian vertebrae. In Diplodocus, postcranial pneumatization is calculated to have lightened the living animal by 7-10%—and that does not include the extraskeletal diverticula, pulmonary air sacs, lungs, or tracheae. If all of these air reservoirs are taken into account, the specific gravity of Diplodocus is 0.80, higher than published values for birds but lower than those for squamates and crocodilians.

Pneumatization of the cervical vertebrae probably facilitated the evolution of long necks in sauropods. Necks longer than nine meters evolved at least four times, in mamenchisaurs, diplodocids, brachiosaurids, and titanosaurs. Increases in the number of cervical vertebrae, their proportional lengths, and their internal complexity occurred in parallel in most of these lineages.

I dedicate this dissertation to

Vicki and London Wedel.

In my work, my laughter and my dreams,

you are my Muses and the ends to my means.

# ACKNOWLEDGMENTS

A person is lucky to find an advisor with whom he gets along, who will inspire him to achieve things that he did not dream for himself, but who still retains the ease and approachability of friendship. It has been my peculiar fortune to have had not one but three such advisors in my graduate career: Kevin Padian and Bill Clemens during my dissertation work at Berkeley, and Rich Cifelli during my master's work at the University of Oklahoma. Kevin and Bill have challenged me to work harder and think more deeply than I ever have before, both directly and by their examples. Their doors, their homes, and their hearts have been open. I am proud to call them my friends, but the word that best describes my feelings toward them is fealty.

David and Marvalee Wake served on my qualifying exam committee and on my dissertation committee. I was very lucky to have had them both as instructors in my first year at Berkeley. I have always been intimidated by the depth of their knowledge and by the conceptual breadth that they bring to the big questions in evolutionary biology, and I asked them to be on my committees because I knew that I would work harder if I had to face them at the end. They are both surpassingly open and kind, and I'm sure that they will be either amused or embarrassed by this confession. The total time that I've spent talking with each of them must add up to only a handful of hours, but those conversations tower over all of the other experiences of my six years here.

Clark Howell served as the outside member on my qualifying exam committee, and served the same role on my dissertation committee for four years. He was a wonderful mentor and friend, and his passing this spring was a shocking loss. The benefit that I gained from him is entirely out of proportion to the time we spent together. I am glad that I got to see him and have one more dazzling, wide-ranging discussion before he went. I wish that I could have shown him this dissertation, and told him how much his input and his approval meant to me.

John Gerhart got me thinking about development as a motor of evolutionary change right after I got to Berkeley, and he generously agreed to serve as my outside member here at the end of the project. I thank him for both inspiration and assistance. I aspire to someday bring his ideas to the study of pneumaticity.

David Lindberg and Brent Mishler broadened my mind in unexpected and breathtaking ways. I am grateful to David for serving on my qualifying exam committee, and I am grateful to both of them for always greeting me with a smile and always leaving me with a head full of new ideas.

Pat Holroyd probably taught me more about how to survive as a grad student and how to conduct myself as a professional than anyone else at Berkeley. I know that she has done the same for a lot of other graduate students, too. Officially she's a museum scientist in the UCMP; unofficially she is one of the best teachers in the department. I am grateful to her for advice, commiseration, and friendship. Mark Goodwin and Jane Mason have been good friends as well.

I have been very fortunate to have friendly and lively labmates during my time at Berkeley. Former Padian lab members Ken Angielczyk and Jim Parham are my Obi-Wan and Yoda, respectively. If I could do it over again I would spend more time working with Drew Lee, writing with Randy Irmis, watching movies with Sarah

Werning, and playing pool with Jackie Moustakas. Katie Brakora and Brian Swartz have also been great fun.

I arrived at Berkeley with a superb cohort, the Ones. Many thanks to all of them for good times and good thoughts, especially Brian Kraatz, Alan Shabel, Joel Abraham, Matt Butler, and Eric Harris. I am also grateful to Nick Pyenson for being a constant source of inspiration and entertainment.

My dissertation research was funded by grants from the Department of Integrative Biology, the UCMP, the Doris O. and Samuel P. Welles Fund, the Jurassic Foundation, and the Berkeley chapter of Sigma Xi. I thank Kent Sanders and Jay Grimaldi for friendship, hospitality, and CT scanning.

My parents, John and Norma Wedel, allowed themselves to be dragged to museums all over the country when I was a kid. They also instilled a love of learning and a work ethic that have never ceased to reward me. Everything they ever taught me has paid off. I have tried to make my career an act of thanks.

# INTRODUCTION

The goal of this dissertation is to explore postcranial pneumaticity in non-avian dinosaurs, both as an interesting phenomenon in its own right, and as the skeletal footprint of the respiratory system. Most previous studies of pneumaticity have focused on theropods and their air sac systems. I have tried to broaden the scope of our knowledge of pneumaticity through inquiries into (1) the origins of pneumaticity in basal saurischians, (2) the evolution of pneumaticity in sauropodomorph dinosaurs, (3) the implications of pneumaticity for the origin of the avian air sac system, and (4) the effect of pneumatization on the mass of dinosaurs, especially as a potential factor in neck elongation in sauropods.

In Chapter One I review the evidence for pneumaticity in sauropod dinosaurs and attempt to describe as many aspects of the system as possible. These aspects include the external traces of pneumaticity on skeletal elements, the internal structure of pneumatic bones, the ratio of bone tissue to air space within a pneumatic element, and the distribution of pneumaticity in the skeleton. I also explore the implications of pneumaticity for mass estimates of sauropods.

Chapter Two is an evaluation of the evidence for pneumaticity in basal sauropodomorphs (or ‘prosauropods’). Most ‘prosauropods’ lack unequivocal evidence of postcranial pneumaticity, but they are phylogenetically bracketed by sauropods and theropods that have extensive postcranial pneumaticity. I discuss the implications of this phylogenetic distribution of pneumaticity for the origin—or origins—of air sacs in Saurischia.

I present evidence for air sacs in non-avian dinosaurs in Chapter Three, and review alternative hypotheses and arguments against the air sac hypothesis. I also introduce new frameworks for testing the hypotheses that most saurischian dinosaurs had cervical and abdominal air sacs like those of birds, and I describe new evidence that supports the air sac hypothesis.

Chapter Four deals with the evolution of long necks in sauropod dinosaurs. Neck elongation occurred in parallel in several sauropod lineages, and increases in the distribution, complexity, and lightness of pneumatic bones appear to have been related to the evolution of long necks. I test these apparent correlations by mapping the relevant characters onto a phylogeny of sauropods, and also in statistical analyses using phylogenetically independent contrasts. These analyses suggest that size and neck length were not correlated in sauropods, but the relationships among these variables and pneumaticity are not clear.

## ABBREVIATIONS

**Institutional abbreviations**—AMNH, American Museum of Natural History, New York, USA; BYU, Earth Sciences Museum, Brigham Young University, Provo, USA; BMNH, The Natural History Museum, London, UK; CCG, Chengdu College of Geology, Chengdu, China; CM, Carnegie Museum of Natural History, Pittsburgh, USA; DGM, Museo de la Divisão Geologia y Mineralogia, Rio de Janeiro, Brazil; DMNH, Denver Museum of Nature and Science, Denver, Colorado; FMNH, Field Museum of Natural History, Chicago, USA; GPIT, Institut für Paläontologie und Geologie der Universität Tübingen, Tübingen, Germany; HM, Humbolt Museum für Naturkunde, Berlin, Germany; IGM, Mongolian Institute of Geology, Ulan Bataar, Mongolia; IVPP, Institute of Vertebrate Paleontology and Paleoanthropology, Beijing, China; LCM, Leicester City Museum, Leicester, UK; LVP-GSC, Laboratory of Vertebrate Paleontology, Geological Survey of China; MAL, Malawi Department of Antiquities Collection, Lilongwe and Nguludi, Malawi; MIWG, Museum of the Isle of Wight, UK; MNN, Musée National du Niger, Niamey, Niger Republic; MPM, Museo Padre Molina, Rio Gallegos, Santa Cruz, Argentina; MSM, Mesa Southwest Museum, Mesa, USA; MVZ, Museum of Vertebrate Zoology, Berkeley, USA; NMMNH, New Mexico Museum of Natural History, Albuquerque, USA; OMNH, Oklahoma Museum of Natural History, Norman, USA; OUMNH, Oxford University Museum of Natural History, Oxford, UK; PMU, Paleontological Museum, Uppsala, Sweden; PVL, Colección de Paleontología de Vertebrados de la Fundación Instituto Miguel Lillo, Tucumán, Argentina; PVSJ, Museo de Ciencias Naturales, Universidad Nacional de San Juan, San Juan, Argentina; SMNS, Staatliches Museum fur Natürkunde, Stuttgart, Germany; UCMP, University of California Museum of Paleontology, Berkeley, USA; USNM, National Museum of Natural History, Washington, D.C., USA; WDC, Wyoming Dinosaur Center, Thermopolis, USA; YPM, Yale Peabody Museum, New Haven, USA.

**Anatomical abbreviations—**ACDL, anterior centrodiapophyseal lamina; AL, accessory lamina; AVF, anteroventral fossa; CIT, canalis intertransversarius; CML, camella; CMR, camera; DSV, diverticulum supervertebrale; FOR, foramen; FOS, fossa; LAM, lamina; NAD, neural arch diverticulum; NAF, neural arch fossa; NCL, neural canal; NCS, neurocentral suture; NCV, neural cavity; NSF, neural spine fossa; PCDL, posterior centrodiapophyseal lamina; PDF, posterodorsal fossa; PODL, postzygodiapophyseal lamina; PPDL, paradiapophyseal lamina; PRDL, prezygodiapophyseal lamina; SPOL, spinopostzygapophyseal lamina; SPRL, spinoprezygapophyseal lamina; VK, ventral keel (lamina abbreviations after Wilson 1999).

# CHAPTER ONE

# POSTCRANIAL SKELETAL PNEUMATICITY IN SAUROPODS AND ITS IMPLICATIONS FOR MASS ESTIMATES

## INTRODUCTION

One of the signal features of sauropods, and one of the cornerstones of our fascination with them, is their apparent efficiency of design. The presacral neural spines of all sauropods have a complex of bony ridges or plates known as vertebral laminae (Fig. 1-1; abbreviations used in the figures are listed below). In addition, the vertebral centra of most sauropods bear deep fossae or have large foramina that open into internal chambers. The laminae and cavities of sauropod vertebrae are often considered to be adaptations for mass reduction (Osborn, 1899; Hatcher, 1901; Gilmore, 1925) and have been important in studies of sauropod evolution (McIntosh, 1990; Wilson, 1999). The possibility that these structures were pneumatic—that they contained or partitioned air-filled diverticula of the lungs or air sacs—has been recognized for over a century (Seeley, 1870; Janensch, 1947). However, pneumaticity in sauropods has received little attention until recently (Britt, 1997; Wilson, 1999; Wedel, 2003a, b).

My goal herein is to review previous work on pneumaticity in sauropods, discuss some outstanding problems, and outline possible directions for future studies. To that end, the paper is organized around three questions. What criteria do we use to infer pneumaticity in sauropod fossils? What characteristics of pneumatic bones have been (or could be) described? Finally, how can we apply data on skeletal pneumaticity to paleobiological problems, such as estimating the masses of sauropods? Before attempting to answer these questions, it will be useful to review skeletal pneumaticity in living vertebrates.

## SKELETAL PNEUMATICITY IN EXTANT TAXA

Pneumatization of the postcranial skeleton in various ornithodiran groups, including sauropods, is just one instance of the more general phenomenon of skeletal pneumatization. Skeletal pneumatization, which includes paranasal, paratympanic, and pulmonary pneumatic spaces, is unique to archosaurs and advanced synapsids (Witmer, 1997, 1999). However, diverticula (epithelium-lined outgrowths) of the pharynx or trachea are present in representative taxa from most major lineages of tetrapods, including frogs (Duellman and Trueb, 1986), snakes (Young, 1991, 1992), birds (King, 1966; McClelland, 1989a), and primates (Janensch, 1947). Pharyngeal and tracheal diverticula are often used to inflate specialized structures used in phonation or visual display. These diverticula do not invade any bones except the hyoid, which is pneumatized by tracheal diverticula in the howler monkey Alouatta (Janensch, 1947; Mycetes of his usage). Diverticula of paranasal and paratympanic air spaces extend down the neck in some species of birds, but these diverticula are subcutaneous or intermuscular and do not pneumatize the postcranial skeleton (King, 1966). Extremely rare examples of cervical pneumatization have been reported in humans, but these are pathological cases related to occipito-atlantal fusion (Sadler et al., 1996). Among extant taxa, only birds have extensive postcranial skeletal pneumaticity (PSP).

Extant birds have relatively small, inflexible lungs and an extensive system of air sacs in the thorax and abdomen. The air sacs are flexible and devoid of parenchymal tissue, and their primary function is to ventilate the lungs (King, 1966; Duncker, 1971; McClelland, 1989b). In most birds, the air sacs also give rise to a network of diverticula. Diverticula pass into the viscera, between muscles, and under the skin in various taxa (Richardson, 1939; King, 1966; Duncker, 1971). If a diverticulum comes into contact with a bone, the bone may become pneumatized. Bremer (1940) described the pneumatization of the humerus in the chicken (Gallus) as follows. The diverticulum enters the bone because osteoclasts break down the bony tissue ahead of it. The bony tissue immediately adjacent to the diverticulum is replaced by mesenchymal tissue, which degenerates or is resorbed and is in turn replaced by the growing diverticulum. As the diverticulum bores through the cortical bone it produces a pneumatic foramen, which must remain open for pneumatization to proceed normally (Ojala, 1957). Once the bone has been penetrated, branches of the diverticulum spread through the marrow cavity by replacing bony trabeculae. The marrow is reduced to small islands of tissue surrounded by the diverticulum. As these islands of marrow degenerate, the branches of the diverticulum anastomose and form a single, epithelium-lined air cavity that occupies most of the internal volume of the bone. The trabecular structure of the bone is greatly reduced, and the inner layers of the cortex are resorbed.

Witmer (1990) pointed out that a pneumatic foramen does not have to be located on the pneumatic bone in question; the intraosseous diverticulum may have spread across a suture from an adjacent pneumatic bone. He called this extramural pneumatization and contrasted it with intramural pneumatization, in which a diverticulum directly invades a bone and produces a pneumatic foramen. Although Witmer (1990) was concerned with cranial pneumatization, extramural pneumatization also occurs in the postcranial skeleton, for example, between fused vertebrae in the chicken (King, 1957; Hogg, 1984a).

The term air sac has been used by some authors for any reservoir of air in an animal that is lined by epithelium and devoid of parenchymal tissue (e.g., Brattstrom, 1959; Cranford et al., 1996). The same term is often used in the ornithological literature to refer specifically to the pulmonary air sacs of birds (e.g., Müller, 1907). In this paper, the term air sac is restricted to indicate the pulmonary air sacs of birds. All other epithelium-lined air reservoirs, including those that develop from the lungs and air sacs, are called diverticula. Another important difference is between a pneumatic diverticulum, which is a soft-tissue structure, and the bony recess that it may occupy (Witmer, 1999). In many cases, the bony recess is produced by the diverticulum through the process of pneumatization. This causal relationship allows us to infer the presence of diverticula from certain kinds of bony recesses. The study of skeletal pneumaticity in fossil taxa is founded upon such inferences.

## WHAT CRITERIA DO WE USE TO INFER PNEUMATICITY IN FOSSILS?

How do we recognize skeletal pneumaticity? More specifically, what are the osteological correlates (sensu Witmer, 1995, 1997) of pneumatic diverticula, such that the presence of the latter can be inferred from the former? Several authors, including Hunter (1774) and Müller (1907), list differences between pneumatic and apneumatic bones. These authors focused on recognizing pneumaticity in extant birds and thus referred to attributes that tend not to fossilize, such as vascularity, oil content, and color. Britt (1993, 1997) provided the most comprehensive list of pneumatic features identifiable in fossil bones: internal chambers with foramina, fossae with crenulate texture, smooth or crenulate tracks (grooves), bones with thin outer walls, and large foramina.

### Internal Chambers With Foramina

The most obvious osteological correlate of pneumaticity is the presence of foramina that lead to large internal chambers. Large chambers, often called ‘pleurocoels,’ are present in the presacral vertebrae of most sauropods. They may also be present in the sacral and caudal vertebrae, as in Apatosaurus and Diplodocus (see Ostrom and McIntosh, 1966:pl. 30 and Osborn, 1899:fig. 13, respectively). In extant birds, such chambers are invariably associated with pneumatic diverticula (Britt, 1993). The presence of similar chambers in the bones of sauropods, theropods, and pterosaurs has been accepted by most authors as prima facie evidence of pneumaticity (Seeley, 1870; Cope, 1877; Marsh, 1877; Janensch, 1947; Romer, 1966; Britt, 1993, 1997; O’Connor, 2002). As far as I am aware, no substantive alternative hypotheses have been advanced; as Janensch (1947:10, translated from the German by G. Maier) said, “There is no basis to consider the pleurocentral cavities in sauropod vertebrae as different from similar structures in the vertebrae of birds.” In short, no soft tissues other than pneumatic diverticula are known to produce large foramina that lead to internal chambers, and these chambers constitute unequivocal evidence of pneumaticity.

One of the primary differences between the pneumatic vertebrae of different sauropod taxa is the subdivision of the internal chambers. Some taxa, such as Camarasaurus, have only a few large chambers, whereas others, such as Saltasaurus, have many small chambers (Fig. 1-1). Vertebrae with many small chambers have been characterized as ‘complex’ (Britt, 1993; Wedel, 2003b), in contrast to ‘simple’ vertebrae with few chambers. The concept of ‘biological complexity’ has several potential meanings (McShea, 1996). In this paper, complexity refers only to the level of internal subdivision of pneumatic bones; complex bones have more chambers than simple ones. This is ‘nonhierarchical object complexity’ in the terminology of McShea (1996).

**Extramural Pneumatization**—The only obvious opportunities for extramural pneumatization in the postcranial skeletons of sauropods are between fused sacral and caudal vertebrae and between the sacrum and ilium. Sacral vertebrae of baby sauropods have deep fossae (Wedel et al., 2000:fig. 14), and, at least in Apatosaurus, a complex of internal chambers is present before the sacral vertebrae fuse (Ostrom and McIntosh, 1966:pl. 30). The co-ossified blocks of caudal vertebrae in Diplodocus often include centra with large pneumatic foramina (Gilmore, 1932:fig. 3). It is possible that co-ossified centra without foramina could be pneumatized by intraosseous diverticula of adjacent pneumatic vertebrae, although this has not been demonstrated.

Sanz et al. (1999) reported that 'cancellous tissue' is present in the presacral vertebrae, ribs, and ilium of Epachthosaurus and Saltasaurus. The presacral vertebrae of Saltasaurus are pneumatic and have camellate internal structure (Fig. 1-1K-N), and pneumatic ribs are known in several titanosaurs (Wilson and Sereno, 1998). Further, spongiosa (sensu Francillon-Vieillot et al., 1990) are present in apneumatic vertebrae of many—possibly all—sauropods (see the section on mass estimates below), so cancellous bone is not limited to titanosaurs. For these reasons, it seems that the 'cancellous tissue' of Sanz et al. (1990) is synonymous with camellate pneumatic bone. If so, then the ilia of some titanosaurs may have been pneumatic. Two possible routes for pneumatization of the ilium are by diverticula of abdominal air sacs or by extramural pneumatization from the sacrum. However, the possibility of ilial pneumatization must remain speculative until better evidence for it is presented.

**Neural Cavities**—In many sauropods, the neural spines of the dorsal vertebrae contain large chambers. These chambers communicate with the outside by way of large foramina beneath the diapophyses. Upchurch and Martin (2003) called such chambers neural cavities and discussed their occurrence in Cetiosaurus, Barapasaurus, and Patagosaurus. According to Upchurch and Martin (2003:218), "In Barapasaurus and Patagosaurus, the neural cavity is linked to the external surface of the arch by a lateral foramen which lies immediately below the base of the transverse process, just in front of the posterior centrodiapophyseal lamina [pcdl]" (see Fig. 1-1A). In some dorsal vertebrae of Barapasaurus, the neural canal is open dorsally and communicates with the neural cavity (Jain et al., 1979). Upchurch and Martin (2003) mentioned that similar cavities are present in some neosauropods, and Bonaparte (1986:fig. 19.7) illustrated neural cavities in Camarasaurus and Diplodocus. Jain et al. (1979) and Upchurch and Martin (2003) also described a second morphology (in Barapasaurus and Cetiosaurus, respectively), in which the neural cavity is divided into two halves by a median septum and does not communicate with the neural canal (Fig. 1-1C). Neural cavities are interpreted as pneumatic for the same reason that the more familiar cavities in vertebral centra are: they are large internal chambers connected to the outside through prominent foramina (Britt, 1993).

**Pneumatic Ribs**—The dorsal ribs of some sauropods have large foramina that lead to internal chambers. The best known examples of costal pneumaticity in sauropods are the pneumatic ribs of Brachiosaurus (Riggs, 1904; Janensch, 1950). Pneumatic dorsal ribs are also present in Euhelopus and some titanosaurs (Wilson and Sereno, 1998). Gilmore (1936) described a foramen that leads to an internal cavity in a dorsal rib of Apatosaurus, and pneumatic dorsal ribs have also been reported in the diplodocid Supersaurus (Lovelace et al., 2003). Pneumatic dorsal ribs have not been found in Haplocanthosaurus, Camarasaurus, or any basal diplodocoids, so the character evidently evolved independently in diplodocids and titanosauriforms. Pneumatic ribs are part of a mounting list of pneumatic characters that evolved in parallel in diplodocids and titanosauriforms, along with complex vertebral chambers and pneumatic caudal vertebrae (see below).

## Fossae and Laminae

**Pneumatic Fossae—**Fossae are ubiquitous in sauropod vertebrae and are often the sole evidence of pneumaticity. For example, basal sauropods such as Barapasaurus have shallow fossae on the presacral centra and neural spines, but lack the large internal chambers typical of later sauropods (Fig. 1-1). Are these fossae pneumatic? The naive assumption that all fossae are pneumatic will surely lead to the overestimation of pneumaticity. On the other hand, to deny that any fossae are pneumatic unless they contain foramina that lead to large internal chambers is equally false. We need criteria to distinguish pneumatic fossae from non-pneumatic fossae.

The best case for a pneumatic fossa is a fossa that contains pneumatic foramina within its boundaries. The Brachiosaurus vertebra shown in Fig. 1-2 has large, sharply-lipped pneumatic foramina in most of the fossae on the lateral sides of the centrum and neural spine (see also Janensch, 1950, and Wilson, 1999). Similar foramina-within-fossae are present in the vertebrae of many other neosauropods, including Diplodocus (Hatcher, 1901:pls. 3 and 7), Tendaguria (Bonaparte et al., 2000:fig. 17 and pl. 8), and Sauroposeidon (Wedel et al., 2000:fig. 8b). The inference that these fossae are pneumatic relies on the presence of unequivocally pneumatic features within the fossae. The inference pneumaticity is less supported in the case of blind fossae that contain no foramina, such as the large fossae on the dorsal centra of Barapasaurus (Fig. 1-1).

Wilson (1999) proposed that ‘subfossae,’ or fossae-within-fossae, might further support the inference of pneumaticity. “These well defined, smooth-walled depressions are present in many sauropods and seem to be analogous to the more pronounced coels [foramina] that characterize Brachiosaurus. Like the coels, these depressions may have housed smaller pneumatic diverticuli [sic] in life" (Wilson, 1999:651). This hypothesis is supported by the complex morphology of some pneumatic diverticula in birds. In the ostrich, the large diverticula that lay alongside the cervical vertebrae consist of bundles of smaller diverticula (Wedel, 2003a:fig. 2). It seems reasonable to expect that when such a bundle comes into contact with a bone, the aggregate would produce a fossa, within which each diverticulum would produce a subfossa. This hypothesis can and should be tested in future computed tomography (CT) studies. Gower (2001:121) argued that the "multipartite fossae" and "deep multi-chambered concavities" in the dorsal vertebrae of Erythrosuchus were more consistent with pneumaticity than with muscular or vascular structures (but see O'Connor, 2002).

Britt (1993) proposed that crenulate texture of the external bone is evidence that some fossae are pneumatic. In Sauroposeidon the difference in texture between the pneumatic fossae and the adjacent bone is striking, and this allows the boundaries of the fossae to be precisely plotted (Wedel et al., 2000:fig. 7). However, there is little doubt that the fossae of Sauroposeidon are pneumatic, because they contain pneumatic foramina. The inference that a blind fossa is pneumatic based on texture alone is less certain. Blind fossae can also contain muscles or adipose tissue (O'Connor, 2002). It is not known if these three kinds of fossae can be reliably distinguished on the basis of bone texture. Until this is tested, inferring pneumaticity on the basis of bone texture alone may not be warranted.

For the time being, I know of no test that can definitively determine whether a blind fossa housed a pneumatic diverticulum or some other soft tissue. Pneumatic diverticula often induce bone resorption when they come into contact with the skeleton, and it is possible that external pneumatic features might be recognized by some distinctive aspect of cortical bone histology. I do not suggest that this must be the case, but it is worth investigating.

To determine if a fossa is pneumatic or not, it is worthwhile to consider other potentially pneumatic features on or in the same bone. Consider the fossa bounded by the podl, prdl, spol, and sprl in Haplocanthosaurus (Fig. 1-3). At least in the cervical vertebrae, these fossae do not contain any pneumatic foramina or subfossae, they do not lead to any obvious pneumatic tracks, and the bone texture is smooth rather than crenulate (pers. obs.). In other words, nothing about the fossae themselves indicates that they were pneumatic (as opposed to containing adipose deposits or other soft tissues). However, the centra of the same vertebrae contain deep, sharp-lipped cavities that penetrate to a narrow median septum. By the criteria discussed herein, the cavities in the centra are unequivocally pneumatic. Their presence demonstrates that pneumatic diverticula were in close contact with all of the preserved cervical vertebrae. Because we already know that pneumatic diverticula contacted the cervical vertebrae, it seems safe to infer that the neural spine fossae are pneumatic in origin. At least, the inference of pneumaticity is better founded than it would be based on the neural spine fossae alone.

(As an aside, the nomenclature for vertebral laminae has been thoroughly reviewed and standardized [Wilson, 1999], but no standard nomenclature for vertebral fossae exists. It is tempting to propose such a nomenclature, if only to avoid circumlocutions like that used above ["the fossae bounded by the podl, prdl, spol, and sprl"]. However, a separate nomenclature for fossae is unnecessary and could be misleading. Hatcher [1901] named several fossae, such as the "infraprezygapophyseal cavity," using the same spatial orientation terms that were commonly used for naming laminae [e.g., Osborn, 1899]. Such a position-based nomenclature for fossae shares all of the faults of the old orientation-based systems for naming laminae [see Wilson, 1999 for further discussion]. Laminae should be defined by the structures they connect [Wilson, 1999]. Similarly, I think that fossae should be defined by the laminae that bound them. To list all of the bounding laminae when referring to a fossa may be awkward, but it is also precise.)

**Vertebral Laminae and the Origins of PSP—**If we order archosaur vertebrae in terms of putatively pneumatic features, the resulting arrangement has no obvious gaps and is roughly congruent with current phylogenies (i.e., Sereno, 1991; Wilson, 2002). At one end of the spectrum are vertebrae that lack laminae, such as those of extant crocodilians. Very shallow depressions may be present on the neural spines or centra, but these depressions are not bounded by an obvious lip and do not contain subfossae or large foramina. The next grade of vertebral construction is represented by Marasuchus, which has low ridges below some of the presacral diapophyses (Sereno and Arcucci, 1994); these ridges may represent rudimentary laminae (Wilson, 1999). At the next level, a series of diapophyseal and zygapophyseal laminae is primitive for Saurischia (Wilson, 1999). These laminae are present in Herrerasaurus and prosauropods (Sereno and Novas, 1994; Bonaparte, 1986), but the fossae they enclose are blind, lack subfossae, and have no obvious textural differences from the adjacent bone (pers. obs.). Vertebral centra of these taxa lack fossae. Shallow fossae are present on the centra of early sauropods such as Isanosaurus, Shunosaurus, and Barapasaurus, and neural chambers may be present in the arch and spine (Jain et al., 1979; Zhang, 1988; Buffetaut et al., 2000). In Jobaria and Haplocanthosaurus the central fossae are bounded by a sharp lip and penetrate to a median septum (Sereno et al., 1999; Wedel, 2003b; pers. obs.). Finally, most neosauropods have prominent pneumatic foramina that open into chambers that ramify within the centrum, and the fossae of the neural arches and spines contain subfossae or pneumatic foramina.

It is not clear where pneumaticity first appears in the preceding series. At one end of the scale are the vertebrae of crocodiles, which are known to be apneumatic. At the other end are the vertebrae of neosauropods, the pneumatic features of which are virtually identical to those of birds (Janensch, 1947). In between, the inference of pneumaticity receives more support as we approach Neosauropoda, but the "break point" between apneumatic and pneumatic morphologies is debatable. The primitive saurischian complex of laminae first appears in small dinosaurs and seems to be structural overkill if pneumatic diverticula were absent (Wilson, 1999). An apneumatic interpretation of these laminae requires that a large number of structures that are clearly related to pneumatization in later forms be primitively present for other reasons, and leaves us (at least for now) without a satisfying hypothesis to explain the origin of vertebral laminae. The blind fossae of early saurischians are, at best, equivocal evidence of pneumaticity. However, any explanation that pushes the origin of PSP forward in time will accumulate a corresponding number of ad hoc hypotheses to explain the early appearance of laminae and fossae. For these reasons, I favor Wilson's (1999) hypothesis that laminae are pneumatic in origin and that the appearance of laminae marks the appearance of PSP, although, as Wilson (1999:651) pointed out, more work is needed.

Gower (2001) posited widespread pneumaticity in Archosauria based on vertebral fossae. If he is right, PSP originated before the divergence between crocodile- and bird-line archosaurs and was present in virtually all of the non-crocodilian taxa in the series discussed above. O'Connor (2002) questioned the reliability of blind fossae as indicators of pneumaticity, but he did not present evidence to falsify Gower's hypothesis. Indeed, hypotheses of pneumaticity are difficult to falsify; although it is often easy to demonstrate that a bone has been pneumatized, it is difficult to demonstrate that it has not (Hogg, 1980). For now, the possibility that the fossae described by Gower are pneumatic cannot be ruled out, but neither can less radical alternative hypotheses.

**Other Osteological Correlates of Pneumaticity**

Pneumatic tracks, thin outer walls, and large foramina are not likely to be falsely interpreted as pneumatic features in sauropods. External tracks are only rarely identified in sauropods. Wedel et al. (2000:fig. 7) illustrated a pneumatic track in Sauroposeidon, but the track was not the basis for the pneumatic interpretation; rather, the track was identified as pneumatic because it led away from a deep, sharply-lipped pneumatic fossa. Many sauropod vertebrae have thin outer walls, especially those of the aforementioned Sauroposeidon (Fig. 1-4). However, the thin outer walls of sauropod vertebrae invariably bound large internal chambers that are clearly pneumatic, so, again, the inference of pneumaticity does not rest on the equivocal feature. Finally, there is the question of foramina that are not pneumatic, such as nutrient or nervous foramina. Britt et al. (1998) proposed that pneumatic foramina could be distinguished from nutrient foramina on the basis of relative size, with pneumatic foramina typically being about an order of magnitude larger, relative to the length of the centrum. The two kinds of foramina could also be distinguished based on the internal structure of the vertebrae. Pneumatic vertebrae typically lack trabecular bone (Bremer, 1940; Schepelmann, 1990), and have compact bone in their outer walls and in the septa between pneumatic cavities (Reid, 1996). The presence of trabecular bone inside a vertebra is evidence that it is either apneumatic, or at least incompletely pneumatized (King, 1957). Distinguishing pneumatic foramina from nutrient foramina is a potential problem in studies of birds and other small theropods, but most sauropods are simply so large that pneumatic and nutrient foramina are unlikely to be confused. Even juvenile sauropods tend to have large pneumatic fossae rather than small foramina (see Wedel et al., 2000:fig. 14).

## DESCRIPTION OF PNEUMATIC ELEMENTS

At least four aspects of skeletal pneumaticity can be described: the external traces of pneumaticity (discussed above); the internal complexity of an element; the ratio of bone to air space within an element; and the distribution of pneumatic features along the vertebral column.

### Internal Complexity of Pneumatic Bones

This variable has received the most attention in previous studies, and is only briefly reviewed here. Longman (1933) recognized that sauropod vertebrae with internal chambers fall into two broad types, those with a few large chambers and those with many small chambers. Longman called the first type phanerocamerate and the second cryptocamerillan (although he did not explicitly discuss them as products of skeletal pneumatization). Britt (1993, 1997) independently made the same observation and used the terms camerate and camellate to describe large-chambered and small-chambered vertebrae, respectively. Wedel et al. (2000) expanded this terminology to include categories for vertebrae with fossae only and vertebrae with combinations of large and small chambers (Table 1-1). Wedel et al. (2000) and Wedel (2003b) also discussed the phylogenetic distribution of different internal structure types. In general, the vertebrae of early diverging sauropods such as Shunosaurus and Barapasaurus have external fossae but lack internal chambers. Camerae are present in the vertebrae of diplodocids, Camarasaurus, and Brachiosaurus. Presacral vertebrae of Brachiosaurus also have camellae in the condyles and cotyles, and camellae are variably present in the neural spine and apophyses. The vertebrae of Sauroposeidon and most titanosaurs lack camerae and are entirely filled with camellae, although some titanosaurs may have vertebral camerae. From published descriptions (Young and Zhao, 1972; Russell and Zheng, 1994), the vertebrae of Mamenchisaurus appear to be camellate.

From the foregoing, it might appear that the internal structures of sauropod vertebrae, their evolution, and their phylogenetic distribution are all well understood. In fact, vertebral internal structure is only known for a small minority of sauropods. Even in those taxa for which the internal structure is known, this knowledge is usually limited to a handful of vertebrae or even a single element, which severely limits our ability to assess serial, ontogenetic, and population-level variation. Despite these limitations, three broad generalizations can be made. First, the vertebrae of very young sauropods tend to have a simple I-beam shape in cross section, with large lateral fossae separated by a median septum (Wedel, 2003b). This is true even for taxa in which the vertebrae of adults are highly subdivided, such as Apatosaurus. In these taxa the internal complexity of the vertebrae increased during ontogeny. The second generalization is that complex internal structures evolved several times, in Mamenchisaurus, diplodocids, and one or more times in Titanosauriformes (Wedel, 2003b). This suggests a general evolutionary trend toward increasing complexity of vertebral internal structure in sauropods, albeit one that took different forms in different lineages (i.e., polycamerate vertebrae in Diplodocidae and somphospondylous vertebrae in Somphospondyli) and that may have been subject to reversals (i.e, camerate vertebrae in some titanosaurs; see Wedel, 2003b). Finally, the largest and longest necked sauropods, such as Mamenchisaurus, the diplodocines, brachiosaurids, Euhelopus, and titanosaurs such as Argentinosaurus and the unnamed taxon represented by DGM Serie A, all have polycamerate, semicamellate, or fully camellate internal structures. I have previously stated that the complex internal structures were correlated with increasing size and neck length (Wedel, 2003a, b). This may or may not be true; I have not performed any phylogenetic tests of character correlation. Nevertheless, the presence of complex internal structures in the vertebrae of the largest and longest necked sauropods suggests that size, neck length, and internal structure are related.

## Volume of Air Within a Pneumatic Bone

The aspect of skeletal pneumaticity that has probably received the least attention to date is the ratio of bone tissue to empty space inside a pneumatic bone. Although many authors have commented on the weight-saving design of sauropod vertebrae (Osborn, 1899; Hatcher, 1901; Gilmore, 1925), no one has quantified just how much mass was saved. The savings in mass could have important paleobiological implications; for example, in determining how much mass to subtract from volumetric mass estimates.

Currey and Alexander (1985) and Cubo and Casinos (2000) reported relevant data on the long bones of birds, which are tubular and may be filled with marrow or air. In both studies, the variable of interest was *K*, the inner diameter of the element divided by its outer diameter. Both studies found mean values of *K* between 0.77 and 0.80 for pneumatic bones. The mean for marrow-filled bird bones is 0.65 (Cubo and Casinos, 2000), and the mean for terrestrial mammals is 0.53 (calculated from Currey and Alexander, 1985:table 1).

The *K* value is a parameter of tubular bones; it is meaningless when applied to bones with more complex shapes or internal structures, such as sauropod vertebrae. I propose the Air Space Proportion (ASP), or the proportion of the volume of a bone—or the area of a bone section—that is occupied by air spaces, as a variable that can be applied to both tubular and non-tubular bones. One problem is that measuring the volumes of objects is difficult and often imprecise. It is usually easier to measure the relevant surface areas of a cross section, but any one cross section may not be representative of the entire bone. For example, the long bones of birds and mammals are usually tubular at mid-shaft, but the epiphyses mostly consist of marrow-filled trabecular bone or pneumatic camellate bone. Nevertheless, it may be easier to take the mean of several cross sections as an approximation of volume than to directly measure volume, especially in the case of large, fragile, matrix-filled sauropod vertebrae.

For the avian long bones described above, data were only presented for a single cross section located at mid-shaft. Therefore, the ASP values I am about to discuss may not be representative of the entire bones, but they probably approximate the volumes (total and air) of the diaphyses. For tubular bones, ASP may be determined by squaring *K* (if *r* is the inner diameter and *R* the outer, then *K* is $r/R$, ASP is $\pi r^2/\pi R^2$ or simply $r^2/R^2$, and ASP=$K^2$). For the *K* of pneumatic bones, Currey and Alexander (1985) report lower and upper bounds of 0.69 and 0.86, and I calculate a mean of 0.80 from the data presented in their table 1. Using a larger sample size, Cubo and Casinos (2000) found a slightly lower mean K of 0.77. The equivalent values of ASP are 0.48 and 0.74, with a mean of 0.64, or 0.59 for the mean of Cubo and Casinos (2000). This means that, on average, the diaphysis of a pneumatic avian long bone is 59-64% air by volume.

How do these numbers compare with the ASPs of sauropod vertebrae? To find out, I measured the area occupied by bone and the total area for several cross-sections of sauropod vertebrae (see Fig. 1-5 for an example). I obtained the cross-sectional images from CT scans, published cross-sections, and photographs of broken or cut vertebrae. For image analysis I used Image J, a free program available online from the National Institutes of Health (Rasband, 2003). Some results are presented in Table 1-2 (this research is in progress and I will present more complete results elsewhere). The results should be approached with caution: I have only analyzed a few vertebrae from a handful of taxa, and only one or a few cross sections for each bone, so the results may not be representative of either the vertebrae, the regions of the vertebral column, or the taxa to which they belong. The sample is strongly biased toward cervical vertebrae simply because cervicals are roughly cylindrical and fit through CT scanners better than dorsal or sacral vertebrae. Despite these caveats, some regularities emerge.

First, ASP values range from 0.32 to 0.89, with a mean of 0.60. Even though the data may not be truly representative, it seems reasonable to conclude that most sauropod vertebrae contained at least 50% air by volume, and probably somewhat more. This assumes that the cavities in sauropod vertebrae were entirely filled with air and the amount of soft tissue was negligible. Chandra Pal and Bharadwaj (1971) found that the air spaces in pneumatic bird bones are lined by simple squamous epithelium, so the assumption is probably valid. The ASP values presented here for sauropod vertebrae are similar to the range and mean found for pneumatic long bones of birds (or at least their diaphyses).

Second, although only a handful of measurements are available for each taxon, it is already clear that ASP can vary widely from slice to slice within a single vertebra and probably also between vertebrae of different regions of the skeleton and between individuals of the same species. As we collect more data we may find more predictable relationships, for example, between the ASP values of cervical and dorsal vertebrae or between certain taxa. The system may also be so variable that such relationships will be impossible to detect, if they even exist. Rampant variation seems to be the rule for skeletal pneumaticity in general (e.g., King, 1957; Cranford et al., 1996; Weiglein, 1999), and it would be surprising if ASP were not also highly variable.

Third, the lowest values of ASP—0.32 in Apatosaurus and 0.39 in Brachiosaurus—are for slices through the cotyle, or bony cup, at the posterior end of the centrum. Here the cortical bone is doubled back on itself to form the cup, and the wall of the cotyle itself is at an angle to the slice and appears wider in cross section. The cotyle is surrounded by pneumatic chambers in both Apatosaurus and Brachiosaurus, but these become smaller and eventually disappear toward the end of the vertebra. For these reasons, the cotyle is expected to have a lower ASP than the rest of the vertebra.

Fourth, Sauroposeidon has the highest values of ASP, up to a remarkable 0.89. The values for Sauroposeidon are even higher than those for the closely related Brachiosaurus, and the ranges for the two taxa do not overlap (although they may come to when a larger sample is considered). A very high ASP is probably an autapomorphy of Sauroposeidon and may have evolved to help lighten its extremely long (~12 m) neck.

Finally, ASP appears to be independent of the internal complexity of the vertebrae. The Saltasaurus vertebra is the most highly subdivided of the sample. The I-beam-like vertebrae of the juvenile Pleurocoelus and Phuwiangosaurus are the least subdivided; the other taxa fall somewhere in the middle. Nevertheless, most values in the table, including those for Saltasaurus, Pleurocoelus, and Phuwiangosaurus, fall between 0.50 and 0.60. The means for all taxa other than Sauroposeidon also fall within the same range, so there is no apparent trend that relates ASP to internal complexity. Cast in evolutionary terms, this indicates that the evolution of complex internal structures from simple ones involved a redistribution rather than a reduction of bony tissue within the vertebrae. The ASP values of the juvenile Pleurocoelus and Phuwiangosaurus imply that a similar redistribution was involved in the ontogenetic derivation of complex chambers from juvenile fossae.

The results presented here are preliminary, and the available data are better suited for suggesting hypotheses than for testing them. Much work remains to be done, both in gathering comparative data from extant forms and in exploring the implications of pneumaticity for sauropod biomechanics.

**Distribution of Pneumaticity Along the Vertebral Column**

The two previous sections dealt with the characteristics of a single pneumatic bone. We must also consider the location of pneumatic features in the skeleton, because these features constrain the minimum extent of the diverticular system. For example, in the USNM 10865 skeleton of Diplodocus, pneumatic foramina are present on every vertebra between the axis and the nineteenth caudal (Gilmore, 1932, and pers. obs.; foramina are only present on caudals 1-18 in the skeleton of Diplodocus described by Osborn, 1899, and on caudals 1-16 in the mounted DMNS skeleton). This means that in life the pneumatic diverticula reached at least as far anteriorly as the axis and as far posteriorly as caudal vertebra 19 (Fig. 1-6). The diverticular system may have been more extensive and simply failed to pneumatize any more bones, but it could not have been any less extensive.

In mapping the distribution of pneumaticity along the vertebral column, it is important to consider where on the vertebrae the pneumatic features are located. In the co-ossified block of Diplodocus caudal vertebrae illustrated by Gilmore (1932:fig. 3), the centra of caudals 15-19 bear large pneumatic foramina, but the neural spines lack laminae and do not appear to have been pneumatic. This is in contrast to the presacral, sacral, and anterior caudal vertebrae, which have heavily sculpted neural spines with deep fossae and scattered foramina (see Osborn, 1899:figs. 7 and 13). In the opposite condition, the neural spines bear laminae and fossae and may have been pneumatic, but the centra lack pneumatic features. Examples include the middle and posterior dorsal vertebrae of Jobaria (see Sereno et al., 1999:fig. 3). Sauropod vertebrae can therefore exist in one of four states: (1) both centrum and neural spine pneumatic, as in the presacral vertebrae of most neosauropods; (2) centrum pneumatic but neural spine apneumatic, as in the middle caudals of Diplodocus; (3) neural spine pneumatic but centrum apneumatic, as in the posterior dorsals of Jobaria (assuming that the laminate neural spines are pneumatic); or (4) no signs of pneumaticity in the centrum or neural spine, as in the distal caudals of most sauropods. Pneumatization of the centrum typically results in large internal cavities with prominent foramina, so the inference of pneumaticity is well supported in conditions (1) and (2). In condition (3) the situation may be less clear. In derived neosauropods such as Brachiosaurus and the diplodocids, the neural spine fossae often bear small subfossae and foramina, which indicate that these fossae are pneumatic (see Janensch, 1950; Curtice and Stadtman, 2001). In more basal sauropods such as Haplocanthosaurus, the neural spine fossae are often blind and lack the heavily sculpted texture seen in later forms. The neural spines of these basal sauropods may have been pneumatic, but the inference is less well founded.

The earliest sauropodomorph with distinctly emarginated pneumatic fossae is Thecodontosaurus caducus (Yates, 2003). In T. caducus, pneumatic fossae are only present on the middle cervical vertebrae. This means that the fossae must have been produced by diverticula of cervical air sacs similar to those of birds (as opposed to diverticula of the lungs proper). A similar pattern of pneumatization in Coelophysis indicates that cervical air sacs were present in both sauropodomorphs and theropods by the Norian (Late Triassic), and cervical air sacs are probably primitive for saurischians (Wedel, 2004).

In general, more derived sauropods tended to pneumatize more of the vertebral column. Except for the atlas, which is always apneumatic, pneumatic chambers (or prominent fossae) are present in the cervical vertebrae of Shunosaurus; in the cervical and anterior dorsal vertebrae of Jobaria; in all of the presacral vertebrae of Cetiosaurus; in the presacral and sacral vertebrae of most neosauropods; and in the presacral, sacral, and caudal vertebrae of diplodocids and saltasaurids (Wedel 2003a, b, and pers. obs.). This caudad progression of vertebral pneumaticity also occurred in the evolution of theropods (Britt, 1993), and occurs ontogenetically in extant birds (Cover, 1953; Hogg, 1984b). At a gross level, the system is both homoplastic and recapitulatory.

In extant birds, diverticula of the cervical air sacs do not extend farther posteriorly than the anterior thoracic vertebrae. If the diverticula of sauropods followed the same pattern of development as those of birds, then the presence of pneumatic sacral vertebrae in most neosauropods indicates the presence of abdominal air sacs (Wedel et al., 2000). There are no strong reasons to doubt that neosauropods had abdominal air sacs. However, the future discovery of a sauropod with a pneumatic hiatus—a gap in the pneumatization of the dorsal vertebrae—would unequivocally demonstrate the presence of abdominal air sacs and their diverticula (Wedel, 2003a).

## APPLICATION TO A PALEOBIOLOGICAL PROBLEM: MASS ESTIMATES

The implications of PSP for sauropod paleobiology are only beginning to be explored. In particular, skeletal pneumaticity may be an important factor in future studies of the biomechanics and respiratory physiology of sauropods. The most obvious implication of extensive PSP in sauropods is that they may have weighed less than is commonly thought. In this section, the problem of estimating the masses of sauropods is used as an example of how information about PSP may be applied to a paleobiological question.

Two distinct questions proceed from the observation that most sauropod skeletons were highly pneumatic. The first is purely methodological: (how) should we take pneumaticity into account in estimating the masses of sauropods? The second question is paleobiological: if we find that pneumaticity significantly lightened sauropods, how does that affect our understanding of sauropods as living animals? If pneumaticity did not significantly lighten sauropods, then the second question is moot, so I will consider the methodological question first.

## Methods

The masses of dinosaurs are generally estimated using allometric equations based on limb bone dimensions (Russell et al., 1980; Anderson et al., 1985) or volumetric measurements using physical or computer models (Colbert, 1962; Paul, 1988, 1997; Henderson, 1999). If allometric equations are used, then pneumaticity need not be taken into account; the limb bones are assumed to have been as circumferentially robust as they needed to be to support the animal's mass, regardless of how the body was constituted. If an animal with a pneumatic skeleton was lighter than it would have been otherwise, this should already be reflected in its limb bone morphology, and no correction is necessary. On the other hand, if volumetric measurements are used, then it is possible to take skeletal pneumaticity into account and failure to do so may result in mass estimates that are too high.

Volumetric mass estimation is performed in three steps (Alexander, 1989). First, the volume of a scale model of the organism is measured. Next, the volume of the model is multiplied by the scale factor to obtain the volume of the organism in life. Finally, the volume of the organism is multiplied by the estimated density to obtain its mass. The presence of air in the respiratory system and pneumatic diverticula can be accounted for in the first two steps, by reducing the estimated volume of model or the organism, or in the third step, by adjusting the density used in the mass calculation. Both methods have been used in published mass estimates of dinosaurs. Alexander (1989) used plastic models in his volumetric study, and he drilled holes to represent the lungs before estimating the center of mass of each model and the proportion of mass supported by the fore and hind limbs (see Alexander, 1989:figs. 4.6 and 5.3).

Curiously, he does not seem to have drilled the holes before performing his mass estimates; at least, the holes are only mentioned in conjunction with the center of mass and limb support studies. Henderson (1999) included lung spaces in his digital models for mass estimation purposes, and later included air sacs and diverticula in a buoyancy study (Henderson, 2004). Paul (1988, 1997) used the alternative method of adjusting the density values for the mass calculations. He assigned a specific gravity (SG) of 0.9 to the trunk to account for lungs and air sacs, and an SG of 0.6 to the neck to account for pneumatization of the vertebrae.

Before attempting to estimate the volume of air in a sauropod, it is important to recognize that the air was distributed among four separate regions: (1) the trachea, (2) the 'core' respiratory system of lungs and, possibly, pulmonary air sacs, (3) the extraskeletal (i.e., visceral, intermuscular, and subcutaneous) diverticula, and (4) the pneumatic bones. These divisions are important for two reasons. First, the volumes of each region are differently constrained by skeletal remains. The volume of air in the skeleton can be estimated with a high degree of confidence because the sizes of the air spaces can be measured from fossils. In contrast, the volume of the trachea is not constrained by skeletal remains and must be estimated by comparison to extant taxa. The lung/air sac system and extraskeletal diverticula are only partly constrained by the skeleton (see below). This leads to the second point, which is that estimates of all four regions can be made independently, so that skeletal pneumaticity can be taken into account regardless of conformation (bird-like, crocodile-like, etc.) and volume of the core respiratory system.

## An Example Using Diplodocus

Consider the volume of air present inside a living Diplodocus. Practically all available mass estimates for Diplodocus (Colbert, 1962; Alexander, 1985; Paul, 1997; Henderson, 1999) are based on CM 84, the nearly complete skeleton described by Hatcher (1901). Uncorrected volumetric mass estimates—i.e, those that do not include lungs, air sacs, or diverticula—for this individual range from 11,700 kg (Colbert, 1962, as modified by Alexander, 1989:table 2.2) to 18,500 kg (Alexander, 1985). Paul (1997) calculated a mass of 11,400 kg using the corrected SGs cited above, and Henderson (1999) estimated 14,912 kg, or 13,421 kg after deducting 10% to represent the lungs. For the purposes of this example, the volume of the animal is assumed to have been 15,000 liters. The estimated volumes of various air reservoirs and their effects on body mass are shown in Table 1-3.

Estimating the volume of air in the vertebral centra is the most straightforward. I used published measurements of centrum length and diameter from Hatcher (1901) and Gilmore (1932) and treated the centra as cylinders. The caudal series of CM 84 is incomplete, so I substituted the measurements for USNM 1065 from Gilmore (1932); comparison of the measurements of the elements common to both skeletons indicates that the two animals were roughly the same size. I multiplied the volumes obtained by 0.60, the mean ASP of the sauropod vertebrae listed in Table 1-2, to obtain the total volume of air in the centra.

The volume of air in the neural spines is harder to calculate. The neural spines are complex shapes and are not easily approximated with simple geometric models. Furthermore, the fossae on the neural arches and spines only partially enclosed the diverticula that occupied them. Did the diverticula completely fill the space between adjacent laminae, did they bulge outward into the surrounding tissues, or did surrounding tissues bulge inward? In the complete absence of in vivo measurements of diverticulum volume in birds it is impossible to say. Based on the size of the neural spine relative to the centrum in most sauropods (see Fig. 1-2), it seems reasonable to assume that in the cervical vertebrae, at least as much air was present in the arch and spine as in the centrum, if not more. In the high-spined dorsal and sacral vertebrae (see Fig. 1-1), the volume of air in the neural arch and spine may have been twice that in the centrum. Finally, proximal caudal vertebrae have large neural spines but the size of the spines decreases rapidly in successive vertebrae. On average, the caudal neural spines of *Diplodocus* may have contained only half as much air as their associated centra. These estimates are admittedly rough, but they are probably conservative and so they will suffice for this example.

As they developed, the intraosseous diverticula replaced bony tissue, and the density of that tissue must be taken into account in estimating how much mass was saved by pneumatization of the skeleton. In apneumatic sauropod vertebrae the internal structure is filled with cancellous bone and presumably supported red (erythropoeitic) bone marrow (Fig. 1-7). Distal caudal vertebrae of the theropod Ceratosaurus have a large central chamber or centrocoel (Madsen and Welles, 2000:fig. 6). This cavity lacks large foramina that would connect it to the outside, so it cannot be pneumatic in origin. The medullary cavities of apneumatic avian and mammalian long bones are filled with adipose tissue that acts as lightweight packing material (Currey and Alexander, 1985), and the same may have been true of the centrocoels in Ceratosaurus caudals. The presence of a similar marrow cavity in sauropod vertebrae prior to pneumatization cannot be ruled out, but to my knowledge no such cavities have been reported. In birds, the intraosseous diverticula erode the inner surfaces of the cortical bone in addition to replacing the cancellous bone (Bremer, 1940), so pneumatic bones tend to have thinner walls than apneumatic bones (Currey and Alexander, 1985; Cubo and Casinos, 2000). The tissues that may have been replaced by intraosseous diverticula have SGs that range from 0.9 for some fats and oils to 3.2 for apatite (Schmidt-Nielsen, 1983:451 and table 11.5). For this example, I estimated that the tissue replaced by the intraosseous diverticula had an average SG of 1.5 (calculated from data presented in Cubo and Casinos, 2000), so air cavities that total 970 liters replace 1455 kg of tissue. The extraskeletal diverticula, trachea, lungs, and air sacs did not replace bony tissue in the body. They are assumed to replace soft tissues (density of one gram/$cm^3$) in the solid model.

Extraskeletal diverticula include visceral, intermuscular, and subcutaneous diverticula. None of these leave traces that are likely to be fossilized. The bony skeleton places only two constraints on the extraskeletal diverticula. First, as previously discussed, the distribution of pneumatic bones in the skeleton limits the minimum extent of the diverticular system. Thus, we can infer that the vertebral diverticula in Diplodocus must have extended from the axis to the nineteenth caudal vertebra (at least in USNM 1065), but the course and diameter of the diverticula are unknown. The second constraint imposed by the skeleton is that the canalis intertransversarius, if it existed, could not have been larger than the transverse foramina where it passed through them, although it may have been smaller or increased in diameter on either side. I am unaware of any studies in which the _in vivo_ volume of the avian diverticular system is measured. This information vacuum prevents me from including a volume estimate for the diverticular system in Table 1-3.

To estimate the volume of the trachea, I used the allometric equations presented by Hinds and Calder (1971) for birds. The length equation, $L = 16.77M^{0.394}$, where L is the length of the trachea in cm and M is the mass of the animal in kg, yielded a predicted tracheal length of 6.8 meters for a 12-ton animal. The cervical series of _Diplodocus_ CM 84 is 6.7 meters long and the trachea may have been somewhat longer, and I judged the correspondence between the neck length and predicted tracheal length to be close enough to justify using the equations, especially for the coarse level of detail needed in this example. The volume equation, $V = 3.724M^{1.090}$, yields a volume of 104 liters.

Finally, the volume of the lungs and air sacs must be taken into account. The lungs and air sacs are only constrained by the skeleton in that they must fit inside the ribcage and share space with the viscera. Based on measurements from caimans and large ungulates, Alexander (1989) subtracted eight percent from the volume of each of his models to account for lungs. Data presented by King (1966:table 3) indicate that the lungs and air sacs of birds may occupy 10-20% of the volume of the body. Hazlehurst and Rayner (1992) found an average SG of 0.73 in a sample of 25 birds from 12 unspecified species. On this basis, they concluded that the lungs and air sacs occupy about a quarter of the volume of the body in birds. However, some of the air in their birds probably resided in extraskeletal diverticula or pneumatic bones, so the volume of the lungs and air sacs may have been somewhat lower. In the interests of erring conservatively, I put the volume of the lungs and air sacs at 10% of the body volume.

The results of these calculations are necessarily tentative. The lungs and air sacs were probably not much smaller than estimated here, but they may have been much larger; the trachea could not have been much shorter but may have been much longer, or it may have been of different or irregular diameter (see McClelland, 1989a for tracheal convolutions and bulbous expansions in birds); the neural spines may have contained much more or somewhat less air; the ASP of Diplodocus vertebrae may be higher or lower; and the tissue replaced by the intraosseous diverticula may have been more or less dense. The extraskeletal diverticula have not been accounted for at all, although they were certainly extensive in linear terms and were probably voluminous as well. Uncertainties aside, it seems likely that the vertebrae contained a large volume of air, possibly 1000 liters or more if the very tall neural spines are taken into account. This air mainly replaced dense bony tissue, so skeletal pneumatization may have lightened the animal by up to 10%—and that does not include the extraskeletal diverticula or pulmonary air sacs. In the example presented here, the volume of air in the body of Diplodocus is calculated to have replaced about 3000 kg of tissue that would have been present if the animal were solid. If the total volume of the body was 15,000 liters and the density of the remaining tissue was one gram per cubic centimeter, the body mass would have been about 12 metric tons and the SG of the entire body would have been 0.8. This is lower than the SGs of squamates and crocodilians (0.81-0.89) found by Colbert (1962), higher than the SGs of birds (0.73) found by Hazlehurst and Rayner (1992), and about the same as the SGs (0.79-0.82) used by Henderson (2004) in his study of sauropod buoyancy. Note that the amount of mass saved by skeletal pneumatization is independent of the estimated volume of the body, but the proportion of mass saved is not. Thus if we start with Alexander's (1985) 18,500 liter estimate for the body volume of Diplodocus, the mass saved is still 1455 kg, but this is only eight percent of the solid mass, not ten percent as in the previous example.

It could be argued that adjusting the estimated mass of a sauropod by a mere 8-10% is pointless. The mass of the living animal may have periodically fluctuated by that amount or more, depending on the amount of fat it carried and how much food it held in its gut (Paul, 1997). Further, the proposed correction is tiny compared to the range of mass estimates produced by different studies, from 11,700 kg (Paul, 1997) to 18,500 kg (Alexander, 1985). However, there are several reasons for taking into account the mass saved by skeletal pneumatization. The first is that estimating the mass of extinct animals is fraught with uncertainty, but we should account for as many sources of error as possible, and PSP is a particularly large source of error if it is not considered. Also, the range of mass estimates for certain taxa may be very wide, but 8-10% of the body mass is still a sizeable fraction when applied to any one estimate. The entire neck and head account for about the same percentage of mass in volumetric studies (Alexander, 1989; Paul, 1997), so failing to account for PSP may be as gross an error as omitting the neck and head from the volumetric model. These are the purely methodological reasons for considering the effect of PSP on body mass. There is also the paleobiological consideration, which is that the living animal was 8-10% lighter because of PSP than it would have been without. Mass reduction of this magnitude almost certainly carries a selective advantage (Currey and Alexander, 1985), and this may explain the presence of extensive PSP in many sauropods.

An alternative possibility is that sauropod skeletons weighed as much as they would have in the absence of PSP, but that pneumatization allowed the elements to be larger and stronger for the same mass. This hypothesis was first articulated by Hunter (1774) to explain skeletal pneumatization in birds. It is supported by the observation that the skeletons of birds are not significantly lighter than the skeletons of comparably sized mammals (Prange et al., 1979). If this hypothesis is correct, pneumatic elements should be noticeably larger and more voluminous than non-pneumatic elements. The transitions from pneumatic to apneumatic regions of the vertebral column in Jobaria (Sereno et al., 1999:fig. 3) and Diplodocus (Osborn, 1899:fig. 13; Gilmore, 1932:fig. 3 and pl. 6) are not marked by obvious changes in size or form of the vertebrae. This supports the hypotheses that pneumatic vertebrae were lighter than apneumatic vertebrae and that PSP really did lighten sauropod skeletons.

**Paleobiological Implications**

The importance of PSP for sauropod paleobiology is still largely unexplored. To date, Henderson’s (2004) study of sauropod buoyancy is the only investigation of the biomechanical effects of PSP. Henderson included pneumatic diverticula in and around the vertebrae in his computer models of sauropods, and found that floating sauropods were both highly buoyant and highly unstable. Pneumaticity may also be important in future studies of neck support in sauropods. Alexander (1985, 1989) calculated that a large elastin ligament would be better suited than muscles to holding up the neck of Diplodocus. His calculations were based on a volumetric estimate of 1340 liters (and, thus, 1340 kg) for the neck and head. Using the values in Table 1-3, one fifth of that volume, or 268 liters, was occupied by air spaces. If Paul (1997) and Henderson (2004) are correct, the SG of the neck may have been as low as 0.6, which would bring the mass of the neck down to about 800 kg (the same result could be obtained by applying the air volumes in Table 1-3 to a more slender neck model than that used by Alexander). As the mass of the neck goes down, so to does the perceived need for a large ‘nuchal’ ligament, the existence of which is controversial (see Wedel et al., 2000; Dodson and Harris, 2001; Tsuihiji, 2004).

Recognition of skeletal pneumaticity in sauropods may also affect physiological calculations. For example, most published studies of thermal conductance in dinosaurs (e.g., Spotila et al., 1973, 1991) have modeled dinosaur bodies using solid cylinders. Air is a better insulator than conductor, but moving bodies of air may cool adjacent tissues by convection or evaporation. The pneumatic diverticula of birds tend to be blind-ended tubes except where they anastomose (Cover, 1953), and most are poorly vascularized (Duncker, 1971), so there appears to be little potential for evaporative cooling. On the other hand, thermal panting is an important homeostatic mechanism for controlling body temperature in birds and depends on evaporation from nasal, buccopharyngeal, and upper tracheal regions (Lasiewski, 1972; Menaum and Richards, 1975). At the very least, the inclusion of tracheae, lungs and pneumatic diverticula in thermal conductance models would decrease the effective radius of some of the constituent cylinders. What effect, if any, this would have on the results of thermal conductance studies is unknown, which is precisely the point: it has not been tested.

## PROBLEMS AND PROSPECTS FOR FURTHER RESEARCH

Despite a long history of study, research on PSP is, in many ways, still in its infancy. Anyone who doubts the accuracy of this statement is directed to Hunter (1774). In the first published study of PSP, Hunter developed two of the major functional hypotheses entertained today: pneumaticity may lighten the skeleton, or it may strengthen the skeleton by allowing bones of larger diameter for the same mass as marrow-filled bones (see Witmer, 1997, for a historical perspective on these and other hypotheses). Although many later authors have documented the presence and extent of PSP in certain birds (e.g., Crisp, 1857; King, 1957), most have focused on one or a few species (O’Connor, 2004), some have produced conflicting accounts (reviewed by King, 1957), and few have attempted to test functional hypotheses (but see Warncke and Stork, 1977; Currey and Alexander, 1985; Cubo and Casinos, 2000; O’Connor, 2004). Evolutionary patterns of PSP in birds are difficult to discern because few species have been studied (King, 1966), usually with little or no phylogenetic context (O’Connor, 2002, 2004). Limits of knowledge of PSP in extant vertebrates necessarily limit what can be inferred from the fossil record. For example, disagreements between various published accounts of the development of pneumatization in birds frustrate attempts to infer the ontogenetic development of PSP in sauropods (Wedel, 2003a).

Another problem for studies of PSP in fossil organisms is small sample sizes. As mentioned above, few taxa have been intensively studied and the importance of serial, ontogenetic, and intraspecific variation is difficult to assess. Sample sizes are mainly limited by the inherent attributes of the fossils: fossilized bones are rare, at least compared to the bones of extant vertebrates; they may be crushed or distorted; and they are often too large, too heavy, or too fragile to be easily manipulated. Even if these difficulties are overcome, most of the pneumatic morphology is still inaccessible, locked inside the bones.

**Sources of Data**

Information on the internal structure of fossil bones comes from three sources: CT studies, cut sections of bones, and broken bones. Although CT studies of fossils are becoming more common, access to scanners is very limited and can be prohibitively expensive. Large fossils, such as sauropod vertebrae, cause logistical problems. Most medical CT scanners have apertures 50 cm or less in diameter, and many sauropod vertebrae are simply too big to fit through the scanners. Furthermore, medical scanners are not designed to image large, dense objects like sauropod bones. The relatively low-energy x-rays employed by medical scanners may fail to penetrate large bones, and this can produce artifacts in the resulting images (Wedel et al., 2000). Industrial CT scanners can image denser materials, but the rotating platforms used in many industrial scanners are too small to accept most sauropod vertebrae. For the near future, CT will likely remain a tool of great promise but limited application.

Cut sections of bones can yield valuable information about pneumatic internal structures. The cuts may be made in the field to break aggregates of bones into manageable pieces, as in the cut Sauroposeidon vertebra shown in Fig. 1-4. Less commonly, bones may be deliberately cut to expose their cross sections or internal structures, such as the cut specimens illustrated by Janensch (1947:fig. 5) and Martill and Naish (2001:pl. 32). Cutting into specimens is invasive and potentially dangerous to both researchers and fossils. Although cut specimens will continue to appear from time to time, they are unlikely to become a major source of data. In contrast, broken bones are ubiquitous. The delicate structure of pneumatic bones, even large sauropod vertebrae, may make them more prone to breakage than apneumatic bones. For these reasons broken bones are an important resource in studies of PSP and could be exploited more in the future. Published illustrations of broken sauropod vertebrae are numerous; notable examples include Cope (1878a:fig. 5), Hatcher (1901:pl. 7), Longman (1933:pl. 16 and fig. 3), and Dalla Vecchia (1999:figs. 2 and 19). A beautiful example from outside Sauropoda is the broken transverse process of Tyrannosaurus illustrated by Brochu (2003:fig. 75).

**Directions for Future Research**

Four attributes of pneumatic bones are listed above under 'Description of pneumatic elements': (1) external pneumatic features, (2) internal structure, (3) ASP, and (4) distribution of pneumaticity in the skeleton. Only the second attribute has been systematically surveyed in sauropods (Wedel, 2003b), although aspects of the first are treated by Wilson (1999). Knowledge of the fourth is mainly limited to the observation that diplodocines and saltasaurines have pneumatic caudal vertebrae and other sauropods do not (Wedel, 2003b). All existing data on the ASPs of sauropod vertebrae are presented in Table 1-2. Not only do all four attributes need further study, the levels of serial, ontogenetic, and intraspecific variation should be assessed whenever possible. Similar data on PSP in pterosaurs, non-avian theropods, and birds are needed to test phylogenetic and functional hypotheses.

The pneumatic diverticula of birds are morphologically and morphogenetically intermediate between the core respiratory system of lungs and air sacs and the pneumatic bones. Understanding the development, evolution, and possible functions of diverticula is therefore crucial for interpreting patterns of PSP in fossil vertebrates. Müller (1907), Richardson (1939), Cover (1953), King (1966), Duncker (1971) and a few others described the form and extent of the diverticular network in the few birds for which it is known, but information on many bird species is lacking or has been inadequately documented (King, 1966). The ontogenetic development of the diverticula is very poorly understood; most of what we think we know is based on inferences derived from patterns of skeletal pneumatization (Hogg, 1984a; McClelland, 1989b). Such inferences tell us nothing about the development of the many visceral, intermuscular, and subcutaneous diverticula that do not contact the skeleton or pneumatize any bones. These diverticula could not have evolved to pneumatize the skeleton. Most diverticula that pneumatize the skeleton must grow out from the core respiratory system before they reach their ‘target’ bones, so they probably also evolved for reasons other than skeletal pneumatization (Wedel, 2003a). Those reasons are unknown, in part because the physiological functions—or exaptive effects (sensu Gould and Vrba, 1982)—of diverticula remain obscure. Three important physiological questions that could be answered with existing methods are: (1) what volume of air is contained in the diverticula in life; (2) what is the rate of diffusion of air into and out of blind-ended diverticula; and (3) in cases where diverticula of different air sacs anastomose, is air actively circulated through the resulting loops?

Finally, more work is needed on the origins of PSP; if nothing else, Gower's (2001) unconventional hypothesis has drawn attention to this need. Potential projects include histological and biomechanical studies to assess the structure and functions of vertebral laminae (Wilson, 1999). In addition, criteria for distinguishing the osteological traces of adipose deposits, muscles, vascular structures, and pneumatic diverticula are badly needed for the interpretation of potentially pneumatic features in fossil bones. This problem is the subject of ongoing research by O'Connor (1999, 2001, 2002).

## CONCLUSIONS

The best evidence for pneumaticity in a fossil element is the presence of large foramina that lead to internal chambers. Based on this criterion, pneumatic diverticula were present in the vertebrae of most sauropods and in the ribs of some. Vertebral laminae and fossae were clearly associated with pneumatic diverticula in most eusauropods, but it is not clear whether this was the case in more basal forms. Measurements of vertebral cross sections indicate that, on average, pneumatic sauropod vertebrae were 50-60% air by volume. Taking skeletal pneumaticity into account may reduce mass estimates of sauropods by up to 10%. Although the functional and physiological implications of pneumaticity in sauropods and other archosaurs remain largely unexplored, most of the outstanding problems appear tractable, and there is great potential for progress in future studies of pneumaticity.

TABLE 1-1. Classification of sauropod vertebrae into morphologic categories based on pneumatic characters. After Wedel et al. (2000:table 3).

| Category | Definition |
|---|---|
| Acamerate | Pneumatic characters limited to fossae; fossae do not significantly invade the centrum. |
| Procamerate | Deep fossae penetrate to median septum, but are not enclosed by ostial margins. |
| Camerate | Large, enclosed camerae with regular branching pattern; cameral generations usually limited to 3. |
| Polycamerate | Large, enclosed camerae with regular branching pattern; cameral generations usually 3 or more, with increased number of branches at each generation. |
| Semicamellate | Camellae present but limited in extent; large camerae may also be present. |
| Camellate | Internal structure entirely composed of camellae; neural arch laminae not reduced. Large external fossae may also be present. |
| Somphospondylous | Internal structure entirely composed of camellae; neural arch laminae reduced; neural spine with inflated appearance. |

TABLE 1-2. The air space proportion (ASP) of transverse sections through vertebrae of sauropods and other saurischians. Only values for published sections are presented. Much more work will be required to determine norms for different taxa and different regions of the vertebral column, and the values presented here may not be representative of either. Nevertheless, these values suggest that pneumatic sauropod vertebrae were often 50-60% air by volume. Abbreviation: O&M, Ostrom and MacIntosh.

| Taxon | Region | ASP | Source |
|---|---|---|---|
| Apatosaurus | Cervical condyle | 0.69 | Wedel (2003b:fig. 6b) |
| | Cervical mid-centrum | 0.52 | Wedel (2003b:fig. 6c) |
| | Cervical cotyle | 0.32 | Wedel (2003b:fig. 6d) |
| Barosaurus | Cervical mid-centrum | 0.56 | Janensch (1947:fig. 8) |
| | Cervical, near cotyle | 0.77 | Janensch (1947:fig. 3) |
| | Caudal mid-centrum | 0.47 | Janensch (1947:fig. 9) |
| Brachiosaurus | Cervical condyle | 0.73 | Janensch (1950:fig. 70) |
| | Cervical mid-centrum | 0.67 | Wedel et al. (2000:fig. 12c) |
| | Cervical cotyle | 0.39 | Wedel et al. (2000:fig. 12d) |
| | Dorsal mid-centrum | 0.59 | Janensch (1947:fig. 2) |
| Camarasaurus | Cervical condyle | 0.49 | Wedel (2003b:fig. 9b) |
| | Cervical mid-centrum | 0.52 | Wedel (2003b:fig. 9c) |
| | Cervical, near cotyle | 0.50 | Wedel (2003b:fig. 9d) |
| | Dorsal mid-centrum | 0.63 | O&M (1966:pl. 23) |

TABLE 1-2. (continued)

| Taxon | Region | ASP | Source |
|---|---|---|---|
| Camarasaurus | Dorsal mid-centrum | 0.58 | O&M (1966:pl. 24) |
| | Dorsal mid-centrum | 0.71 | O&M (1966:pl. 25) |
| Pleurocoelus | Cervical mid-centrum | 0.55 | Lull (1911:pl. 15) |
| Phuwiangosaurus | Cervical mid-centrum | 0.55 | Martin (1994:fig. 2) |
| Saltasaurus | Dorsal mid-centrum | 0.55 | Powell (1992:fig. 16) |
| Sauroposeidon | Cervical prezyg. ramus | 0.89 | Fig. 1-4 |
| | Cervical mid-centrum | 0.74 | Wedel et al. (2000:fig. 12g) |
| | Cervical postzygapophysis | 0.75 | Wedel et al. (2000:fig. 12h) |
| Theropoda | Cervical prezygapophysis | 0.48 | Janensch (1947:fig. 16) |
| | Dorsal mid-centrum | 0.50 | Janensch (1947:fig. 15) |
| Mean of sauropod measurements (13.17/22) | | 0.60 | |

TABLE 1-3. The volume of air in Diplodocus. See the text for methods of estimation.

| System | Total Volume (L) | Air Volume (L) | Mass Savings (kg) |
|---|---|---|---|
| Trachea | | 104 | 104 |
| Lungs and air sacs | | 1500 | 1500 |
| Extraskeletal diverticula | | ? | ? |
| Pneumatic vertebrae | | | |
| Centra | | | |
| Cervicals 2-15 | 136 | 82 | |
| Dorsals 1-10 | 208 | 125 | |
| Sacrals 1-5 | 75 | 45 | |
| Caudals 1-19 | 329 | 198 | |
| Subtotal for centra | 748 | 450 | |
| Neural spines | | | |
| Cervicals 2-15 | 136 | 82 | |
| Dorsals 1-10 | 416 | 250 | |
| Sacrals 1-5 | 150 | 90 | |
| Caudals 1-19 | 165 | 99 | |
| Subtotal for spines | 867 | 520 | |
| Subtotal for vertebrae | 1615 | 970 | 1455 |
| Total volume of air spaces | | 2574 | |
| Total mass replaced by air spaces | | | 3059 |

FIGURE 1-1. Pneumatic features in dorsal vertebrae of Barapasaurus (A-D), Camarasaurus (E-G), Diplodocus (H-J), and Saltasaurus (K-N). Anterior is to the left; different elements are not to scale. A, a posterior dorsal vertebra of Barapasaurus. The opening of the neural cavity is under the transverse process. B, a midsagittal section through a mid-dorsal vertebra of Barapasaurus showing the neural cavity above the neural canal. C, a transverse section through the posterior dorsal shown in A (position 1). In this vertebra, the neural cavities on either side are separated by a narrow median septum and do not communicate with the neural canal. The centrum bears large, shallow fossae. D, a transverse section through the mid-dorsal shown in B. The neural cavity opens to either side beneath the transverse processes. No bony structures separate the neural cavity from the neural canal. The fossae on the centrum are smaller and deeper than in the previous example. A-D redrawn from Jain et al. (1979:pls. 101 and 102). E, an anterior dorsal vertebra of Camarasaurus. F, a transverse section through the centrum (E, position 1) showing the large camerae that occupy most of the volume of the centrum. G, a horizontal section (E, position 2). E-G redrawn from Ostrom and McIntosh (1966:pl. 24). H, a posterior dorsal vertebra of Diplodocus. Modified from Gilmore (1932:fig. 2). I, transverse sections through the neural spines of other Diplodocus dorsals (similar to H, position 1). The neural spine has no body or central corpus of bone for most of its length. Instead it is composed of intersecting bony laminae. This form of construction is typical for the presacral neural spines of most sauropods outside the clade Somphospondyli. Modified from Osborn (1899:fig. 4). J, a horizontal section through a generalized Diplodocus dorsal (similar to H, position 2). This diagram is based on several broken elements and is not intended to represent a specific specimen. The large camerae in the mid-centrum connect to several smaller chambers at either end. K, a transverse section through the top of the neural spine of an anterior dorsal vertebra of Saltasaurus (L, position 1). Compare the internal pneumatic chambers in the neural spine of Saltasaurus with the external fossae in the neural spine of Diplodocus shown in J. L, an anterior dorsal vertebra of Saltasaurus. M, a transverse section through the centrum (L, position 2). N, a horizontal section (L, position 3). In most members of the clade Somphospondyli the neural spines and centra are filled with small camellae. K-N modified from Powell (1992:fig. 16).

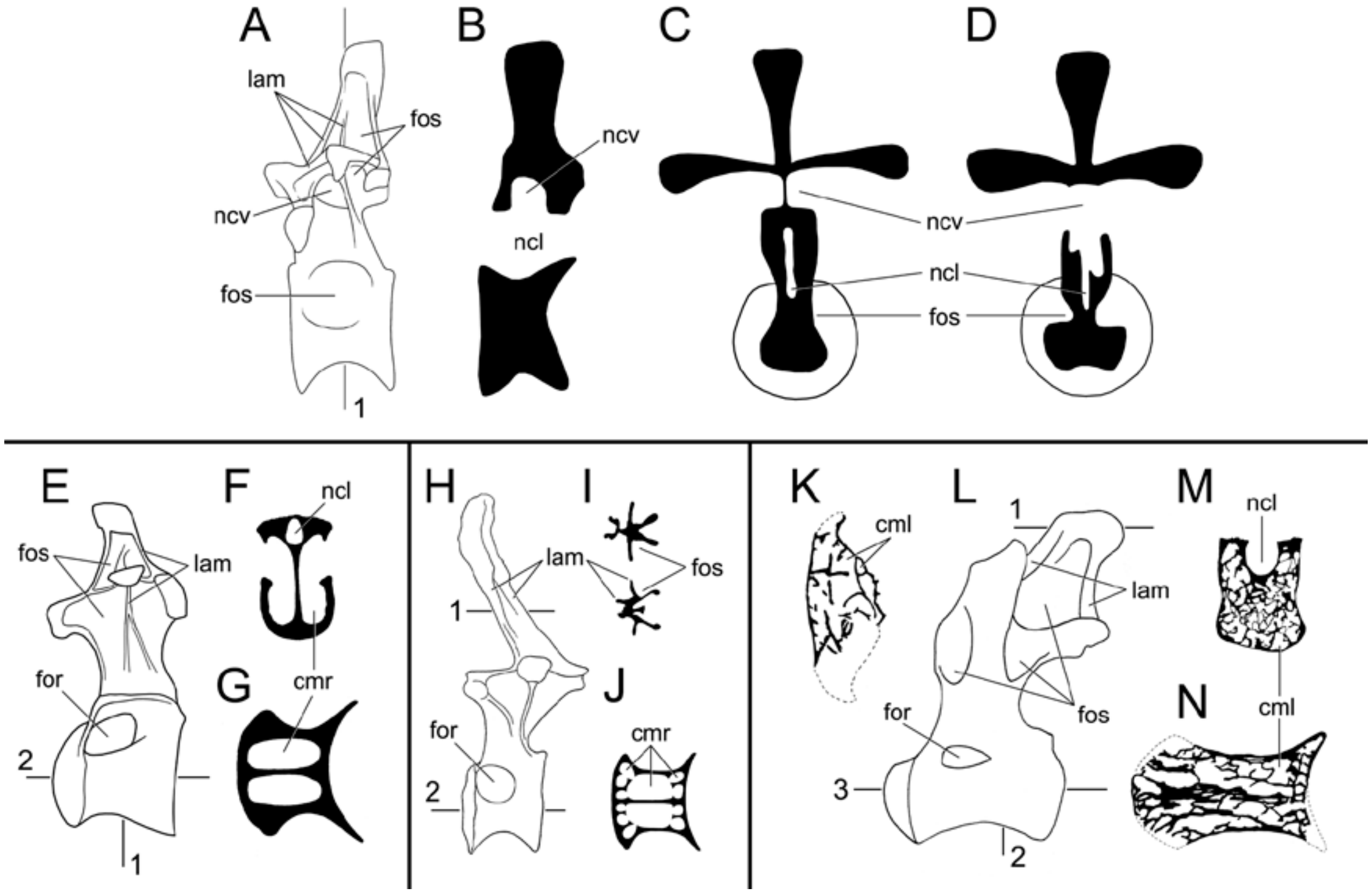

FIGURE 1-2. A cervical vertebra of Brachiosaurus and a hypothetical reconstruction of the pneumatic diverticula. A, BYU 12866, a mid-cervical vertebra of Brachiosaurus, in left lateral view. The neural spine fossae are bounded on all sides by the four laminae that connect the pre- and postzygapophyses to the neurapophysis and diapophysis. Some of the neural spine fossae contain large, sharp-lipped foramina. B, possible appearance of the pneumatic diverticula, shown in black. We can be fairly certain that pneumatic diverticula occupied the fossae on the neural arch, neural spine, and centrum, but the connections between various diverticula and their order of appearance during ontogeny remain speculative. Here the diverticula have been restored based on those of birds, with the canalis intertransversarius running alongside the centrum and the diverticulum supervertebrale occupying the neural spine fossae (see Müller [1907:figs. 3-5, 7, 11, and 12] for the appearance of these diverticula in the pigeon). Any connections between the canalis intertransversarius and diverticulum supervertebrale probably passed intermuscularly, because the laminae bounding the neural spine fossae are uninterrupted by tracks or grooves. C, a transverse section through the mid-centrum (A, position 1) traced from a CT image (Wedel et al. 2000:fig. 12C) and corrected for distortion. The volume of air filling the fossae and camellae in the neural arch and spine is unknown, but it may have equaled or exceeded the volume of air in the centrum. Lamina terminology after Wilson (1999). Scale bar is 20 cm.

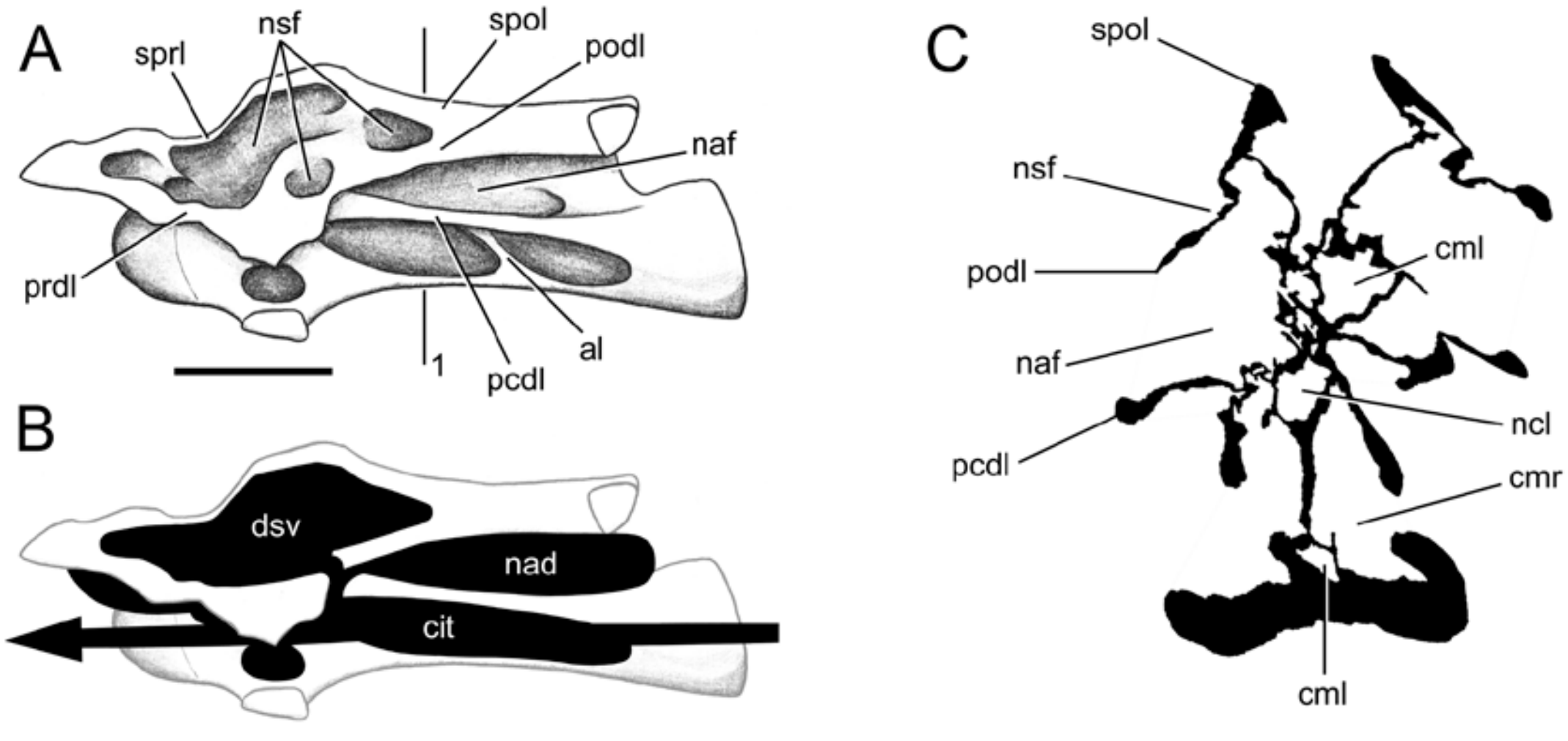

FIGURE 1-3. Pneumatic features in a cervical vertebra of Haplocanthosaurus. A, a posterior cervical of Haplocanthosaurus in right lateral view (CM 879-7; this specimen was erroneously referred to as CM 572 in Upchurch, 1998:fig. 8, and as CM 897-7 in Wedel et al., 2000:fig. 2, Wedel, 2003a:fig. 3, and Wedel, 2003b:fig. 1). Modified from Hatcher (1903:pl. 2). B-E, cross-sections traced from CT slices. B, section at A, position 1. C, section at A, position 2. The opening of the neural canal and the absence of the neurocentral suture on one side are due to a break in the specimen. D, section at A, position 3. E, section at A, position 4. The neurocentral sutures are unfused over most of their length, indicating that this animal was not fully mature. Scale bar is 5 cm.

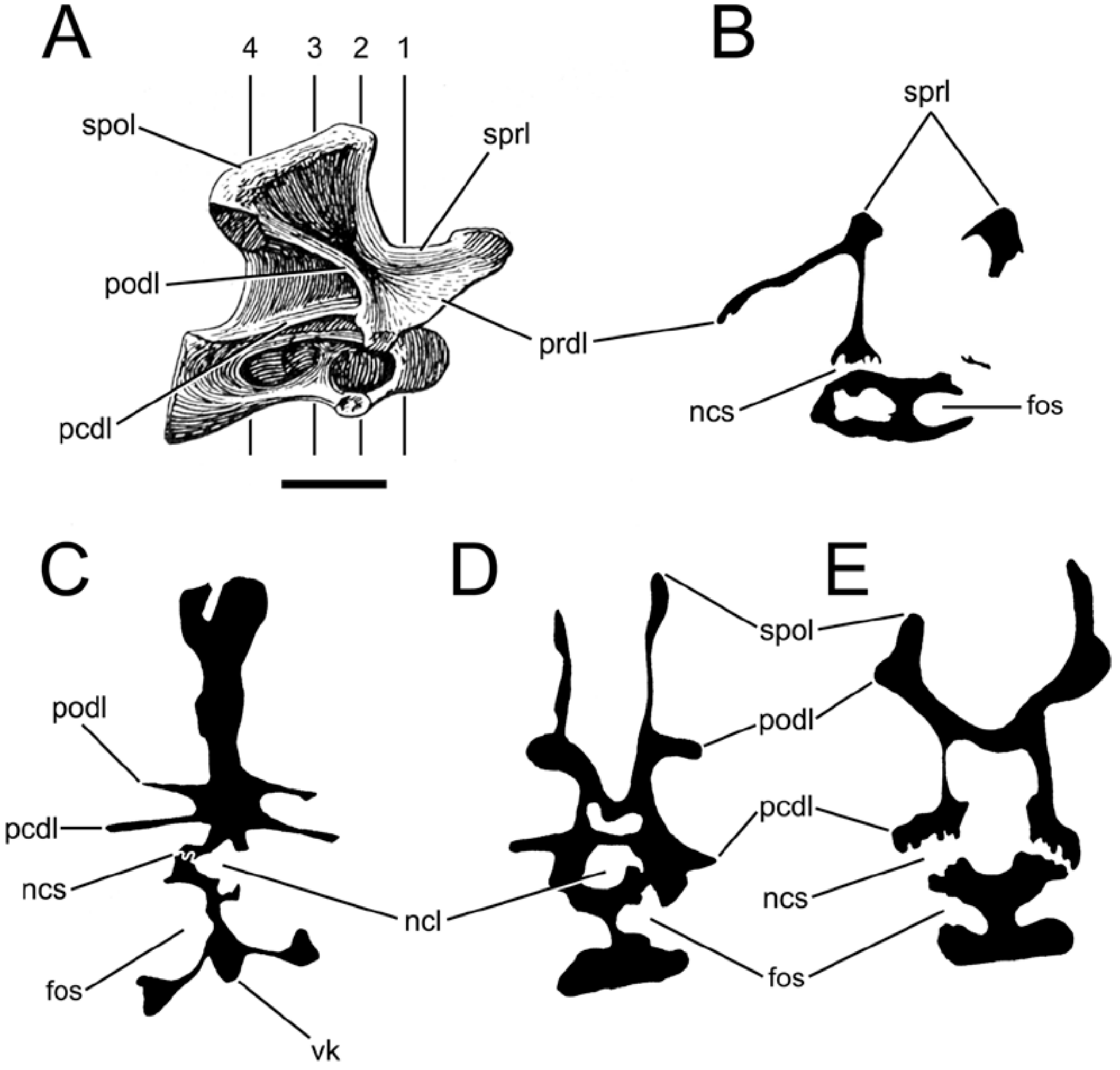

FIGURE 1-4. Internal structure of a cervical vertebra of Sauroposeidon, OMNH 53062. A, the posterior two-thirds of C5 and the condyle and prezygapophysis of C6 in right lateral view. The field crew cut though C6 to divide the specimen into manageable pieces. B, cross section of C6 at the level of the break, traced from a CT image (A, position 1) and photographs of the broken end. The left side of the specimen was facing up in the field and the bone on that side is badly weathered. Over most of the broken surface the internal structure is obscured by plaster or too damaged to trace, but it is cleanly exposed in the ramus of the right prezygapophysis (outlined). C, the internal structure of the prezygapophyseal ramus, traced from a photograph. The arrows indicate the thickness of the bone at several points, as measured with a pair of digital calipers. The camellae are filled with sandstone.

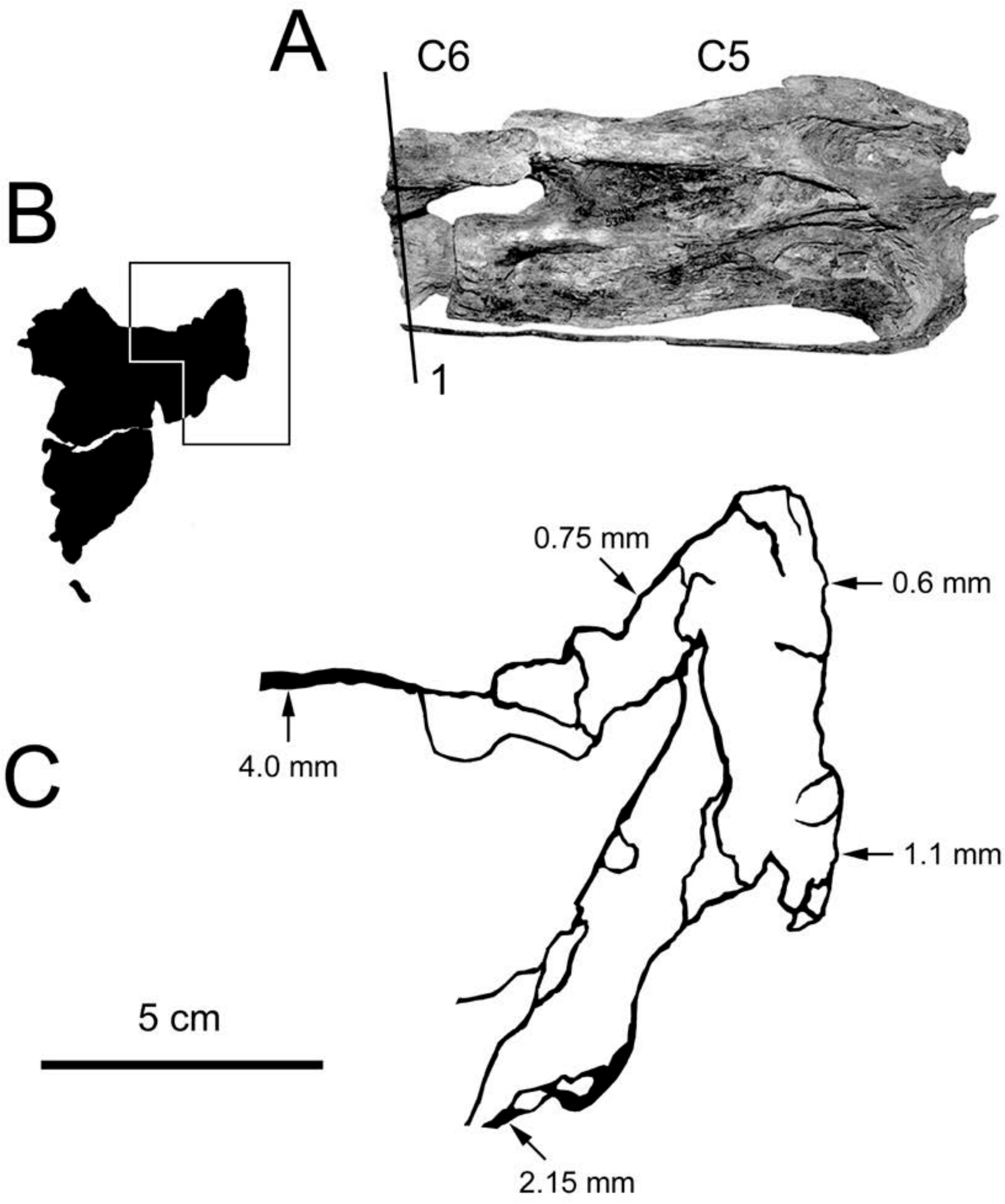

FIGURE 1-5. How to determine the air space proportion (ASP) of a bone section. A, a section is traced from a photograph, CT image, or published illustration; in this case, a transverse section of a Barosaurus africanus cervical vertebra from Janensch (1947:fig. 3). B, imaging software is used to fill the bone, air space, and background with different colors. The number of pixels of each color can then be counted using Image J (or any program with a pixel count function) and used to compute the ASP. In this case, bone is black and air is white, so the ASP is (white pixels) / (black pixels + white pixels).

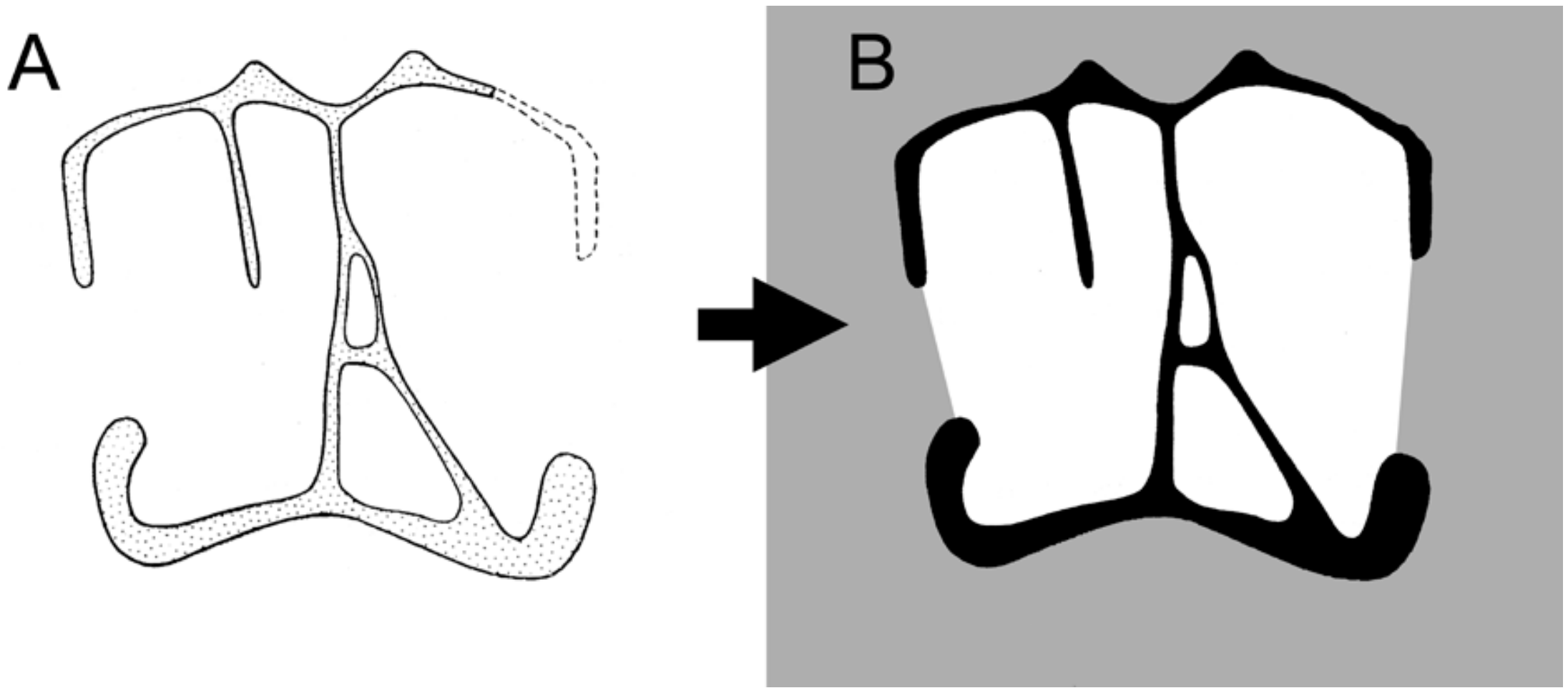
A
B

FIGURE 1-6. Hypothetical conformation of the respiratory system of a diplodocid sauropod. The left forelimb, pectoral girdle, and ribs have been removed for clarity. The lung is shown in dark grey, air sacs are light grey, and pneumatic diverticula are black. Only some of the elements shown in this illustration can be determined with certainty: the minimum length of the trachea, the presence of at least some air sacs, and the minimum extent of the pneumatic diverticula. The rest of the respiratory system has been restored based on that of birds, but this remains speculative. The skeleton is modified from Norman (1985:83).

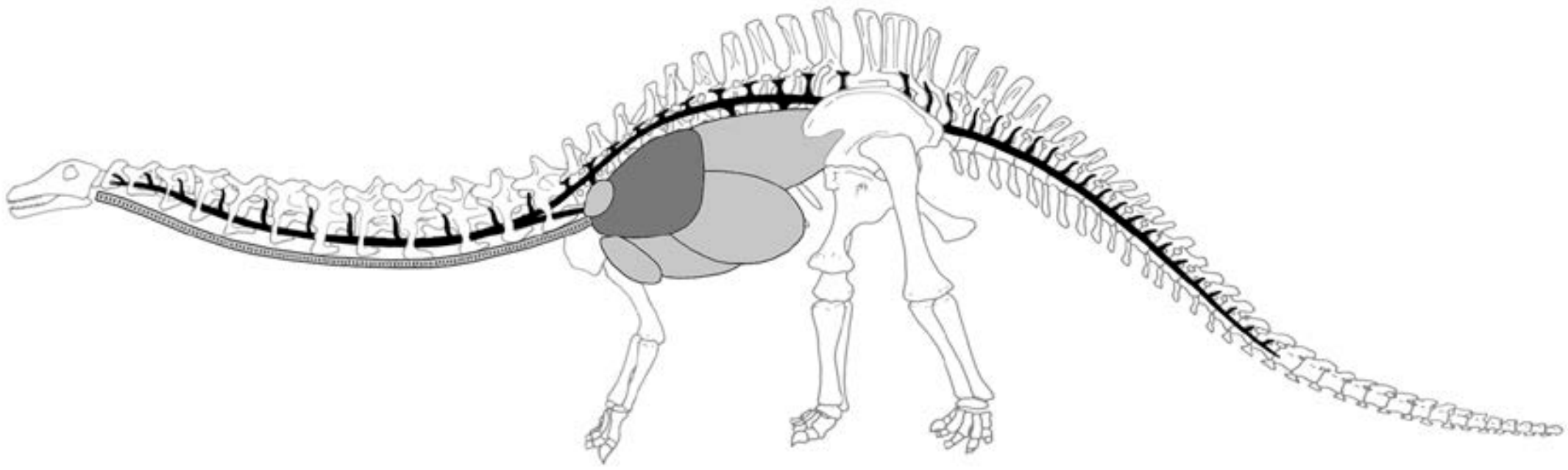

FIGURE 1-7. Internal structure of OMNH 27794, a partial distal caudal vertebra of a titanosauriform. The internal structure is composed of apneumatic cancellous bone, and no medullary cavity is present. Scale bar is 1 cm.

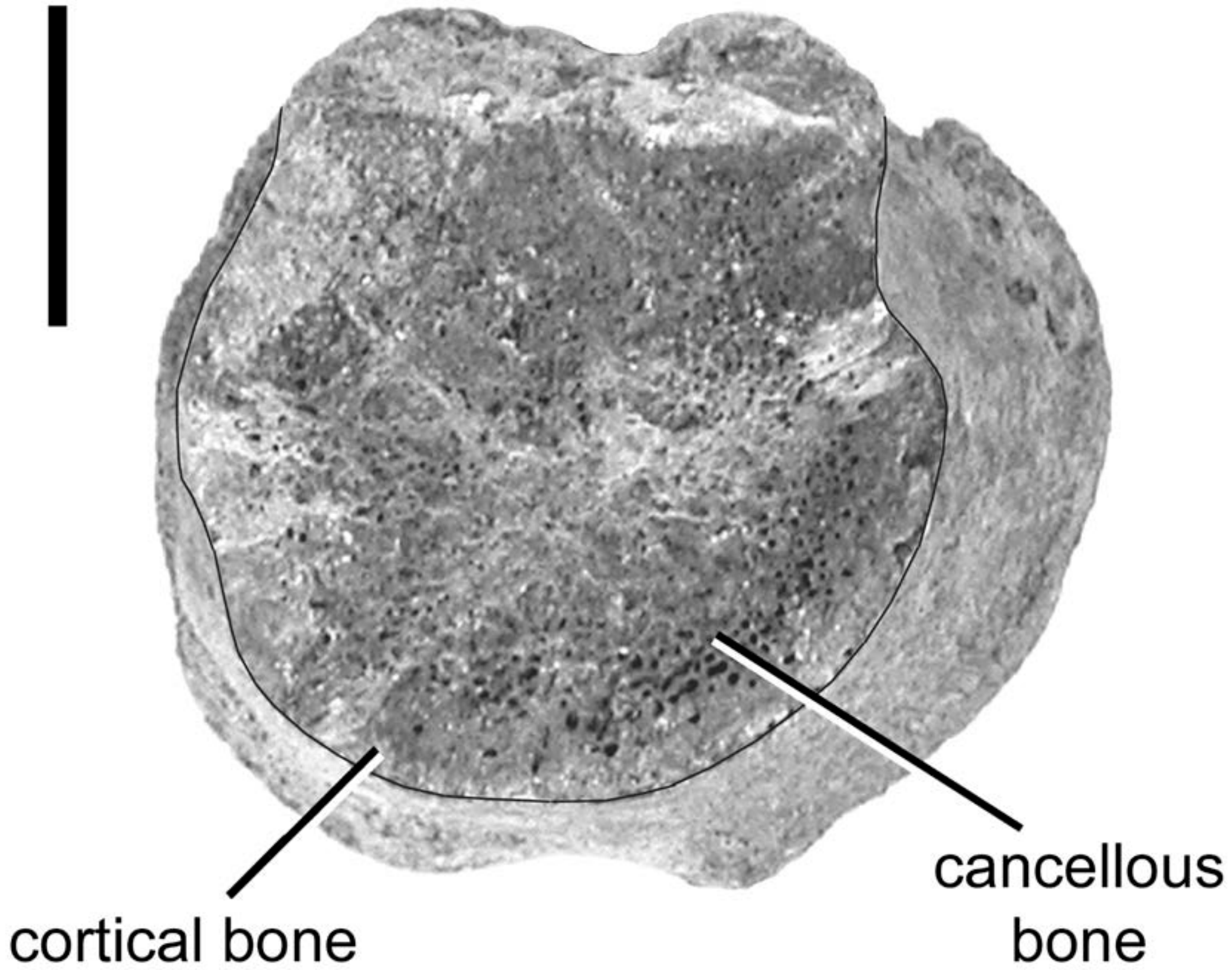
cortical bone
cancellous
bone

# CHAPTER TWO

# WHAT PNEUMATICITY TELLS US ABOUT "PROSAUROPODS", AND VICE VERSA

## INTRODUCTION

Pneumaticity is a prominent feature of the postcranial skeleton in theropod and sauropod dinosaurs. In contrast, there is little evidence for postcranial pneumaticity in basal sauropodomorphs (informally referred to as 'prosauropods' in this paper), although from time to time some aspects of prosauropod osteology have been posited as evidence of pneumaticity (Britt 1997) or compared to unequivocal pneumatic structures in sauropods (Yates 2003; Galton and Upchurch 2004). My goals in this paper are to review the evidence for postcranial skeletal pneumaticity (PSP) in prosauropods and to discuss the origin of air sacs and pneumaticity in early dinosaurs and their relatives.

Prosauropod taxonomy is currently in a state of flux, as other papers in this volume attest (Sereno 2007; Upchurch et al. 2007; Yates 2007). Prosauropods were traditionally considered a paraphyletic assemblage that gave rise to sauropods. Sereno (1998) recovered a monophyletic Prosauropoda, defined this clade (anchored upon Plateosaurus) as a monophyletic sister taxon to Sauropoda, and united the two in a node-based Sauropodomorpha. A similar phylogenetic hypothesis was described by Galton and Upchurch (2004). However, other recent phylogenetic analyses (Yates 2003, 2004; Yates and Kitching 2003) have found that some prosauropods are closer to Saltasaurus than to Plateosaurus; thus under current phylogenetic definitions they should be regarded as basal sauropods. Some other taxa (e.g. Saturnalia) lie outside Sauropodomorpha as defined by Sereno (1998) altogether (Yates 2003, 2004; Yates and Kitching 2003; Langer 2004). However, the monophyly or paraphyly of the group of taxa traditionally called prosauropods is not critical to the purposes of this paper. What is important is that all traditional 'prosauropods' have two things in common: they lack unequivocal evidence of pneumatic cavities in their vertebrae and ribs, and they are phylogenetically bracketed by sauropods and theropods (Figure 2-1).

## POSTCRANIAL PNEUMATICITY IN THEROPOD AND SAUROPOD DINOSAURS

Before examining the evidence for PSP in 'prosauropods', I will review the conditions present in other saurischian dinosaurs. The sister taxon of Sauropodomorpha is Theropoda; consequently, 'prosauropods' are phylogenetically bracketed in part by birds, the only clade of extant vertebrates with extensive PSP. The relationship between the respiratory system and pneumatic postcranial bones in birds has been described many times (e.g. Müller 1908; King 1966; Duncker 1971; O'Connor 2004), and is briefly summarized here. The relatively small, constant volume, unidirectional flow-through lungs of birds are ventilated by the attached air sacs, which are large, flexible and devoid of parenchymal tissue. The lungs and air sacs also produce air-filled tubes called diverticula that pass between the viscera, between the muscles, and under the skin. Where a diverticulum comes into contact with a bone, it may (but does not always) induce bone resorption, which can produce pneumatic tracks, fossae, or foramina. If resorption of the cortex produces a foramen, the diverticulum may enter the medullary space and replace the existing internal structure with a series of air-filled chambers of varying complexity. The best description of this process is provided by Bremer (1940).

The extent of PSP varies in different avian clades. Almost any postcranial bones can become pneumatized; in large soaring birds such as pelicans, almost the entire skeleton is pneumatic, including the distal limb elements (O'Connor 2004). Although many large volant and flightless birds have highly pneumatic skeletons, the correlation between body size and the extent of PSP in birds is weak (O'Connor 2004). PSP tends to be reduced or absent in diving birds (Gier 1952; O'Connor 2004). Different parts of the skeleton become pneumatized by diverticula of different air sacs in extant birds (Table 2-1); this is important because it allows us to make inferences regarding the evolution of air sacs in fossil taxa. PSP in non-avian theropods generally follows the avian model (Britt 1993, 1997; O'Connor and Claessens 2005; O'Connor 2006). Patterns of pneumatization along the vertebral column indicate that both anterior and posterior air sacs (presumably cervical and abdominal) had evolved by the time of the ceratosaur-tetanuran divergence (O'Connor and Claessens 2005).

Fossae are present in the presacral vertebrae of basal sauropods such as Shunosaurus and Barapasaurus (Britt 1993; Wilson and Sereno 1998). These fossae are similar to the unequivocally pneumatic foramina and camerae of more derived sauropods, both in their position on individual vertebrae and in their distribution along the vertebral column, and because of these similarities they have usually been regarded as pneumatic in origin (Britt 1993, 1997; Wedel 2003a). However, similar fossae are present in other tetrapods that lack PSP, so the presence of fossae alone is at best equivocal evidence for PSP (O'Connor 2006; see below). The vertebrae of more derived sauropods have foramina that communicate with large internal chambers; the combination of foramina and large internal chambers is an unambiguous indicator of PSP (O'Connor 2006). There is a general trend in sauropod evolution for PSP to spread posteriorly along the vertebral column, albeit to different extents in different clades and with a few reversals (Wedel 2003b; Figure 2-2). In both sauropods and theropods, fossae in basal forms were replaced by large-chambered (camerate) vertebrae and eventually small-chambered (camellate) vertebrae in more derived taxa (Britt 1993, 1997; Wedel 2003a).

The evolution of PSP in sauropods mirrors in detail that of non-avian theropods. At the level of individual elements (e.g. vertebrae and ribs), pneumatic features in sauropods compare very closely with those of both avian and non-avian theropods (Figure 2-3). In terms of the ratio of bony tissue to air space within a pneumatic element, sauropod vertebrae are, on average, comparable with the limb bones of many extant birds: about 60 per cent air by volume (Wedel 2004, 2005; Woodward 2005; Schwarz and Fritsch 2006). At the level of the skeleton, osteological indicators of pneumaticity spread as far back as the mid-caudal vertebrae in at least two groups of sauropods, the diplodocines and saltasaurines (Osborn 1899; Powell 1992). Among non-avian theropods, extensive pneumatization of the caudal series evolved only in oviraptorosaurs (Osmólska et al. 2004). Finally, limited appendicular pneumaticity was probably present in both sauropods and non-avian theropods. The dromaeosaur Buitreraptor has a pneumatic furcula (Makovicky et al. 2005), and a large foramen in the proximal femur of the oviraptorid Shixinggia is probably also pneumatic in origin (Lü and Zhang 2005). Large chambers have been reported in the ilia of the basal diplodocoid Amazonsaurus (Carvalho et al. 2003) and in several titanosaurs (Powell 1992; Sanz et al. 1999, Xu et al. 2006). Although these chambers are similar to unequivocally pneumatic spaces in the other saurischians, it has not yet been shown that the ilial chambers are connected to foramina, which are necessary for pneumatization to occur (see O'Connor 2006).

## EVIDENCE OF PNEUMATICITY IN ‘PROSAUROPODS’

Historically, postcranial pneumaticity in ‘prosauropods’ has received little attention, which is to be expected given the paucity of available evidence. Janensch (1947) posited that a foramen in a dorsal vertebra of Plateosaurus might have been pneumatic, but he attached no great weight to this hypothesis. Britt (1997) considered vertebral laminae evidence of pneumaticity in ‘prosauropods’. Most recently, Yates (2003, p. 14, fig. 12) identified ‘pleurocoel-like pits’ in the mid-cervical vertebrae of Pantydraco caducus, and Galton and Upchurch (2004, p. 245) referred to fossae in the posterior dorsals of some prosauropods as ‘pleurocoelar indentations’. The ‘pleurocoel-like’ structures were not explicitly described as pneumatic in either work. Although fossae are not unambiguous indicators of pneumaticity (O'Connor 2006), vertebral fossae seem to be an early step toward full pneumatization, both ontogenetically and phylogenetically (Wedel 2003a). Putative pneumatic characters in ‘prosauropods’ can be divided into three categories: vertebral laminae, foramina and fossae, which will be discussed in this order, below.

**Vertebral Laminae**

Vertebral laminae are struts or plates of bone that connect the various apophyses of a vertebra to each other and to the centrum. The landmarks that are usually connected in this way are the pre- and postzygapophyses, the diapophyses and parapophyses, and the neurapophysis. The form and occurrence of the major laminae in saurischian dinosaurs were reviewed by Wilson (1999). In addition to a basic set of laminae common to all saurischians, many sauropods and theropods have other irregularly developed laminae that are usually not named but are collectively called accessory laminae. Laminae tend to be more numerous and more sharply defined in camerate than camellate vertebrae (Wilson and Sereno 1998; Wedel 2003a). Camellate vertebrae evolved relatively early in the radiation of non-avian theropods (Britt 1993, 1997), and most derived theropods have less elaborate systems of laminae than do neosauropods. This may explain why the literature on laminae has tended to focus on sauropods (e.g. Osborn 1899; Osborn and Mook 1921; Janensch 1929a, 1950; Wilson 1999).

Two problems with the identification of laminae that are relevant to the question of pneumaticity are how well-developed a ridge of bone must be before we call it a lamina, and whether laminae are primarily additive structures formed by the deposition of new bone, or are simply bone that is left over following the formation of fossae. The first problem is important because, as shown below, incipient laminae are broadly distributed among archosaurs. To what extent are the distinct laminae of saurischian dinosaurs new (= apomorphic) structures, rather than modifications of pre-existing ones? This question has ramifications for the evolution of laminae and for coding of laminae as characters in phylogenetic analyses.

The second question can be stated: do laminae grow out from the corpus of the vertebra to define the fossae that they bound, or do we only recognize laminae as distinct structures because the bone between them has been removed? For example, the cervical vertebra of Nigersaurus illustrated by Sereno and Wilson (2005, fig. 5.8) has on the lateral face of the neural spine two fossae divided by an accessory lamina (Figure 2-4). At its edges, the anteroventral fossa approaches both the prezygapophysis and the diapophysis. This region is flat or convex in most other neosauropods, which have a lateral fossa in roughly the same position as the posterodorsal fossa in Nigersaurus. It seems likely, therefore, that the anteroventral fossa in Nigersaurus is a new morphological feature, and that the accessory lamina can only be recognized as a lamina because a fossa has been excavated below it. Conversely, the vertebrae of most tetrapods do not have straight bars of bone that connect the zygapophyses to the neurapophysis, but this is exactly what the spinopre- and spinopostzygapophyseal laminae of some sauropods do (Figure 2-4). In comparison with the condition in other tetrapods, including prosauropods, these laminae appear to be additive structures. These potentially opposing processes of lamina formation should be kept in mind while reading the following descriptions.

The laminae of sauropods often form the boundaries of fossae that have been interpreted as pneumatic, either because they contain foramina that lead to internal chambers or because they are heavily sculpted, with numerous subfossae (sensu Wilson 1999) and a distinct bony texture (although texture alone is not necessarily a good indicator of pneumaticity; see O'Connor 2006). Wilson (1999) considered whether sauropod laminae existed to provide mechanical support or to subdivide pneumatic diverticula, and concluded that they probably served both functions simultaneously. Following from the aforementioned discussion, we might also ask if sauropod laminae exist because the pneumatic diverticula are subdivided, as they often are in birds (e.g. Wedel 2003b, fig. 2), and these subdivisions are impressed into the bone, leaving laminae between them. Rather than try to determine which structure has morphogenetic precedence, it may be more useful to view sauropod vertebrae in light of Witmer's (1997) hypothesis that the form of a pneumatic bone can be viewed as the outcome of a struggle between bone tissue, which grows partly in response to biomechanical stress, and pneumatic diverticula, which are opportunistic and invasive and spread wherever possible (see Sadler et al. [1996] and Anorbe et al. [2000] for examples of proliferating diverticula).

The laminae of 'prosauropods' differ from those of sauropods in three important ways. The first is that prosauropods have fewer laminae. The laminae that connect the diapophysis to the centrum, parapophysis and zygapophyses are usually present (Wilson 1999), but those that connect the neurapophysis to other landmarks are absent (Figure 2-5; but see Bonaparte 1999, figs. 13–16 on Lessemsaurus). The second is that laminae are confined to the presacral vertebrae in "prosauropods", whereas the sacral vertebrae of neosauropods and the caudal vertebrae of diplodocids also bear laminae.

The third and most important difference between the laminae of sauropods and 'prosauropod' is that the fossae bounded by the latter are blind. These fossae do not contain foramina or subfossae and they do not have a distinctive bone texture. Consequently, there is no strong reason to suspect that these fossae contained pneumatic diverticula. O'Connor (2006) found that similar fossae in extant crocodilians and birds may contain cartilage or adipose tissue. Considering whether the laminae are additive structures or remnants of fossa formation sheds little light on the problem. Some laminae, such as the PRDLs of Plateosaurus cervicals, are straight-line structures that appear to have been added, compared to the condition in vertebrae that lack laminae (Figure 2-5). Others, such as the PODLs in the same vertebrae, are only detectable because they have been undercut by a fossa. The form of the fossae themselves provides no obvious clues to their contents in vivo.

Laminae like those of 'prosauropods' occur in many other archosaurs. Desojo et al. (2002) and Parker (2003) recognized that many of the laminae described by Wilson (1999) for saurischian dinosaurs are also present in basal archosaurs and pseudosuchians. The full complement of diapophyseal laminae is present in dorsal vertebrae of the basal archosauriform Erythrosuchus and in those of poposaurs such as Sillosuchus and Arizonasaurus, including the PCDL, PODL, PPDL and PRDL (Figure 2-6; see Alcober and Parrish 1997; Nesbitt 2005). At least in Erythrosuchus, the fossae bounded by these laminae contain aggregates of vascular foramina; obvious foramina like these are not present or at least not common in the interlaminar fossae of 'prosauropods' (pers. obs.). Incipient laminae are also present in some neosuchian crocodyliforms. Most dorsal vertebrae of Goniopholis stovalli have rudimentary PCDLs and PODLs (Figure 2-7). The PODL is bounded dorsally by a shallow fossa on the lateral face of the neural spine and ventrally by a deep infrapostzygapophyseal fossa. In at least some of the vertebrae, the fossa on the side of the neural spine has a distinct margin (Figure 2-7b). Although most neosuchian crocodyliforms have extensive skull pneumatization (Witmer 1997; Tykoski et al. 2002), PSP is absent in the clade (O'Connor 2006).

Vertebral laminae also occur in non-amniotes. The best example is probably the plethodontid salamander Aneides lugubris, in which plate-like shelves of bone connect the parapophyses of dorsal vertebrae to the ventrolateral margins of the centra (Wake 1963). These shelves of bone are absent in other species of Aneides and in other plethodontid genera, and they are thus additive structures that are apomorphic for A. lugubris (compare Wake [1963], fig. 9 with Wake and Lawson [1973], fig. 6). A. lugubris has the most prolonged ontogeny of any plethodontid, and it is peramorphic relative to other species in the genus, with a more extensively ossified skeleton (Wake 1963; Wake et al. 1983). The development of laminae in the species is probably an epiphenomenon of the extensive ossification of the skeleton, which in turn is related to adaptations for arboreality and feeding (Larson et al. 1981). As such, the laminae of A. lugubris are not homologous with those of archosaurs in a taxic sense, and they are probably produced by different developmental processes. Still, A. lugubris demonstrates that laminae can evolve in vertebrates that are far removed from basal dinosaurs in both genealogy and body size, and it provides a potential system in which to investigate the development of laminae in an extant tetrapod.

On one hand, it is possible that the laminae and fossae of basal archosauriform and pseudosuchian vertebrae appear pneumatic because they are pneumatic (Gower 2001), but the morphology is not compelling. Like the fossae of ‘prosauropod’ neural spines, those of Erythrosuchus and Arizonasaurus lack subfossae, foramina that lead to large internal chambers, or altered texture. The presence of similar features in crocodyliforms and salamanders is strong evidence that the morphologies in question can be produced in the absence of pneumaticity. Unlike ‘prosauropods’, basal archosauriforms and pseudosuchians are not bracketed by taxa with unequivocal evidence of pneumaticity, so inferring that they had pneumatic vertebrae would require pulmonary diverticula and possibly also air sacs to have evolved much earlier than otherwise supposed (see Palaeobiological Implications, below).

Regardless of when the capacity for PSP evolved, the laminae of ‘prosauropods’ bound fossae that are not unequivocally pneumatic. Similar laminae are present in crocodilians, a group in which postcranial pneumaticity is entirely absent. Some of the osteological traces of diverticula are subtle, and the possibility that the neural arch fossae of “prosauropods” accommodated pneumatic diverticula cannot be ruled out, but there is no strong evidence for it.

**Foramina**

The only putative pneumatic foramen in a ‘prosauropod’ is that described by Janensch (1947) in a cervical vertebra of Plateosaurus. Janensch argued that the size (11x4 mm) and form of the foramen were more consistent with a pneumatic than a vascular interpretation. The identity of the foramen can only be settled by firsthand observation, preferably with a computed tomographic (CT) scan to determine if the foramen leads to any large internal chambers. Unfortunately, such an examination has yet to be conducted. However, the caudal vertebrae of some whales are similar in size to Plateosaurus dorsal vertebrae (~20 cm in maximum linear dimension) and have vascular foramina up to 30 mm in diameter (pers. obs.), so large foramina do not necessarily indicate the presence of pneumaticity. In general, the prominent foramina and internal chambers that are typical of sauropod vertebrae are absent in the vertebrae of ‘prosauropods’.

**Fossae**

The first step in recognizing pneumatic fossae is to distinguish between vertebrae that have distinct fossae and those that are merely waisted (narrower in the middle than at the ends). The vertebrae of most vertebrates are waisted to some extent. In humans the effect is barely noticeable, but in some archosaurs the ‘waist’ of the vertebra may be only half the diameter of the ends of the centrum (e.g. Nesbitt 2005, fig. 16). Some degree of waisting is to be expected based on the early development of vertebrae. Cell-dense regions of the embryonic axial column become intervertebral discs, and lower-density regions become vertebral bodies. This produces centra that are inherently waisted (Verbout 1985, pl. 10; Wake 1992, figs. 6.5 and 6.7). The degree of waisting has occasionally been used as a taxonomic character (Case 1907), but to date there is no clear explanation of why some vertebrae are more waisted than others. In any case, vertebral waisting is widespread in vertebrates and is not evidence for pneumaticity.

Waisting aside, fossae still suffer from a problem of definition. Consider a spectrum of morphological possibilities (Figure 2-8). At one end is a vertebra that is waisted but lacks distinct fossae: for example, a thoracic vertebra of an artiodactyl. At the other end is a vertebra with large foramina that open into internal chambers, such as a dorsal vertebra of Saltasaurus. The ‘chamber morphospace’ between these endpoints is filled with a continuum of deeper and more distinct fossae and camerae. Adjacent to the artiodactyl vertebra we might put a vertebra that has fossae with a distinct margin on one side but not the other, like those in the cervical centra of Arizonasaurus, which are bounded dorsally by the PCDL. Next a fossa that has a distinct bony rim on all sides, but that is not enclosed by a bony lip, like those in dorsal centra of adult Barapasaurus or juvenile Apatosaurus. The penultimate example is a fossa that is enclosed by a bony lip, but that is little expanded beyond the boundaries of the opening, such as the fossae in presacral centra of Haplocanthosaurus (Britt [1993] referred to these chambers as camerae, whereas Wedel termed them fossae [Wedel et al. 2000; Wedel 2003a]. The morphology of these features is intermediate between that of fossae and camerae, and either term could reasonably be applied). Finally, in neosauropods such as Camarasaurus and Saltasaurus the space beyond the bony lip is greatly expanded, so that the result is a foramen that leads to camerae or camellae.

The fossae along this spectrum vary in geometry and they are not all pneumatic. Although Goniopholis is extinct and not part of the crown-group Crocodylia, it is highly unlikely that the caudal vertebrae of this semi-aquatic neosuchian were pneumatic. Nevertheless, they bear lateral fossae with distinct margins that are very similar to structures that are sometimes interpreted as pneumatic in dinosaurs, such as the sacral 'pleurocoels' of ornithomimosaurs. However, distinct margins alone are not compelling evidence of pneumaticity. Conversely, truly pneumatic fossae need not have distinct margins. For example, the fossae behind the prezygapophyses of ratites lack clear margins, but CT scans show that they house pneumatic diverticula, and they sometimes contain pneumatic foramina (Figure 2-9). In extant birds, the pneumatic canalis intertransversarius lies alongside the centrum (Müller 1908), but many birds have cervical centra that are laterally convex and lack any fossae (the pneumatic foramina are usually located inside the cervical rib loop or ansa costotransversaria).

The foregoing discussion implies that where chambers lack a distinct lip of bone, geometry alone is a poor clue to whether or not a given fossa has a pneumatic origin or not. Other lines of evidence must be used, such as position in the body, the presence or absence of adjacent pneumatic foramina, subfossae, or textural differences (and even the last two may be misleading; see O'Connor 2006).

Vertebral centra of 'prosauropods' can be quite narrow-waisted, and some have lateral grooves or fossae that are bounded on one side by a lamina. As with neural arch laminae, these features are sometimes associated with pneumaticity but they are not diagnostic for it. The 'pleurocoelar indentations' mentioned by Galton and Upchurch (2004) do not have a distinct boundary or lip in any of the specimens that I have examined (e.g. Moser 2003, pl. 16). The only known 'prosauropod' with distinctly emarginated lateral fossae is Pantydraco caducus (Yates 2003). Cervical vertebrae 6–8 of BMNH P24, the holotype of P. caducus, have small, distinct fossae just behind the diapophyses (Figure 2-10). The fossae are high on the centra and may have crossed the neurocentral sutures, which are open. The fossa on the eighth cervical looks darker than it should because it is coated with glue. The ninth cervical has a very shallow, teardrop-shaped hollow in the same region of the centrum. The bone texture in this hollow is noticeably smoother than on the rest of the centrum (this is especially apparent under low-angle lighting). That the fossa on the ninth cervical is shallower and less distinct than those on cervicals 6–8 is reminiscent of the diminution of pneumatic features observed at the transition from pneumatic to apneumatic vertebrae, as seen in the anterior dorsal vertebrae of Jobaria (Sereno et al. 1999, fig. 3) and the middle caudal vertebrae of Diplodocus (Osborn 1899, fig. 13). The holotype specimen of P. caducus represents an immature individual (Yates 2003), however, so the shallow fossa on the ninth cervical may be incompletely developed.

Are the fossae of P. caducus pneumatic? If so, they are the only good evidence for invasive pneumatic features in the postcrania of 'prosauropods'. Previously, I have assumed that they were pneumatic, based in part on the distinct margins of the fossae in cervicals 6–8, and also on the fact that the fossae only occur on cervicals 6–9 (Wedel 2006). The first line of evidence is inadequate to unequivocally diagnose pneumaticity. The second is also problematic. Cervical vertebrae 5–9 are the only ones that are always pneumatized in the chicken (Hogg 1984a), and the cervical and anterior thoracic vertebrae are the first parts of the axial skeleton to be pneumatized during the ontogeny of birds (Cover 1953; Hogg 1984b). The spread of pneumaticity posteriorly along the vertebral column in the ontogeny of birds appears to faithfully recapitulate the evolution of pneumaticity in theropods and sauropods (Wedel 2003b, 2005). The presence of fossae on the midcervical vertebrae of P. caducus is easily explained if the fossae are pneumatic; their appearance in that part of the skeleton mirrors early ontogeny in birds and is also consistent with later trends in the evolution of PSP in sauropodomorphs (Figure 2-2). On all four vertebrae, the fossae are not closely associated with laminae and cannot be dismissed as epiphenomena of lamina formation (see O'Connor 2006); a specific soft-tissue influence was causally related to the formation of the fossae. The geometry of the fossae is not sufficient to specify that soft-tissue influence, because adipose, muscular and pulmonary tissues have all been found to occupy similar fossae in other tetrapods (O'Connor 2006). On the other hand, the presence of the fossae only on the midcervical vertebrae is difficult to explain if they were not produced by pneumatic diverticula like those of more derived sauropods.

**Summary**

Vertebral laminae and shallow depressions on the centra are widespread in archosauriforms and not diagnostic of pneumaticity, although it is difficult to rule out the possibility that they may have been associated with pneumatic diverticula. ‘Prosauropods’ have fewer laminae than most sauropods, fewer vertebrae with laminae, and the fossae adjacent to the laminae are almost always blind (with no large foramina or chambers). A foramen in a vertebra of Plateosaurus and distinct fossae in the cervical vertebrae of Pantydraco caducus are the best evidence for potential pneumaticity in ‘prosauropods’, but neither is an unambiguous indicator of PSP and both would benefit from further study. In any case, the diagnostic osteological correlates of pneumaticity that are common in sauropods and theropods are absent or extremely rare in ‘prosauropods’, and the putative pneumatic features that are widespread in ‘prosauropods’ (laminae and shallow fossae) are not compelling evidence of pneumaticity. To leave aside for a moment the question of ‘prosauropod’ monophyly, ‘prosauropods’ are unusual as the only sizeable group (or grade) of saurischian dinosaurs that lack extensive PSP.

## PALAEOBIOLOGICAL IMPLICATIONS

Pneumatic bones are of palaeobiological interest in two ways. We may be interested in the pneumatic bones themselves: in their external and internal morphology, in the ratio of bone to air space, and in the ways that they develop. Pneumatic bones are also important, arguably more important, as osteological markers of the pulmonary system. In this section I discuss the origins of pneumaticity and of air sacs, and the implications for the respiratory physiology of sauropodomorphs.

### Origin of the diverticular lung and PSP

The first part of the postcranial skeleton to be pneumatized in any saurischian dinosaur is the cervical column. The fossae in the mid-cervical vertebrae of Pantydraco caducus are not definitely pneumatic on the basis of geometry alone. However, their placement in the skeleton is suspiciously similar to the early stages of pneumatization in birds. The same is true of fossae in the cervical column of the basal sauropod Shunosaurus (Wilson and Sereno 1998). Among basal theropods, Coelophysis bauri is the earliest well-represented taxon with evidence of pneumaticity. The postaxial cervical vertebrae of C. bauri have pneumatic cavities that occupy most of the neural spine and that communicate with the outside through several large foramina (Colbert 1989).

The pattern of pneumatization in these early-diverging saurischians indicates the presence of cervical air sacs like those of birds. It is true that in sauropsids diverticula may develop from practically any portion of the respiratory system. However, it does not follow that the diverticula that pneumatize the skeleton can come from anywhere (contra Hillenius and Ruben 2004), for two reasons. First, in extant birds the cervical vertebrae are only pneumatized by diverticula of cervical air sacs. Diverticula of the cranial air spaces, larynx and trachea are never known to pneumatize the postcranial skeleton (King 1966), and diverticula of the parenchymal portion of the lung only pneumatize the vertebrae and ribs adjacent to the lungs (O'Connor 2004). Second, as discussed below, pneumatization of the posterior half of the body is accomplished only by diverticula of abdominal air sacs (O'Connor and Claessens 2005). These observations of extant taxa provide valuable guidelines for interpreting patterns of skeletal pneumatization in fossil taxa. Pneumatization by diverticula of cervical air sacs is the only mechanism for pneumatizing the neck that is (1) known to occur in extant taxa, and (2) consistent with the pattern of pneumatization found in basal saurischians (Wedel 2006).

Pneumaticity in basal saurischians is extremely limited. The bone removed by pneumatization of the postcranial skeleton (or fossa formation, if the fossae of Pantydraco are not pneumatic) accounted for much less than one per cent of the total body volume in both Coelophysis and Pantydraco (see Appendix I), compared to several per cent for more derived sauropods and theropods (Wedel 2004, 2005). PSP probably did not evolve as an adaptation for lightening the skeleton, although it seems to have been exapted for that purpose later in saurischian evolution (Wedel 2003b).

Furthermore, diverticula did not evolve to pneumatize the skeleton. In the first place, many of the diverticula of birds are visceral, subcutaneous or intermuscular, and do not pneumatize any bones (Duncker 1971). Skeletal pneumatization cannot be invoked to explain the presence of these diverticula. In the second place, the presence of diverticula is a prerequisite for pneumatization of the skeleton. The immediate ancestors of Coelophysis and Pantydraco must have already had cervical diverticula (assuming that the fossae of the latter are pneumatic in origin). Pneumatization of the cervical series could not happen until these diverticula were already in place, so the diverticula must have evolved for some other reason.

Alternatively, the origins of paravertebral diverticula and of PSP may have been coincident. The first step may have been a developmental accident that allowed the diverticula to push beyond the coelom and these 'unleashed' diverticula may have pneumatized the vertebral column immediately. This sort of morphogenetic behavior on the part of diverticula is plausible on the basis of cases in the human clinical literature (e.g. Sadler et al. 1996; Anorbe et al. 2000). The main argument against this near-saltational scenario is that the first vertebrae to be pneumatized in both sauropodomorphs (Pantydraco, Shunosaurus) and theropods (Coelophysis) are cervicals that are not adjacent to the lungs (Wedel 2006).

**Origin of flow-through ventilation**

Flow-through ventilation requires that air sacs be present both anterior and posterior to the parenchymal portion of the lung. Given the pattern of pneumatization found in pterosaurs, sauropods and theropods, we may infer that cervical air sacs were present in the ancestral ornithodiran (or evolved independently in pterosaurs and saurischians). The next problem is to determine when abdominal air sacs originated and how many times.

In extant birds, the posterior thoracic, synsacral and caudal vertebrae, pelvic girdle and hindlimb are only pneumatized by diverticula of abdominal air sacs (O'Connor and Claessens 2005; contra Ruben et al. 2003; Hillenius and Ruben 2004; Chinsamy and Hillenius 2004). So if a fossil archosaur is discovered with pneumatic vertebrae posterior to the mid-thorax, we have a compelling case for inferring that the animal had abdominal air sacs. Pneumatic vertebrae in the 'posterior compartment' are present in pterosaurs, diplodocid and macronarian sauropods, and in most clades of neotheropods, but are absent in non-dinosaurian dinosauromorphs, ornithischians, herrerasaurids, 'prosauropods', basal sauropods, dicraeosaurids, and in basal members of most neotheropod clades (e.g. Baryonyx, Ceratosaurus and Allosaurus; pers. obs.).

How many times did abdominal air sacs evolve? Possibly just once, before the ornithodiran divergence; possibly twice, in pterosaurs and saurischians; or possibly three times, in pterosaurs, sauropods and theropods (Figure 2-1). We could take this to its logical conclusion and assume that abdominal air sacs evolved afresh in every group with posterior compartment pneumaticity; this would require the independent origin of abdominal air sacs in ceratosaurs, allosauroids and coelurosaurs, for example (not to mention several independent derivations within coelurosaurs).

The alternative is that some or all of the groups listed above had abdominal air sacs but failed to pneumatize any elements in the posterior compartment. The same condition pertains in many extant birds (O'Connor 2004, table 2). O'Connor and Claessens (2005) posited an origin of abdominal air sacs by the time of the ceratosaur-tetanuran divergence, based on the presence of posterior compartment pneumatization in Majungatholus, and despite its absence in basal ceratosaurs and basal tetanurans.

In terms of evolutionary change, ventilation mechanisms are highly conserved, PSP is highly labile, and diverticula seem to lie between these extremes. All birds have essentially the same lung architecture; the biggest difference among living forms is the presence or absence of a neopulmo (Duncker 1971). On the other hand, PSP varies widely within small clades and even within populations (King 1966; Hogg 1984a; O'Connor 2004). Diverticula appear to be more conserved than PSP, although a dedicated study comparing the evolution of the two is needed. For example, most birds have femoral and perirenal diverticula, but the femur and pelvis are only pneumatized in a subset of these taxa (Müller 1908; King 1966; Duncker 1971). These observations are necessarily tentative, given the paucity of phylogenetically-based comparative studies of pneumatic diverticula and PSP (but see O'Connor 2004). Further, our knowledge of variation in the pulmonary system and its diverticula is based entirely on extant birds, and may not be applicable to other saurischians.

Nevertheless, the evolutionary malleability of lungs, diverticula, and PSP in birds should not be ignored in reconstructing the pulmonary systems of fossil archosaurs. The absence of unequivocal PSP in most 'prosauropods' does not mean that they lacked air sacs. Depending on the preferred phylogenetic hypothesis, Sauropodomorpha is only one or two nodes away from Neotheropoda. Most neosauropods have pneumatic vertebrae in the posterior compartment. If these sauropods found some way to pneumatize the posterior compartment without abdominal air sacs, then surely the same could be true of some or all non-avian theropods. Likewise, if posterior compartment pneumaticity is prima facie evidence of abdominal air sacs in theropods, then abdominal air sacs must also have been present in sauropods (and, by extension, pterosaurs). It is more parsimonious to infer that cervical and abdominal air sacs were present in the ancestral saurischian, but did not pneumatize the skeleton in ‘prosauropods’, than to infer independent origins of air sacs in sauropods and theropods.

Most pterosaurs have extensively pneumatized skeletons, although it is not clear whether pneumaticity is present in any of the Triassic forms (Bonde and Christiansen 2003). The presence of PSP in pterosaurs, sauropodomorphs and theropods suggests that air sacs may have been present in the ancestral ornithodiran. An apparent problem with pushing the origin of air-sac-driven breathing back before the origin of Saurischia is the utter absence of PSP in ornithischians. PSP appeared in pterosaurs, sauropodomorphs and theropods relatively quickly after the divergence of each clade: by the Norian in theropods (Colbert 1989) and no later than the Early Jurassic in pterosaurs and sauropodomorphs (Bonde and Christiansen 2003; Wedel 2005). If ornithischians had air sacs and diverticula then it is odd that they never evolved PSP during the 160 million years of their existence. However, this problem may be more illusory than real. The invasion of bone by pneumatic epithelium is essentially opportunistic (Witmer 1997). Although pneumatic diverticula may radically remodel both the exterior and interior of an affected bone, this remodeling cannot occur if the diverticula never come into contact with the bone, and may not occur even if they do. Furthermore, for all of the potential advantages it conveys, PSP is still an exaptation of a pre-existing system: in an adaptive sense, lineages that lack PSP don’t know what they’re missing. Recall that PSP in basal saurischians did little to lighten the skeleton (see above). Ornithischians may have had air sacs without diverticula, or diverticula without PSP. It is pointless to consider the advantages that ornithischians ‘lost’ by never evolving PSP, because that evolution would have hinged on the incidental contact of bone and air sac and could not have been anticipated or sought by natural selection.

The problem of determining when abdominal air sacs evolved is challenging because it forces us to decide between events of unknown probability: the possibility that ornithischians had an air sac system and never 'discovered' PSP (if abdominal air sacs are primitive for Ornithodira), versus the possibility that a system of cervical and abdominal air sacs evolved independently in pterosaurs and saurischians. Currently, available evidence is insufficient to falsify either hypothesis.

**Sauropodomorph palaeobiology**

It is likely that ‘prosauropods’ had cervical and abdominal air sacs, given the strong evidence for both in sauropods and theropods. We may not be able to determine for certain whether ‘prosauropods’ had a bird-like flow-through lung, but the requisite air sacs were almost certainly present. Our null hypothesis for the respiratory physiology of ‘prosauropods’ should take into account some form of air-sac-driven ventilation.

The air sacs of birds mitigate the problem of tracheal dead space (Schmidt-Nielsen 1972), and some birds have improbably long tracheae (i.e. longer than the entire body of the bird; see McClelland 1989). In addition, birds can ventilate their air sacs without blowing air through the lungs, which allows them to avoid alkalosis during thermoregulatory panting (Schmidt-Nielsen et al. 1969). Finally, flow-through breathing allows birds to extract much more oxygen from the air than mammals can (Bernstein 1976). In general, sauropods were larger and longer-necked than ‘prosauropods’, and the aforementioned capabilities of a bird-like ventilation system may have helped sauropods overcome the physiological challenges imposed by long necks and large bodies, including tracheal dead space, heat retention and oxygen uptake.

The one obvious advantage that ‘prosauropods’ did not share with sauropods is the very lightweight skeletal construction afforded by pneumaticity. In life, the average pneumatic sauropod vertebra was approximately 60 per cent air by volume (Wedel 2005; Woodward 2005; Schwarz and Fritsh 2006). All else being equal, a sauropod could have a neck two-thirds longer than that of a prosauropod for the same skeletal mass. Pneumaticity helped sauropods overcome constraints on neck length, and thereby opened feeding opportunities that were not available to ‘prosauropods’. The importance of that difference is unknown, but it is worth considering in reconstructions of sauropodomorph evolution and palaeobiology.

TABLE 2-1. Parts of the postcranial skeleton that are pneumatized by diverticula of different parts of the respiratory system in extant birds. Pneumaticity varies widely within populations and clades, and not all elements are pneumatized in all taxa. Based on Duncker (1971) and O'Connor (2004).

| Respiratory structure | Skeletal elements |
|---|---|
| Lung (parenchymal portion) | Adjacent thoracic vertebrae and ribs |
| Clavicular air sac | Sternum, sternal ribs, pectoral girdle and humerus |
| Cervical air sac | Cervical and anterior thoracic vertebrae and associated ribs |
| Anterior thoracic air sac | Sternal ribs |
| Posterior thoracic air sac | (none reported) |
| Abdominal air sac | Posterior thoracic, synsacral and caudal vertebrae, pelvic girdle and femur |
| Subcutaneous diverticula | Distal limb elements |

FIGURE 2-1. A phylogeny of archosaurs showing the evolution of postcranial skeletal pneumaticity and air sacs. Pantydraco is shown as having limited postcranial pneumaticity. The evidence for this is ambiguous; see text for discussion. Based on the phylogenetic framework of Brochu (2001) and Yates (2003).

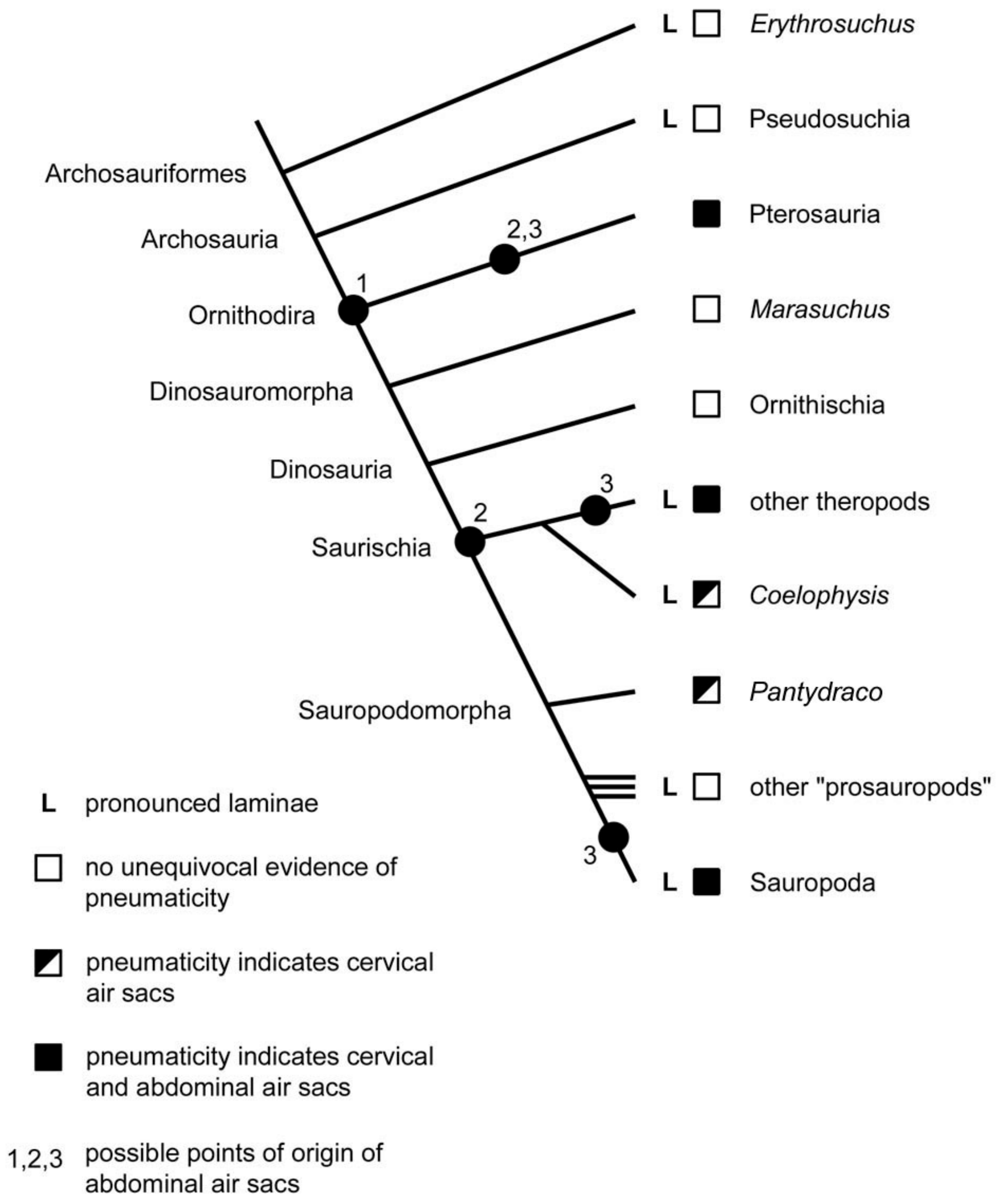
L *Erythrosuchus*
L Pseudosuchia
Pterosauria
*Marasuchus*
Ornithischia
L other theropods
L *Coelophysis*
*Pantydraco*
L other "prosauropods"
L Sauropoda
Archosauriformes
Archosauria
Ornithodira
Dinosauromorpha
Dinosauria
Saurischia
Sauropodomorpha
1
2,3
2
3
3
L pronounced laminae
no unequivocal evidence of pneumaticity
pneumaticity indicates cervical air sacs
pneumaticity indicates cervical and abdominal air sacs
1,2,3 possible points of origin of abdominal air sacs

FIGURE 2-2. A diagram showing the distribution of fossae and pneumatic chambers (black boxes) along the vertebral column in sauropods. Only the lineage leading to diplodocines is shown here. The same caudal extension of pneumatic features also occurred independently in macronarian sauropods, culminating in saltasaurines, and several times in theropods. The format of the diagram is based on Wilson and Sereno (1998, fig. 47). Phylogeny based on Wilson (2002), Yates (2003), and Upchurch et al. (2004).

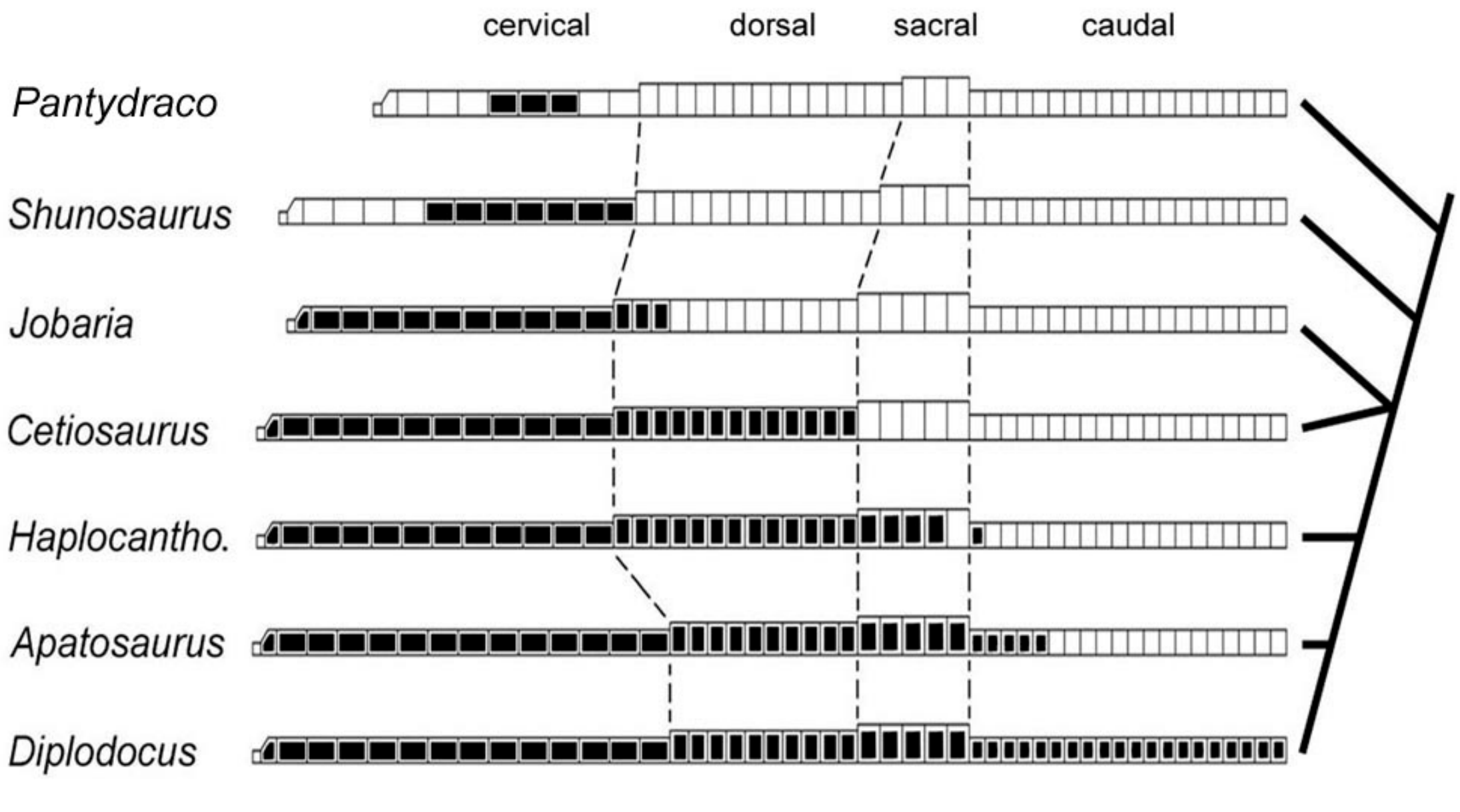

FIGURE 2-3**.** Pneumatic foramina (black) in thoracic or dorsal vertebrae of an extant bird, a non-avian theropod, and a sauropod, in right lateral view (above) and posterior view (below). A, a crane, Grus. B, an abelisaurid, Majungatholus. C, a diplodocid, Apatosaurus. A and B traced from O'Connor and Claessens (2005, fig. 3). C traced from a photograph of OMNH 1382. Scale bars represent 1 cm (A), 3 cm (B) and 20 cm (C), respectively.

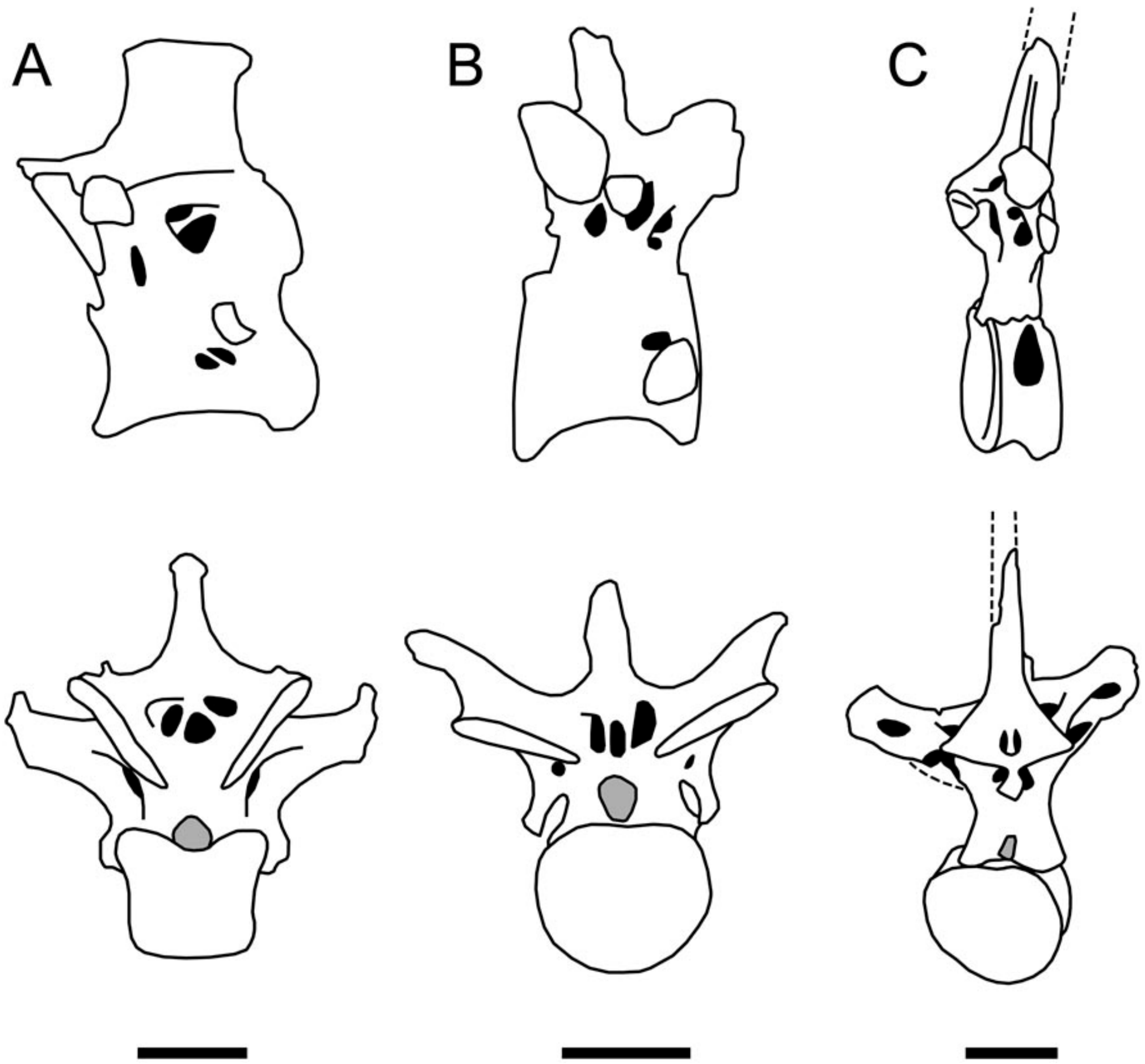
A
B
C

FIGURE 2-4**.** Laminae, fossae and foramina in cervical vertebrae of Nigersaurus and Apatosaurus. A, fifth cervical vertebra of Nigersaurus, traced from Sereno and Wilson (2005, fig. 5.8). B, tenth cervical vertebra of Apatosaurus, traced from Gilmore (1936, pl. 24). Scale bars represent 5 cm (A) and 20 cm (B), respectively.

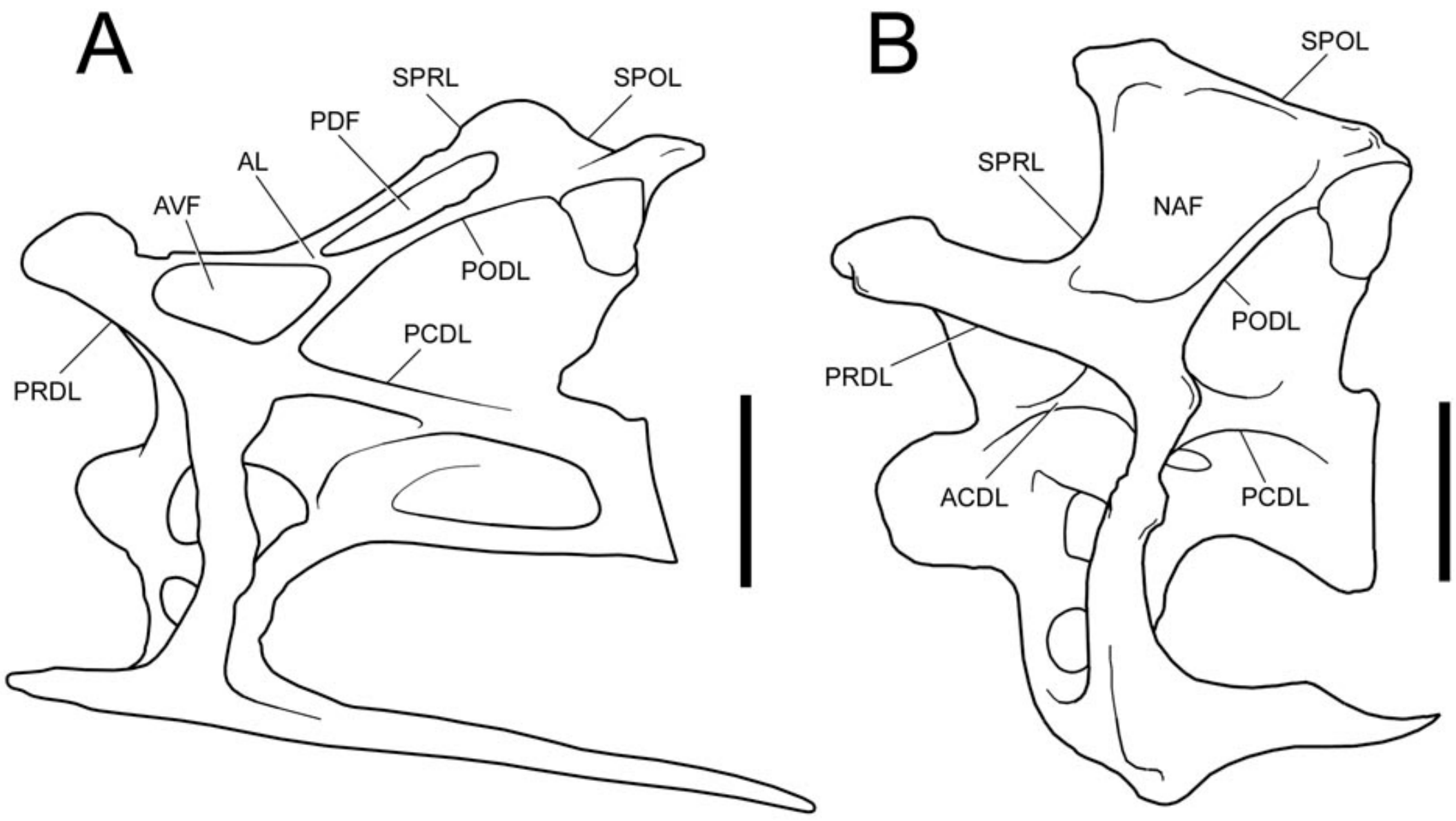

FIGURE 2-5**.** Vertebrae of Plateosaurus trossingensis (SMNS 13200) in left lateral view. A, the eighth cervical vertebra. B, dorsal vertebrae 1–4. Scale bars represent 5 cm.

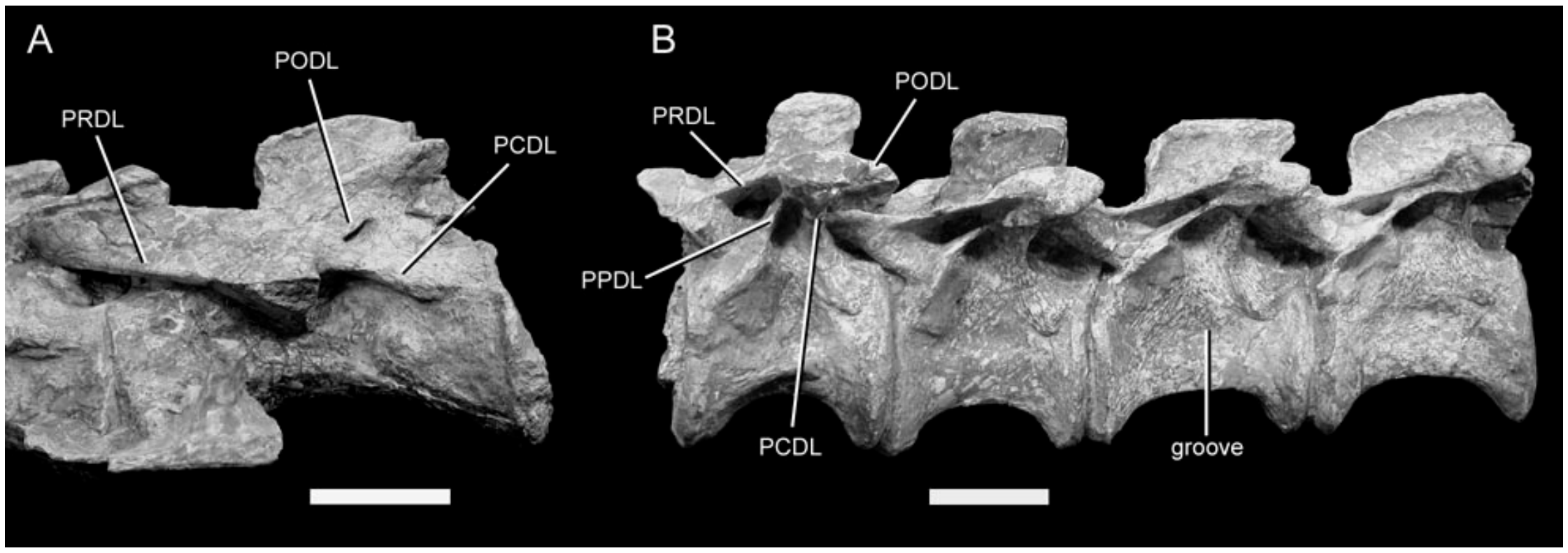

FIGURE 2-6**.** Dorsal vertebrae of Erythrosuchus africanus (BMNH R533). A, right lateral and B, ventrolateral views. Scale bar represents 5 cm.

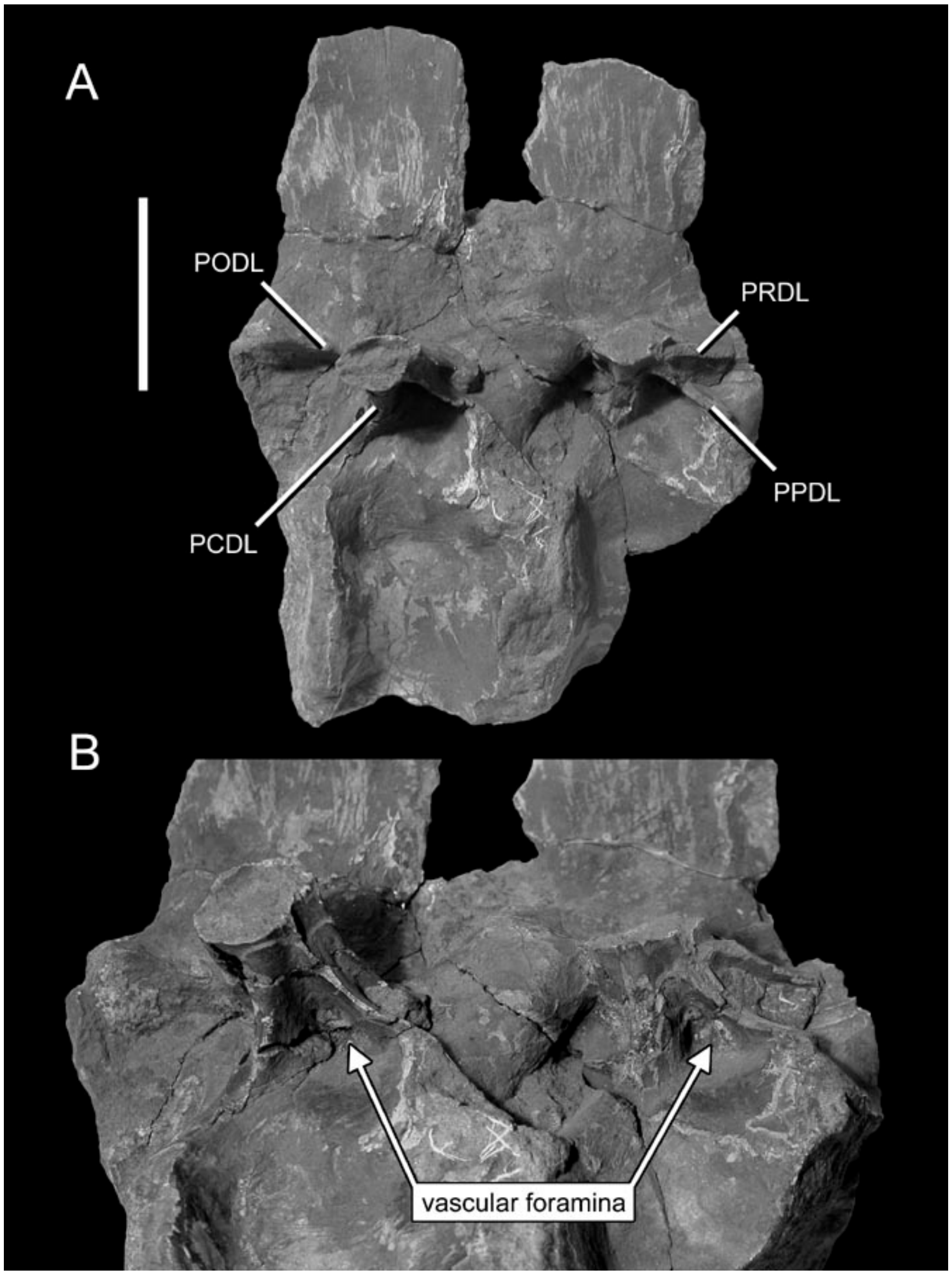

FIGURE 2-7**.** Dorsal and caudal vertebrae of Goniopholis stovalli. These vertebrae are part of an associated collection of several individuals from the type locality of the species. A, a dorsal vertebra (OMNH 2504) in left posterolateral view. B, a dorsal vertebra (OMNH 2470) in right lateral view. C, a middle caudal centrum (OMNH 2448) in right lateral view. D, a distal caudal centrum (OMNH 2454) in left lateral view. White arrows in B and D highlight the margins of fossae. Scale bar represents 1 cm.

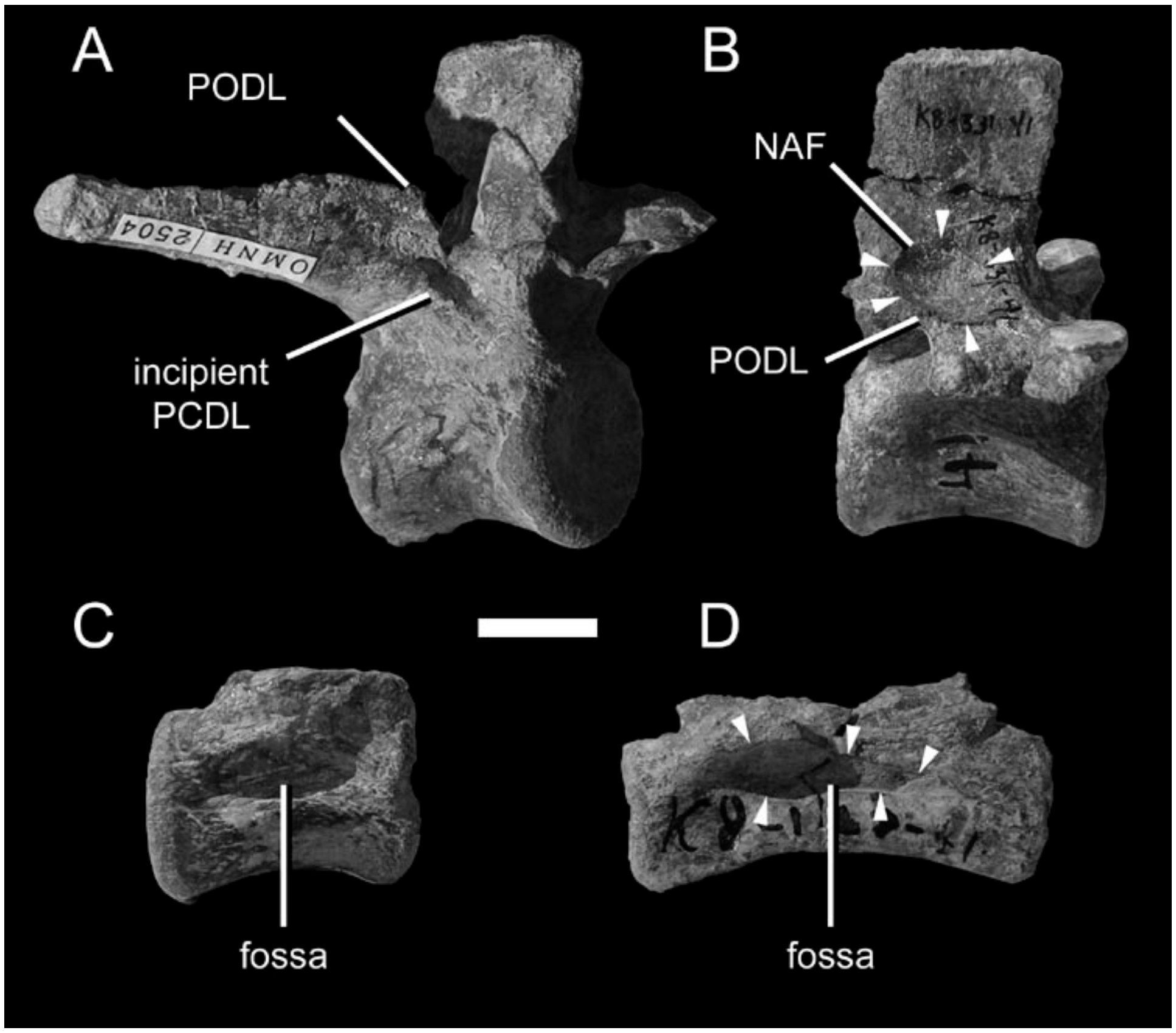

FIGURE 2-8**.** A diagram showing the evolution of fossae and pneumatic chambers in sauropodomorphs and their outgroups. Vertebrae are shown in left lateral view with lines marking the position of the cross-sections, and are not to scale. The omission of 'prosauropods' from the figure is deliberate; they have no relevant apomorphic characters and their vertebrae tend to resemble those of many non-dinosaurian archosaurs. Cross sections are based on firsthand observation (Giraffa and Arizonasaurus), published sections (Barapasaurus, Camarasaurus and Saltasaurus), or CT scans (Apatosaurus and Haplocanthosaurus). Giraffa based on FMNH 34426. Arizonasaurus based on MSM 4590 and Nesbitt (2005, fig. 17). Barapasaurus based on Jain et al. (1979, pls. 101-102). Apatosaurus based on CM 11339. Haplocanthosaurus based on CM 572. Camarasaurus based on Ostrom and McIntosh (1966, pl. 24). Saltasaurus modified from Powell (1992, fig. 16).

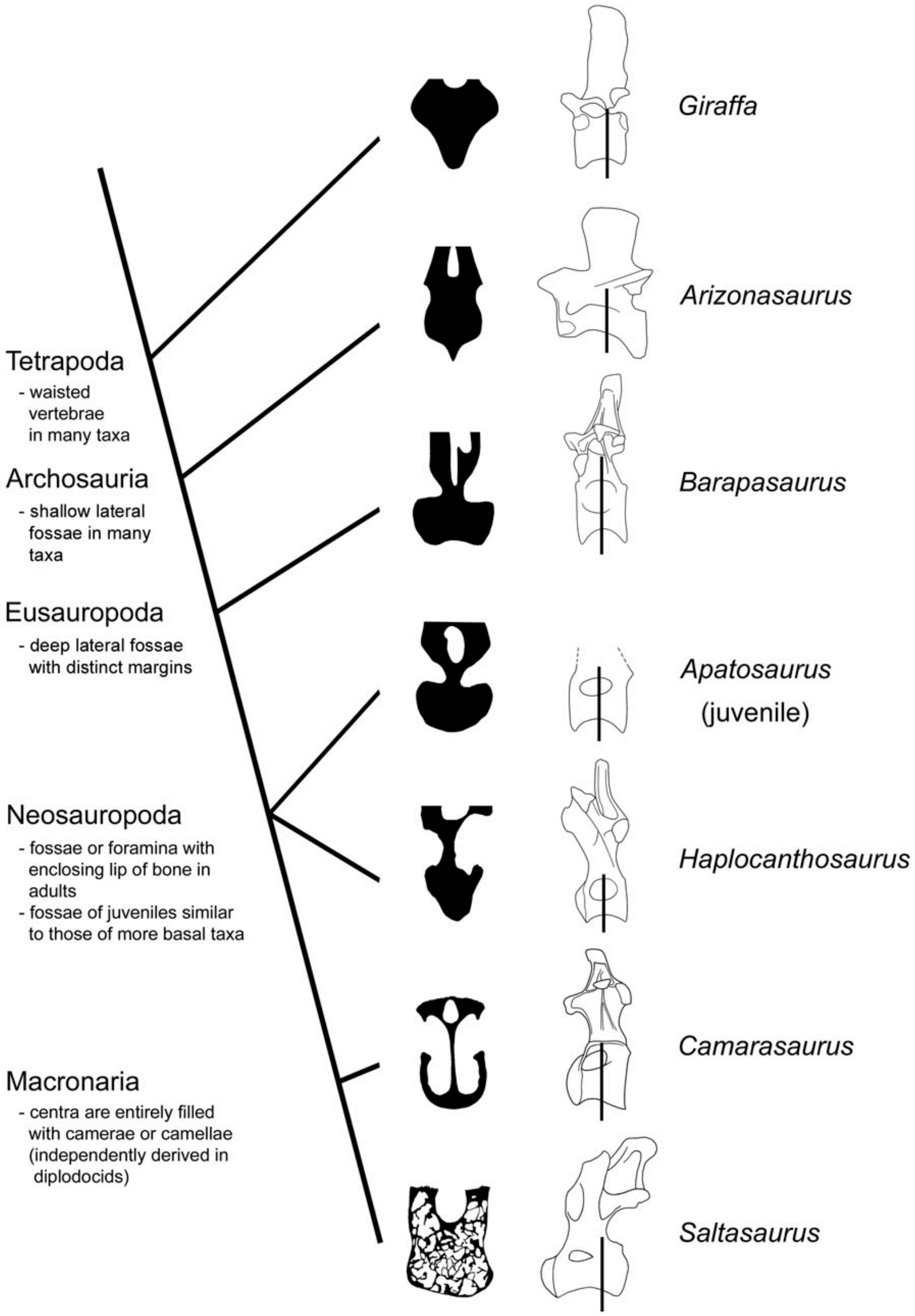
Giraffa
Arizonasaurus
Barapasaurus
Apatosaurus
(juvenile)
Haplocanthosaurus
Camarasaurus
Saltasaurus
Tetrapoda
- waisted vertebrae in many taxa
Archosauria
- shallow lateral fossae in many taxa
Eusauropoda
- deep lateral fossae with distinct margins
Neosauropoda
- fossae or foramina with enclosing lip of bone in adults
- fossae of juveniles similar to those of more basal taxa
Macronaria
- centra are entirely filled with camerae or camellae (independently derived in diplodocids)

FIGURE 2-9**.** An uncatalogued cervical vertebra of an emu (Dromaius novaehollandiae) from the OMNH comparative collection. Scale bar represents 2 cm.

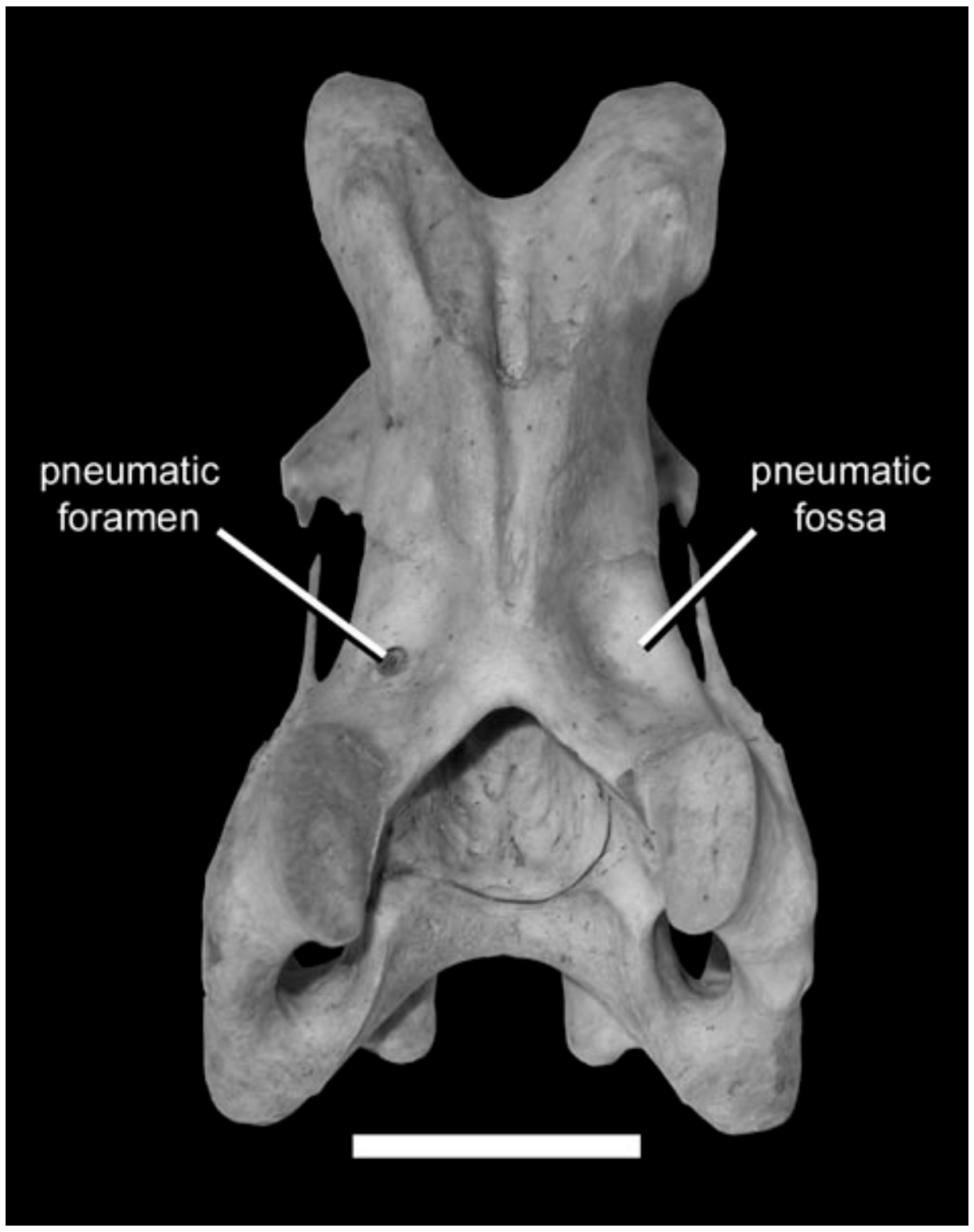

pneumatic
foramen
pneumatic
fossa

FIGURE 2-10**.** Vertebrae of Pantydraco caducus, BMNH P24. A, cervical vertebrae 6–8 in left lateral view. B, cervical vertebra 9 in right lateral view. Scale bars represent 1 cm.

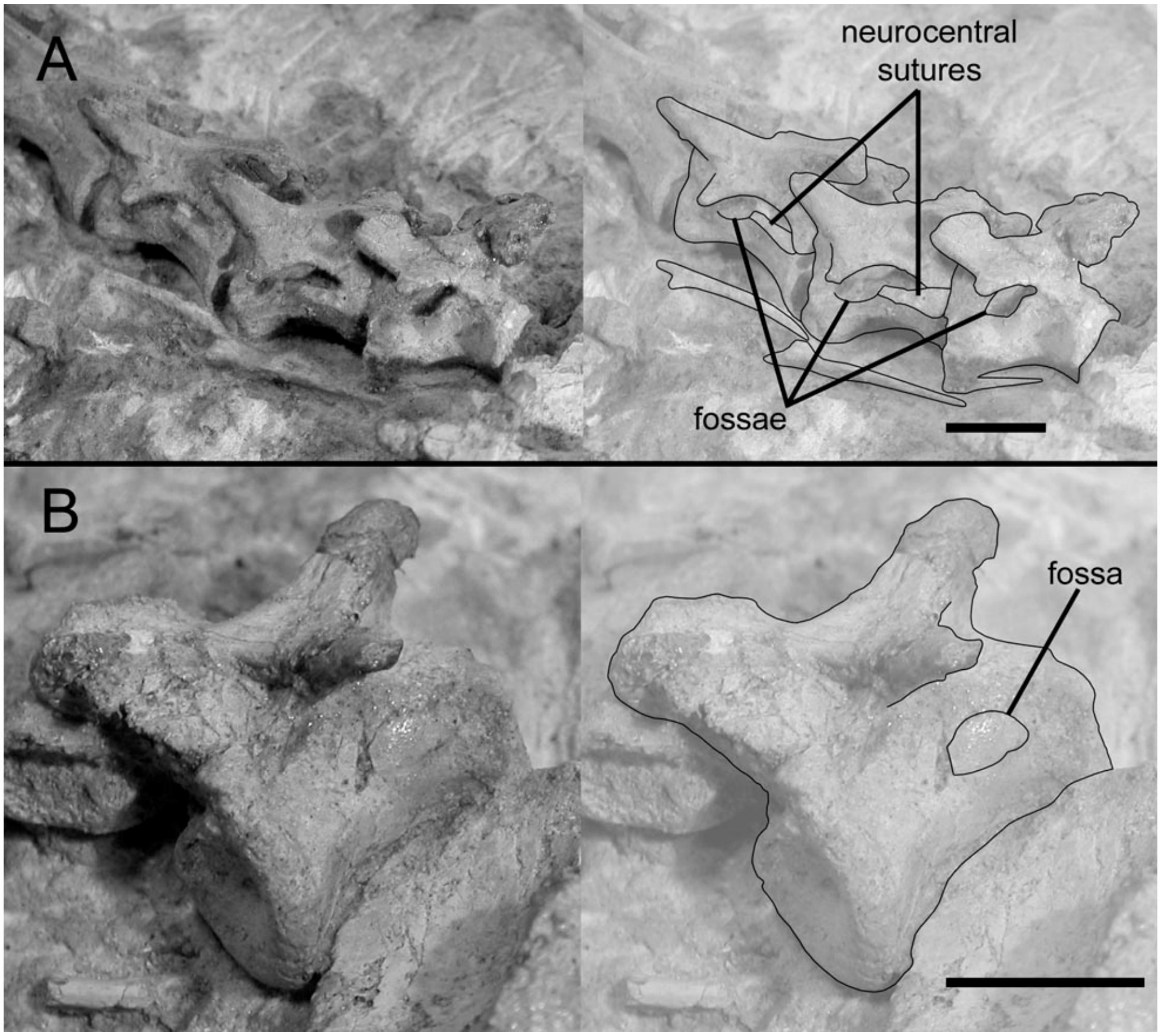
A
neurocentral sutures
fossae
B
fossa

## CHAPTER THREE

# THE EVOLUTION OF PULMONARY AIR SACS IN DINOSAURS AND THE ORIGIN OF THE AVIAN LUNG

## INTRODUCTION

Obtaining oxygen from the environment is a challenge that all animals share. Respiration may rely on passive diffusion of oxygen across a moist membrane, or the respiratory medium (air or water) may be actively driven across the gas exchange surface, either by pumping the medium or moving the gas exchanger. Movement of air or water over the gas exchanger opens the possibility of moving the blood or other oxygen-bearing fluid across or against the flow, to produce cross- or counter-current exchange and greatly enhance diffusion. For example, in the gills of sharks and actinopterygians the blood flows anteriorly through the gill lamellae, against the flow of water, and this allows high levels of gas exchange (Hughes and Morgan 1973). However, these aquatic animals are limited by the amount of diffused oxygen that water can hold—about one-thirtieth as much oxygen as is present in air. Terrestrial arthropods, mammals, non-avian reptiles, amphibians, and many fishes take advantage of the higher oxygen concentration in air. In most cases this means breathing with lungs, but buccal and cutaneous respiration are important in some taxa such as the lungless plethodontid salamanders. However, all these animals rely on static diffusion for gas exchange.

Birds are unique among extant animals in having a respiratory system that uses flow-through ventilation with air as the respiratory medium. The lungs of birds consist of minute, parallel tubes called parabronchi, which are surrounded by dense networks of even smaller passages called air capillaries (Duncker 1971). Air is blown through the parabronchi on both inspiration and expiration by air sacs, which are attached to the lungs both anteriorly and posteriorly. Blood flow in the capillaries of the lungs is at right angles to the direction of airflow in the parabronchi and against the direction of airflow in the air capillaries (Scheid 1979). This combination of cross- and countercurrent exchange allows birds to extract up to 160% more oxygen from the air than mammals can (i.e., up to 260% of mammalian levels; Brown et al. 1997).

The lungs and air sacs of birds also give rise to a network of blind-ended air-filled tubes of epithelium, or diverticula. These diverticula may be present throughout the body, among the viscera, between muscles, and under the skin (King 1966, Duncker 1971). Some diverticula enter the bones of the postcranial skeleton. The marrow of these bones is resorbed and replaced with air spaces, the dense trabeculae are reorganized into an interconnected network of larger air cells, and the walls of the bones typically become thinner as the inner layers of bone are resorbed (Bremer 1940a, b). These changes reduce the density of the pneumatic bones, which in birds are typically only half as dense as apneumatic bones (see Appendix II). Pneumatization of the postcranial skeleton is an epiphenomenon of the formation of lung and air sac diverticula. Gas exchange in the diverticula is minimal, and in some birds postcranial skeletal pneumaticity (PSP) is entirely absent (e.g., Apteryx, Gavia; Owen 1841, Gier 1952).

PSP is not necessary to the function of the respiratory system. However, it provides a skeletal fingerprint of the lungs and air sacs, and forms the basis for inferences about respiration in the fossil relatives of birds. PSP is present in most saurischian dinosaurs and pterosaurs and was recognized in these animals from very early discoveries (Owen 1856, Seeley 1870). After a century of infrequent study (Janensch 1947), PSP in fossil archosaurs has received increasing attention in the past decade and a half (Britt 1993, 1997, Britt et al. 1998, Wedel et al. 2000, Wedel 2003a, b, 2005, 2006a, 2007, Christiansen and Bonde 2000, Bonde and Christiansen 2003, O’Connor and Claessens 2005, O’Connor 2006). Most of these studies have supported the hypothesis that at least some saurischian dinosaurs and pterosaurs had some components of an air sac system like that of birds. In contrast, other authors have argued that the air sacs of dinosaurs were limited in extent and irrelevant to lung ventilation (Ruben et al. 2003), or that PSP is completely uninformative on the respiratory anatomy of extinct taxa (Farmer 2006).

My goals in this paper are, first, to outline the history of arguments for and against the hypothesis that saurischian dinosaurs had air sac systems and flow-through lung ventilation like that of birds; second, to present new data on PSP in dinosaurs, especially sauropods, that is relevant to the problem; and third, to discuss the origin of the avian lung-air sac system in a comparative framework.

CT protocols follow Wedel et al. (2000).

## RECENT HYPOTHESES ABOUT DINOSAUR RESPIRATION

The hypothesis that at least some dinosaurs had an air sac system like that of birds has a complex history. The debates over dinosaurian physiology and the dinosaurian origin of birds have made dinosaurian respiration a contentious topic. Some authors (Ruben et al. 2003, Chinsamy and Hillenius 2004, Hillenius and Ruben 2004) have argued that dinosaurs were similar to extant lizards, turtles, or crocodilians in their respiratory anatomy, constrained to ectothermic, "reptilian" activity levels, and ruled out as possible ancestors of birds. Others have suggested, based on shared derived similarities among birds and non-avian dinosaurs, that the latter had air sac driven respiration (Wedel 2003, O'Connor and Claessens 2005) and elevated metabolic rates (Padian and Horner 2002, 2004). The conformation of the dinosaurian respiratory system is an anatomical problem, and it should be answered on the basis of anatomical observations and physiological correlates.

Three main hypotheses have been proposed: (1) dinosaurs had lungs like those of extant non-avian reptiles (e.g., lizards, turtles, or crocodilians), which were ventilated by movement of the anterior thoracic ribs (Spotila et al. 1991, Hengst et al. 1996); (2) dinosaurs had lungs like those of extant crocodilians, which were ventilated by a hepatic piston mechanism (Ruben et al. 1997, 1998, 1999, Chinsamy and Hillenius 2004, Hillenius and Ruben 2004); and (3) dinosaurs or some subset thereof (i.e., saurischians, theropods, or coelurosaurs) had a lung/air sac system similar to that of extant birds, which was ventilated primarily by rib movement and may have been assisted in some taxa by an accessory aspiration pump that involved the gastralia (Bakker 1972, 1974, Daniels and Pratt 1992, Britt 1993, 1997, Wedel et al. 2000,

Wedel 2003a, b, 2005, 2006a, 2007, Carrier and Farmer 2000a, Claessens 2004, O'Connor and Claessens 2005, O'Connor 2006). These competing hypotheses have inspired new research in many areas, including the respiratory mechanics of extant vertebrates (Ruben et al. 1987, Carrier and Farmer 2000b, Claessens 2004a), the osteological correlates of different ventilation systems (Claessens 2004a, O'Connor 2006), and the distribution of PSP in fossil archosaurs (e.g., Britt 1997, Gower 2001, Bonde and Christiansen 2003, Wedel 2003b, O'Connor 2006). In this section I review the history of arguments for the air sac hypothesis, against the air sac hypothesis, and for alternative hypotheses.

### Arguments for the air sac hypothesis

**Large dinosaurs required efficient, bird-like lung ventilation**—Some authors have concluded that sauropod dinosaurs must have had a bird-like system of lungs ventilated by air sacs simply to overcome the debits imposed by large size, especially the tracheal dead space associated with a long neck (Daniels and Pratt 1992; Paladino et al. 1997). Tracheal dead space is the volume of 'spent' air left in the trachea at the end of exhalation that must be re-inhaled before any fresh air reaches the lungs. In birds, tracheal dead space is overcome by the air sac system (Duncker 1971, Schmidt-Nielsen 1972), and some birds have coiled trachea that may be longer than the entire body of the animal (Hinds and Calder 1971, Clench 1978, Prange et al. 1985, McLelland 1989a). Tracheal dead space in long-necked sauropods was almost certainly large (>100 L; Daniels and Pratt 1992, Hengst et al. 1996, Wedel 2005). Nevertheless, it did not stop sauropods from evolving very long necks. Members of at least four sauropod clades independently acquired necks longer than 9 meters, and Supersaurus had a neck at least 14 meters long (Wedel 2006b).

These observations are consilient with the presence of a ventilatory air sac system in sauropods, but it is not clear that sauropods required air sac ventilation on this basis alone. The largest sperm whales (Physeter) have skulls more than five meters long (e.g., USNM 301634 = 514 cm; N. Pyenson, pers. comm.); the blowhole is located anterior to the rostral end of the skull (Cranford 1999) and the lungs are some distance behind the head. The distance between the blowhole and lungs of the sperm whale may therefore approach seven meters (Paul and Leahy 1994, Paul 1998). This distance is comparable to the minimum length of the trachea in Diplodocus (Wedel 2005). Despite the large respiratory dead space, the sperm whale supports a high metabolic rate and an active lifestyle using a mammalian bellows lung. Physiological modeling can, at best, explore the utility of air sac breathing in dinosaurs. The question of whether or not air sacs were actually present must be decided on anatomical grounds.

**The presence of PSP in saurischian dinosaurs indicates bird-like lung ventilation—**The presence of PSP in saurischian dinosaurs has been recognized since the nineteenth century (Owen 1856, Seeley 1870, Cope 1877, Marsh 1877). Birds are the only extant amniotes with PSP, and the postcranial skeletons of birds are pneumatized exclusively by diverticula of the lungs and air sacs (O’Connor 2006). On this basis, some authors have posited that some or all saurischian dinosaurs had bird-like lungs (Bakker 1972, 1974). Note that I am separating this hypothesis, based only on the presence of PSP, from those that rely on the distribution of PSP in the skeleton; the latter are discussed in the next section.

The idea that the mere presence of PSP indicates bird-like lung ventilation has been criticized as speculative and “entirely hypothetical” (Feduccia 1973:p. 167). However, it is not as weak as it may appear. We may hypothesize various mechanisms for pneumatizing the postcranial skeleton, but the fact remains that the only mechanism that has actually been demonstrated in extant animals—pneumatization by diverticula of the lungs and air sacs—is present only in birds, and it is an epiphenomenon of the lung/air sac system.

A weakness of this hypothesis is that it assumes that PSP is a corollary of a fully avian lung/air sac system. The ventilatory mechanics of extant crocodilians and birds are more derived than in many of their extinct outgroups. For example, mobile pubes evolved in Crocodyliformes but not in other pseudosuchian clades (Claessens 2004a), and ossified sternal ribs with synovial articulations are present in birds but not in most other ornithodirans. As they diverged from a common ancestor, the linear ancestors of birds and crocodilians must have passed through functionally intermediate stages. Non-avian dinosaurs did not necessarily have the same pulmonary anatomy as crocodilians or extant birds. As hypotheses of pulmonary anatomy in dinosaurs, “croc lungs” versus “bird lungs” is a false dichotomy. It is more informative to identify the derived features that non-avian dinosaurs share with their extant relatives, and to determine the hierarchical distribution of these characters in archosaurian phylogeny.

**The distribution of PSP in saurischian dinosaurs indicates that specific components of a bird-like lung/air sac system were present—**The air sac hypothesis, as it has been developed over the past several years, does not depend on the mere presence of PSP. Rather, pneumatization of different parts of the avian skeleton is diagnostic for specific pulmonary structures. The unvarying relationships among air sacs and regions of the skeleton form the basis for inferences about the pulmonary anatomy of extinct taxa.

In extant birds, the cervical and anterior thoracic vertebrae are pneumatized by diverticula of cervical air sacs; the furcula, sternum, humeri, and pectoral girdle are pneumatized by diverticula of the clavicular air sacs; thoracic vertebrae and ribs adjacent to the lungs are pneumatized by diverticula of the lungs; sternal ribs are pneumatized by diverticula of the anterior thoracic air sacs; and posterior thoracic, synsacral, and caudal vertebrae, femora, and pelvic girdle elements are pneumatized by diverticula of abdominal air sacs (Duncker 1971, O'Connor 2006). Distal limb elements are pneumatized by subcutaneous diverticula (O'Connor 2006). Not all elements are pneumatic in all taxa, and some birds lack PSP altogether (e.g., loons; Gier 1952).

The relationships among the different parts of the pulmonary system and their respective skeletal 'domains' are invariant in all birds that have been studied to date (O'Connor and Claessens 2005, O'Connor 2006). Diverticula of cervical air sacs never pneumatize elements posterior to the middle of the thoracic series, diverticula of abdominal air sacs never penetrate anterior to the middle of the thoracic series, and cervical and synsacral vertebrae are never pneumatized by diverticula of the lungs.

These distinctions are important because the relationships among pulmonary regions and their skeletal domains have been historically obscure (Wedel 2003a, O'Connor 2006).

The invariant relationships among components of the respiratory system and the regions of the skeleton that they pneumatize form the basis for inferences about the pulmonary anatomy of extinct taxa. Even opponents of the air sac hypothesis (e.g., Ruben et al. 2003) have conceded that cervical air sacs must have present in most or all saurischian dinosaurs, based on pneumatization of the neck. Pneumatic sacral and caudal vertebrae are also present in several clades of both sauropods and theropods (see below). The hypothesis that pneumatic sacral and caudal vertebrae indicate the presence of abdominal air sacs in non-avian dinosaurs was first advanced by Wedel et al. (2000). This suggestion is supported by recent studies of birds (O'Connor and Claessens 2005, O'Connor 2006). The presence of PSP in both the anterior (cervical, anterior thoracic) and posterior (posterior thoracic, sacral, and caudal) regions of the vertebral column in sauropods and theropods is strong evidence that air sacs were present both anterior and posterior to the lung, and that these non-avian dinosaurs had at least the anatomical structures necessary for flow-through lung ventilation (Figure 3-1). The hypothesis that both anterior and posterior air sacs were present in many saurischians is further supported when development and phylogeny are taken into account, as described below.

**The evolution of PSP in saurischian dinosaurs parallels the development of PSP in extant birds—**The cervical and anterior thoracic vertebrae are the first parts of the axial skeleton to be pneumatized during the ontogeny of birds, and they are pneumatized by diverticula of the cervical air sacs (Figure 3-2; Cover 1953; Hogg 1984b). Later in ontogeny diverticula of the abdominal air sacs pneumatize the posterior thoracic vertebrae and synsacrum. The sequence of pneumatization of the vertebral column in avian ontogeny closely parallels the evolutionary sequence of vertebral pneumatization in both sauropods and theropods (Figure 3-3). PSP is present only in the cervical series in basal members of both saurischian lineages. Dorsal, sacral, and caudal vertebrae become pneumatized in successively more derived taxa (Britt 1993, Wedel 2007). Thus the spread of pneumaticity posteriorly along the vertebral column in the ontogeny of birds appears faithfully to recapitulate the evolution of pneumaticity in theropods and sauropods (Wedel 2003a, 2005).

The posterior progression of vertebral pneumatization in birds is not caused by diverticula developing from a single, anteriorly located source. Rather, diverticula of different sources (cervical air sacs, lungs, abdominal air sacs) pneumatize their respective skeletal domains at different times. It is not clear why vertebral pneumatization in birds proceeds front-to-back, as opposed to back-to-front or in both directions starting from the middle. Nevertheless, the identical sequence of pneumatization in avian ontogeny and saurischian phylogeny is further support for the hypothesis that extinct saurischian dinosaurs had lungs and air sacs like those of extant birds.

**Arguments against the air sac hypothesis**

**Non-avian dinosaurs lacked the musculoskeletal mechanisms necessary to ventilate an air sac system**—The dorsal (rib-bearing) vertebrae, ribs, and sternum of birds form a bony box that encloses the viscera and air sacs. Movement of the axial skeleton is highly constrained by ossification of elements that are often cartilaginous or absent in other tetrapod clades (e.g., sternal ribs, uncinate processes). Some of these derived features are absent in non-avian dinosaurs, and this has led some workers to argue that dinosaurs could not have ventilated an air sac system. To analyze this claim I focus on three questions. First, what are the ventilatory motions of the avian musculoskeletal system? Second, are the derived features in question necessary components of the ventilatory mechanism, and are they present in all extant birds? Third, were the axial skeletons of non-avian dinosaurs capable of producing equivalent motions?

The thoracic cage of birds includes the thoracic vertebrae, vertebral ribs (often with ossified uncinate processes), sternal ribs, and sternum. The vertebral ribs are bicipital and articulate with dorsoventrally separated diapophyses and parapophyses on the thoracic vertebrae. These attachments constrain the ribs to move fore and aft. The ends of the vertebral ribs are connected to the sternum by sternal ribs, at least some of which are ossified. The sternum is also connected to the shoulder joints by the large coracoids. During inspiration, the vertebral ribs swing forward and laterally expand the thorax. The angle between the vertebral ribs and sternal ribs widens, and the distal ends of the sternal ribs are displaced ventrally. The sternum is hinged at the shoulder joints by the coracoids, and the posterior end of the sternum is displaced ventrally. Together, these skeletal movements expand the thorax laterally and posteroventrally. As the thorax expands, air is drawn through the trachea into the posterior air sacs (posterior thoracic and abdominal), and air that was already in the respiratory system is drawn through the lung into the anterior air sacs (cervical, clavicular, anterior thoracic; Duncker 1971).

Opponents of the air sac hypothesis have argued that certain components of the avian musculoskeletal system are necessary for ventilation and that dinosaurs that lacked these components could not have ventilated an air sac system. These arguments have focused on three components: the sternum, sternal ribs, and uncinate processes.

Many authors have argued that a large sternum is necessary to ventilate the air sac system, especially the abdominal air sacs (Hengst et al. 1996, Ruben et al. 1997, 1998, Chinsamy and Hillenius 2004). It is true that the sternum moves during ventilatory movements of the ribs, and that costal movements may be transmitted to air sacs by a large sternum. However, in some birds the sternum is very small and does not come into contact with any of the posterior air sacs. Examples of birds with particularly small sterna include extant Apteryx and Dromaius (Owen 1841, Mivart 1877, Paul 2001) and extinct Aepyornis and Dromornis (Murray and Vickers-Rich 2004:fig. 154). The morphology of these flightless birds shows that the sternum does not have to be particularly large or come into contact with the posterior air sacs to ventilate them. Sternal elements in non-avian dinosaurs consist of paired plates. Ossified sternal plates are present in dromaeosaurs (Norell and Makovicky 1997, 1999), oviraptorids (Barsbold 1981, 1983; Clark et al. 1999), sauropodomorphs, and the major clades of Ornithischia (Romer 1956). Other than dromaeosaurs and oviraptorids, sternal plates have not been found in non-avian theropods, and these theropods presumably had cartilaginous sternal elements. Furthermore, the sterna of non-avian saurischian dinosaurs were proportionally as large as or larger than those of flightless birds (Table 3-1).

Some authors have also argued that the connection of the vertebral ribs to the sternum by ossified bicipital sternal ribs is also a necessary condition for the ventilation of the avian lung (Ruben et al. 1997, 1998; Chinsamy and Hillenius 2004). However, the condition in large flightless birds suggests that this is not the case. Most ratites have free vertebral ribs both anterior and posterior to the vertebral ribs that articulate with the sternum (Mivart 1877, pers. obs.). Furthermore, it has never been explained why the sternal ribs would have to be ossified to function in ventilation. Most mammals, including humans, ventilate large lungs with cartilaginous sternal ribs. The limb and girdle articulations of many dinosaurs, especially large sauropods, are relatively simple and the bones do not fit tightly together (Coombs 1975). Cartilage must have made up a large part of the locomotor apparatus in these animals; it would not be surprising if the same were true of the axial skeleton. Ossified sternal ribs have been found in Apatosaurus (Marsh 1896, Claessens 2004b), Velociraptor (Norell and Makovicky 1999), and an unnamed oviraptorid (Clark et al. 1999), but are otherwise unknown in non-avian dinosaurs.

Hengst et al. (1996) hypothesized that the uncinate processes of birds are integral to their derived respiration because they guide and stabilize the movements of the ribs. No evidence was adduced to support this interpretation. In contrast, Codd et al. (2005, p. 856) found that whereas “any putative stiffening function of the uncinate processes cannot be completely ruled out”, the uncinate processes function primarily as attachment points for muscles involved in respiration (e.g., external intercostal, appendicocostalis, and external oblique mm.). Ossified uncinate processes are present in dromaeosaurs and oviraptorids (reviewed by Hwang et al. 2002) but unknown in other non-avian dinosaurs.

Uncinate processes are also absent in some extant birds (e.g., Dromaius, Casuarius, anhimids and megapodids: Mivart 1877, Bellairs and Jenkin 1960, Baumel and Witmer 1993), and thus they are at best an accessory, and not a necessary, component of avian lung ventilation.

Instead of focusing on particular apomorphies that are either not present in all birds (large sternum, uncinate processes) or not clearly necessary for air sac ventilation (ossified rather than cartilaginous sternal ribs), it may be more productive to identify the skeletal movements take place during avian respiration and the effects of these movements on the shape and volume of the thoracic cage, and then to ask whether the skeletons of non-avian dinosaurs were able to produce similar movements. As discussed above, during inspiration the vertebral ribs of birds swing forward and laterally expand the thoracic cage. At the same time, the sternal ribs depress the sternum, and this ventral movement increases toward the posterior end of the sternum. The cumulative effect of these movements is that the space bounded by the thoracic skeleton becomes wider and deeper, and this expansion is greatest posteriorly. The space bounded by the thoracic cage can be thought of as a truncated cone, the narrow end of which is bounded by the anterior vertebral ribs, coracoids, and sternum, and the base of which is located at the posterior end of the bony thorax. During inspiration the cone increases in diameter, but this dilation does not take place evenly along the length of the cone; the increase is much greater at the base than at the narrow end.

Could the skeletons of non-avian dinosaurs have produced similar movements? Like most amniotes, all dinosaurs have bicipital vertebral ribs that articulate with widely separated diapophyses and parapophyses. Furthermore, the plane of articulation changes along the vertebral column (Figure 3-4). In anterior dorsal vertebrae the parapophysis is located on the centrum or the ventral part of the neural arch, and the axis of rotation of the rib is nearly vertical. This would constrain the rib to move anteroposteriorly, with little dorsoventral excursion. In successively posterior vertebrae the parapophysis migrates up the neural arch, and the axis of rotation changes from being nearly vertical to nearly horizontal. The posterior dorsal ribs were constrained to move dorsoventrally, with little anteroposterior excursion. Similar serial changes in the angle of the rib articulation occur in many non-avian dinosaurs (e.g., Tenontosaurus, Tyrannosaurus; Ostrom 1970, Brochu 2003), although the degree to which the angle approaches the horizontal in the posterior dorsal vertebrae varies among clades. In some clades the parapophyses remain ventromedial to the diapophyses throughout the dorsal series, as in birds (e.g., Haplocanthosaurus, ceratopsians; Hatcher 1903, Hatcher et al. 1907).

If we again imagine the thoracic cavity as a truncated cone, the posterior ribs would have moved at right angles to the cone, whereas more anterior ribs would have moved at acute angles to the surface of the cone. Therefore the respiratory movements in non-avian dinosaurs would have had a similar effect on the volume of the thorax as those of extant birds. The main difference is that in birds the dilation of the thorax is greatest posteroventrally, whereas in non-avian dinosaurs it may have been greater posterodorsally, or the posterior expansion may have been more even in the dorsal and ventral directions. Without preserved sternal ribs for almost all taxa, neither possibility can be ruled out.

The role of skeletal movements in avian respiration is only beginning to be understood, and some movements proposed on the basis of anatomy do not actually contribute to ventilation (Claessens 2004a). The ideas presented here about how non-avian dinosaurs may have ventilated an air sac system are hypothetical, and they could be falsified by more detailed functional studies. Nevertheless, there is no basis for inferring that non-avian dinosaurs could not have ventilated an air sac system, based simply on the absence of some avian features.

**Abdominal air sacs could not have been present because their development would have herniated the diaphragm—**Ruben et al. (1997, 1998, 1999) argued that theropod dinosaurs ventilated their lungs with a hepatic piston diaphragm. They further claimed that birds could not be theropod descendants because abdominal air sacs could not have evolved without herniating the diaphragm and such hernias would have been selected against.

The hypothesis that a hepatic piston diaphragm was present in theropods is problematic for many reasons; see the section below on 'Alternative Hypotheses'. The anatomical and phylogenetic distribution of PSP supports the hypothesis that abdominal air sacs were present in most theropods (O'Connor and Claessens 2006), and they may have been present in the ancestral saurischian (Wedel 2007). The taxa that Ruben et al. (1997, 1998, 1999) use as evidence for the muscular diaphragm (Scipionyx and Sinosauropteryx) are deeply nested in clades in which abdominal air sacs are inferred on anatomical grounds to have been primitively present. So even if the inference of a muscular diaphragm in Scipionyx and Sinosauropteryx were correct (and it is contradicted by several lines of evidence), the question would seem to be how and why plesiomorphic air sacs were lost in these particular theropods.

**Vertebral pneumaticity develops from cervical air sacs alone and does not indicate the presence of abdominal air sacs—**Some authors have argued that the vertebral column of birds is only pneumatized by diverticula of the cervical air sacs.

> "Pneumatization of the vertebrae and ribs is invariably accomplished by diverticuli [sic] of the cervical air sacs (McLelland 1989a [1989b herein]), which are located outside the trunk and contribute little, if anything, to the respiratory air flow (Scheid and Piiper 1989). Presence of pneumatized vertebrae in non-avian dinosaurs therefore only speaks of the possible presence of such nonrespiratory diverticuli [sic], and cannot be regarded as indicative of an extensive, avian-style abdominal air-sac system" (Ruben et al. 2003, p. 153).

These points are reiterated by Chinsamy and Hillenius (2004) and Hillenius and Ruben (2004), but many of them are misleading or incorrect. McLelland (1989b) did not claim that the vertebral column was "invariably" pneumatized by diverticula of cervical air sacs; in fact, he stated that "the abdominal air sac aerates the synsacrum, pelvis and femur" (p. 272); by listing the synsacrum and pelvis separately he clearly meant that the synsacral vertebrae are pneumatized by the abdominal air sac, and this is confirmed by the sources he cited: Hogg (1984a, b). Pneumatization of the posterior compartment by diverticula of the abdominal air sacs has been known for a century (Muller 1907, Cover 1953, King 1966, 1975, Duncker 1971, Hogg 1984a, b, Bezuidenhout et al. 1999). Cover (1953) and King (1975) did not claim that the abdominal air sacs never pneumatized the vertebral column; in fact they stated the opposite. The cervical air sacs are located adjacent to the anterior thoracic vertebrae and posterior to the coracoids and glenoid fossae (O'Connor 2004, fig. 1), so it is not clear in what sense they are "outside the trunk". In any case, it is the functional relationships of the air sacs to the lungs that are important, not their positions relative to an arbitrarily defined "trunk". The fact that the vertebral diverticula are "nonrespiratory" (presumably this means 'not contributing to ventilation or gas exchange', because they are developmentally and topologically part of the respiratory system) is beside the point; the diverticula are of interest not because they function in ventilation or gas exchange, but because they indicate the presence of specific air sacs that do function in ventilation.

**PSP is completely uninformative—**Still other authors have argued that PSP does not inform us about the structure of the respiratory system at all.

> "Without integrating functional data into the study, the most that can be inferred from post-cranial pneumaticity in extinct animals is that, as pointed out by Owen (1856), the pneumatized bones received parts of the lung in the living animal… Because pneumaticity has no known functional role in ventilation or thermoregulation or metabolic rates, its usefulness as a hard-part correlate for lung structure and metabolism is, unfortunately, questionable" (Farmer 2006, pp. 91-92).

Farmer has failed to distinguish between inferences based on the presence of postcranial pneumaticity and inferences based on the distribution of postcranial pneumaticity. If all we know about a bone is that it is pneumatic, then she is correct in stating that the most we can conclude is that it was connected to the respiratory system. (The thermoregulatory function of pneumaticity discussed by Seeley [1870] has been demonstrated for cranial pneumaticity [Warncke and Stork 1977] but not for PSP [Witmer 1997]). But the inference of cervical and abdominal air sacs in non-avian dinosaurs does not depend simply on the existence of postcranial pneumaticity. Rather, these inferences are based on patterns of postcranial pneumaticity that are diagnostic for specific air sacs. Similarly, the paleobiological implications of PSP are not based on its mere presence, but rather on the probable capabilities of the air sac system, of which PSP is simply the osteological footprint.

**Alternative hypotheses**

**Assumptions based on the presumed pace of evolution—**It has frequently been assumed that dinosaurs had lungs like those of extant non-avian reptiles. Dinosaurs have often been perceived as essentially "reptilian" because that is how they are classified in the Linnean system. All members of Reptilia as conceived by Linnaeus (i.e., excluding birds) have septate lungs, and therefore that is the default assumption for dinosaurs (Chinsamy and Hillenius 2004). However, this assumption does not take into account the actual distribution of primitive and derived characters within sauropsids (Padian and Horner 2002). Some authors have assumed that non-avian dinosaurs had lungs like those of turtles (Spotila et al. 1991) or lizards (Gale 1997, 1998), but to date these assumptions have not been justified by evidence.

Some authors have argued that Mesozoic dinosaurs must have been relatively primitive in their respiratory anatomy (i.e, similar to basal reptiles) because they originated a long time ago and had little time to evolve derived respiratory systems (Hengst et al. 1996, Paladino et al. 1997). This argument makes no reference to the derived character states that are actually present in various dinosaurian groups, nor does it correlate these features with specific modes of ventilation. It is simply an unjustified assumption about the rate of evolution (Padian and Horner 2002). Paladino et al. (1997) contrasted the physiological attributes of reptiles on one hand and mammals and birds on the other, explained that because they lived a long time ago non-avian dinosaurs must have been more like reptiles, and then paradoxically concluded that Apatosaurus must have had a fully avian lung air sac system. If non-avian dinosaurs evolved a fully avian pulmonary system by the Late Jurassic, it is not clear why the rest of their physiology could not have been equally derived.

**Costal ventilation of a crocodilian-like lung**—The idea that dinosaurs had costally ventilated lungs like those of non-avian reptiles has only been developed as a substantial hypothesis by Hengst and Rigby (1994) and Hengst et al. (1996). In both papers, Hengst and his coauthors use Apatosaurus as a model for dinosaurian respiration. Their analysis is specifically based on the mounted skeleton of Apatosaurus in the Field Museum (FMNH P25112; this is a composite of at least two individuals; see Riggs 1903 and McIntosh 1990) and a flawed skeletal reconstruction by Marsh (1891) that includes one third too many dorsal vertebrae (13 instead of 10—see Gilmore 1936; Marsh had used the correct number in earlier restoration from 1883).

The hypothesis of Hengst and his coauthors includes three major points that are relevant to lung structure and ventilation:

1. The lung structure of crocodilians is primitive to both clades of archosaurs, the crocodilian lineage and the bird lineage.
2. "Because crocodilians coexisted with dinosaurs and birds are phylogenetic descendants of dinosaurs, it is likely that a form similar to the crocodilian lung was present in the earliest dinosaurs and that this lung structure was perpetuated" (Hengst et al. 1996:p. 335).
3. Only the first five ribs were used to ventilate the lungs, based on the size and robustness of the ribs and the transverse processes of the associated vertebrae. The lungs were confined to the space between the first and fifth ribs and consequently they were small.

I will discuss each of these points in turn.

The first point, that the general structure of crocodilian lungs is primitive for archosaurs, may be partly accurate, but not because crocodilians and dinosaurs are both ancient lineages. Rather, Perry (1998, 2001) has pointed out that despite their different ventilatory mechanics birds and crocodilians share a suite of pulmonary characters that are not present in other reptiles. These features are most parsimoniously interpreted as synapomorphies of Archosauria. In general, crocodilians are not good models for ancestral archosaurs. Extant crocodilians are most likely descended from small (skull length ~5cm), long-legged, probably insectivorous sphenosuchians (Clark et al. 2004). Together with anatomical and developmental evidence, the fossil history of crocodilians suggests that their ancestors were active terrestrial cursors. Non-crocodilian pseudosuchians may have shared some pulmonary synapomorphies with crocodilians and birds, but we cannot assume that they were similar to extant crocodilians in either pulmonary anatomy or aerobic capacity.

The second point is simply a bald assumption that non-avian dinosaurs had not progressed from the primitive state or evolved a bird-like lung yet. This presumes knowledge of the pace of evolution, denies the existence of intermediate forms, and ignores the hierarchical distribution of derived characters in dinosaurs and other archosaurs.

The third hypothesis, that one can predict the posterior extent of the lungs by locating the most robust vertebral rib, has never received any baseline testing on extant animals. It is not supported by dissections of lizards, crocodilians, birds, or mammals, in which the serial position of the most robust rib varies among taxa and is often not correlated with the posterior ends of the lungs (pers. obs.). In extant birds the most robust ribs are usually in the middle of the series, with more gracile ribs both anterior and posterior. The robust ribs do not even mark the end of the lung proper, and the air sacs often project well beyond the posterior margin of the ribcage (see Duncker 1971).

In Apatosaurus, the first five ribs may have been more robust than the rest not because they were the only ones used in respiration but because they supported the scapula and forelimb. This anatomical relationship has been recognized in skeletal reconstructions for more than a century (Marsh 1891, Stevens and Parrish 2005).

In summary, the assumptions of Hengst and coauthors regarding pulmonary anatomy and skeletal function are contradicted by comparative anatomy and phylogeny. There is no reason to assume that the lungs of non-avian dinosaurs were more like those of crocodilians than those of birds, or that they were constrained to a small anterior part of the bony thorax. In fact, osteological evidence suggests the opposite, that sauropods and theropods had bird-like lungs with large air sacs that were present throughout and beyond the bony thorax.

**Hepatic piston ventilation of a crocodilian-like lung—**The holotype specimens of the small theropods Sinosauropteryx and Scipionyx have dark stains in the area of the ribcage (Ji and Ji 1996, Dal Sasso and Signore 1998). On the basis of these stains, Ruben et al. (1997, 1998, 1999) made the following inferences, which have subsequently been defended by Ruben et al. (2003), Chinsamy and Hillenius (2004), and Hillenius and Ruben (2004): (1) the stains represent the dinosaurs' livers, (2) the preservation of the stains faithfully captures the anatomical positions of the livers, (3) the presence of such large livers shows that these theropods ventilated their lungs with hepatic piston diaphragms like those of extant crocodilians, (4) the presence of such diaphragms would have precluded the evolution of abdominal air sacs, and (5) theropod dinosaurs cannot be ancestral to birds.

This chain of inferences is exhaustively rebutted elsewhere (Christiansen and Bonde 2000, Hutchinson 2001, Padian 2001, Paul 2001, Padian and Horner 2002, Witmer 2002). The strongest piece of evidence against the interpretation of Ruben and his colleagues is that the liver of the carrion crow (Corvus corone) is anteriorly arched and spans the entire distance from the sternum to the vertebral column (Duncker 1971). In lateral view, the bird's liver has exactly the same profile as the stains associated with the Sinosauropteryx and Scipionyx holotypes (Paul 2001). The dorsoventrally expansive liver does not interfere with the placement or function of the air sacs, which are lateral to it. So even if we allow that Ruben and colleagues have interpreted the fossils correctly (and there are strong reasons to suspect otherwise), their central inference—that the size and form of the livers is inconsistent with an avian air sac system—is simply incorrect.

## POSTCRANIAL PNEUMATICITY AND THE AIR SACS OF NON-AVIAN DINOSAURS

### Contributions of this study in the context of previous work

The most important work to date on PSP in non-avian dinosaurs and birds is that of O'Connor (2004, 2006) and O'Connor and Claessens (2005). The major contributions of these studies have been (1) to establish the relationships among pulmonary components and regions of the skeleton in birds, (2) to show that patterns of PSP present in non-avian theropods allow us to infer the presence and relative position of the lungs, cervical air sacs, and abdominal air sacs in these animals, and (3) to show that most non-avian theropods had all of the pulmonary components necessary for flow-through lung ventilation.

In this paper I build on the results of O'Connor (2004, 2006) and O'Connor and Claessens (2006) by taking a broader phylogenetic perspective. In particular, this study deals with the origins of PSP in basal saurischians, and with the evidence for air sacs in sauropodomorph dinosaurs. Sauropodomorpha is the sister group to Theropoda within Saurischia, and sauropodomorphs are the only other clade of dinosaurs with extensive PSP (although PSP is also present and extensive in pterosaurs). Documenting the origin and evolution of PSP in sauropodomorphs provides new information that helps to phylogenetically constrain inferences about the evolution of PSP and the presence of pulmonary air sacs in Saurischia.

**The origin of postcranial pneumaticity in dinosaurs: evidence for cervical air sacs**

Skeletal pneumaticity, or the phenomenon of having air-filled bones, is not broadly distributed in vertebrates. Among extant taxa, pneumatization of the skull occurs only in archosaurs (crocodilians and birds) and mammals. Pneumatization of the postcranial skeleton is even more limited, and occurs only in birds. Skeletal pneumatization in birds is often assumed to be an adaptation for reducing mass and improving flight performance (although on average the skeletons of birds are not lighter than those of similar-sized mammals: Prange et al. 1979). It is less commonly recognized that birds inherited postcranial skeletal pneumaticity (PSP) from their non-flying dinosaurian ancestors. PSP in dinosaurs has received little attention until very recently. This section addresses the origin of PSP in dinosaurs, with the goal of determining what kind of soft tissue system was responsible for bringing the air into the skeleton.

When we find evidence of pneumaticity in fossil forms, it is fair to ask how the air got into the bones. Left to itself, an enclosed volume of air inside the body will usually dissipate by diffusion into nearby blood vessels. If it is to persist, the air reservoir must be maintained, either by perfusion from a vascular rete mirabile, as in the swim bladders of fish (Scholander 1954), or by constant connection to the outside (Ojala 1957). In the case of skeletal pneumatization, a patent sinus ostium or pneumatic foramen is required for normal development (Ojala 1957). Not only must there be a hole in the bone, but there must be an extension of the respiratory or tympanic passages, called a diverticulum, to bring the air from the outside to the pneumatic cavity within the bone.

Pneumatic diverticula may develop from any of four regions of the body: (1) the cranial air spaces, (2) the larynx and trachea, (3) the lungs, and, in birds, (4) the pulmonary air sacs. Conceivably, diverticula from any of these sources could pneumatize portions of the postcranial skeleton. Each of the four regions is therefore a hypothetical source of the diverticula that pneumatized the skeletons of basal dinosaurs. To test these competing hypotheses, I briefly survey the kinds of diverticula that develop from each source in extant sauropsids, predict the pattern of pneumatization that diverticula from each region might be expected to produce, and compare the predicted patterns with the observed patterns of postcranial skeletal pneumatization in basal dinosaurs.

**Pneumatic diverticula in extant sauropsids—**Diverticula of nasal and tympanic air spaces are present in many extant sauropsids. In testudines and squamates, small cranial diverticula may develop from the nasal cavities or nasopharyngeal ducts (Witmer 1999). In no cases, however, do these diverticula extend into the neck or pneumatize any of the bones of the skull. In contrast, the skulls of crocodilians and most birds are extensively pneumatized by diverticula of both the nasal and tympanic cavities. In a few species of birds, diverticula of the cranial air spaces pass into the neck and may reach as far as the shoulder (King 1966). These cervicocephalic diverticula are intermuscular or subcutaneous. Although they may anastomose with diverticula of the pulmonary air sacs, they do not pneumatize any postcranial bones. If similar diverticula did pneumatize the postcranial skeleton in dinosaurs, they might invade the cervical column at any point, although they might be expected to pneumatize the anterior cervical vertebrae first.

Diverticula of the larynx and trachea are present in some squamates and birds, but absent in crocodilians. Most snakes have a ‘tracheal lung;’ that is, an expanded tracheal membrane that bears highly vascular parenchymal tissue and functions in gas exchange (Wallach 1998). More conventional tracheal diverticula are also present in some snakes, and they are used to inflate the neck or acoustically modify the hiss (Young 1991, 1992). Laryngeal and tracheal diverticula are also present in a few birds. These diverticula typically form large sacs that lie between the trachea and esophagus or between the trachea and the skin, and they are primarily used in phonation (McLelland 1989a). Laryngeal and tracheal diverticula do not pneumatize any bones in extant sauropsids. In fact, given that the membranes of air sacs and diverticula are relatively inelastic and the trachea is mobile (at least in some birds), it may be impossible for tracheal diverticula to invade the vertebral column. If it is possible for tracheal diverticula to pneumatize the skeleton, they could conceivably do so anywhere along the neck.

Among extant vertebrates, diverticula of the lung itself are only present in birds. In certain lizards and many snakes the posterior portion of the lung may be devoid of parenchymal tissue (Perry 1998). Such regions of the lungs are sometimes called ‘air sacs,’ but they do not ventilate the parenchymal portions of the lungs or give rise to diverticula as do the pulmonary air sacs of birds. In birds, small diverticula develop from the bronchial branches of the lungs. These diverticula may pneumatize the adjacent thoracic vertebrae and ribs (Müller 1907, O’Connor 2004, O’Connor and Claessens 2005). The expected pattern of pneumatization in dinosaurs would be the same: diverticula of the lungs should invade the thoracic vertebrae and ribs.

Finally, we must consider the diverticula of the pulmonary air sacs of birds. In most birds there are nine air sacs: a single clavicular sac and paired cervical, anterior thoracic, posterior thoracic, and abdominal sacs. All of the air sacs give rise to diverticula in at least some species of birds (Duncker 1971). Skeletal pneumatization varies widely within clades and even within populations (Hogg 1984a; O’Connor 2004), and not all of the elements listed below are pneumatized in all taxa. Diverticula of the cervical air sac pneumatize the cervical and anterior thoracic vertebrae and their attendant ribs. The sternum, sternal ribs, coracoids, clavicles, scapulae, and humeri are pneumatized by diverticula of the clavicular air sac. The anterior and posterior thoracic air sacs give rise to visceral diverticula that lie between the esophagus and pericardium and between the esophagus and liver. Diverticula of the anterior thoracic air sac occasionally pneumatize adjacent sternal ribs. Finally, diverticula of the abdominal air sacs pneumatize the posterior thoracic vertebrae, synsacrum, pelvic girdles, and femora.

The predicted ontogenetic pattern of pneumatization produced by diverticula of pulmonary air sacs depends on which diverticula develop first. The first postcranial bones to be pneumatized in the turkey (Meleagris gallopavo) are the sternum and thoracic vertebrae (Cover 1953). In the chicken (Gallus gallus), pneumaticity first appears in the humeri and cervical vertebrae (Hogg 1984b). It therefore appears that diverticula of the cervical and clavicular sacs pneumatize the presacral vertebrae and humeri or sternum, respectively, at about the same time in development. The humeri and sternum are not pneumatized in any non-avian dinosaurs, so we need only consider the relative timing of vertebral pneumatization. In this case, it is clear that the cervical or anterior thoracic vertebrae are pneumatized first; these regions of the spine are pneumatized by diverticula of the cervical air sac. Hogg (1984a) reported that in the chicken cervical vertebrae five through nine are always pneumatized; the rest of the postatlantal cervical vertebrae are pneumatized in most, but not all, individuals. This suggests that cervicals five through nine are the first to be pneumatized, and that in a few individuals pneumatization of the neck does not proceed any further.

**Patterns of pneumatization in basal dinosaurs—**Postcranial skeletal pneumaticity (PSP) is present in some “prosauropods” and most sauropods and theropods, but absent in ornithischians and many “prosauropods” (Britt 1993, 1997). (Traditionally, Prosauropoda included all non-sauropod sauropodomorphs [Sereno 1999], but recent analyses suggest that this group is paraphyletic [Yates 2003, 2007; Upchurch et al. 2007]. In this paper, I use "prosauropods" to refer to all non-sauropod sauropodomorphs). Further, pneumaticity is well developed in most pterosaurs. This distribution of the character requires either multiple origins or multiple losses (3- 1).

The Late Triassic (Norian) forms Pantydraco caducus (formerly Thecodontosaurus caducus; see Galton and Yates 2007) and Coelophysis bauri are the earliest-diverging sauropodomorph and theropod, respectively, that show convincing evidence of PSP. P. caducus and C. bauri may have inherited systems of pneumatic diverticula from a common ancestor, but they are the first representatives of their separate lineages in which the diverticula invaded the postcranial skeleton. P. caducus differs from all other prosauropods in having deep pneumatic fossae in the centra of its cervical vertebrae (Yates 2003). The placement of these cavities on the vertebrae, their invasive nature, and the presence of a distinct margin of bone bounding each cavity all argue for the interpretation of these cavities as pneumatic in origin. Furthermore, the fossae are only present in cervicals six through eight; thus, the pattern of pneumatization in Pantydraco is remarkably similar to the earliest stages of skeletal pneumatization in the chicken (Figure 3-5). Pneumatic cavities of this sort do not reappear in sauropodomorphs until the advent of basal sauropods, such as Shunosaurus, which have pneumatic cervical vertebrae. Coelophysis is the most basal theropod with evidence of PSP. The postaxial cervical vertebrae of C. bauri have pneumatic cavities that occupy most of the neural spine and that communicate with the outside through several large foramina (Colbert 1989). C. bauri is similar to more derived theropods, including birds, in having pneumatic cervical vertebrae.

**Evaluating the hypotheses—**Now that the patterns of pneumatization in the earliest dinosaurs with PSP have been established, we can evaluate the four hypotheses outlined above (Table 3-2). The hypothesis that the patterns of skeletal pneumatization observed in P. caducus and C. bauri were produced by diverticula of cranial air spaces is not well supported. Cranial diverticula do not pneumatize postcranial bones in any extant sauropsids, and the most anterior vertebrae of P. caducus and C. bauri are not pneumatized. Practically any pattern of cervical pneumatization could be produced by diverticula of the larynx and trachea, but these diverticula also do not pneumatize any postcranial bones in extant forms. Diverticula of the lungs sometimes pneumatize the thoracic vertebrae and ribs in birds, but these elements are not pneumatized in P. caducus or C. bauri.

That leaves the fourth hypothesis, that the PSP of basal saurischians was produced by diverticula of pulmonary air sacs. Diverticula of the air sacs pneumatize many elements of the postcranial skeleton in extant birds. The patterns of pneumatization observed in Pantydraco and Coelophysis are exactly what we should expect to see if diverticula of cervical air sacs were responsible. In effect, P. caducus and C. bauri resemble snapshots of the ontogenetic development of PSP in birds (Figure 3-5). The patterns of pneumatization present in these taxa show that cervical air sacs and their diverticula were present in both saurischian lineages in the Late Triassic.

**Evidence for clavicular air sacs**

The furcula of the dromaeosaur Buitreraptor is pneumatic (Makovicky et al. 2005). The furculae is pneumatized in some birds, such as swans (pers. obs), by diverticula of the clavicular air sac (Duncker 1971). This suggests that either the clavicular air sac evolved before the divergence of dromaeosaurs and birds, or that the two groups evolved clavicular air sacs in parallel.

A broken humerus of the basal tyrannosauroid Eotyrannus shows several large, irregular chambers. Although the form of the chambers is reminiscent of pneumatic internal structure, the proximal part of the humerus is crushed and it is not clear if any pneumatic foramina are present (Hutt et al. 2001). The situation with Eotyrannus is different from the iliac chambers mentioned above for sauropods. In the latter case, iliac chambers are known to be absent in many or most sauropods, and they are found only in taxa that are known (titanosaurs) or suspected (Amazonsaurus) to have had sacral pneumaticity and, therefore, abdominal air sac diverticula. In contrast, hollow appendicular elements are ubiquitious in theropods (Gauthier 1986, Colbert 1989), so the hollow humerus of Eotyrannus is expected. The question then becomes, are the chambers pneumatic or not? From the evidence in hand, both hypotheses are viable. If Eotyrannus had a pneumatic humerus, it was the only known non-avian dinosaur that did; pneumatic humeri have not been reported in the more derived taxa that lie close to the origin of birds and have been confused with birds (e.g., Caudipteryx, Sinornithosaurus: Ji et al. 1998; Xu et al. 1999). Those facts weigh against the pneumatic interpretation for Eotyrannus, but it cannot be ruled out on the basis of available evidence. New and better specimens will be required to resolve this problem.

**Evidence for abdominal air sacs, I. Pneumatization of the post-thoracic vertebral column**

Wedel et al. (2000) first proposed the hypothesis that pneumatic post-thoracic vertebrae in non-avian dinosaurs implied the presence of abdominal air sacs. The posterior thoracic vertebrae and synsacrum of birds are pneumatized by diverticula of the abdominal air sacs. The posterior thoracic, sacral, and anterior caudal vertebrae are pneumatic in many sauropods and non-avian theropods. Therefore it seems likely that these taxa had abdominal air sacs. All of those statements are still accurate, but further work has clarified several important points. (For the sake of simplicity, the term 'posterior compartment' is used herein to refer to the portions of the vertebral column that are pneumatized by diverticula of the abdominal air sacs.)

The crucial inference, that posterior compartment PSP implies the presence of abdominal air sacs, has been obscured by inconsistency in published accounts. Diverticula of the cervical air sacs, lungs, and abdominal air sacs may anastomose to produce a continuous network of diverticula that spans most or all of the vertebral column (Cover 1953). In some older accounts (e.g., Cover 1953), the entire network of vertebral diverticula is called an extension of the cervical diverticulum, and so some authors have inferred that diverticula of the cervical air sacs alone can pneumatize the entire vertebral column (King 1975, Britt et al. 1998). If that were true, then it would not be possible to infer the presence of abdominal air sacs based on post-thoracic PSP (Britt et al. 1998). As discussed above, some authors have used these uncertainties to argue that PSP does not provide evidence for abdominal air sacs (e.g., Ruben et al. 2003) or for the presence of any air sacs (e.g., Farmer 2006), although in both cases the authors have ignored numerous published accounts to the contrary (e.g., Muller 1907, Cover 1953, King 1966, 1975, Duncker 1971, Hogg 1984a, b, Bezuidenhout et al. 1999).

All of the uncertainties and contrary claims listed above have now been dispelled or falsified by O'Connor and Claessens (2005) and O'Connor (2006). Diverticula of the cervical vertebrae never pass farther posteriorly than the middle thoracic vertebrae, diverticula of the lungs pneumatize only the thoracic vertebrae and vertebral ribs immediately adjacent to the lungs, and the posterior compartment is only pneumatized by diverticula of abdominal air sacs. These relationships were found to be invariant in more than 200 individual birds representing 19 extant orders, and they support the hypothesis that posterior compartment PSP in sauropods and theropods indicates the presence of abdominal air sacs. No other hypothesis is consistent with known patterns of pneumatization in extant tetrapods.

**Evidence—**In theropods, pneumatization of the posterior compartment is present in at least some taxa in Abelisauroidea, Spinosauroidea, Allosauroidea, Tyrannosauroidea, Oviraptorosauria, and Dromaeosauridae (O'Connor and Claessens, 2005).

Among sauropods, posterior compartment pneumaticity was recognized very early. Marsh (1878, 1879, 1881, 1884) described and figured cavities in the sacral vertebrae of many sauropods, and he considered hollow sacral vertebrae a diagnostic character of Sauropoda. Pneumatic caudal vertebrae were first illustrated by Marsh (1890) for Barosaurus. Osborn (1899), Hatcher (1901), and Gilmore (1932) also illustrated pneumatic caudal vertebrae in Diplodocus. Although Marsh described the sacral vertebrae of Morosaurus (= Camarasaurus) as hollow, the evidence for sacral pneumaticity in Camarasaurus is equivocal (McIntosh et al. 1996a). In Camarasaurus lewisi (BYU 9047) the sacral vertebrae have large lateral fossae but no foramina, and the internal structure of the vertebrae is composed of apneumatic spongiosa (pers. obs.). However, unequivocally pneumatic sacral vertebrae are present in Brachiosaurus altithorax (Riggs 1904) and B. brancai (Janensch 1947). In Malawisaurus dixeyi the neural spine of the first caudal vertebra is pneumatic but the centrum is not (Figure 3-6). Posterior compartment pneumaticity is therefore present in both lineages of Neosauropoda (Diplodocoidea and Macronaria) and in some non-neosauropods, such as Mamenchisaurus youngi (Pi et al. 1996).

**Evidence for abdominal air sacs, II. Pneumatization of the pelvic girdle and hindlimb**

Pneumatization of the pelvic girdle and hindlimb in birds is straightforward: it is accomplished by diverticula of the abdominal air sacs, and this fact has never been debated (Muller 1907, Cover 1953, King 1966, 1975, Duncker 1971, Hogg 1984a, b, Bezuidenhout et al. 1999, O’Connor and Claessens 2005, O’Connor 2006). Pneumatization of the pelvic girdle and hindlimb elements in non-avian dinosaurs would be further evidence for the presence of abdominal air sacs.

**Evidence—**Among sauropods, large chambers have been reported in the ilia of the diplodocoid Amazonsaurus (Carvalho et al. 2003), the titanosaur Sonidosaurus (Xu et al. 2006), and the saltasaurines Saltasaurus and Neuquensaurus (Powell 1992, Sanz et al. 1999). From published descriptions, these internal chambers appear to have the same morphology as those in the pneumatic vertebrae of sauropods, and some authors (e.g. Carvalho et al. 2003, Xu et al. 2006) have interpreted the chambers as pneumatic. However, the case would be stronger if we knew how the air got into the bones. Pneumatization cannot take place and pneumatic chambers cannot persist without a patent (open) foramen (Ojala 1957, Witmer 1997). The case for appendicular pneumaticity in sauropods would be strengthened by the discovery of pneumatic foramina on the outside of the ilium, or a series of chambers connecting the ilium to the sacral vertebrae.

Although we should keep that caveat in mind, there is no strong reason to doubt that the chambers reported in the ilia of the sauropods listed above are pneumatic. We already have compelling evidence of sacral pneumaticity in both Diplodocoidea and Macronaria (see above). Iliac chambers are so far only found in clades in which sacral pneumatization is known, so the phylogenetic distribution of iliac chambers is consistent with the hypothesis that they are pneumatic. At the very least, we know from broken specimens (e.g., BYU 9047, Jensen 1988) that iliac chambers are absent in most sauropods, so we can characterize the presence of iliac chambers as a derived character that was independently acquired in Diplodocoidea and Macronaria in taxa for which sacral pneumaticity was present.

In theropods, a large foramen is present in the proximal femur of the oviraptorosaur Shixinggia (Lu and Zhang 2005). In its size and location this foramen is similar to pneumatic foramina in the femora of extant birds. It would be helpful to know what connections, if any, this foramen shares with spaces inside the femur. Still, as with the iliac chambers of sauropods, there is no strong reason to doubt that this foramen is pneumatic. If it is, then femoral pneumaticity evolved independently in oviraptorosaurs and birds (although femoral diverticula may have been present in the common ancestor of both clades).

### Evidence for abdominal air sacs, III. Pneumatic hiatuses

**Definition and occurrence in birds—**Diverticula of cervical air sacs, lungs, and abdominal air sacs invade the vertebral column at different points. Diverticula of the cervical air sacs first contact the posterior cervical and anterior thoracic vertebrae; diverticula of the lungs may invade the vertebrae adjacent to the lungs; and diverticula of the abdominal air sacs may invade the synsacrum at several points (King 1957, Duncker 1971, O'Connor 2006). Paravertebral diverticula derived from these sources may grow along the column until they contact each other and anastomose. The growth and anastomosis of the paravertebral diverticula may produce an uninterrupted pattern of vertebral pneumatization, so that every vertebra from the second or third cervical back to the free caudals is pneumatic.

However, in some cases the vertebral column is not continuously pneumatized. One or more pneumatic hiatuses may be produced as a result. A pneumatic hiatus is an apneumatic portion of the vertebral column that is bordered by pneumatic vertebrae both anteriorly and posteriorly (Figure 3-7; Wedel 2003a). There are at least three kinds of pneumatic hiatus, which I define here based on their positions. A Type I hiatus is a gap in pneumatization between diverticula of the cervical air sacs and lungs, and it appears in the most posterior cervical vertebrae or the most anterior dorsal vertebrae. A Type II hiatus is a gap in pneumatization between diverticula of the cervical air sacs or lungs and diverticula of the abdominal air sacs. It may occur in the middle or posterior thoracic vertebrae (Figure 3-8). Finally, a Type III hiatus is a gap in pneumatization between different diverticula of the abdominal air sac, and it may occur in the (syn)sacrum or caudal vertebrae.

King (1957) and Hogg (1984a) described the distribution of PSP in chickens, Gallus gallus. Although neither author used the term ‘pneumatic hiatus’, both studies document the existence of all three types of pneumatic hiatus (Table 3-3). In both studies, the majority of the birds examined had Type II hiatuses; Type I and Type III hiatuses were rare. King (1957) found no Type I hiatuses but his study had a much smaller sample size (six birds) than Hogg’s (1984a) study (51 birds). The two authors also found differences in the location of the hiatuses in different breeds. In his study of Rhode Island Reds, King (1957) found that the fourth thoracic vertebra was infrequently pneumatized and the fifth thoracic vertebra was never pneumatized. King and Kelly (1956) obtained similar results in a study of 50 chickens of unknown breed. In contrast, Hogg (1984a) found that in Golden Comets the fourth and fifth thoracic vertebrae were almost always pneumatized, and the second and third thoracic vertebrae were frequently apneumatic.

Each type of pneumatic hiatus is informative. The most anterior thoracic vertebrae may be pneumatized by diverticula of cervical air sacs or by diverticula of the lungs. Based on osteological evidence alone, we cannot determine whether diverticula of the cervical air sacs or lungs were involved if the posterior cervical and anterior thoracic vertebrae are all pneumatic. However, if a Type I hiatus is present, then the anterior thoracic vertebrae posterior to the hiatus could only have been pneumatized by diverticula of the lung. Similarly, vertebrae posterior to a Type II or Type III hiatus could only have been pneumatized by diverticula of abdominal air sacs.

**Importance—**If the development of pneumaticity in non-avian dinosaurs followed that of birds, then pneumatization of the cervical, dorsal, and sacral vertebrae in some sauropods and theropods shows that they had both cervical and abdominal air sacs—and, therefore, all of the components necessary for flow-through lung ventilation (O'Connor and Claessens 2005). However, this inference only holds if the development of pneumaticity in non-avian dinosaurs followed that of birds. Although diverticula of the cervical air sacs or lungs never pneumatize the posterior compartment in extant birds, we can at least hypothesize the possibility. The inference that abdominal air sacs were present in sauropods and non-avian theropods is already robust (see above); it would be stronger still if we found pneumatic hiatuses in these groups, because the vertebrae posterior to the hiatus would have to have been pneumatized independently, by diverticula of abdominal air sacs (Figure 3-9; Wedel 2003a). One such hiatus is present in the sauropod Haplocanthosaurus.

**Evidence: Haplocanthosaurus**—Haplocanthosaurus is a small (femur length = 128 cm) sauropod from the Upper Jurassic Morrison Formation of Western North America. Haplocanthosaurus has some characters in common with both Diplodocoidea and Macronaria, the two clades that make up Neosauropoda. This mosaic of features causes it be unstable in phylogenetic analyses; in various analyses it has been recovered as a eusauropod basal to Neosauropoda (Upchurch 1998, Rauhut et al. 2005), the most basal diplodocoid (Wilson 2002, fig. 13A), the most basal macronarian (Wilson and Sereno 1998), or a macronarian more derived than Camarasaurus (Upchurch et al. 2004).

The CM 879 specimen of Haplocanthosaurus has a mostly complete vertebral column. The first few cervical vertebrae are missing or incompletely preserved, so the precise presacral count is unknown, but it probably included 13 cervicals and 13 dorsals (Wilson and Sereno, 1998). The sacrum is incomplete, and includes five coossified spines and the fourth and fifth sacral centra. The first seven caudal vertebrae are present and complete. The dorsal vertebrae and most of the cervical vertebrae are preserved with the neural spines attached to the centra, although the neurocentral sutures are still visible. The neural spines and centra are preserved as separate elements in at least one cervical vertebra, in the sacral vertebrae, and in caudal vertebrae three through seven. Interestingly, the neurocentral sutures of the first two caudal vertebrae are fused. Hatcher (1903:38) wrote of the caudal vertebrae that "Owing to the age of the individual the neural arches and transverse processes are not coossified with their respective centra although those of the first two are still held in place." The first two caudal vertebrae are not illustrated with any visible sutures, however (Hatcher 1903, pl. 2), and in my examination of the elements I found no evidence of suture lines between the neural spines and centra.

All of the preserved cervical and dorsal centra have prominent lateral cavities that penetrate to a median septum (Figure 3-10). A CT scan of a dorsal vertebra of the CM 572 specimen of Haplocanthosaurus shows that the lateral fossae do not invade the condyle or the ventral half of the centrum and that they are only partially bounded by a distinct lip of bone. Fossae in the dorsal vertebrae of CM 879 have the same morphology. In both specimens the ventral margins of the lateral fossae are more clearly delimited than the dorsal margins, so that the fossae open dorsolaterally.

The sacral neural spines of CM 879 have no distinct pneumatic fossae or foramina. The only well-developed laminae are the spinodiapophyseal laminae, which are present only in the first three spines (Figure 3-11). In particular, the fossae present on the neural arch and spine of the first caudal vertebra (described below) are absent in the sacral neural spines.

The fourth sacral vertebra of CM 879 is asymmetrically pneumatized. The right side of the centrum bears a large, distinct fossa that extends upward underneath the facet for the sacral rib. This fossa is 78 mm long, 33 mm tall, and 27 mm deep. The fossa differs from the apneumatic fossae of extant crocodilians in being proportionally larger and deeper and in having a distinct margin. The dorsal margin is much more pronounced than the ventral, so in cross section the fossa is similar to the fossae of the dorsal vertebrae, only flipped upside down. The left side of the centrum is strongly waisted (i.e., narrower in the middle than at either end) but has no distinct fossa below the articular surface for the sacral rib.

On both sides of the centrum there are smaller cavities above the sacral rib facets (Figure 3-12). On each side this space is bounded ventrally and anteriorly by the sacral rib facet, posteriorly by the rim of the cotyle, and dorsally by a lamina of bone. The laminae may be either infrapostzygapophyseal or infrahyposphenal; their exact identity is obscured by the thorough coossification of the sacral spines. Each space is also divided into anterodorsal and posteroventral compartments by an accessory lamina. The anterodorsal compartments on both sides consist of shallow fossae only a few millimeters deep. The posteroventral fossae are much deeper. On the left side the posteroventral fossa is cone-shaped and 32 mm deep. The fossa on the right side is similar in size and shape, but it still contains some matrix so its depth cannot be determined. Hatcher (1903:figs. 15 and 20) illustrated both the lateral and dorsal fossae on the right side of the centrum.

The fifth sacral vertebra has no evidence of pneumaticity. The sides of the centrum are shallowly waisted, but there are no invasive fossae anywhere on the element. The facets for the sacral ribs are more dorsally extensive than in the fourth sacral and cover the area occupied by the dorsal fossae in the preceding vertebra. It is possible that the large sacral rib facets simply left no room for the dorsal fossae to form. The sacral rib facets are no more ventrally extensive than those of the fourth sacral. In other words, there is room for lateral fossae on the sides of the centrum, but the fossae are not present.

The first caudal vertebra has deep, distinct fossae on both the centrum and the neural spine (Figure 3-13). The lateral fossae of the centrum are similar in size and form to the fossa on the right side of the centrum of sacral vertebra four. The fossa on the right side of the centrum is 69 mm long, 41 mm tall, and 31 mm deep. Like the right-hand fossa of the fourth sacral, it extends upward under the attachment of the transverse process and the dorsal margin is more sharply delineated than the ventral. The fossa on the left side of the centrum is 54 mm long and 29 mm tall and mostly filled with matrix. It is 12 mm deep from the rim to the surface of the matrix. It is not clear if the matrix-filled part of the fossa extends up under the attachment of the transverse process.

The neural spine fossae of the first caudal are all located just posterior to the prezygapophyses. The vertebra lacks a true intraprezygapophyseal lamina. Instead, a low rampart of bone connects the prezygapophyseal rami below and behind the prezygapophyses. A bilobate fossa is situated behind this bony rampart at the base of the prespinal ligament scar. The fossa is 17 mm long, 16 mm wide, and at least 17 mm deep. As currently preserved, the bottom of the fossa contains some glue and bits of broken bone, so the bony recess may be a few millimeters deeper. The fossa is too deep and too narrow to accept the postzygapophyses of the preceding vertebra.

In dorsal view, with anterior to the top, the prezygapophyseal laminae of the first caudal vertebra resemble the letter M. The middle legs of the M are formed by short spinoprezygapophyseal laminae that converge on the lower portion of the neural spine. The bilobate fossa described in the preceding paragraph sits just above the convergence of the middle legs. The lateral legs of the M are formed by posterolaterally directed laminae that connect the prezygapophyses to the transverse processes. On the right side of the vertebra these laminae are poorly developed, and there is only a shallow depression between the right spinoprezygapophyseal lamina and the posterolateral lamina. On the left side the laminae are much more pronounced and the same space is occupied by a prominent fossa. The bottom of this fossa contains a shallow pool of dried glue. The fossa is 19 mm deep from the edge of the posterolateral lamina to the surface of the glue, and was probably no more than 21 mm deep in life.

None of the fossae present in the first caudal vertebra are present in the second. The lateral faces of the centra are shallowly waisted but have no distinct fossae. Two small nutrient foramina are present on the right side of the centrum, both less than 2 mm in diameter. Most of the right prezygapophysis is missing. The remainder of the right prezygapophysis, the left prezygapophysis, and the neural spine form three sides of a rectangular trough. This trough does not extend posteriorly or ventrally past the margin of the right prezygapophysis. It is no larger or deeper than it needed to be to accept the postzygapophyses of the first caudal vertebra.

In summary, the fourth sacral and first caudal vertebrae have a variety of large, distinct fossae that compare well to those found on the dorsal vertebrae. These fossae are compelling evidence of pneumaticity. These fossae are absent in the centra and neural spines of the fifth sacral and second caudal vertebra. The apneumatic fifth sacral vertebra is bordered anteriorly and posteriorly by pneumatic vertebrae, and constitutes a Type III pneumatic hiatus. The first caudal vertebra could only have been pneumatized by diverticula of abdominal air sacs (Figure 3-14).

## DISCUSSION

### Shared Developmental Pathways and the Origin(s) of Air Sacs and PSP

Until now, the hypothesis that non-avian saurischian dinosaurs had cervical and abdominal air sacs has been supported by the presence of pneumaticity in the parts of the skeleton that are pneumatized by those air sacs in extant birds. The evidence presented above shows that PSP in non-avian dinosaurs also shared some aspects of development with PSP in birds. In basal theropods and sauropodomorphs PSP is present only in the cervical vertebrae. This shows that diverticula of the cervical air sacs must have developed anteriorly from the thorax before they pneumatized the skeleton, just as in extant birds. The pneumatic hiatus in Haplocanthosaurus shows that vertebral diverticula developed from more than one part of the abdominal air sacs. The parallel between the evolution of PSP in non-avian dinosaurs and its development in birds also suggests that similar generative mechanisms were responsible.

Evidence for PSP in Archaeopteryx is equivocal (O'Connor 2006, Mayr et al. 2007, contra Britt et al. 1998, Christiansen and Bonde 2000). Foramina in the vertebrae and pelvic elements are not clearly pneumatic. Some may be vascular foramina, and some may actually be breaks in the specimens (O'Connor 2006). Furthermore, radiographs of at least one specimen show that the vertebrae are dense and solidly constructed, which is more consistent with apneumatic bone than with pneumaticity (Mayr et al. 2007). However, PSP was definitely present in other basal birds (e.g., Ichthyornis, Jeholornis; Marsh 1880, Zhou and Zhang 2003), and many other basal birds have humeral and vertebral fossae that may be pneumatic (Sanz et al. 1995). Therefore it is not clear if PSP in non-avian theropods is taxically homologous with that of birds. On the basis of currently available evidence, the absence of unequivocal PSP in Archaeopteryx could represent a loss in that taxon alone.

Archaeopteryx is not the only lacuna in the phylogenetic distribution of PSP in Saurischia. Sauropodomorphs evolved PSP independently from theropods; PSP is absent in the most basal known sauropodomorph (Saturnalia; Langer et al. 1999) and equivocal in most basal sauropodomorphs (Wedel 2007). Furthermore, posterior compartment pneumaticity evolved independently in diplodocoids, titanosauriforms, ceratosaurians, and coelurosaurians. The skeletal traces of abdominal air sacs are not phylogenetically continuous throughout Saurischia, and therefore also not taxically homologous.

Although PSP does not have a continuous phylogenetic distribution in Saurischia, it seems to have been produced by similar developmental pathways in sauropodomorphs, non-avian theropods, and birds, and none of the gaps in the phylogenetic distribution of pneumatic characters are very large. It is possible that both cervical and abdominal air sacs were present in the ancestral saurischian, but did not pneumatize the postcranial skeleton in some descendants of that ancestor—for example, most basal sauropodomorphs. A parallel situation exists in birds, which all have an air sac system even though PSP has been lost in some clades.

On the other hand, air sacs may have evolved independently in sauropodomorphs and theropods. If air sacs were absent in the ancestral saurischian, they must have evolved in both lineages very soon after the divergence of theropods and sauropodomorphs, because PSP is present in the basal members of both clades (Coelophysis and Pantydraco, respectively). The ancestral saurischian must have had at least the potential to evolve air sacs, but this potential may have been realized independently in theropods and sauropodomorphs. Whether air sacs are primitive for Saurischia or evolved independently in theropods and sauropodomorphs cannot be answered on the basis of currently available evidence.

## The Origin of Flow-Through Lung Ventilation

Flow-through lung ventilation like that of birds minimally requires four things: (1) lungs that are organized into tubes rather than sacs, (2) air sacs anterior to the lungs, (3) air sacs posterior to the lungs, and (4) a musculoskeletal system capable of driving ventilation. Evidence from the fossil record, particularly PSP, can extend our understanding of the origin of this system, but it may not be sufficient to pinpoint the origin of avian-style lung ventilation.

Fossil evidence and phylogenetic inferences suggest that all of the components necessary for flow-through lung ventilation were present in basal saurischians. But the hypothesis that bird-like lung ventilation was common to all saurischians comes with two important caveats. The first is obvious: at least for now, we have no way of knowing the path of inspired air in non-avian dinosaurs. The fossil evidence can only show that most saurischians had all of the anatomical components necessary for flow-through lung ventilation. So it is possible that basal sauropodomorphs and basal theropods had air sacs but not flow-through ventilation.

The other caveat extends uncertainty in the opposite direction. PSP is present in pterosaurs (Bonde and Christiansen 2003, O'Connor 2006) but absent in the closest outgroups of Saurischia—Ornithischia and non-dinosaurian dinosauromorphs. We know from avian development that the air sac system and its diverticula are present before they leave skeletal traces (Locy and Larsell 1916a, b; Hogg 1984a, b), and that the fully avian air sac system can be present without producing PSP, as in loons and penguins. Therefore we cannot rule out the possibility that an air sac system and possibly even flow-through lung ventilation evolved much earlier in archosaurs, and that this system only became detectable in the fossil record when it started leaving diagnostic traces in the skeletons of basal saurischians and pterosaurs.

The evolution of PSP in saurischians is marked by trends in the invasiveness and physical scale of the pneumatic traces. Pneumatic fossae in the vertebrae of basal sauropods and theropods are antecedent to the simple, camerate (large-chambered) vertebrae of more derived taxa, which in turn give way to the complex, camellate (small-chambered) vertebrae of the most derived taxa in both lineages. If in our quest to find the origin of PSP we go down the tree through progressively more basal taxa, the problem is not that evidence of PSP disappears entirely. It is that the shallow, unbounded fossae of basal dinosaurs are no longer diagnostic for pneumaticity (Wedel 2007). Similar fossae are present in the vertebrae of many tetrapods, and they may be associated with many soft tissues, including muscles, cartilage, and fat (O'Connor 2006).

One potential step forward is to search for criteria that would allow us to distinguish pneumatic fossae from those associated with other soft tissues. Such criteria might be present in the microscopic surface texture of pneumatic bones, or in their histology, which has been little studied except for a few brief treatments (Reid 1996, Woodward 2005). Perhaps the ancestors of Saurischia had air sacs and diverticula but the skeletal traces of the respiratory system are so faint or so non-diagnostic that we have not recognized them. Perhaps they had air sacs but no diverticula, or no air sacs at all. There is no guarantee that diagnostic criteria for pneumatic fossae exist to be found, or that if found they will be present in the outgroups to Saurischia. Nevertheless, if we want to improve our understanding of the origin of the avian air sac system, that is where and how we will have to look.

TABLE 3-1. Sternum length compared to trunk length in various saurischians. Trunk length is measured from the first dorsal vertebra to the posterior end of the ilium. The sternal elements of non-avian saurischian dinosaurs are proportionally as large or larger than those of some flightless birds.

| Taxon | Sternum Length (cm) | Trunk Length (cm) | Sternum L/ Trunk L | Source |
|---|---|---|---|---|
| **Birds** | | | | |
| Apteryx | 3.1 | 18 | 0.17 | MVZ Hild 1480 |
| Dromornis | 30 | 135 | 0.22 | Murray & Vickers-Rich 2004 |
| Megalapteryx | 6.3 | 47 | 0.13 | Worthy & Holdaway 2002 |
| **Non-avian theropods** | | | | |
| Velociraptor | 6.1 | 30[a] | 0.20 | Norell & Makovicky 1997, 1999 |
| **Sauropods** | | | | |
| Apatosaurus | 51 | 261 | 0.20 | Upchurch et al. 2004 |
| Brachiosaurus | 110 | 480[e] | 0.23 | Janensch 1950, 1961 |
| Saltasaurus | 61 | 207 | 0.29 | Powell 2003 |

[a] based on IGM 100/985 and IGM 100/986

[e] estimated from incomplete dorsal and sacral series

TABLE 3-2. Pneumatic diverticula in extant sauropsids and their potential to pneumatize the postcranial skeleton.

| Source of diverticula | Pneumatize postcranial skeleton in extant forms? | Predicted pattern of PSP in non-avian dinosaurs |
|---|---|---|
| Cranial air spaces | No | Anterior cervical vertebrae |
| Larynx and trachea | No | Cervical vertebrae |
| Lungs | Yes | Thoracic vertebrae and ribs |
| Pulmonary air sacs | Yes | |
| Cervical | | Cervical or ant. thoracic verts. |
| Clavicular | | Sternum and humerus |
| Thoracic | | Sternal ribs |
| Abdominal | | Synsacrum, pelvis, femur |

TABLE 3-3. Vertebral pneumaticity in the posterior compartment in sauropods. Abbreviations: PT, posterior thoracic vertebrae; S, sacral vertebrae; AC, anterior caudal vertebrae; MC, middle caudal vertebrae.

| Taxon | PT | S | AC | MC | Source or specimen |
|---|---|---|---|---|---|
| *Omeisaurus tianfuensis* | X | - | - | - | He et al. 1988 |
| *Mamenchisaurus youngi* | X | - | - | - | Pi et al. 1996 |
| *M. hochuanensis* | X | - | - | - | CCG V 20401 |
| *Apatosaurus louisae* | X | X | - | - | CM 3018 |
| *Apatosaurus* sp. | na | na | X | na | OMNH 1436 |
| *Diplodocus carnegii* | X | X | X | X | CM 84 |
| *Barosaurus lentus* | X | X | X | X | AMNH 6341 |
| *Camarasaurus supremus* | X | - | - | - | AMNH 5761 |
| *Camarasaurus lewisi* | X | - | - | - | BYU 9047 |
| *Haplocanthosaurus priscus* | X | X | X | - | CM 572, 879 |
| *Brachiosaurus altithorax* | X | X | - | na | FMNH P 25701 |
| *Brachiosaurus brancai* | X | X | - | - | Janensch 1950 |
| *Euhelopus zdanskyi* | X | X | na | na | Wiman 1929 |
| *Malawisaurus dixeyi* | X | X | X | - | MAL holotype series |

TABLE 3-4. Frequencies of pneumatic hiatuses in two breeds of chickens. Data on Rhode Island Reds from King (1957); data on Golden Comets from Hogg (1984a). See text for descriptions of different types of hiatuses.

| | Pneumatic Hiatuses | | |
|---|---|---|---|
| | Type I | Type II | Type III |
| Rhode Island Red | | | |
| Males (1) | 0 / 0% | 1 / 100% | 0 / 0% |
| Females (5) | 0 / 0% | 5 / 100% | 2 / 40% |
| Total (6) | 0 / 0% | 6 / 100% | 2 / 33% |
| Golden Comet | | | |
| Males (3) | 1 / 33% | 2 / 67% | 0 / 0% |
| Females (48) | 3 / 6% | 34 / 71% | 2 / 4% |
| Total (51) | 4 / 8% | 36 / 71% | 2 / 4% |

FIGURE 3-1. Evidence for air sacs in fossil archosaurs. Letters next to each taxon indicate that they have patterns of PSP that are diagnostic for certain air sacs: C, cervical air sacs; A, abdominal air sacs; Cl, clavicular air sacs. See text for diagnostic criteria. In both sauropodomorphs and theropods, cervical air sacs pneumatized the skeleton before abdominal air sacs. Evidence for PSP in Archaeopteryx is equivocal (O'Connor 2006). A naïve reading of the fossil record would suggest that different air sacs evolved independently several times—for example, independent origins of abdominal air sacs in Mamenchisaurus and Neosauropoda. However, it is more reasonable to infer that cervical and abdominal air sacs, at least, were present in all members of Eusaurischia, and simply failed to pneumatize the postcranial skeleton in some taxa. It is possible that an air sac system is primitive for Ornithodira, but the total absence of PSP in Ornithischia, a diverse and long-lived clade, is problematic (see Wedel 2007). Phylogeny based on Gauthier (1986), Wilson (2002), Zhou and Zhang (2002), Upchurch et al. (2004, 2007), Makovicky et al. (2005), and Yates (2007).

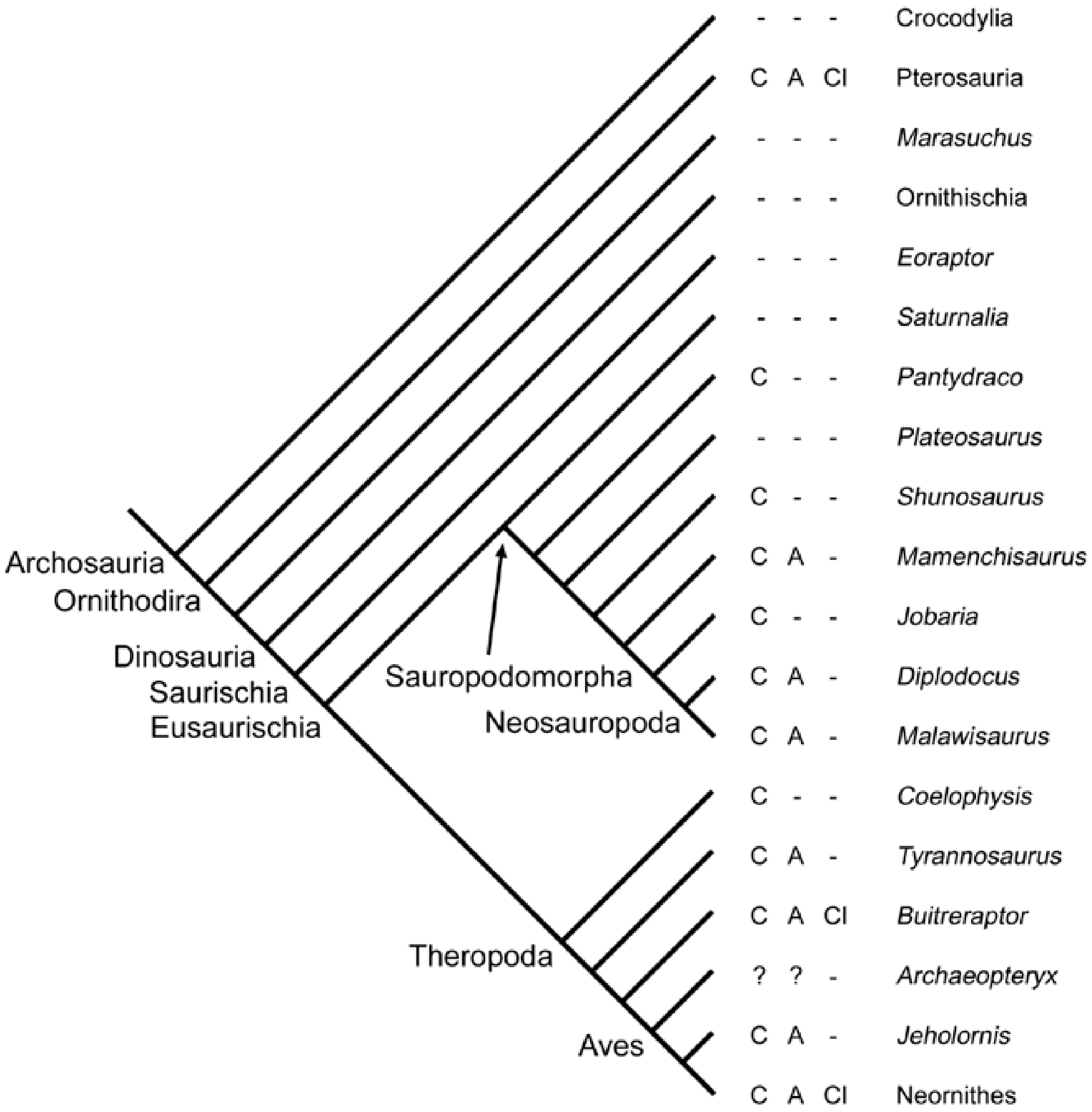
- - - Crocodylia
C A Cl Pterosauria
- - - *Marasuchus*
- - - Ornithischia
- - - *Eoraptor*
- - - *Saturnalia*
C - - *Pantydraco*
- - - *Plateosaurus*
C - - *Shunosaurus*
C A - *Mamenchisaurus*
C - - *Jobaria*
C A - *Diplodocus*
C A - *Malawisaurus*
C - - *Coelophysis*
C A - *Tyrannosaurus*
C A Cl *Buitreraptor*
? ? - *Archaeopteryx*
C A - *Jeholornis*
C A Cl Neornithes
Archosauria
Ornithodira
Dinosauria
Saurischia
Eusaurischia
Sauropodomorpha
Neosauropoda
Theropoda
Aves

FIGURE 3-2. Pneumatization of the vertebral column in the chicken. Pneumatic vertebrae are shown in black. Data are from Hogg (1984b); vertebrae are shown at earliest date of complete pneumatization. Some rare variations are not shown; for example, the second and third thoracic vertebrae were pneumatized in one individual (from a total of 44) examined by Hogg (1984b). The spread of PSP along the vertebral column in the chicken parallels the evolution of PSP in non-avian theropods and sauropods; compare to Figure 3-3.

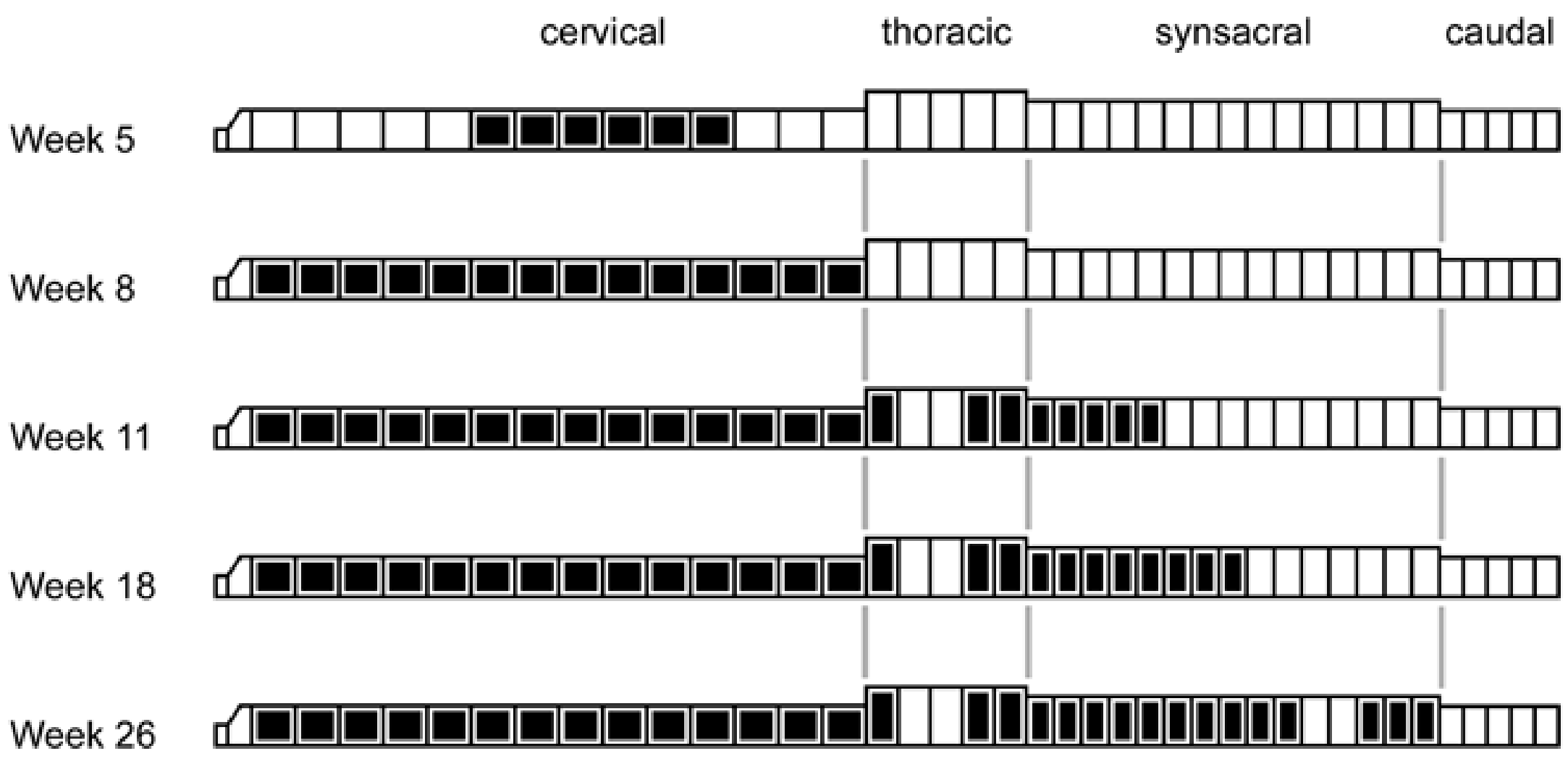

FIGURE 3-3. The distribution of fossae and pneumatic chambers (black boxes) along the vertebral column in sauropodomorphs. Only the lineage leading to diplodocines is shown here. A similar caudal extension of pneumatic features occurred independently in macronarian sauropods (Table 3-3) and several times in theropods. It also parallels the development of vertebral pneumaticity in birds; compare to Figure 3-2. The format of the diagram is based on Wilson and Sereno (1998, fig. 47). Phylogeny based on Wilson (2002), Yates (2003), and Upchurch et al. (2004). Data on Shunosaurus from Zhang (1988). All others based on personal observations of specimens; see list in Appendix III.

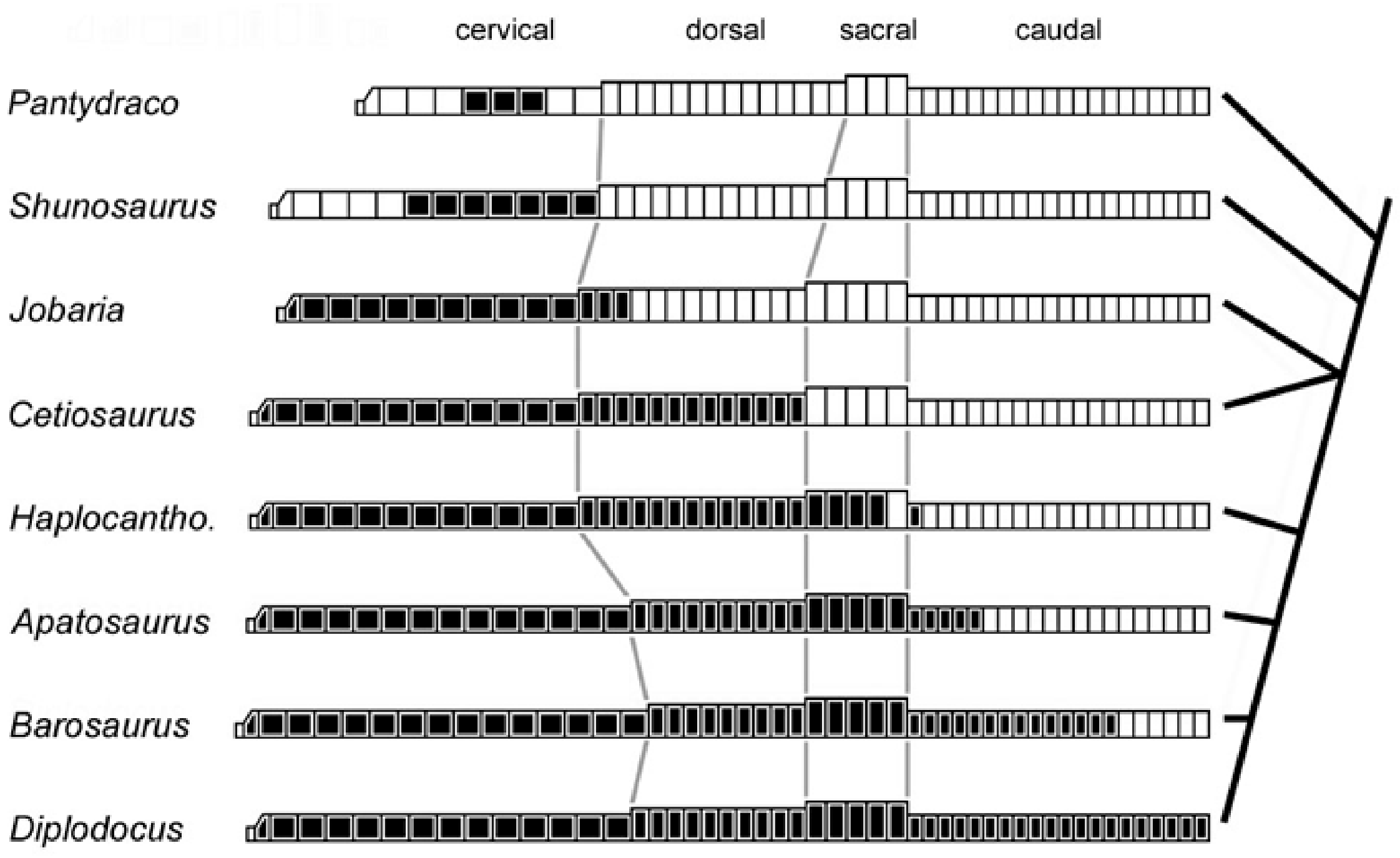

FIGURE 3-4. Rib attachments in saurischian dinosaurs. Dorsal vertebrae of (A) Lufengosaurus, (B) Diplodocus, and (C) Tyrannosaurus in right lateral view. Positions of diapophyses and parapophyses are represented by dots. In (A), straight lines represent axes of rotation for vertebral ribs, and arrows show the direction of rib movement during inhalation. A traced from Young (1941, fig. 6), B traced from Hatcher (1901, pl. 7), C traced from Brochu (2003, fig. 56).

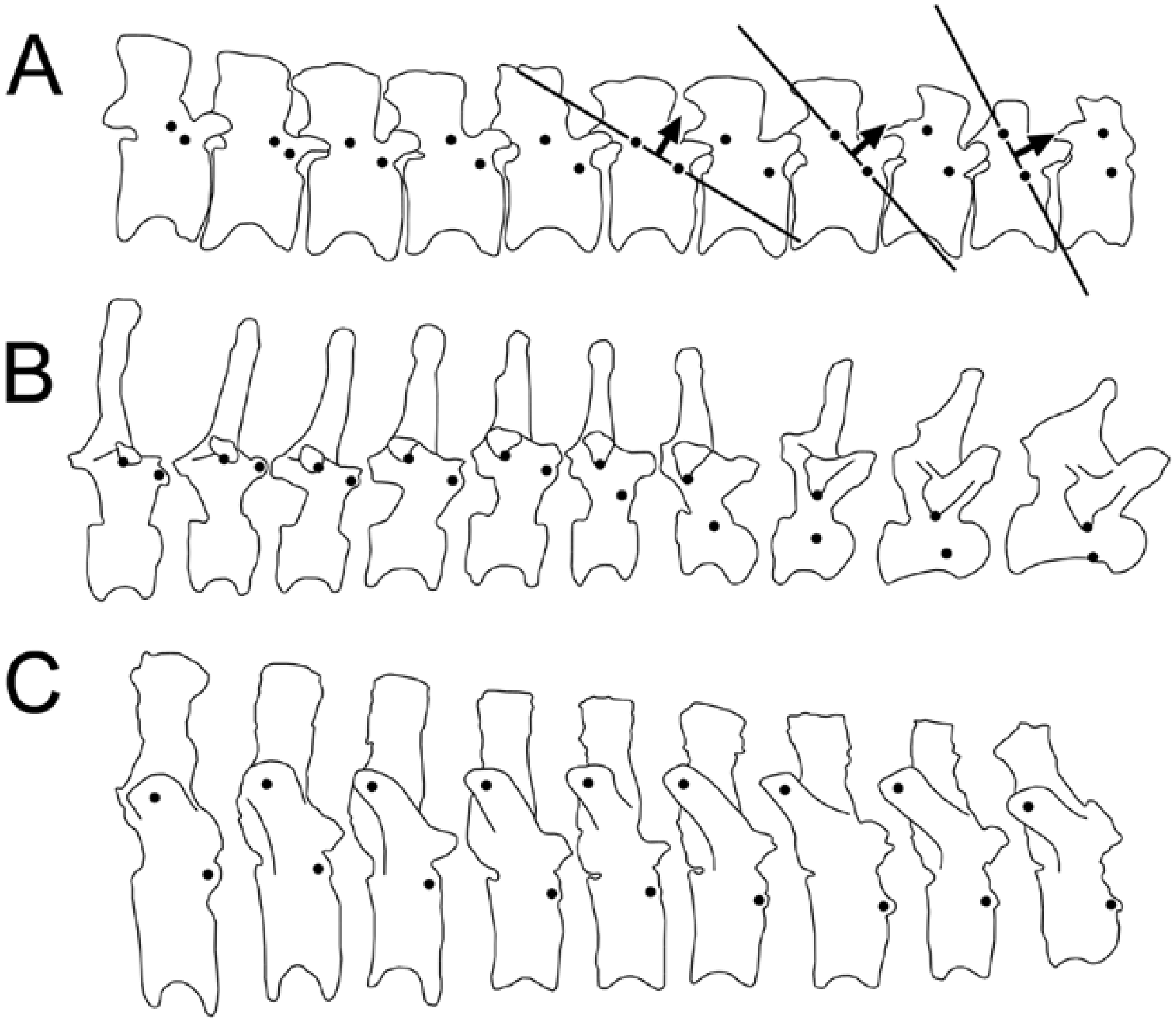
A
B
C

FIGURE 3-5. Pneumatic vertebrae (black) in (A) a 7-week-old chicken, Gallus gallus; (B) Coelophysis bauri; and (C) Pantydraco caducus. Modified from Storer (1951:fig. 31-5) (A), Colbert (1989:fig. 88) (B), and Benton et al. (2000:fig. 19) (C).

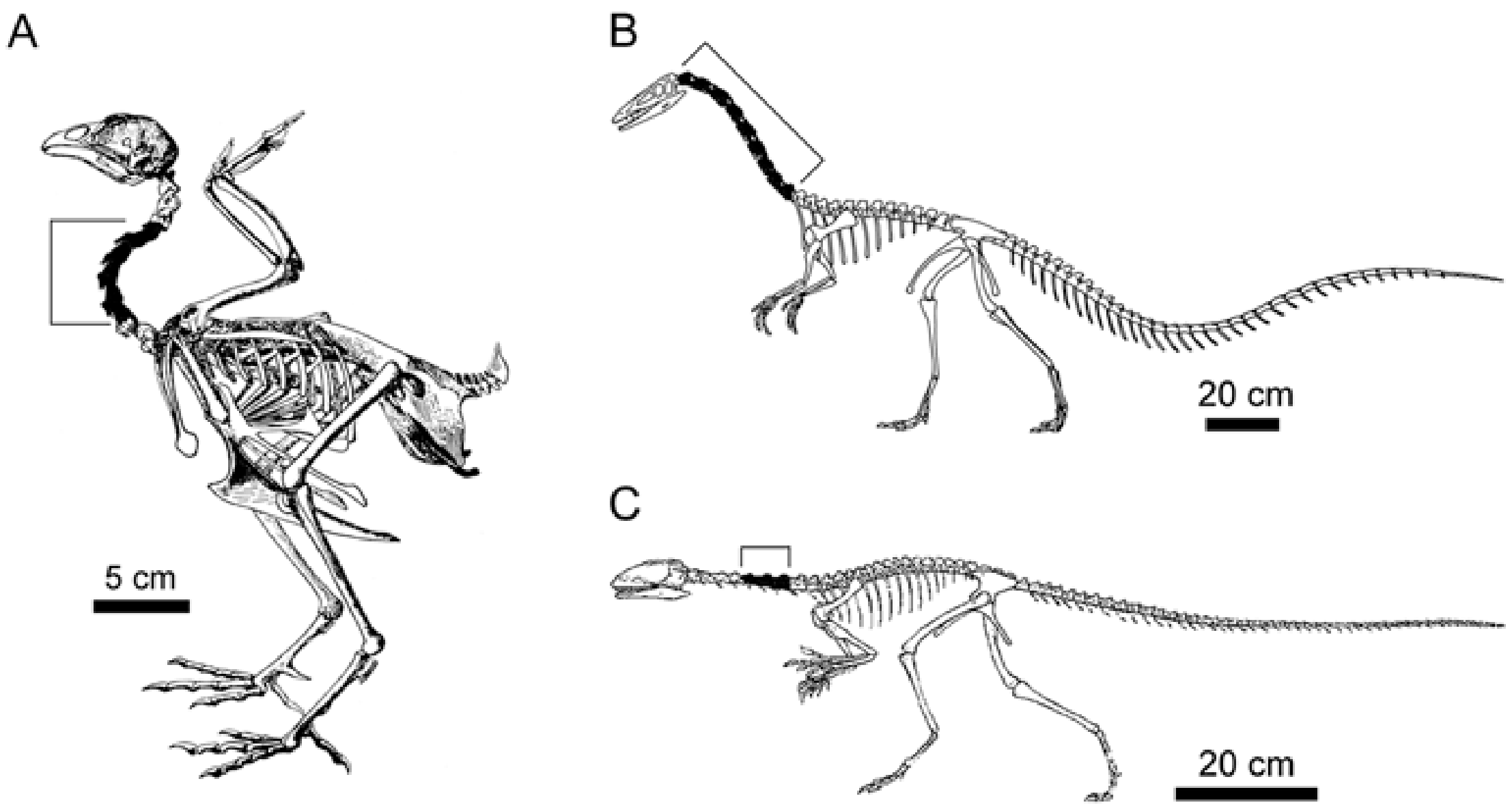

FIGURE 3-6. MAL-200, an anterior caudal vertebra of Malawisaurus dixeyi. A. The vertebra in left lateral view showing the position of CT slices. B.-D. CT cross-sections. Matrix was erased from the internal chambers using Photoshop 5.5. Pneumatic foramina on the neural arch and spine are connected to a network of internal chambers, but the centrum is apneumatic.

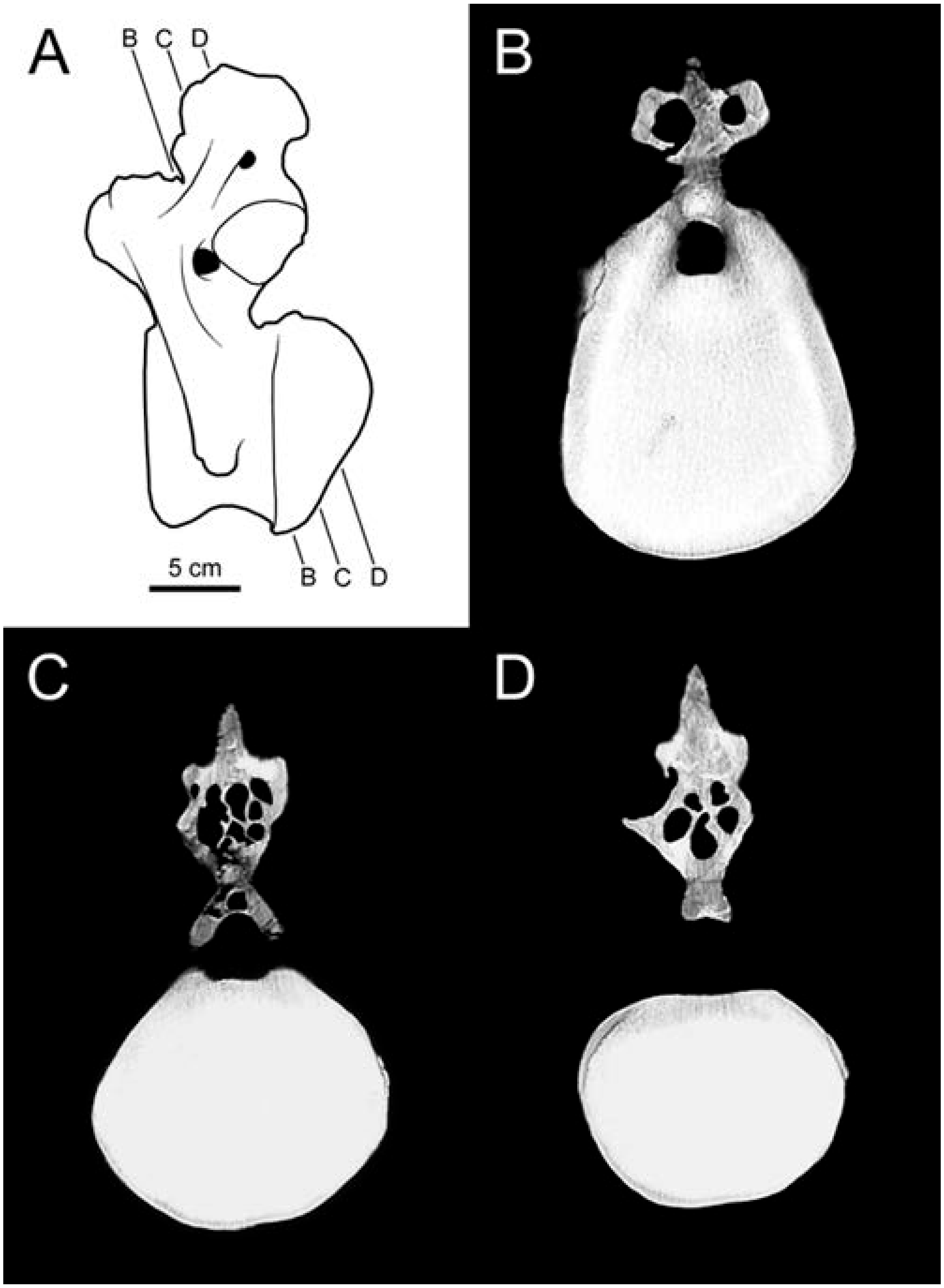

FIGURE 3-7. Pneumatization of the vertebral column in the chicken, Gallus gallus. Pneumatic vertebrae are stippled. The vertebral column is pneumatized by diverticula of the cervical air sacs, lungs, and abdominal air sacs. A pneumatic hiatus is one or more apneumatic vertebrae that are bordered anteriorly and posteriorly by pneumatic vertebrae. These hiatuses are produced if the diverticula from the different parts of the respiratory system do not meet and anastomose. Supporting data come from King (1957), King and Kelly (1956), Hogg (1984a, b), and from personal examination of museum specimens. Inspired by King (1957:fig. 1).

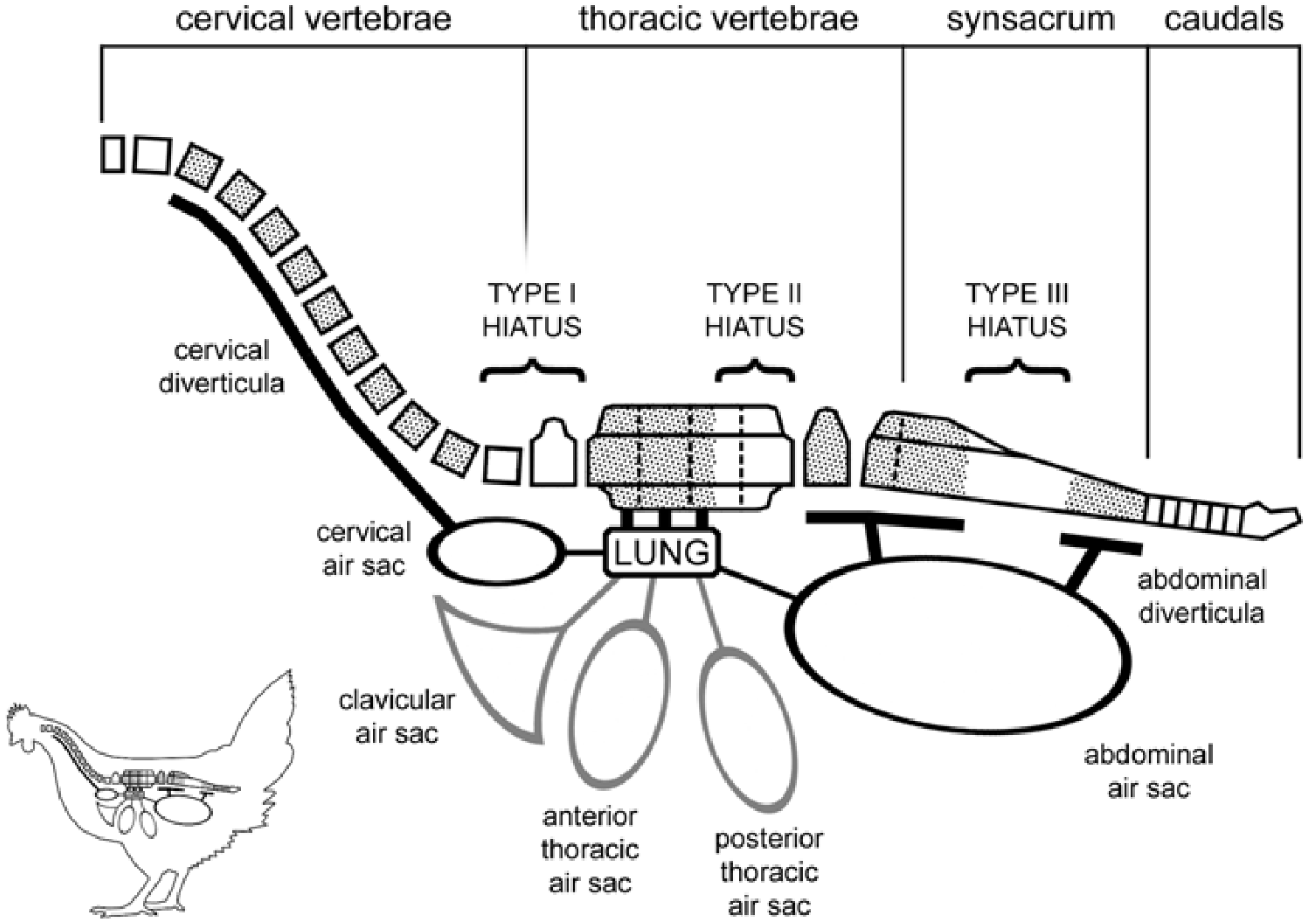

FIGURE 3-8. A pneumatic hiatus in a chicken. The notarium of UCMP 119225 is composed of four vertebrae. The three anterior vertebrae are pneumatic, but the fourth is not. A. The specimen in left lateral view under normal lighting. B. The specimen lit from behind to show the pneumatic (translucent) and apneumatic (opaque) regions. C. A micro CT slice through a pneumatic vertebra. D. A micro CT slice through the apneumatic vertebra. Note the density of the trabeculae in D compared to C. The anterior synsacral vertebrae of this individual are pneumatic. The apneumatic vertebra is bordered anteriorly and posteriorly by pneumatic vertebrae, and constitutes a Type II pneumatic hiatus.

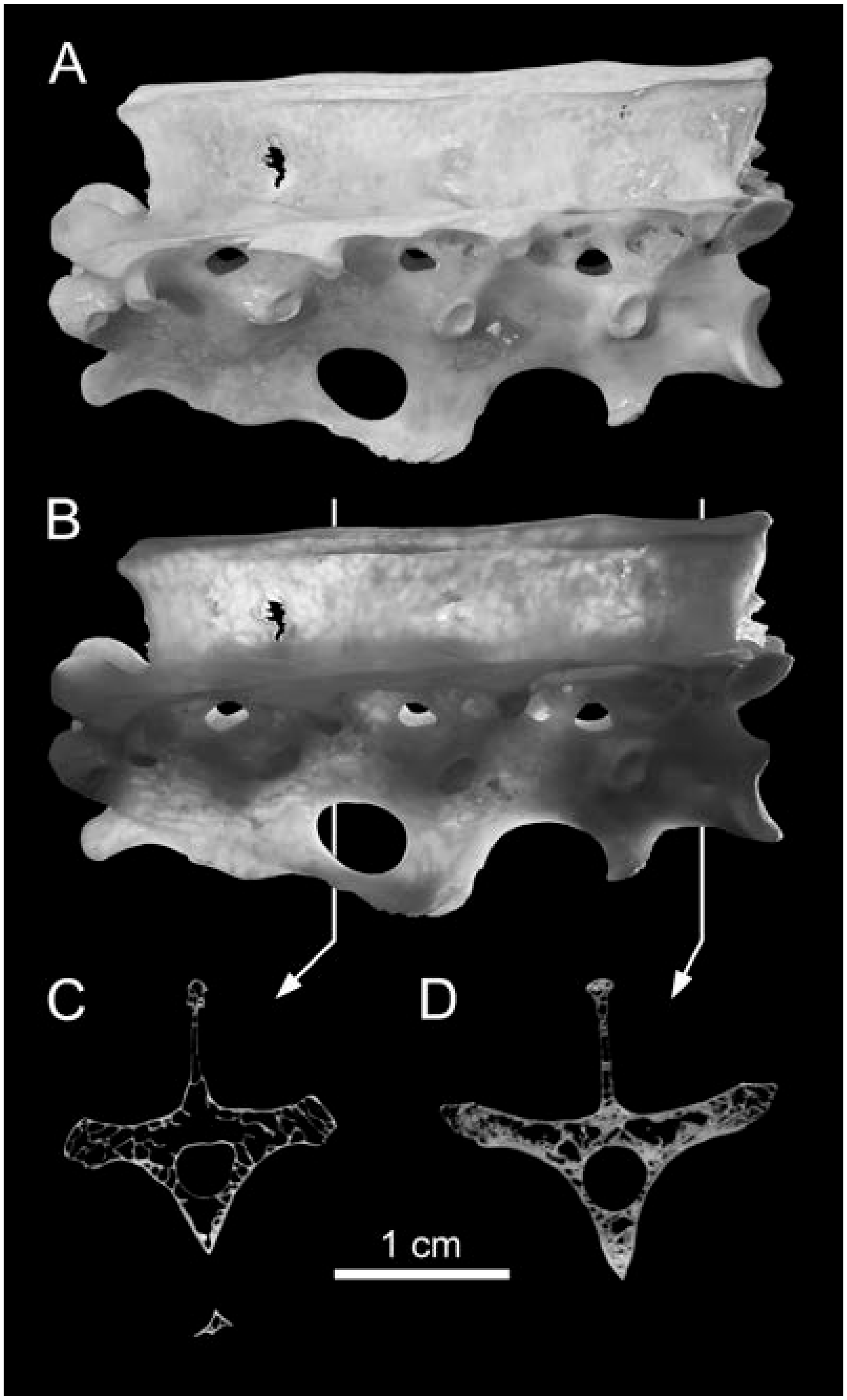
A
B
C
D
1 cm

FIGURE 3-9. Criteria for inferring the presence of abdominal air sacs in non-avian dinosaurs. A sauropod is shown, but the same logic applies to theropods. Pneumatized vertebrae are shown in black. Small arrows show the spread of pneumatic diverticula, and large arrows represent ontogenetic trajectories. A. Pneumatization of the vertebrae by diverticula of cervical air sacs (ca), lungs (L), and abdominal air sacs (aa). Other air sacs (light gray) may be present, but are not known to pneumatize the vertebral column. B. Pneumatization of the vertebrae by diverticula of cervical air sacs alone. C. A hypothetical sauropod with pneumatic hiatuses. Vertebrae posterior to a Type II or Type III hiatus could only be pneumatized by diverticula of abdominal air sacs. Therefore, the presence of these hiatuses demonstrates that abdominal air sacs were present. D. Pneumatization of the posterior dorsal, sacral, and caudal vertebrae does not necessarily demonstrate the presence of abdominal air sacs, because continuous pneumatization of the vertebral column could be produced by anastomosing diverticula of the cervical and abdominal air sacs (as in A) or by cervical air sacs alone (as in B). The A-C and A-D transformations are known to happen in extant birds. The B-D transformation does not happen in extant birds (O'Connor 2006), but cannot be ruled out in non-avian dinosaurs (because it depends on the behavior of epithelial diverticula that do not themselves fossilize). Modified from Wedel (2003a:fig. 4).

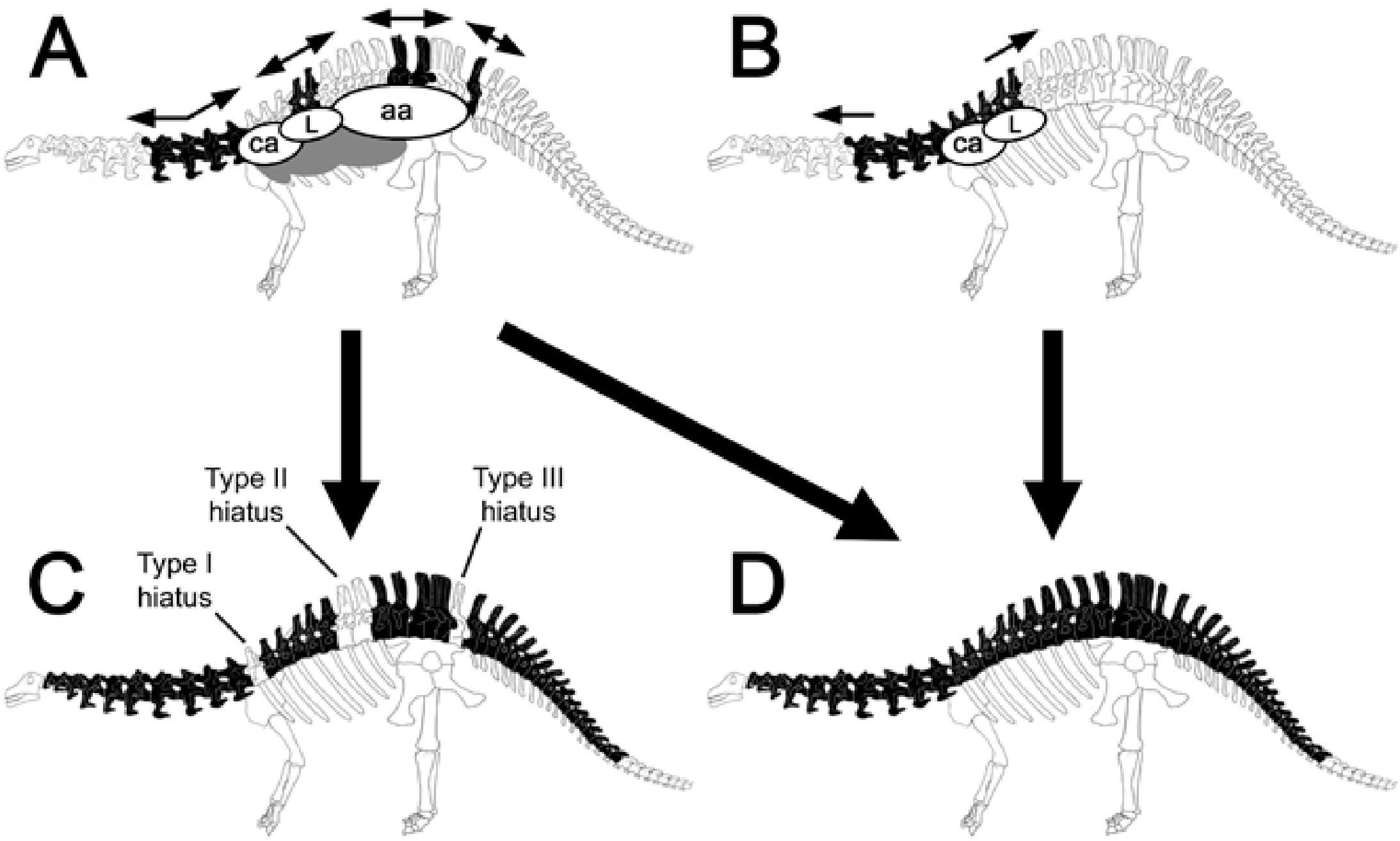

FIGURE 3-10. Pneumatization of the presacral vertebrae in Haplocanthosaurus. A. X-ray image of a posterior cervical vertebra of CM 879 in right lateral view. B. A CT slice through the same vertebra. C. X-ray image of an anterior dorsal vertebra of CM 572 in left lateral view. D. X-ray image of the same vertebra in anterior view. All of the preserved presacral vertebrae of both specimens have large, sharp-lipped fossae that penetrate to a narrow median septum.

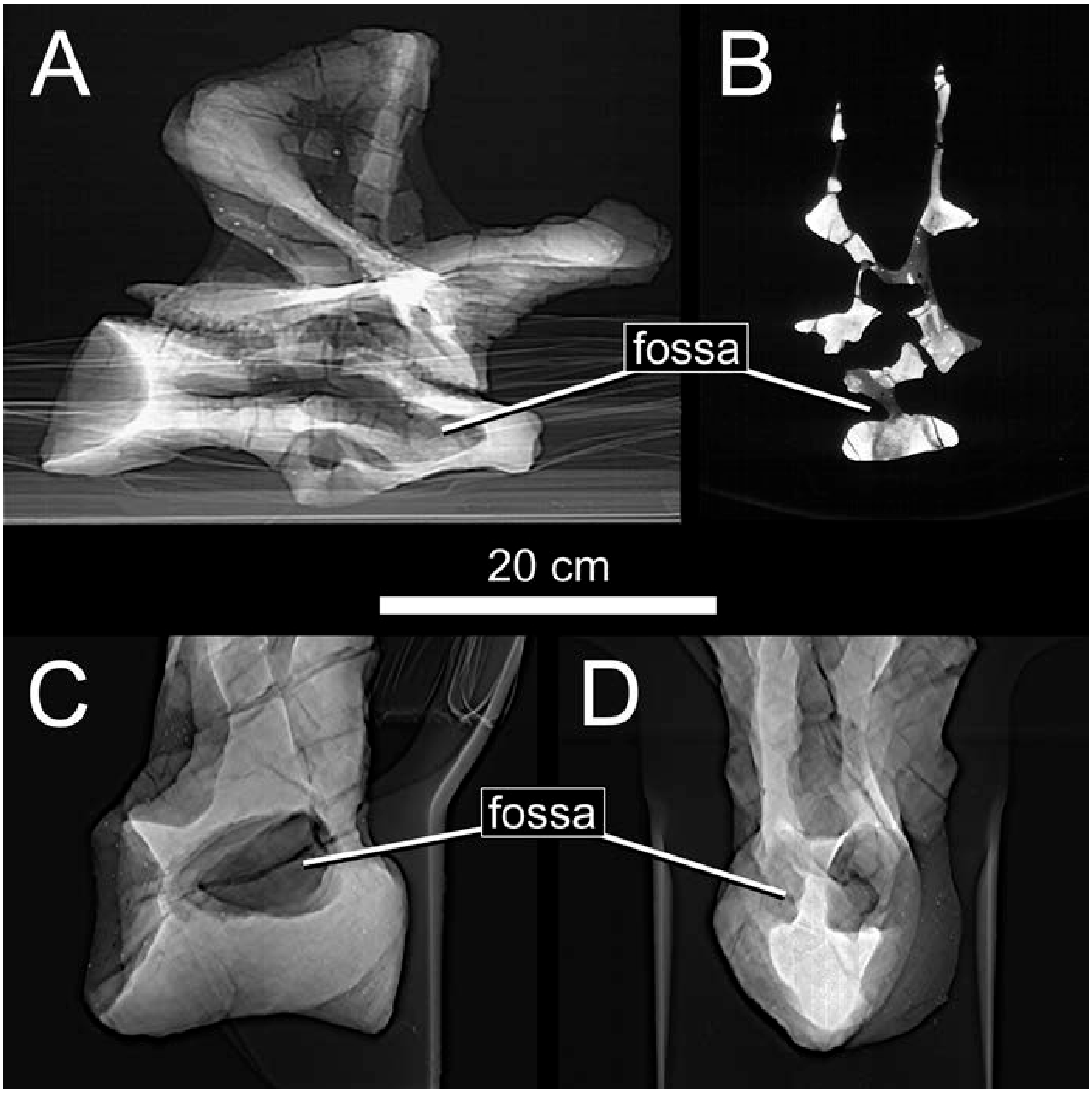

FIGURE 3-11. A pneumatic hiatus in a sauropod dinosaur. The preserved portions of the sacrum (S1-S5) and anterior caudal vertebrae (Ca1-Ca3) of Haplocanthosaurus CM 879 are shown in right lateral (top) and left lateral (bottom) views. All of the preserved cervical and dorsal vertebrae have large, distinct fossae. Distinct fossae are also present on the right sides of the fourth sacral and first caudal vertebrae, and on the left side of the first caudal. The left side of the fourth sacral and both sides of the fifth sacral are waisted but lack distinct fossae (see text for discussion), and constitute a Type III pneumatic hiatus.

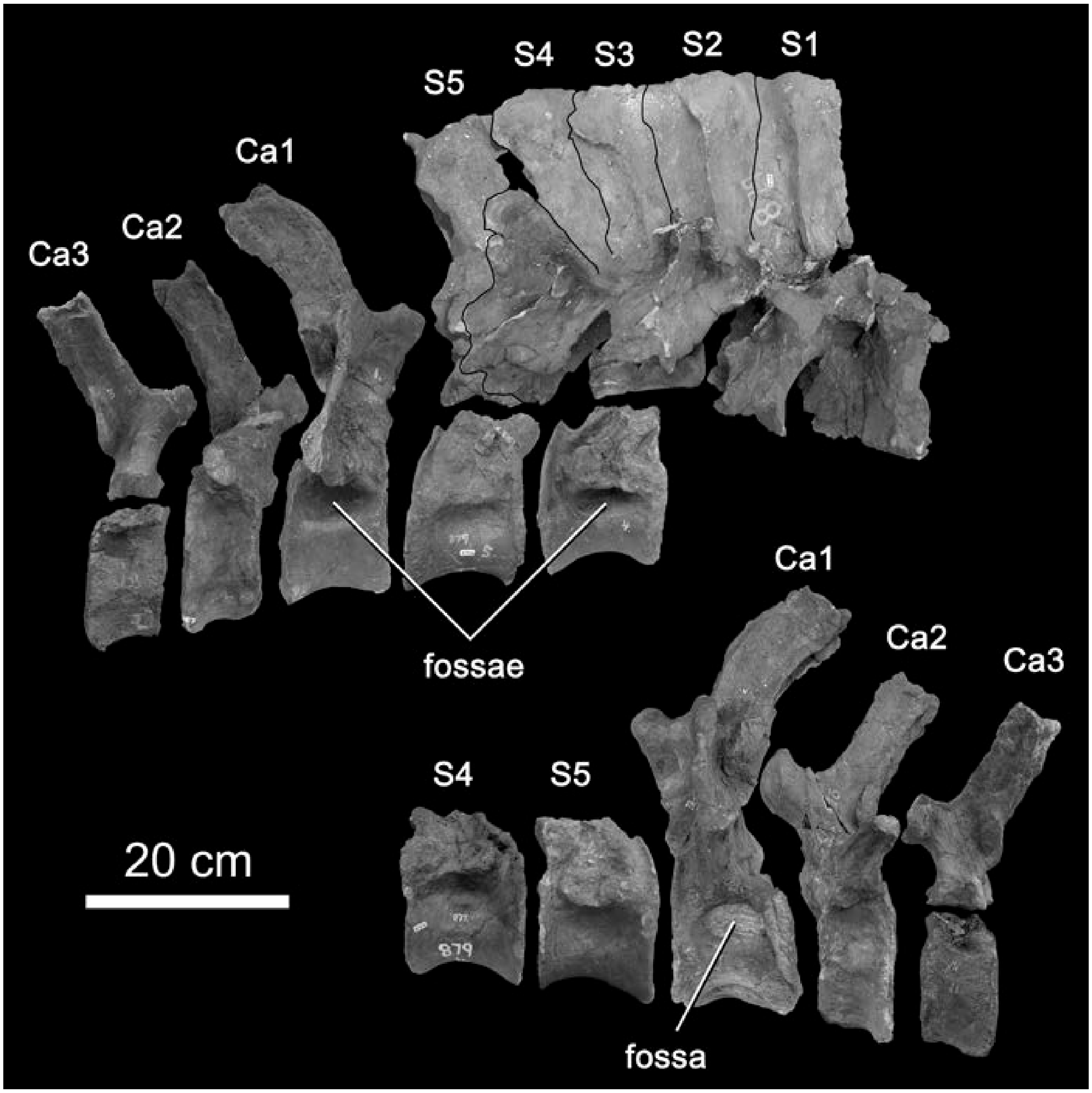

FIGURE 3-12. The fourth and fifth sacral centra of Haplocanthosaurus CM 879. Above, the centrum of the fourth sacral vertebra in right posterolateral (A), right lateral (B), left lateral (C), and left posterolateral (D) views. Below, the centrum of the fifth sacral vertebra in right lateral (E) and left lateral (F) views. The centrum of S4 has dorsolateral fossae (dlf) on both sides and a lateral fossa (lf) on the right side. The left side of S4 is waisted but lacks a lateral fossa. The centrum of S5 is waisted and lacks both lateral and dorsolateral fossae.

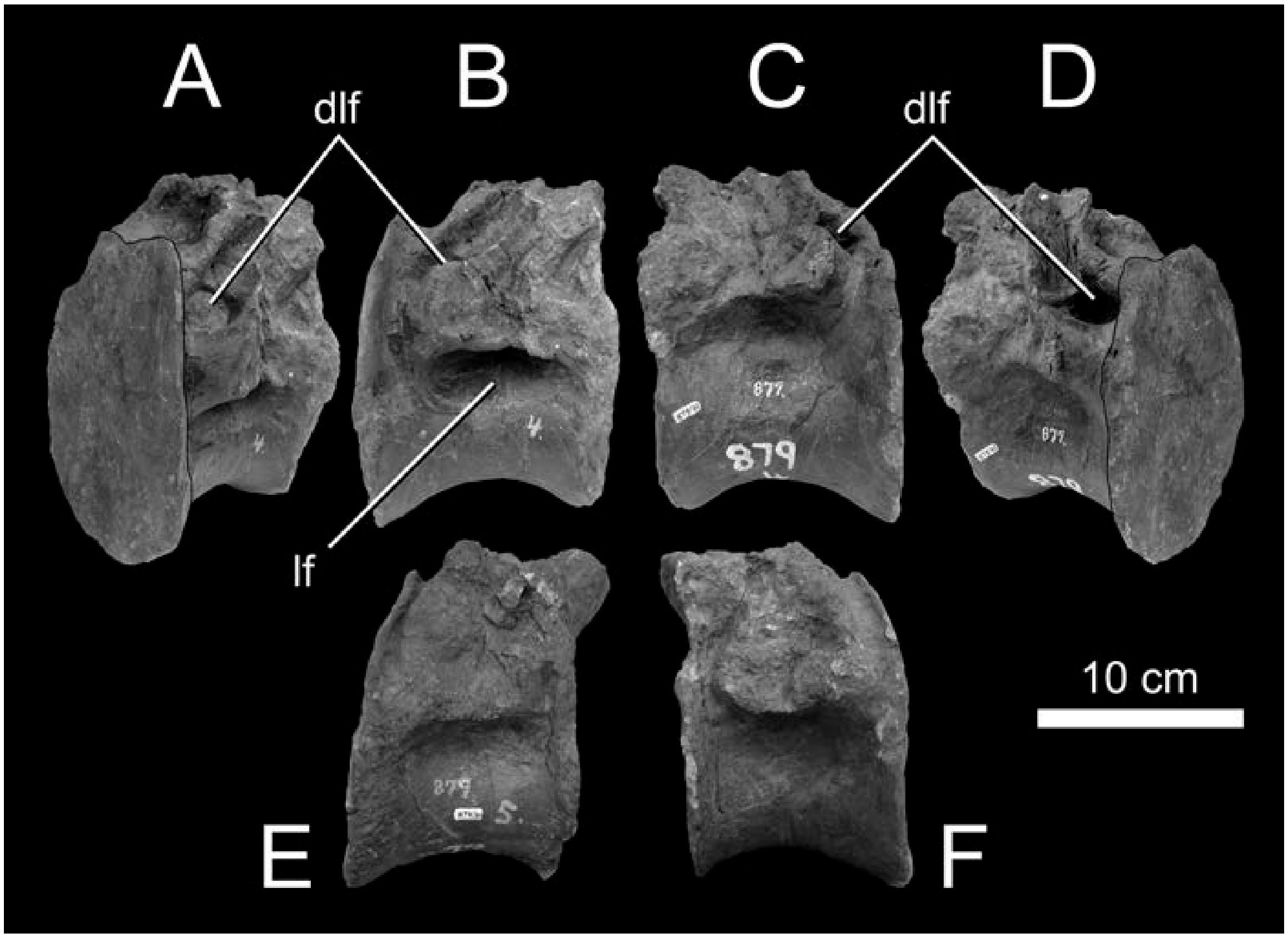

FIGURE 3-13. Anterior caudal vertebrae of Haplocanthosaurus in dorsal view. A. The first caudal vertebra. B. The second caudal vertebra.

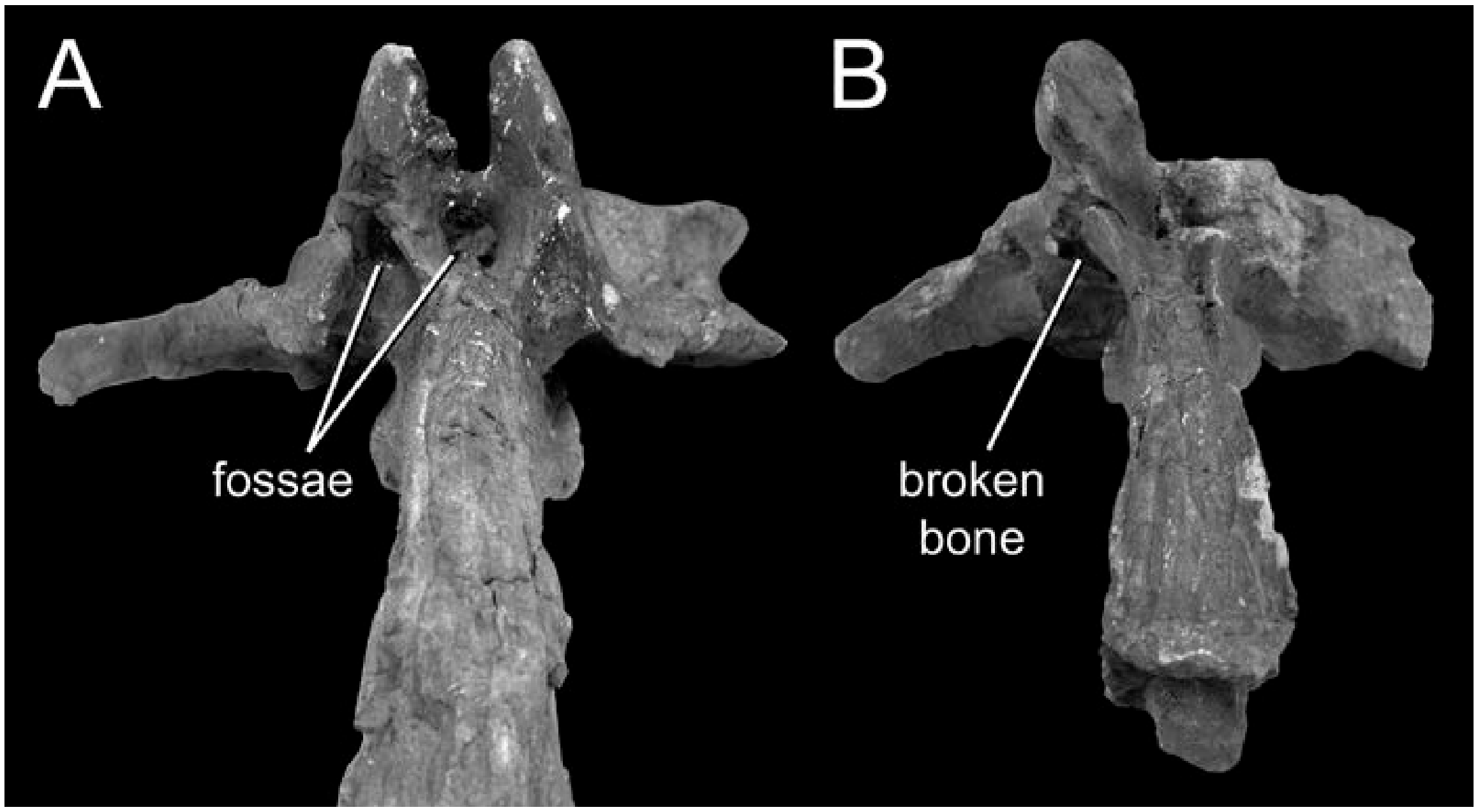
A
fossae
B
broken
bone

FIGURE 3-14. The air sacs of Haplocanthosaurus. Preserved elements of CM 879 are shown in right lateral view. The cervical and anterior thoracic vertebrae were pneumatized by diverticula of cervical air sacs (C). Middle thoracic vertebrae were pneumatized by diverticula of the lung (L). Diverticula of the abdominal air sac (A) pneumatized the posterior thoracic, sacral, and first caudal vertebrae. Other, intermediate air sacs were probably present (I), but their presence is not detectable from the preserved elements.

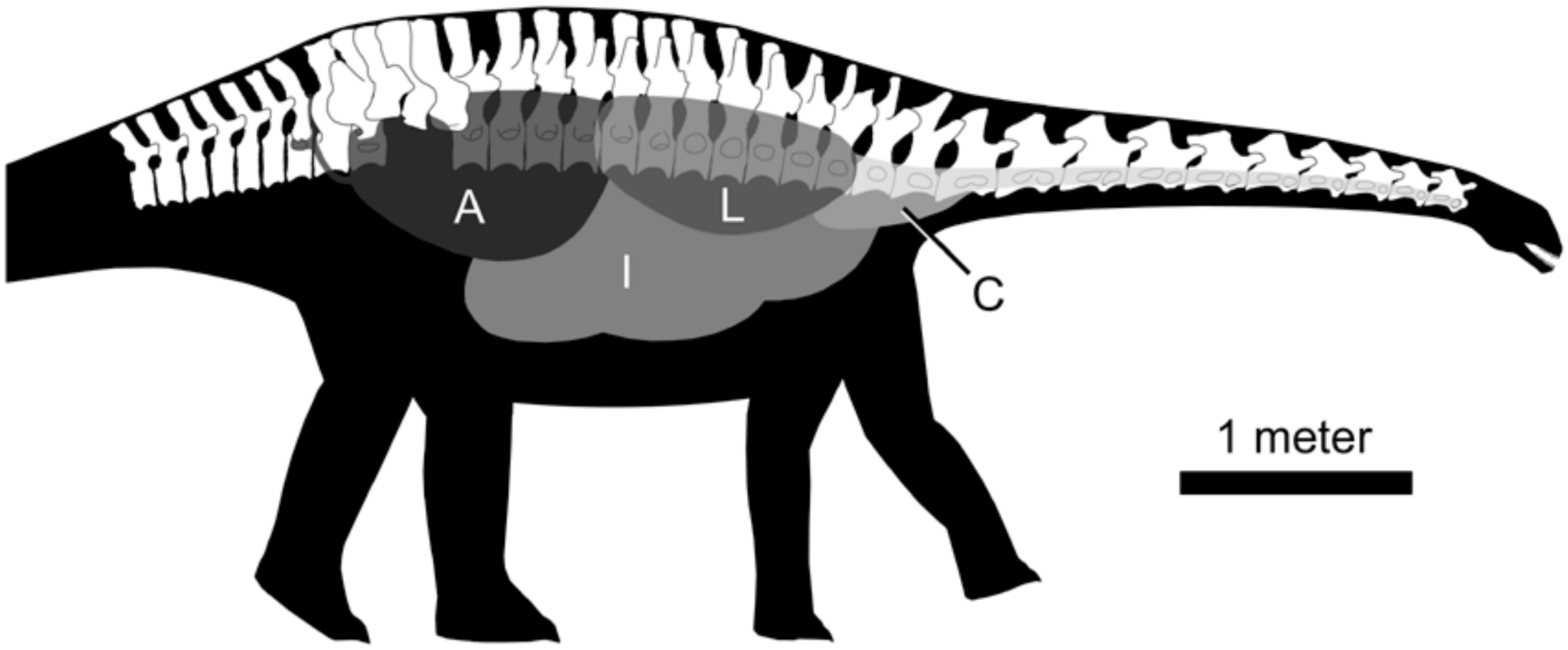
A
L
I
C
1 meter

# CHAPTER FOUR

# THE EVOLUTION OF LONG NECKS IN SAUROPOD DINOSAURS

## INTRODUCTION

At their extremes, dinosaurs were completely outside the bounds of our experience with living things. Curiosity about dinosaur sizes goes beyond popular fascination. The biggest dinosaurs seem to push the limits of what is possible for terrestrial animals, and they are therefore of interest to biomechanists and physiologists who seek to understand those limits. Sauropod dinosaurs are the natural targets of this curiosity; the largest sauropods were up to three times as long and ten times as heavy as any other terrestrial vertebrates, including representatives of other dinosaurian clades.

Sauropods were not only the largest terrestrial animals of all time, they also had the longest necks of any animals. Complete sauropod necks up to 9.5 meters long have been recovered (Young and Zhao 1972), and isolated vertebrae show that the necks of some sauropods were much longer. The longest sauropod vertebrae were more than 1.4 meters long, and vertebrae more than one meter long are known from at least three different sauropod clades.

Sauropod necks were not only long, they were also lightly built. The presacral vertebrae of most sauropods were filled with pneumatic spaces. The mass-reducing effects of pneumatization have been recognized for over a century, but the relationship of pneumaticity to large size and long necks in sauropods has never been investigated with more than anecdotal rigor.

The goals of this paper are to answer the following questions:

1. How long were the necks of the largest and longest-necked sauropods, and what was the maximum neck length achieved by each of the major sauropod clades?
2. How many times did characters related to neck length and pneumaticity evolve independently in sauropods?
3. How important was pneumaticity in reducing the mass of sauropod vertebrae?
4. Are variables related to pneumaticity, neck length, and body size statistically correlated?

## MAXIMUM NECK LENGTHS OF SAUROPODS

In the following description, the taxa are discussed in an approximate phylogenetic order, from most basal to most derived. Most taxa for which a rigorous estimate of neck length is possible are shown in Figure 4-1. Measurements of individual vertebrae are provided in Table 4-1, and variables related to neck length are presented in Table 4-2.

This list does not include every sauropod for which the neck length is known or can be estimated, but only those taxa that are of interest in determining the maximum neck length attained by each clade. Most of the measurements of individual vertebrae discussed herein are in terms of functional length. Functional length is defined here as the length that each vertebra contributes to the total neck length, and it is measured from the posterior edge of the condyle to the posterior rim of the cotyle (Figure 4-2). Many authors report both centrum length and functional length in tables of measurements (e.g., McIntosh 2005); in these cases functional length is usually reported as the length of the centrum without the condyle.

Sauropods evolved longer necks by increasing the number of cervical vertebrae, recruiting dorsal vertebrae into the cervical series, and increasing the proportional length of the individual vertebrae. Cervical and dorsal counts for the taxa discussed below are provided in Table 4-2. Proportional length of the vertebrae is quantified as the elongation index (EI), which is the length of a vertebra divided by the vertical diameter of the cotyle (Figure 4-2; Wedel et al. 2000).

When metric measurements are converted to English in popular reports, there is a mysterious tendency to round up. All of the measurements and estimates of neck length are given in both meters and feet to prevent unintentional exaggeration in the future. Most of the neck lengths discussed below are estimates based on partial cervical series, or on a single cervical vertebra, or, in the absence of preserved cervical vertebrae, on cross-scaling from related forms. Throughout, I have made sources of certainty and uncertainty as explicit as possible, but approach the more speculative estimates with appropriate skepticism.

**Sauropoda**

<u>Omeisaurus</u> <u>tianfuensis</u>

Cervical count: 17

Neck length: 8.5 meters (28 feet)

Key reference: He et al. (1988)

Omeisaurus has the longest neck relative to body size of any sauropod (Table 4-2). In the most complete specimen, CCG T5701, several vertebrae are incomplete and two are missing. However, overlapping material from another, similar-sized specimen, CCG T5703, allows us to estimate the length of the missing or incomplete vertebrae. Combining information from the two specimens yields a total neck length of 8.5 meters. The cervical series of most sauropods are less than 2.5 times the length of the dorsal series, but in Omeisaurus the neck is just over four times the length of the dorsal series.

Mamenchisaurus hochuanensis

Cervical count: 19

Neck length: 9.5 meters (31.2 feet)

Key reference: Young and Zhao (1972)

The longest sauropod neck represented by a complete, articulated cervical series is that of CCG V 20401, the holotype of Mamenchisaurus hochuanensis. Longer-necked sauropods certainly existed. In fact, the neck of M. hochuanensis was only about two-thirds the length of the longest neck for which we have a reasonable estimate (see below). Although the neck length of M. hochuanensis has frequently been reported (e.g., Norman 1985) as 10 meters (33 feet), the articulated cervical series is actually 9.5 meters (31.2 feet). This measurement was made from a mounted cast of the skeleton in the Homogea Museum in Trzic, Slovenia (M.P. Taylor, pers.

comm.), and it is consistent with the summed functional lengths of the individual vertebrae reported by Young and Zhao (1972).

Mamenchisaurus sinocanadorum

Cervical count: ?

Neck length: ~12 meters (39 feet)

Key reference: Russell and Zheng (1993)

Another Chinese sauropod, M. sinocanadorum, may have had an even longer neck. Only the first few vertebrae in the neck were recovered intact, but they are 25% longer on average than the equivalent vertebrae of M. hochuanensis (not including the axis, the proportions of which do not vary much among sauropods—see Wilson and Mohabey 2006). The neck of M. sinocanadorum was probably about 12 meters (~39 feet) long, assuming that the rest of the neck showed the same proportional increase over M. hochuanensis as the anterior portion. This estimate should be treated with extra caution because the species taxonomy of Mamenchisaurus is very uncertain (Upchurch et al. 2004). At least one species referred to the genus, M. youngi, has only 18 cervicals (Pi et al. 1996), whereas the better-known M. hochuanensis has 19.

Turiasaurus riodevensis

Cervical count: ?

Neck length: ?

Key reference: Royo-Torres et al. (2006)

The recently-described Turiasaurus is the largest known sauropod that is not a member of the clade Neosauropoda. Six cervical vertebrae are known, of which the largest is 520 mm long. This is comparable to the largest cervical vertebrae of Camarasaurus supremus, which had a neck less than four meters long (see below). Even if Turiasaurus was built like Euhelopus (and there is no evidence that it was), in which the longest cervical vertebra is only 7% of the length of the neck, its neck would have been less than seven meters long. Despite its large size, Turiasaurus was apparently a rather short-necked sauropod.

Jobaria tiguidensis

Cervical count: 12

Neck length: 4.0 meters (13.1 feet)

Key reference: Sereno et al. (1999)

Jobaria is a large, relatively primitive sauropod from the Early Cretaceous of Niger. It is close to the neosauropod divergence (phylogenetically, but not temporally) and it may be a good model for what the ancestral neosauropod was like. Jobaria is similar to Turiasaurus and Camarasaurus in being large but relatively short-necked. Measurements of the individual vertebrae have not been published, but Sereno et al. (1999) report a total neck length of 403 cm.

**Neosauropoda - Diplodocoidea**

Apatosaurus louisae

Cervical count: 15

Neck length: 4.9 meters (16.1 feet)

Key reference: Gilmore (1936)

Apatosaurus louisae, represented by the holotype skeleton CM 3018, is by no means one of the largest sauropods. However, because the skeleton is so complete, it has been the basis for many reconstructions and mass estimates (e.g., Christiansen 1997; Paul 1997; Henderson, 1999). It also serves as a baseline for estimating the size of larger, more fragmentary individuals of Apatosaurus. The complete cervical series of CM 3018 is 4.9 meters long.

Apatosaurus sp.

Cervical count: 15

Neck length: ~7 meters (23 feet)

Key reference: Stovall (1938)

J. Willis Stovall discovered remains of a truly enormous Apatosaurus from the Oklahoma panhandle in the 1930s. A dorsal vertebra (OMNH 1670), probably D5 based on the shape and proportions of the neural spine, was illustrated by Stovall (1938:fig. 3). At 144 cm tall, it is 36% larger than the fifth dorsal of CM 3018 (106 cm tall). Other elements also indicate an animal about 40% larger than CM 3018. These include a cervical rib (OMNH 1368) 41 cm along the parapophyseal ramus, and a partial caudal vertebra with an articular face that is 43 cm tall and would have been 50 cm wide when complete (OMNH 1331; preserved width is 46 cm). Interestingly, the giant cervical rib is unfused, and at least one of the dorsal vertebrae (OMNH 1382) has open neurocentral sutures. This indicates that despite its size the giant Oklahoma Apatosaurus was not fully mature. If CM 3018 was scaled up by 40%, the neck would be about 7 meters long.

Diplodocus carnegii

Cervical count: 15

Neck length: 6.1 meters (20 feet)

Key reference: Hatcher (1901)

Like Apatosaurus louisae, Diplodocus carnegii is mainly of interest here as a point for comparison to larger, more fragmentary taxa. The complete articulated neck is 6.1 meters long.

Seismosaurus hallorum

Cervical count: ?

Neck length: ?7.3-10.2 meters (24-33.5 feet)

Key reference: Herne and Lucas (2006)

Gillette (1994) estimated the neck of Seismosaurus to be 19.8-21.3 meters (65-70 feet) long, longer than all but the most liberal estimates for Amphicoelias (see below), despite the fact that most elements from the holotype skeleton (NMMNH P-3690) are only 20% longer than those of Diplodocus. He based this on the height of the sacral neural spines, which he supposed must have supported an inordinately long neck and tail. However, tall sacral spines do not necessarily indicate a long neck. The sacral vertebrae of Apatosaurus CM 3018 are comparable in size to those of Diplodocus CM 84, yet the neck of the former is only 80% as long as that of the latter (Table 4-1). Furthermore, Barosaurus has a longer neck than Diplodocus but a shorter tail (McIntosh 2005), so neck and tail lengths are not necessarily correlated and cannot be read from the sacrum alone.

Cervical material of Seismosaurus is apparently fragmentary (Herne and Lucas 2006) and has not been described to date. The longest vertebrae from a 20-meter neck would be more than 2 meters long, and no vertebrae of that length have ever been reported. The cervical vertebrae shown in the skeletal reconstruction of Seismosaurus from Gillette (1994) appear to have been shaded in on a grossly oversized outline of the neck.

According to Herne and Lucas (2006), Seismosaurus had a total length of only 33 m. Based on this and on statements in Gillette (1994) that the individual elements of NMMNH P-3690 are about 20% longer than corresponding bones of Diplodocus (presumably CM 84), I estimate the neck of Seismosaurus to have been 7.3 meters (24 feet) long, or perhaps 10.2 meters (33.5 feet) if it was built like that of Barosaurus (i.e., 40% longer than the neck of Diplodocus).

Barosaurus lentus

Cervical count: ?16

Neck length: ?8.5m (28 feet)

Key reference: McIntosh (2005)

In its general morphology, Barosaurus can be thought of as a longer-necked version of Diplodocus (the two also differ in other, less obvious ways), although no complete cervical series of Barosaurus has been found to date. The individual cervical vertebrae are both proportionally and absolutely longer than those of Diplodocus. Diplodocus and Apatosaurus both have 15 cervical and 10 dorsal vertebrae. Some more basal sauropods also have 25 presacral vertebrae, of which 12 or 13 are dorsals (Wilson and Sereno 1998). This pattern suggests that the diplodocines lengthened their necks by recruiting cervical vertebrae from the dorsal series. Barosaurus appears to have continued this trend. The specimen with the most complete presacral column, AMNH 6341, has only nine dorsal vertebrae. McIntosh (2005) hypothesized that the anteriormost dorsal had been recruited into the neck, and that the complete skeleton would have had 16 cervicals. AMNH 6341 includes the nine most posterior cervical vertebrae, which McIntosh (2005) interpreted as C8-C16. According to this interpretation, C9 is the most anterior vertebra in which the neural spine is bifurcated.

Another specimen of Barosaurus, AMNH 7535, includes C2-C8. The neural spine of the eighth cervical vertebra is not bifurcated, which supports McIntosh's hypothesis that Barosaurus had sixteen cervical vertebrae, although it is not definitive. If the vertebrae of AMNH 7535 were scaled up and combined with AMNH 6341, the total neck length would be about 8.5 meters.

Supersaurus vivianae

Cervical count: ?

Neck length: ?13.3-16.2 meters (43.6-53.2 feet)

Key reference: Jensen (1987)

Supersaurus is without question the longest-necked animal with preserved cervical material. Jim Jensen recovered a single cervical vertebra of Supersaurus from Dry Mesa Quarry in western Colorado. The vertebra, BYU 9024, was originally referred to "Ultrasauros". Later, both the cervical and the holotype dorsal of "Ultrasauros" were shown to belong to a diplodocid, and they were separately referred to Supersaurus by Jensen (1987) and Curtice et al. (1996), respectively.

BYU 9024 has a centrum length of 1378 mm, and a functional length of 1203 mm (Figure 4-3). At 1400 mm, the longest vertebra of Sauroposeidon is marginally longer in total length. However, that length includes the prezygapophyses, which overhang the condyle, and which are missing from BYU 9024. The centrum length of the largest Sauroposeidon vertebra is about 1250 mm, and the functional length is 1190 mm. BYU 9024 therefore has the largest centrum length and functional length of any vertebra that has ever been discovered for any animal. Furthermore, the Supersaurus vertebra is much larger than the Sauroposeidon vertebrae in diameter, and it is a much more massive element overall.

Neck length estimates for Supersaurus vary depending on the taxon chosen for comparison and the serial position assumed for BYU 9024. The vertebra shares many similarities with Barosaurus that are not found in other diplodocines, including a proportionally long centrum, dual posterior centrodiapophyseal laminae, a low neural spine, and ventrolateral flanges that connect to the parapophyses (and thus might be considered posterior centroparapophyseal laminae, similar to those of Sauroposeidon). The neural spine of BYU 9024 is very low and only very slightly bifurcated at its apex. In these characters, it is most similar to C9 of Barosaurus. However, the proportions of the centrum of BYU 9024 are more similar to those of C14 of Barosaurus, which is the longest vertebra of the neck in AMNH 6341. BYU 9024 is 1.6 times as long as C14 of AMNH 6341 and 1.9 times as long as C9. If it was built like that of Barosaurus, the neck of Supersaurus was at least 13.7 meters (44.8 feet) long, and may have been as long as 16.2 meters (53.2 feet).

Based on new material from Wyoming, Lovelace et al. (2005) noted potential synapomorphies shared by Supersaurus and Apatosaurus. BYU 9024 does not closely resemble any of the cervical vertebrae of Apatosaurus. Instead of trying to assign its serial position based on morphology, I conservatively assume that it is the longest vertebra in the series if it is from an Apatosaurus-like neck. At 2.7 times longer than C11 of CM 3018, BYU 9024 implies an Apatosaurus-like neck about 13.3 meters (43.6 feet) long.

Amphicoelias fragillimus

Cervical count: ?

Neck length: ~11.7-20.8 meters (38.4-68.2 feet)

Key reference: Cope (1878b)

A. fragillimus is based on a single dorsal neural spine that has been lost for over a century, and was only reported in a single two-page paper and illustrated with one sketch, the accuracy of which cannot now be assessed. Nevertheless, the missing holotype of A. fragillimus represents the biggest sauropod ever reported, and there are no serious reasons to suspect that the published figure is inaccurate (Carpenter 2006).

According to the illustration provided by Cope (1878b), the height of the preserved portion of the vertebra is 1500 mm and the total height would have been 2210 mm. The most appropriate comparisons are probably with Diplodocus or Barosaurus (McIntosh, 2005). The tallest preserved dorsal vertebra of Diplodocus CM 84 is D9, at 1047 mm, although Hatcher (1901) reconstructed the fragmentary D10 as being 1147 mm tall. The tallest dorsal of Barosaurus AMNH 6341 is D8, at 906 mm. Using these specimens as the bases for comparison, the neck length of Diplodocus is 5.8 or 5.3 times the height of the tallest dorsal (compared to the intact D9 or the reconstructed D10), and that of Barosaurus is 9.4 times the tallest dorsal. The neck of Amphicoelias was 11.7-12.8 meters (38.4-42 feet) long if built like that of Diplodocus, or 20.8 meters (68.2 feet) long if like that of Barosaurus. Note that if A. fragillimus is reconstructed as a larger version of Diplodocus, its neck is still shorter than the most conservative estimate for Supersaurus.

Carpenter (2006) scaled up the dorsal vertebra of the much smaller Amphicoelias altus to restore the missing parts of A. fragillimus, which results in a total height of 2.7 m. If we perform the calculations given above using this larger estimate, the neck of Amphicoelias was 14.3-15.7 meters (46.9-51.5 feet) long if built like that of Diplodocus, or 25.4 meters (83.3 feet) long if built like that of Barosaurus. Carpenter (2006) stated that if A. fragillimus had the same proportions as Diplodocus then its neck length would have been 16.75 meters (55.0 feet), but he did not provide details on how this estimate was made.

It is important to bear in mind that all of this speculation is based on a single sketch of an incomplete vertebra that no one has seen for more than a century.

**Neosauropoda - Macronaria**

Camarasaurus supremus

Cervical count: 12

Neck length: 3.9 meters (12.8 feet)

Key reference: Osborn and Mook (1921)

Camarasaurus is one of the shortest-necked large sauropods. Even the very large Camarasaurus supremus (AMNH 5761) had a neck less than four meters long. A complete cervical series from a single individual has not been described, but a composite cervical series drawn from several similar-sized specimens has a total length of 3.9 meters (Osborn and Mook 1921)

**Brachiosauridae**

Brachiosaurus brancai

Cervical count: 13

Neck length: 8.5-9.6 meters (28-31.5 feet)

Key reference: Janensch (1950)

The 8.5 meters (28 feet) cervical series of HM SII, the holotype and best known skeleton of Brachiosaurus brancai, is the second-longest reasonably complete sauropod neck (missing only the atlas and axis; the total length given above includes the atlas and axis of the smaller HM SI scaled up), behind Mamenchisaurus hochuanensis. However, larger individuals of Brachiosaurus are known. HM XV2 is a fibula 134 cm long, 12.5% larger than that the fibula of HM SII (Janensch 1961). If the entire animal was 12.5% larger than HM SII, the neck would have been 9.6 meters long (31.5 feet).

Unnamed brachiosaurid MIWG.7306

Cervical count: ?

Neck length: ~7.3-8.1 meters (24-26.7 feet)

Key reference: Naish et al. (2004)

A large brachiosaurid vertebra from the Isle of Wight, MIWG.7306 has a functional length of 745 mm. It probably represents a C6, which would give the animal a neck length of 8.1 meters (26.7 feet) if it shared the same proportions as Brachiosaurus brancai. Even if it is from a more posterior position, it is still at least 85% as long as the longest vertebrae from HM SII, so the neck was at least 7.3 meters (24 feet) long. MIWG.7306 is the longest cervical vertebra of a sauropod discovered in Europe to date, and represents the longest-necked dinosaur yet known from the continent.

Sauroposeidon proteles

Cervical count: ?

Neck length: ~11.5 meters (37.7 feet)

Key reference: Wedel et al. (2000)

Sauroposeidon is only known from a series of four articulated cervical vertebrae (OMNH 53062). The vertebrae are very similar to those of Brachiosaurus in their external appearance. OMNH 53062 probably represents C5-C8 based on a transition point in the neural spine shape between C6 and C7 that is shared with Brachiosaurus. The vertebrae of Sauroposeidon are on average 37% longer than C6-C8 of HM SII (C5 of OMNH 53062 is incomplete and its functional length cannot be determined), which indicates a neck 11.5 meters (37.7 feet) long—not 12 meters (39.4 feet) or 12.2 meters (40 feet) as it is often reported. The functional length of C8 of HM SII is 99% of that of the longest vertebra in the series. In light of this, the estimated neck length of Sauroposeidon would change little if the inferred serial positions of the Sauroposeidon vertebrae are incorrect (i.e., the vertebrae do not represent C5-C8).

**Somphospondyli**

Euhelopus zdanskyi

Cervical count: 17

Neck length: 4.0 meters (13.2 feet)

Key reference: Wiman (1929)

Although it is a fairly small sauropod, Euhelopus has a very long neck relative to its body size, and it has the longest complete neck known for any member of the clade Somphospondyli. Like the CM 84 specimen of Diplodocus, it is useful as a reference point for estimating possible neck lengths of other taxa.

Erketu ellisoni

Cervical count: ?

Neck length: ?

Key reference: Ksepka and Norell (2006)

Erketu is known from an articulated lower hindlimb, a sternal plate, and the first five cervical vertebrae. The holotype and only known specimen does not represent a particularly large animal, but the neck vertebrae are among the longest, proportionally, known for any sauropod. The largest uncrushed vertebra, C4, has an elongation index of 5.5. However, in most sauropods C5 and C6 have the highest elongation indices. C5 of Erketu is intact but distorted. An exact determination of the in vivo EI is impossible, but it was probably between 6 and 7—as long or longer, proportionally, than the most attenuated vertebrae of Sauroposeidon. The individual vertebrae of Erketu are 1.7-2.1 times longer than the corresponding vertebrae of Euhelopus. This suggests a total neck length of 6.8-8.4 meters (22.3-27.6 feet). This is remarkable; the tibia of Erketu is 71 cm long, which implies a hip (acetabulum) height of about 2 m, or roughly the same size as a very large draft horse or a small elephant.

**Titanosauria**

Malawisaurus dixeyi

Cervical count: 13

Neck length: 3.4 meters (11.1 feet)

Key reference: Gomani (2005)

Malawisaurus is another small sauropod with a mostly complete neck (missing, as usual, the atlas and axis)—the longest relatively complete cervical series that has been described for any member of the clade Titanosauria.

Rapetosaurus krausei

Cervical count: ≥15

Neck length: ?

Key reference: Curry-Rogers and Forster (2001)

A mostly complete, articulated neck is available for Rapetosaurus, but no measurements of vertebrae have been published. The cervical count has been given as 16 (Curry-Rogers and Forster 2001) and "at least 15" (Curry-Rogers 2005:p. 62); in either case, Rapetosaurus has more cervical vertebrae than any other known titanosaur.

Argentinosaurus huinculensis

Cervical count: ?

Neck length: ?9 meters (29.5 feet)

Key reference: Bonaparte and Coria (1993)

The holotype of Argentinosaurus consists of dorsal and sacral vertebrae, ribs, and a fibula, although one partial and one complete femur have been referred to the genus (Bonaparte 1996, Mazzetta et al. 2004). At 250 cm long, the complete femur is the largest intact limb bone known from any dinosaur. The fact that all of the available elements are anatomically distant from the neck and vary in size relative to the neck in different taxa only underscores how flimsy is the chain of inference in these cases.

Nevertheless, Argentinosaurus and Bruhathkayosaurus (see below) were larger than any other known sauropods whose remains are available for study (i.e., not Amphicoelias fragillimus), and it may be of interest to estimate how long their necks were by comparison to better-known taxa.

The following series of calculations is based on the 250-cm referred femur. In Malawisaurus, the neck is 3.6 times as long as the femur (Gomani, 2005), and in Euhelopus, 4.2 times as long (Wiman, 1929). The neck of Argentinosaurus was 9 meters (29.5 feet) if like that of Malawisaurus, and 10.5 meters (34.4 feet) if like that of Euhelopus. The comparison to Malawisaurus is probably more apt. Euhelopus has more cervical vertebrae than any known titanosaurs, and it apparently evolved its unusually long neck independently. Malawisaurus is similar in cervical morphology to Puertasaurus (Novas et al. 2005) and Trigonosaurus (Campos et al. 2006), which are at least from the same continent and period as Argentinosaurus.

Puertasaurus reuili

Cervical count: ?

Neck length: ~ 9m (29.5 feet)

Key reference: Novas et al. (2005)

Puertasaurus is known from one cervical, one dorsal, and two caudal vertebrae, but they are immense. The cervical vertebra is 118 cm long (including zygapophyses) and 140 cm wide (including the cervical ribs), and the dorsal vertebra is 106 cm tall and 168 cm across the transverse processes.

The Puertasaurus cervical vertebra is very similar to the posterior cervical vertebrae of Malawisaurus, especially MAL-280-4 (see Gomani 2005). MAL-280-4 has a centrum length of 41.5 cm, a functional length (i.e., minus condyle) of 36.5 cm, and accounts for 11% of the 340 cm long neck. It is also one of the longest vertebrae in the neck; the two just anterior to it are less than 1 cm longer in either centrum length or functional length. The centrum of the Puertasaurus cervical is incomplete, but based on comparisons to similar vertebrae (Malawisaurus, Trigonosaurus) I estimate the centrum length at 105 cm and the functional length at 93 cm. Based on its proportions the Puertasaurus cervical is probably from the proximal third of the neck and therefore one of the longest vertebrae in the neck. Scaling up from Malawisaurus indicates a total neck length of about 9 meters (29.5 feet).

Bruhathkayosaurus matleyi

Cervical count: ?

Neck length: ?9.6-15.2 meters (31.5-49.9 feet)

Key reference: Yadagiri and Ayyasami (1989)

Although it is known from fossils that apparently still exist, there is simply very little known about this animal. The only elements that have been described are a tibia and parts of the pelvis. The tibia is reported to be 2 meters long. If we accept the reported tibia length and the referral of Bruhathkayosaurus to Titanosauria, then we can perform a similar series of extrapolations as for Argentinosaurus. In Malawisaurus, the neck is 6.2 times as long as the tibia (Gomani 2005), and in Euhelopus, 7.6 times as long (Wiman 1929). The neck of Bruhathkayosaurus was 12.4 meters (40.7 feet) long if like that of Malawisaurus, and 15.2 meters (49.9 feet) if like that of Euhelopus. With the information currently available, it is impossible to tell which, if either, of these comparisons is more appropriate. It is worth noting that the most conservative and most extreme estimates of neck length for Bruhathkayosaurus are still shorter than the corresponding estimates for Supersaurus.

## HOMOPLASY IN THE EVOLUTION OF LONG NECKS IN SAUROPODS

Resolving characters on a phylogeny is a powerful tool for exploring evolutionary patterns. In particular, repeated evolution of characters in closely related lineages can elucidate shared developmental processes and evolutionary trends. I collected character data by CT scanning and personal observation when possible, and from the literature when necessary (Table 4-2). The phylogeny shown here (Fig. 4-1) is broadly congruent with several recent cladistic analyses (Wilson 2002, Upchurch et al. 2004, Rauhut et al. 2005, Harris 2006), but agnostic about the relationships of several taxa that have proven to be unstable in those analyses (e.g., Cetiosaurus, Jobaria, and Haplocanthosaurus).

The evolution of long necks in sauropods is marked by the repeated evolution of several important characters, including the number of cervical vertebrae, the length of the individual vertebrae, the complexity of vertebral internal structure, and the ratio of bone to air space.

**Cervical Count**—If the primitive number of cervical vertebrae in Titanosauria is 13 (Gomani 2005), then increases to 15 or more cervicals occurred at least four times: in the Mamenchisauridae, Diplodocidae, Euhelopus, and Rapetosaurus.

Alternatively, 15 or more cervicals may be primitive for Somphospondyli, the lower counts in Malawisaurus and Saltasaurus may be reversals, and increases to 15 or more cervicals may have only happened three times. The relatively primitive count of 13 cervicals in Brachiosaurus is noteworthy; no other known sauropod had such a long neck with so few vertebrae. It will be interesting to see if the same will hold true for the other brachiosaurids once better material is discovered.

**Elongation index—**The cervical vertebrae of basal sauropodomorphs and basal sauropods did not exceed an EI of 3.2. Elongate vertebrae with EIs greater than 4.0 evolved independently at least three times, in Mamenchisauridae, Diplodocidae, and Titanosauriformes. The sauropods with the most elongate vertebrae are the brachiosaurid Sauroposeidon and the basal somphospondylan Erketu; in both taxa the maximum EI is greater than 6.0.

**Vertebral internal structure—**Most early-diverging sauropods and some neosauropods have very simple internal structures in their presacral vertebrae. In these taxa, the centrum of each vertebra contains a pair of large chambers and has a cross-section similar to an I-beam. In more derived sauropods the number of pneumatic cavities increases and the vertebrae become honeycombed with small cavities (i.e., the polycamerate and camellate conditions of Wedel et al. 2000).

The mamenchisaurids, diplodocines, brachiosaurids, and long-necked somphospondylians all have complex (i.e., many-chambered: polycamerate, semicamellate, or camellate) vertebrae (Wedel 2003b). Complex internal structure evolved independently in each of those clades. Camellae are probably synapomorphic for Titanosauriformes, but Sauroposeidon evolved a fully camellate internal structure independently of Somphospondyli (Wedel et al. 2000). The longest-necked sauropod (in absolute terms) with procamerate or camerate cervical vertebrae is probably Jobaria. The MNN TIG3 skeleton has a neck 4.03 meters long (Sereno et al. 1999). Atlasaurus and Camarasaurus also had camerate vertebrae, and both had necks less than 4 meters long (Osborn and Mook 1921; Monbaron et al. 1999).

**Air space proportion—**The air space proportion (ASP) of a bone is the proportion of its volume taken up by pneumatic cavities (Wedel 2004, 2005). Dicraeosaurids (Dicraeosaurus, Amargasaurus, and related taxa) had reduced postcranial pneumaticity compared to other neosauropods, both in terms of the number of presacral vertebrae that were pneumatized, and in the air space proportion (Table 4-2; Schwarz and Fritsch 2006). The mean ASP for neosauropods is 0.52 if the dicraeosaurids are included (Table 4-2), and 0.61 if they are not (Table 4-3). The presacral vertebrae of most neosauropod taxa had ASPs between 0.50 and 0.70—as lightly built as the bones of most birds (Wedel 2005). Basal sauropods outside or near the base of Neosauropoda, such as Cetiosaurus, Jobaria, and Haplocanthosaurus, had much lower ASPs, around 0.40 (ASPs of Cetiosaurus and Jobaria are estimates based on personal observations of the holotypes and referred specimens). Unfortunately, these three taxa have proven unstable in recent phylogenetic analyses (e.g., Wilson 2002, Upchurch et al. 2004, Rauhut et al. 2005, Harris 2006), especially with respect to the fundamental neosauropod divergence. It is therefore unclear whether high ASPs evolved independently in Diplodocidae and Macronaria, or if a high ASP is synapomorphic for Neosauropoda and reversed in Dicraeosauridae.

To further complicate matters, no relevant data have been published for any member of the Mamenchisauridae. In published descriptions, the internal structure of mamenchisaurid vertebrae is described as a honeycomb of small cavities (Young and Zhao 1972, Russell and Zheng 1993). In sauropods for which ASP values are available, complex internal structures are always associated with ASPs greater than 0.50. It therefore seems likely that mamenchisaurids had ASPs similar to those of most neosauropods. If that is the case, then high ASPs (> 0.50) evolved at least twice, in Mamenchisauridae and Neosauropoda.

**Summary—**Most characters of interest evolved independently in Mamechisauridae, Diplodocidae, and Titanosauriformes. Within Titanosauriformes, brachiosaurids are characterized by a relatively small number of very elongate cervical vertebrae. The vertebrae of most somphospondylans are proportionally shorter, but increases in cervical count occurred at least twice within Somphospondyli.

Many authors have speculated about the hemodynamic and respiratory debits imposed by the long necks of sauropods (Hohnke 1973; Seymour 1976; Choy & Altman 1992; Daniels & Pratt 1992; Gunga et al. 1995; Badeer & Hicks 1996; Seymour & Lillywhite 2000). Whatever problems came along with a 9-m neck, sauropods were certainly adept at achieving and exceeding that mark, as evidenced by Mamenchisaurus, Supersaurus, Sauroposeidon, and Puertasaurus. It is of particular interest that all of the studies just mentioned used Diplodocus, Barosaurus, Brachiosaurus, or Mamenchisaurus as models. At a probable 16.2 m, the neck of Supersaurus was 70% longer than that of any sauropod whose blood pressure or respiratory dead space has been calculated. In other words, no previous investigation of the physiological limits of sauropod necks has come within a literal stone's throw of what we now know to be possible.

Except Supersaurus and Argentinosaurus, which are both known from at least two localities, all of the largest and longest-necked representatives of the major sauropod clades—Mamenchisaurus sinocanadorum in the Mamenchisauridae, Sauroposeidon in the Brachiosauridae, and Puertasaurus and Bruhathkayosaurus in the Titanosauria—are each known from a single, very incomplete specimen. The chances are negligible that these individuals were world-record representatives of their species. There seems to be an inverse correlation between size and preservation for the largest sauropods.

## PNEUMATICITY AND MASS REDUCTION

In contrast to prosauropods and basal sauropods, mamenchisaurids and neosauropods had extensive vertebral pneumaticity. Percent pneumatization of the presacral vertebrae—the number of pneumatic elements divided by the total presacral count—rose to a maximum early in sauropod evolution and was subject to only one notable reversal, in dicraeosaurids (Table 4-2). The effects of pneumatization on the mass of the cervical series have been little explored.

The presacral vertebrae of most neosauropods (excluding dicraeosaurids) were on average 60% air by volume (Table 4-3). The centrum walls, laminae, septae, and struts that comprised the vertebrae were primarily made of compact bone (Reid 1996). The specific gravity (SG) of compact bone is 1.8-2.0 in most tetrapods (Spector 1956), so an element with an ASP of 0.60 would have an in vivo SG of 0.7-0.8. Some sauropod vertebrae were much lighter. For example, Sauroposeidon has ASP values up to 0.89 and therefore SG as low as 0.2. On the other hand, many basal sauropods had ASPs of 0.30-0.40 and SG of 1.1-1.4.

An important effect of postcranial pneumaticity is to broaden the range of available densities in skeletal construction. Animals without postcranial pneumaticity, including mammals and ornithischian dinosaurs, are constrained to build their skeletons out of bone tissue (SG = 1.8-2.0) and marrow (SG = 0.95; Currey and Alexander 1985). Therefore, the whole-element densities of their postcranial bones will always be between 1.0 and 2.0. The pneumatic bones of pterosaurs and saurischian dinosaurs are made of bone tissue (SG = 1.8-2.0) and air space (SG = 0), which allows them to have whole-element densities that are much lower. The lightest postcranial bones for which data are available are those of Sauroposeidon and some pterosaurs, which had SG as low as 0.2 (calculated from data in Currey and Alexander 1985). The cranial bones of some birds are even lighter. Seki et al. (2005) report an SG of 0.05 for the "bone foam" inside the beak of the toucan (Rhamphastos toco), and an SG of 0.1 for the entire beak. To date, this is the lightest form of bone known in any vertebrate.

To explore the effect of pneumaticity on skeletal construction, I estimated the in vivo mass of the cervical skeleton of Brachiosaurus. From CT scans it is possible to calculate the volume of bone tissue in a single vertebra and thus determine the mass of the element. I multiplied the mass of a single vertebra by scale factors to determine the masses of the other vertebrae in the neck, and added the resulting values to estimate the mass of the whole cervical series. The mass of the cervical column of Brachiosaurus is about 590 kilograms. For the sake of comparison, I used graphic double integration (Hurlburt 1999) to estimate the mass of a Brachiosaurus humerus at 290 kg. The animal's humeri are each 2 meters long but together they weighed almost as much as the 8.5-meter cervical series (Figure 4-4). The presacral vertebrae of Brachiosaurus have an average ASP of 0.65 and SG of 0.63-0.70, depending on the density of the compact bone. The cervical vertebrae of the giraffe are almost exactly twice as dense (SG = 1.3; van Schalkwyk et al. 2004). If the cervical vertebrae of Brachiosaurus were built like those of a giraffe, they would have weighed twice as much as they apparently did.

The cervical vertebrae of Sauroposeidon are 37% longer than those of Brachiosaurus in functional length and about 15% larger in diameter. However, they were also 40% lighter, with a mean ASP of 0.79. The complete cervical series of Sauroposeidon had a volume almost twice that of Brachiosaurus (1.15 x 1.15 x 1.37 = 1.81), but its mass would have been only slightly greater (1.81 x 0.21 / 0.35 = 1.09). Sauroposeidon probably evolved from a Brachiosaurus-like ancestor, and the biomechanical debits implied by its longer neck were largely offset by its increased ASP.

## SIZE, NECK LENGTH, AND PNEUMATICITY

Potential relationships between size and pneumaticity have been recognized for more than a century. Cope (1878b:p. 564) wrote, "But so far as the vertebrae are concerned the following rule is without exception among the Saurians of the Dakota epoch: It is, that the size of the vertebra is in direct proportion to the attenuation of its walls" (original emphasis).

Most statistical methods assume that the data points are independent. The relatedness of all living things means that lineages are not independent, however, and that a certain amount of correlation is to be expected on the basis of common descent. This similarity caused by relationship must be factored out before ordinary statistical methods can be applied to character correlations across a phylogeny. Felsenstein (1985) introduced the method of phylogenetically independent contrasts to make phylogenetic character data amenable to conventional statistics.

Independent contrasts gives us character correlations with evolution factored out. If we are interested in the coevolution of characters across the phylogeny—for example, characters that form an adaptive complex, or cases in which the evolution of one character facilitates or drives the evolution of others—independent contrasts is the opposite of what we want. Two characters that always evolve in concert would be expected to share little correlation in an independent contrasts analysis. Nevertheless, their evolutionary co-occurrence may be of more interest than their statistical correlation. Independent contrasts analyses are presented below, but first it will be informative to describe qualitatively the patterns present in the data (Figure 4-1, Table 4-2).

## Qualitative Relationships

**Body size and neck length—**Even within the limited sample of sauropods for which complete data are available, widely varying combinations of body size and neck length are realized. There are small sauropods with short necks (Amargasaurus, Dicraeosaurus, Saltasaurus), small sauropods with long necks (Euhelopus), large sauropods with short necks (Jobaria, Camarasaurus), and large sauropods with long necks (Brachiosaurus).

**Vertebral internal structure—**The pneumatic vertebrae of basal sauropods have a pair of fossae or camerae, whereas almost all neosauropods have more complex, 'honeycombed' internal structures, typically with dozens of chambers. The vertebrae of Camarasaurus have a handful of small chambers, mostly in the condyle, but they are still relatively simple compared to those of other neosauropods (Wedel et al. 2000, Wedel 2003b). Not all sauropods with complex internal structures are long-necked (e.g., Saltasaurus). However, no sauropod with a simple internal structure had a neck longer than 4 meters, or a proportional neck length (cervical length/dorsal length) much greater than 1.6. In contrast, taxa with complex internal structures evolved necks up to 9 meters long at least four times; most had proportional neck lengths greater than 2.0, and a few had proportional neck lengths greater than 3.5. Studies of birds show that internal struts significantly strengthen pneumatic bones (Rogers and LaBarbera 1993). The biomechanical implications of the various internal structure types in sauropods have not been investigated, but the fact that sauropods with simple internal structures are all short-necked (both absolutely and proportionally) suggests an important relationship between vertebral internal structure and neck elongation.

**Air space proportion—**Comparative hypotheses or inferences about ASP are hampered by the limited amount of available data. Nevertheless, some generalizations can be made, and the situation can only improve with additional study. ASP data based on CT scans (Schwarz and Fritsch 2006), published cross-sections, histological sections (Woodward 2005), or some combination of methods (Table 4-3; Wedel 2005) are mostly drawn from neosauropods with complex internal structures. The only exceptions are Camarasaurus and Haplocanthosaurus (this study) and Dicraeosaurus (Schwarz and Fritsch 2006). The under-representation of basal sauropods and sauropods with simple internal structures obscures evolutionary patterns and may have biased the overall picture of ASPs in sauropods.

Measurements based on CT or other sectioning methods are unavailable for most basal sauropods, but estimates are possible based on examination of broken bones and comparisons to related or similar taxa. The presacral vertebrae of Jobaria are very similar to those of Haplocanthosaurus in morphology and robustness (pers. obs.); Amargasaurus is similar to Dicraeosaurus (Salgado and Bonaparte 1991); an estimate for Cetiosaurus is based on examination of broken vertebrae in the OUNHM collection; and the estimate for Shunosaurus is based on the small size of the pneumatic cavities apparent in published images (Zhang 1988). All of these estimates are approximate and could be off by as much as 10%. However, they are not likely to be off by more than that, and even with this uncertainty in mind it appears that the dicraeosaurids and many basal sauropods had much lower ASPs than most neosauropods. The mean ASP based on the measurements in Table 4-3 is 0.61. The inclusion of the lower estimates for the taxa listed above brings the mean ASP down to 0.52 for neosauropods and 0.41 for all of the taxa included in Table 4-2.

That does not mean that most sauropods had pneumatic vertebrae that were only 40-50% air by volume. Rather, the distribution is split between basal sauropods and dicraeosaurids, which had ASPs of 0.40 or less, and neosauropods (other than dicraeosaurids) which had ASPs of 0.50 or higher.

**The "average" sauropod—**If the mean values presented in Table 4-2 are truly representative of the groups from which they were drawn, then there was no such thing as an "average" sauropod, in that no known taxa embody the mean values. Jobaria and Camarasaurus are too large and their necks are too short (proportionally), and the diplodocids and Brachiosaurus are all too large or too long-necked or both. The closest correspondence is probably between Apatosaurus and the mean for neosauropods apart from dicraeosaurids, and even in this case the fit is only as good as the scope of the table. In terms of variables that are not included in this study, Apatosaurus is one of the most extreme sauropods: it has very robust limb bones and very short forelimbs relative to its hindlimbs (Upchurch et al. 2005). All sauropods shared a similar bauplan—a small head; long neck, tail, and limbs (at least relative to other tetrapods); and a compact body—but within that common plan they explored a large range of shapes and sizes. The fit of these disparate morphologies to possible ecological roles is only beginning to be explored (Barrett and Upchurch 2005; Stevens and Parrish 2005).

## Statistical Analysis

**Materials and Methods—**I used the PDAP module in Mesquite v1.12 (Midford et al. 2005, Maddison & Maddison 2006) to test correlations among variables related to body size, neck length, and pneumaticity. Between 70 and 100 sauropod genera are currently recognized as valid, but most of these are based on very incomplete remains. My goal in these analyses is to examine the largest number of variables for the largest number of taxa. The number of variables could not have been expanded without shrinking the number of included taxa even further; similarly, broader taxonomic sampling would have sharply reduced the number of variables that could have been included.

The analyses include three variables related to absolute size:

- ANL, absolute neck length, the total length of the cervical series
- DL, dorsal length, the total length of the dorsal series
- FL, femur length

Four variables related to neck and body proportions:

- CC, cervical count, the number of cervical vertebrae
- DC, dorsal count, the number of dorsal vertebrae
- MEI, maximum elongation index, the maximum length-to-diameter ratio of the cervical vertebrae
- PNL, proportional neck length, the absolute neck length divided by the dorsal length (ANL/DL)

Three variables related to pneumaticity:

- MC, maximum number of cavities, the maximum number of cavities in a transverse section through the centrum of a presacral vertebra
- PPP, proportion of presacral pneumaticity, the number of pneumatic

presacral vertebrae divided by the total number of presacral vertebrae

- ASP, air space proportion, the volume of air space in presacral vertebrae divided by their total volume

Of all the variables, the fewest data are available for those related to pneumaticity. Furthermore, the gaps in our knowledge are concentrated on the Mamenchisauridae. Except for verbal descriptions of their vertebral internal structure as ‘honeycombed’ (see above), nothing is known about the internal structure or ASP of Omeisaurus and Mamenchisaurus. And yet these genera are among the longest-necked of all sauropods, both proportionally and absolutely. To assess the seriousness of this problem, I ran two sets of analyses. The first set included all of the taxa listed in Table 4-2 but excluded two variables related to pneumaticity (Table 4-4). The second set included all of the variables listed in Table 1 but excluded Omeisaurus and Mamenchisaurus (Table 4-5).

Another problem is that the phylogenetic relationships of several taxa are resolved differently in recent cladistic analyses. To explore the importance of phylogeny, I ran four analyses: both combinations of taxa and variables described above were assessed using two different tree topologies. The first topology (Figure 4-5) is that of Wilson (2002) with Plateosaurus and Lufengosaurus added according to Yates (2007) and Cetiosaurus added according to Upchurch and Martin (2002). The second topology (Figure 4-6) is that of Upchurch et al. (2004) with Plateosaurus and Lufengosaurus added according to Upchurch et al. (2007).

In all cases branch lengths were set to 1. Nonsignificant correlations ($p > 0.05$) between the absolute value and standard deviation of contrasts for each character were found for all characters under both topologies using both sets of taxa, which suggests that this branch length assumption is adequate (Diaz-Uriarte and Garland 1998).

**Results—**Results of the analyses are shown in Tables 4-4 and 4-5 and Figure 4-7. In all of the analyses, the proportional measures of neck length share some correlations with absolute neck length but not with other measures of absolute size (dorsal length and femur length). This is consistent with the observation that both large and small sauropods have a range of proportional neck lengths. Proportional neck length is correlated with maximum elongation index and cervical count, which is also consistent with expectations. These relationships are consistent in all four combinations of taxa, variables, and tree topologies.

In the analyses with all 18 taxa and a reduced set of variables, tree topology has little effect on the results. However, in the analyses with all 10 variables and a reduced set of taxa, tree topology has a large effect on the correlations of the pneumatic characters. ASP is strongly correlated with proportional neck length and cervical count using the topology of Upchurch et al. (2004), weakly correlated with elongation index and absolute neck length, and not correlated with dorsal length or femur length. However, using the topology of Wilson (2002), ASP is strongly correlated with absolute neck length, weakly correlated with dorsal length, femur length, and proportional neck length, and not correlated with elongation index or cervical count. Proportional presacral pneumatization (PPP) is strongly correlated with ASP in the 10-variable analyses but not correlated with other variables. In the 8-variable analyses, PPP is weakly correlated with cervical count and proportional neck length.

The results are relatively unaffected by taxon sampling. Other than the inclusion of ASP, the results of the 16 taxon, 10 variable analyses are similar to those of the 18 taxon, 8 variable analyses.

Parrish (2006) examined neck elongation in sauropods and other ornithodirans and found that neck length was positively correlated with femoral dimensions. However, that study included a broad sample of non-sauropod taxa, and the disparity in size among sauropods and their outgroups may have driven the results. The study also did not include independent contrasts. The strength of Parrish's work is in demonstrating the increase in neck length in sauropods relative to their outgroups. In contrast, this study focuses on the relationships between size and neck length within sauropods. When evolutionary relatedness is removed by the use of independent contrasts, body size and proportional measures of neck length are not correlated.

The implications of these analyses for the importance of pneumaticity are not clear. ASP is more strongly correlated with measures of absolute size under Wilson's (2002) topology, and with proportional measures of neck length under the topology of Upchurch et al. (2004). The lack of unambiguous results concerning pneumaticity is frustrating, but it may accurately reflect reality—either the complicated relationship of pneumaticity to body size and neck length, or our very limited data on pneumaticity, or both.

Furthermore, all of the results of this quantitative study should be viewed with caution. The analyses are limited to those taxa for which relatively complete remains are available. The largest sauropods for which well-studied fossils are available—Supersaurus and Argentinosaurus—were about one quarter again as large as any of the taxa included in the analysis. Bruhathkayosaurus was apparently even larger, and Amphicoelias may have been twice as large as any sauropod known from relatively complete remains. However, none of these very large sauropods could be included because they are all represented by very fragmentary remains. At best, these analyses did not sample the top 20% of sauropod size; they may not have sampled the top 50%. Our very limited knowledge of the largest sauropods hampers evolutionary inferences as much as it does physiological ones.

## CONCLUSIONS

Neck elongation was a pervasive feature of sauropod evolution. Mamenchisaurids, diplodocoids, brachiosaurids, and somphospondylans all independently evolved necks at least 9 meters long. The evolution of long necks was facilitated by increases in the number and proportional lengths of the cervical vertebrae, which also occurred several times independently. All of the longest-necked sauropods had vertebrae with complex, many-chambered internal structures, but the biomechanical importance of these pneumatic features has not been investigated. However, pneumatization lightened the presacral vertebrae of almost all sauropods and probably facilitated the evolution of long necks. For example, the neck of Sauroposeidon is much longer than that of Brachiosaurus but the mass of the cervical series would have been about the same because of the higher ASP in Sauroposeidon. In this way, many of the increases in neck length in sauropods may have been biomechanically “free”, if increases in the volume of the cervical series were offset by increases in ASP.

The largest sauropods had necks at least 13-14 meters long, and they may have had necks longer than 16 meters. The physiological implications of very long necks in sauropods are unknown. Although numerous publications on the topic have appeared, no published study includes taxa with necks more than 10 meters long—a mark that sauropods met at least three times, and probably four. The paucity of data on the largest and longest-necked sauropod is a stumbling block to the paleontologist and physiologist alike. Nevertheless, the repeated evolution of very long necks (i.e., ≥9 meters) in sauropods suggests that either the physiological challenges associated with such necks were not as great as we assume, or that the physiological limits on body size and neck length are more remote than we suppose.

Table 4-1. Measurements of sauropod cervical vertebrae. All measurements are functional lengths; the summed functional lengths of the vertebrae equal the length of the articulated cervical series. Missing vertebrae are indicated by dashes. Jobaria is not included because measurements of the individual vertebrae are not available (although the length of the articulated neck and the cervical count have been reported—see text). Abbreviation: in, incomplete vertebra.

| Taxon | Specimen | c1 | c2 | c3 | c4 | c5 | c6 | c7 | c8 | c9 | c10 | c11 | c12 | c13 | c14 | c15 | c16 | c17 | c18 | c19 |
|---|---|---|---|---|---|---|---|---|---|---|---|---|---|---|---|---|---|---|---|---|
| Plateosaurus trossingensis | GPIT 1 | - | 77 | 99 | 113 | 115 | 108 | 104 | 104 | 92 | 89 | | | | | | | | | |
| Lufengosaurus huenei | LVP-GSC V15 | - | 77 | 113 | 119 | 121 | 128 | 124 | 125 | 107 | 105 | | | | | | | | | |
| Omeisaurus tianfuensis | CCG T5701 | in | 185 | 275 | 335 | in | in | 620 | in | - | - | in | 700 | 705 | 689 | 680 | 551 | 330 | | |
| Omeisaurus tianfuensis | CCG T5703 | 32+ | 170 | 241 | 368 | 495 | 595 | 670 | 673 | 770 | 690 | 640 | in | - | - | - | - | - | | |
| Mamenchisaurus hochuanensis | CCG 20401 | 60 | 160 | 215 | 320 | 415 | 480 | 580 | 590 | 610 | 660 | 730 | 730 | 690 | 690 | 660 | 640 | 550 | 400 | 325 |

Table 4-1. (continued)

| Taxon | Specimen | c1 | c2 | c3 | c4 | c5 | c6 | c7 | c8 | c9 | c10 | c11 | c12 | c13 | c14 | c15 | c16 | c17 | c18 | c19 |
|---|---|---|---|---|---|---|---|---|---|---|---|---|---|---|---|---|---|---|---|---|
| Mamenchisaurus sinocanadorum | IVPP V10603 | - | 170 | 290 | 370 | - | - | - | - | - | - | - | - | - | - | - | - | - | - | - |
| Cetiosaurus oxoniensis | LCM G468.1968 | - | 145 | 193 | 243 | 265 | 265 | 315 | 337 | 356 | 360 | in | 254 | in | | | | | | |
| Dicraeosaurus hansemanni | HM m | - | 160 | 185 | 215 | 225 | 230 | 230 | 235 | 220 | 210 | 155 | 140 | | | | | | | |
| Apatosaurus louisae | CM 3018 | 45 | 190 | 230 | 320 | 360 | 365 | 365 | 420 | 400 | 435 | 445 | 370 | 360 | 330 | 285 | | | | |
| Diplodocus carnegii | CM 84 | 35 | 145 | 210 | 240 | 335 | 410 | 450 | 470 | 505 | 560 | 560 | 550 | 565 | 550 | 505 | | | | |
| Barosaurus lentus | AMNH 6341 | - | - | - | - | - | - | - | 590 | 630 | 660 | 715 | 715 | 760 | 745 | 730 | 620 | | | |
| Barosaurus lentus | AMNH 7535 | - | 74 | 89 | 128 | 152 | 184 | 251 | 292 | 326 | - | - | - | - | - | - | - | | | |
| Supersaurus vivianae | BYU 9024 | - | - | - | - | - | - | - | - | 1193 | - | - | - | - | - | - | - | | | |

Table 4-1. (continued)

| Taxon | Specimen | c1 | c2 | c3 | c4 | c5 | c6 | c7 | c8 | c9 | c10 | c11 | c12 | c13 | c14 | c15 | c16 | c17 | c18 | c19 |
|---|---|---|---|---|---|---|---|---|---|---|---|---|---|---|---|---|---|---|---|---|
| Haplocanthosaurus priscus | CM 879 | - | in | 140 | in | in | 190 | 185 | 175 | 180 | 205 | 220 | 255 | in | | | | | | |
| Camarasaurus supremus | AMNH 5761-1 | - | 180 | 210 | 275 | 315 | in | 445 | 520 | 410 | in | in | in | | | | | | | |
| Camarasaurus lewisi | BYU 9047 | 102 | 155 | 150 | 182 | 250 | 325 | 340 | 360 | 320 | - | - | - | | | | | | | |
| Brachiosaurus brancai | HM SI | - | 232 | 306 | 457 | 560 | 691 | 705 | - | - | - | - | - | - | | | | | | |
| Brachiosaurus brancai | HM SII | - | - | 390 | 580 | 720 | 780 | 820 | 860 | 850 | 870 | 870 | 810 | 670 | | | | | | |
| Unnamed brachiosaurid | MIWG 7306 | - | - | - | - | - | - | - | 745 | - | - | - | - | - | | | | | | |
| Sauroposeidon proteles | OMNH 53062 | - | - | - | - | - | 1070 | 1080 | 1180 | - | - | - | - | - | | | | | | |
| Euhelopus zdanski | PMU R233 | 30 | 95 | 130 | 220 | 235 | 240 | 260 | 260 | 275 | 280 | 285 | 275 | 270 | 265 | 260 | 205 | 180 | | |

Table 4-1. (continued)

| Taxon | Specimen | c1 | c2 | c3 | c4 | c5 | c6 | c7 | c8 | c9 | c10 | c11 | c12 | c13 | c14 | c15 | c16 | c17 | c18 | c19 |
|---|---|---|---|---|---|---|---|---|---|---|---|---|---|---|---|---|---|---|---|---|
| *Erketu* | IGM | - | 160 | 268 | 387 | 489 | in | - | - | - | - | - | - | - | - | - | - | - | | |
| *Ellisoni* | 100/1803 | | | | | | | | | | | | | | | | | | | |
| *Malawisaurus* | MAL | - | - | 185 | 220 | 280 | 310 | 320 | 350 | 370 | 370 | 365 | 230 | 200 | | | | | | |
| *dixeyi* | nos.[a] | | | | | | | | | | | | | | | | | | | |
| *Puertasaurus* | MPM | - | - | - | - | - | - | - | - | 930 | - | - | - | - | | | | | | |
| *reuili* | 10002 | | | | | | | | | | | | | | | | | | | |
| *Saltasaurus* | PVL | - | 70 | 100 | 155 | 160 | 160 | - | 160 | 160 | 175 | 15 | in | 13 | | | | | | |
| *loricatus* | 4017 | | | | | | | | | | | | | | | | | | | |

[a] Each vertebra has a different specimen number. See Gomani (2005) for a complete list.

SOURCES: *Plateosaurus*: Moser (2003); *Lufengosaurus*: Young (1941); *Omeisaurus*: He et al. (1988); *Mamenchisaurus hochuanensis*: Young and Zhao (1972); *Mamenchisaurus sinocanadorum*: Russell and Zheng (1993); *Cetiosaurus*: Upchurch and Martin 2002; *Dicraeosaurus*: Janensch (1929b); *Apatosaurus*: Gilmore (1936); *Diplodocus*: Hatcher (1901); *Barosaurus*, AMNH 6341: McIntosh (2005; *Barosaurus*, AMNH 7535: pers. obs.; *Supersaurus*: pers. obs.; *Haplocanthosaurus*: Hatcher (1903); *Camarasaurus supremus*: Osborn and Mook (1921); *Camarasaurus lewisi*: McIntosh et al. (1996b); *Brachiosaurus*: Janensch (1950); unnamed brachiosaurid, MIWG.7306: Naish et al. (2004); *Sauroposeidon*: pers. obs.; *Euhelopus*: Wiman (1929); *Erketu*: Ksepka and Norell (2006); *Malawisaurus*: Gomani (2005); *Puertasaurus*: Novas et al. (2005); *Saltasaurus*: Powell 1992.

Table 4-2. Variables related to body size and neck length in sauropods and their close outgroups. Abbreviations: CC, cervical count; DC, dorsal count; FL, femur length in cm; DL, length of the dorsal series in meters; ANL, absolute neck length in meters; PNL, proportional neck length, equals ANL/DL; MC, maximum number of pneumatic cavities in vertebrae in transverse section; PPP, proportion of presacral vertebrae that are pneumatic; ASP, air space proportion of pneumatic vertebrae; MEI, maximum elongation index.

| Taxon | CC | DC | FL | DL | ANL | PNL | MC | PPP | ASP | MEI | Source |
|---|---|---|---|---|---|---|---|---|---|---|---|
| Plateosaurus | 10 | 15 | 80 | 1.4 | 1.0 | 0.71 | 0 | 0 | 0 | 2.8 | Huene 1926[p] |
| Lufengosaurus | 10 | 15 | 56 | 1.2 | 1.1 | 0.92 | 0 | 0 | 0 | 2.8 | Young 1941 |
| Shunosaurus | 13 | 13 | 87 | 1.5 | 1.3 | 0.87 | 2 | 0.28 | 0.10[e] | 3.2 | Zhang 1988 |
| Omeisaurus | 17 | 12 | 134 | 2.2 | 8.5 | 4.09 | - | 1.00 | - | 4.9 | He et al. 1988 |
| Mamenchisaurus | 19 | 12 | 128 | 2.7 | 9.5 | 3.52 | - | 1.00 | - | 4.2 | Young & Zhao 1972[p] |
| Jobaria | 12 | 13 | 180 | 3.2 | 4.0 | 1.25 | 2 | 0.58 | 0.40[e] | 2.1 | Sereno et al. 1999[p] |
| Cetiosaurus | 13 | 12 | 166 | 1.8 | 2.9 | 1.61 | 2 | 1.00 | 0.30[e] | 3.2 | Upchurch & Martin 2002[p] |
| Haplocanthosaurus | 13 | 13 | 128 | 1.9 | 2.3 | 1.21 | 2 | 1.00 | 0.40 | 3.0 | Hatcher 1903[p] |

Table 4-2. (continued)

| Taxon | CC | DC | FL | DL | ANL | PNL | MC | PPP | ASP | MEI | Source |
|---|---|---|---|---|---|---|---|---|---|---|---|
| <u>Amargasaurus</u> | 13 | 9 | 105 | 1.3 | 1.5 | 1.15 | 2 | 0.67 | 0.20[e] | 2.1 | Salgado & Bonaparte 1991 |
| <u>Dicraeosaurus</u> | 11 | 13 | 122 | 1.8 | 2.2 | 1.22 | 2 | 0.61 | 0.20 | 2.2 | Janensch 1929b, Schwarz & Fristch 2006 |
| <u>Apatosaurus</u> | 15 | 10 | 179 | 2.2 | 4.9 | 2.23 | 24 | 1.00 | 0.58 | 3.3 | Gilmore 1936[P] |
| <u>Diplodocus</u> | 15 | 10 | 154 | 2.8 | 6.1 | 2.18 | 20 | 1.00 | 0.53 | 4.9 | Hatcher 1901[P] |
| <u>Barosaurus</u> | 16 | 9 | 144 | 2.4 | 8.5 | 3.54 | 20[e] | 1.00 | 0.70 | 5.3 | McIntosh 2005[P] |
| <u>Camarasaurus</u> | 12 | 12 | 180 | 2.5 | 4.0 | 1.60 | 5 | 1.00 | 0.57 | 3.5 | Osborn & Mook 1921[P] |
| <u>Brachiosaurus</u> | 13 | 12 | 203 | 3.8 | 8.5 | 2.24 | 18 | 1.00 | 0.65 | 5.4 | Janensch 1950, 1961[P] |
| <u>Euhelopus</u> | 17 | 13 | 96 | 1.6 | 4.0 | 2.50 | 20+ | 1.00 | 0.60[e] | 4.0 | Wiman 1929, Britt 1993 |
| <u>Malawisaurus</u> | 13 | 10 | 95 | 1.7 | 3.4 | 2.00 | 88 | 1.00 | 0.58 | 4.7 | Gomani 2005[P] |
| <u>Saltasaurus</u> | 13 | 10 | 150 | 1.5 | 1.8 | 1.20 | 54 | 1.00 | 0.69 | 2.7 | Powell 1992 |

Table 4-2. (continued)

| Taxon | CC | DC | FL | DL | ANL | PNL | MC | PPP | ASP | MEI | Source |
|---|---|---|---|---|---|---|---|---|---|---|---|
| Mean for all taxa | 14 | 12 | 133 | 2.1 | 4.2 | 1.89 | 16 | 0.79 | 0.41 | 3.6 | |
| Neosauropod mean | 14 | 11 | 141 | 2.1 | 4.3 | 1.92 | 23 | 0.93 | 0.52 | 3.7 | |
| Neosauropod mean sans dicraeosaurids | 14 | 11 | 150 | 2.3 | 5.2 | 2.19 | 31 | 1.00 | 0.61 | 4.2 | |

[e] estimated from published descriptions (Shunosaurus, Amargasaurus, Euhelopus) or personal examination of specimens (Jobaria, Cetiosaurus)

[p] also includes personal observations

Table 4-3. Air Space Proportion (ASP) of sections through sauropod vertebrae. Measurements are taken from CT sections, photographs, and published images. Sections are transverse unless otherwise noted. Although this dataset is almost three times as large as that reported by Wedel (2005), the mean is the about same, 0.61 compared to 0.60. Abbreviations: C, cervical; Cd, caudal; D, dorsal; P, presacral.

| Taxon | Region | | ASP | Source |
|---|---|---|---|---|
| Apatosaurus | C | condyle | 0.69 | OMNH 01094 |
| | | mid-centrum | 0.52 | " |
| | | posterior centrum | 0.73 | " |
| | | cotyle | 0.32 | " |
| | C | condyle | 0.63 | OMNH 01340 |
| | | mid-centrum | 0.69 | " |
| | | cotyle | 0.49 | " |
| | C | condyle | 0.52 | CM 555 |
| | | mid-centrum | 0.75 | " |
| | | posterior centrum | 0.59 | " |
| | | cotyle | 0.34 | " |
| | C | parapophysis | 0.60 | BYU 11998 |
| | C | cotyle | 0.70 | BYU 11889 |
| Barosaurus | C | mid-centrum | 0.56 | Janensch (1947:fig. 8) |
| | C | posterior centrum | 0.77 | Janensch (1947:fig. 3) |
| | D | condyle (sagittal) | 0.78 | Janensch (1947:fig. 9) |
| | Cd | mid-centrum | 0.47 | Janensch (1947:fig. 7) |

Table 4-3. (continued)

| Taxon | Region | | ASP | Source |
|---|---|---|---|---|
| Brachiosaurus | C | condyle | 0.55 | BYU 12866 |
| | | mid-centrum | 0.67 | “ |
| | | posterior centrum | 0.81 | “ |
| | C | condyle | 0.73 | Janensch (1950:fig. 70) |
| | C | condyle (sagittal) | 0.57 | Janensch (1947:fig. 4) |
| | D | mid-centrum | 0.59 | Janensch (1947:fig. 2) |
| Brachiosauridae | C | mid-centrum | 0.89 | MIWG 7306 |
| | P | | 0.65 | Naish & Martill (2001:pl. 32) |
| | P | | 0.85 | Naish & Martill (2001:pl. 33) |
| | P | | 0.85 | MIWG uncatalogued |
| Camarasaurus | C | condyle | 0.51 | OMNH 01109 |
| | | mid-centrum | 0.68 | “ |
| | | cotyle | 0.54 | “ |
| | C | condyle | 0.49 | OMNH 01313 |
| | | mid-centrum | 0.52 | “ |
| | | cotyle | 0.50 | “ |
| | D | mid-centrum | 0.58 | Ostrom & McIntosh (1966:pl. 23) |
| | D | mid-centrum | 0.63 | Ostrom & McIntosh (1966:pl. 23) |

Table 4-3. (continued)

| Taxon | Region | | ASP | Source |
|---|---|---|---|---|
| Camarasaurus | D | mid-centrum | 0.71 | Ostrom & McIntosh (1966:pl. 23) |
| Chondrosteosaurus | P | centrum (horiz.) | 0.70 | Naish & Martill (2001:fig. 8.5) |
| Diplodocus | C | condyle | 0.56 | BYU 12613 |
| | | mid-centrum | 0.54 | “ |
| | | posterior centrum | 0.66 | “ |
| Haplocanthosaurus | C | condyle | 0.39 | CM 879-7 |
| | | mid-centrum | 0.56 | “ |
| | | posterior centrum | 0.42 | “ |
| | | cotyle | 0.28 | “ |
| | D | mid-centrum | 0.36 | CM 572 |
| Malawisaurus | C | condyle | 0.56 | MAL-280-1 |
| | | mid-centrum | 0.62 | “ |
| | C | condyle | 0.57 | MAL-280-4 |
| | | mid-centrum | 0.56 | “ |
| Phuwiangosaurus | C | mid-centrum | 0.55 | Martin (1994:fig. 2) |
| Pleurocoelus | C | mid-centrum | 0.55 | Lull (1911:pl. 15) |
| Saltasaurus | D | centrum (horiz.) | 0.62 | Powell (1992:fig. 16) |
| | | mid-centrum | 0.55 | “ |
| | | neural spine (horiz.) | 0.82 | “ |
| Saltasaurus | D | prezygapophysis | 0.78 | Powell (1992:fig. 16) |

Table 4-3. (continued)

| Taxon | Region | | ASP | Source |
|---|---|---|---|---|
| Sauroposeidon | C | prezyg. ramus | 0.89 | OMNH 53062 |
| | | postzygapophysis | 0.74 | “ |
| | | anterior centrum | 0.75 | “ |
| Supersaurus | C | mid-centrum | 0.64 | WDC-DMJ021 |
| Sauropoda indet. | C | mid-centrum | 0.54 | OMNH 01866 |
| | C | posterior centrum | 0.46 | OMNH 01867 |
| | C | mid-centrum | 0.55 | OMNH 01882 |
| | | MEAN | 0.61 | |

TABLE 4-4. Character correlations using phylogenetically independent contrasts. Each cell contains the Pearson product-moment correlation coefficient (r2) above and the p-value of the two-tailed t-test below. Only significant ($p \leq 0.05$) p-values are listed. This set of analyses includes all 18 taxa from Table 4-2, but does not include two of the variables related to pneumaticity (MC and ASP). Values in the upper right half of the table were derived using a version of Wilson's (2002) tree topology, and those in the lower left half of the table were derived using a version of the tree topology from Upchurch et al. (2004). For a graphical presentation of these results, see Figure 4-7. Abbreviations follow Table 4-2.

| | CC | DC | FL | DL | ANL | PNL | PPP | MEI |
|---|---|---|---|---|---|---|---|---|
| CC | | 0.312 | 0.184 | 0.119 | 0.470 | 0.723 | 0.593 | 0.465 |
| | | | | | 0.042 | <0.001 | 0.007 | 0.045 |
| DC | 0.426 | | 0.089 | 0.072 | 0.197 | 0.265 | 0.358 | 0.231 |
| | | | | | | | | |
| FL | 0.140 | 0.051 | | 0.757 | 0.420 | 0.017 | 0.355 | 0.059 |
| | | | | <0.001 | | | | |
| DL | 0.026 | 0.120 | 0.710 | | 0.657 | 0.068 | 0.184 | 0.157 |
| | | | <0.001 | | 0.023 | | 0 | 0 |
| ANL | 0.558 | 0.209 | 0.405 | 0.725 | | 0.763 | 0.533 | 0.632 |
| | 0.013 | | | <0.001 | | <0.001 | 0.019 | 0.0037 |
| PNL | 0.775 | 0.331 | 0.063 | 0.258 | 0.827 | | 0.587 | 0.734 |
| | <0.001 | | | | <0.001 | | 0.008 | <0.001 |
| PPP | 0.522 | 0.421 | 0.364 | 0.103 | 0.403 | 0.526 | | 0.412 |
| | 0.022 | | | | | 0.021 | | |
| MEI | 0.557 | 0.254 | 0.037 | 0.238 | 0.641 | 0.753 | 0.388 | |
| | 0.013 | | | | 0.003 | <0.001 | | |

TABLE 4-5. Character correlations using phylogenetically independent contrasts. Each cell contains the Pearson product-moment correlation coefficient (r2) above and the p-value of the two-tailed t-test below. Only significant ($p \leq 0.05$) p-values are listed. This set of analyses includes all 10 variables taxa from Table 4-2, but does not include two of the taxa (Omeisaurus and Mamenchisaurus). Values in the upper right half of the table were derived using a version of Wilson's (2002) tree topology, and those in the lower left half of the table were derived using a version of the tree topology from Upchurch et al. (2004). For a graphical presentation of these results, see Figure 4-7. Abbreviations follow Table 4-2.

| | CC | DC | FL | DL | ANL | PNL | MC | PPP | ASP | MEI |
|---|---|---|---|---|---|---|---|---|---|---|
| CC | | 0.248 | 0.159 | 0.165 | 0.206 | 0.564<br>0.023 | 0.097 | 0.407 | 0.405 | 0.259 |
| DC | 0.359 | | 0.151 | 0.062 | 0.105 | 0.156 | 0.424 | 0.325 | 0.284 | 0.110 |
| FL | 0.263 | 0.457 | | 0.742<br>0.001 | 0.478 | 0.018 | 0.147 | 0.492 | 0.525<br>0.037 | 0.046 |
| DL | 0.186 | 0.179 | 0.698<br>0.003 | | 0.764<br><0.001 | 0.204 | 0.041 | 0.231 | 0.533<br>0.033 | 0.353 |
| ANL | 0.292 | 0.099 | 0.407 | 0.729<br>0.001 | | 0.753<br><0.001 | 0.050 | 0.306 | 0.659<br>0.006 | 0.716<br>0.002 |
| PNL | 0.629<br>0.009 | 0.231 | 0.034 | 0.192 | 0.788<br><0.001 | | 0.138 | 0.341 | 0.590<br>0.016 | 0.666<br>0.005 |
| MC | 0.331 | 0.325 | 0.351 | 0.164 | 0.137 | 0.412 | | 0.112 | 0.270 | 0.315 |
| PPP | 0.549<br>0.028 | 0.359 | 0.220 | 0.053 | 0.281 | 0.488 | 0.230 | | 0.664<br>0.005 | 0.232 |
| ASP | 0.684<br>0.003 | 0.256 | 0.162 | 0.234 | 0.627<br>0.009 | 0.764<br><0.001 | 0.485 | 0.648<br>0.007 | | 0.371 |
| MEI | 0.431 | 0.167 | 0.035 | 0.367 | 0.788<br><0.001 | 0.781<br><0.001 | 0.440 | 0.408 | 0.614<br>0.011 | |

FIGURE 4-1. The evolution of long necks in sauropods. At least four clades evolved necks longer than 9 meters: Mamenchisauridae, Diplodocoidea, Brachiosauridae, and Titanosauridae. See text for details and methods of estimation.

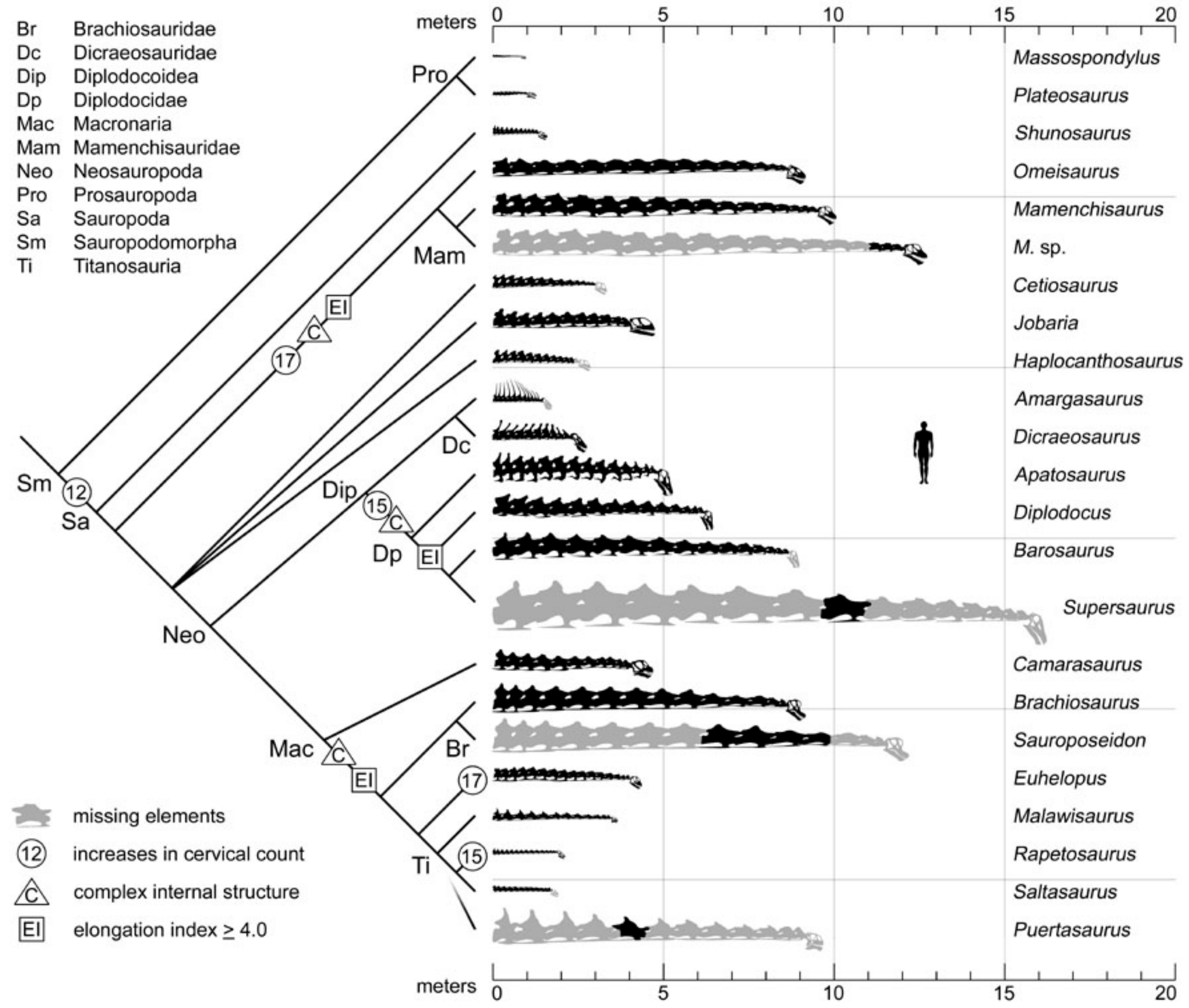

FIGURE 4-2. Measurement protocols used in this paper. A. The relationships among total length, centrum length, and functional length. B. Measurements used to calculate the elongation index (EI). EI equals centrum length divided by cotyle height.

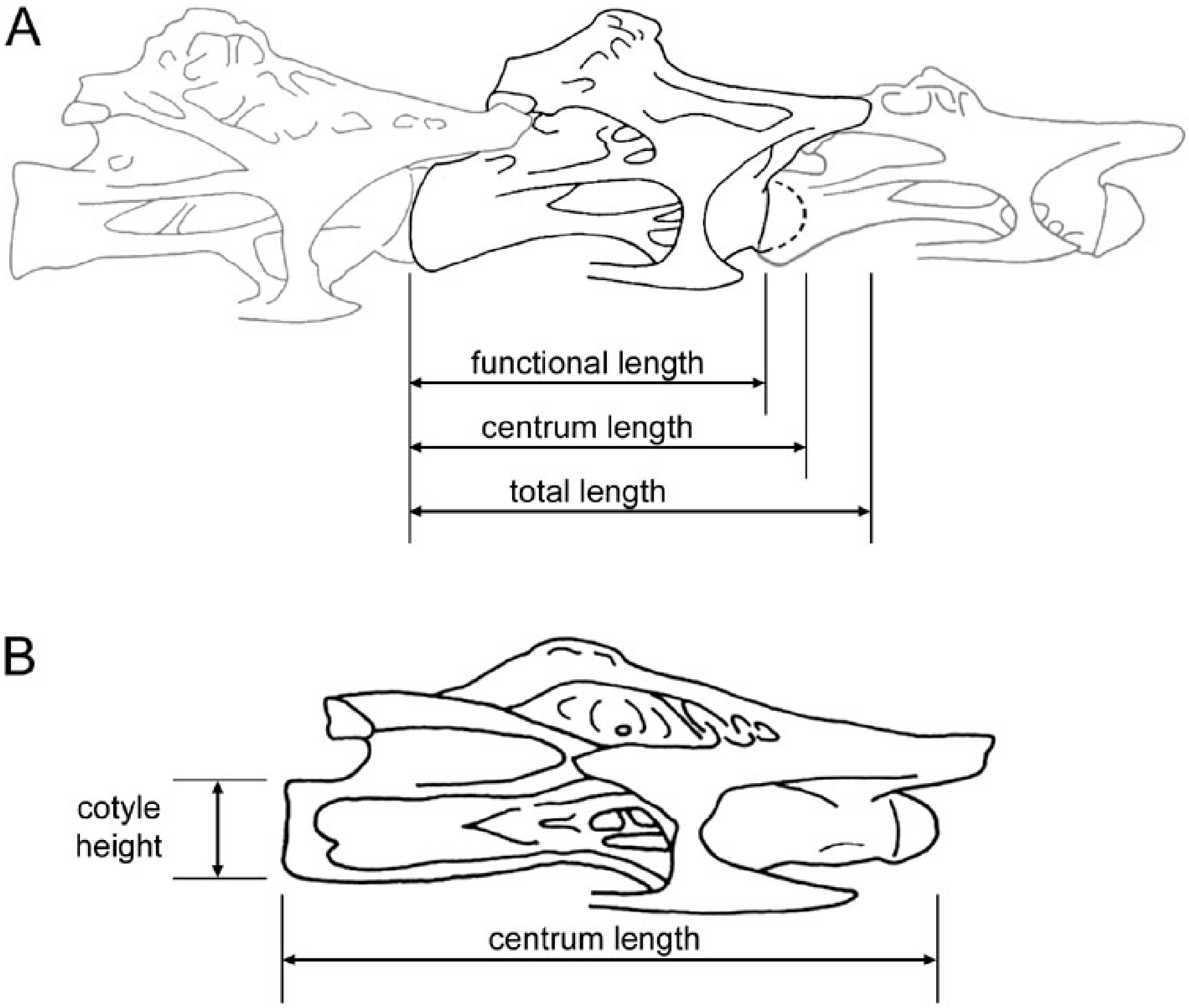

FIGURE 4-3. The three longest vertebrae ever described. All are midcervical vertebrae and all are substantially longer than one meter. They also represent three separate clades: Puertasaurus is a titanosaurian, Sauroposeidon a brachiosaurid, and Supersaurus a diplodocid. Puertasaurus after Novas et al. 2005; Sauroposeidon and Supersaurus photographed by the author.

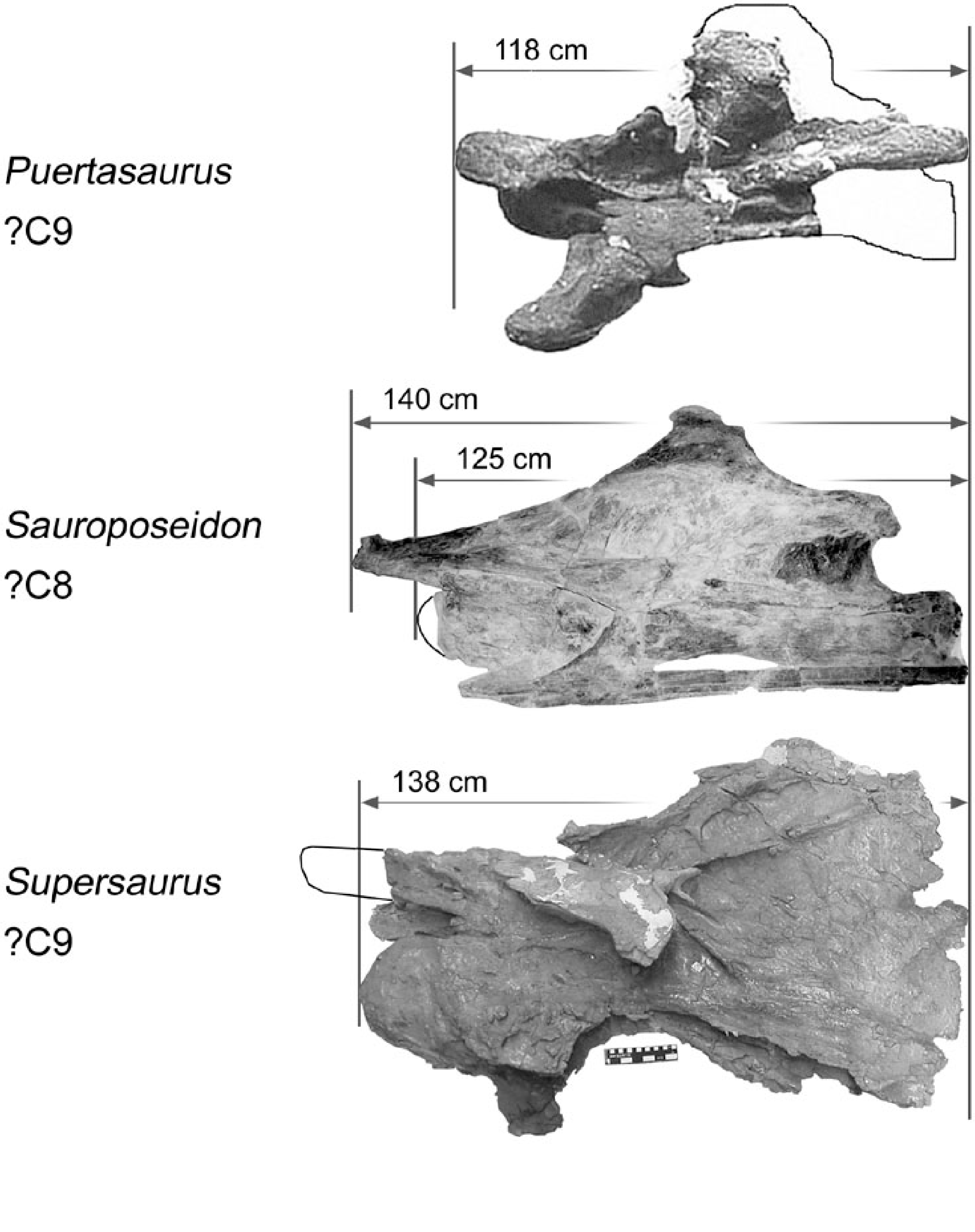

FIGURE 4-4. Pneumatic bones in Brachiosaurus. At roughly 600 kg, the 8.5-meter cervical series of Brachiosaurus had about the same mass as the animal's paired 2-meter humeri. The Brachiosaurus skeleton is scaled to HM SII, the mounted skeleton in Berlin. The human figure is 1.8 meters tall.

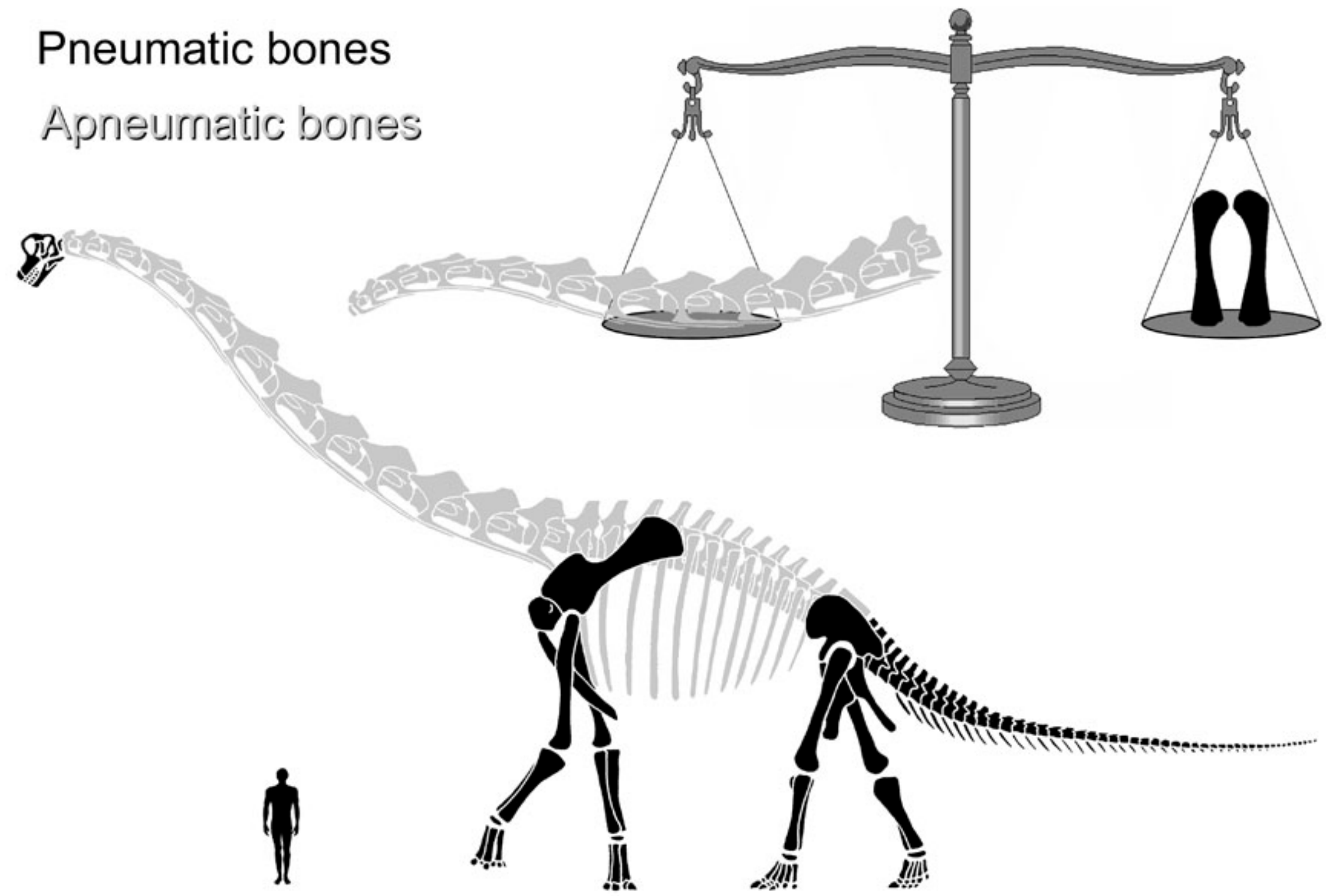
Pneumatic bones
Apneumatic bones

FIGURE 4-5. Taxa used in this study, arranged according to Wilson (2002). Plateosaurus and Lufengosaurus were added based on Yates (2007), and Cetiosaurus was added based on Upchurch and Martin (2002).

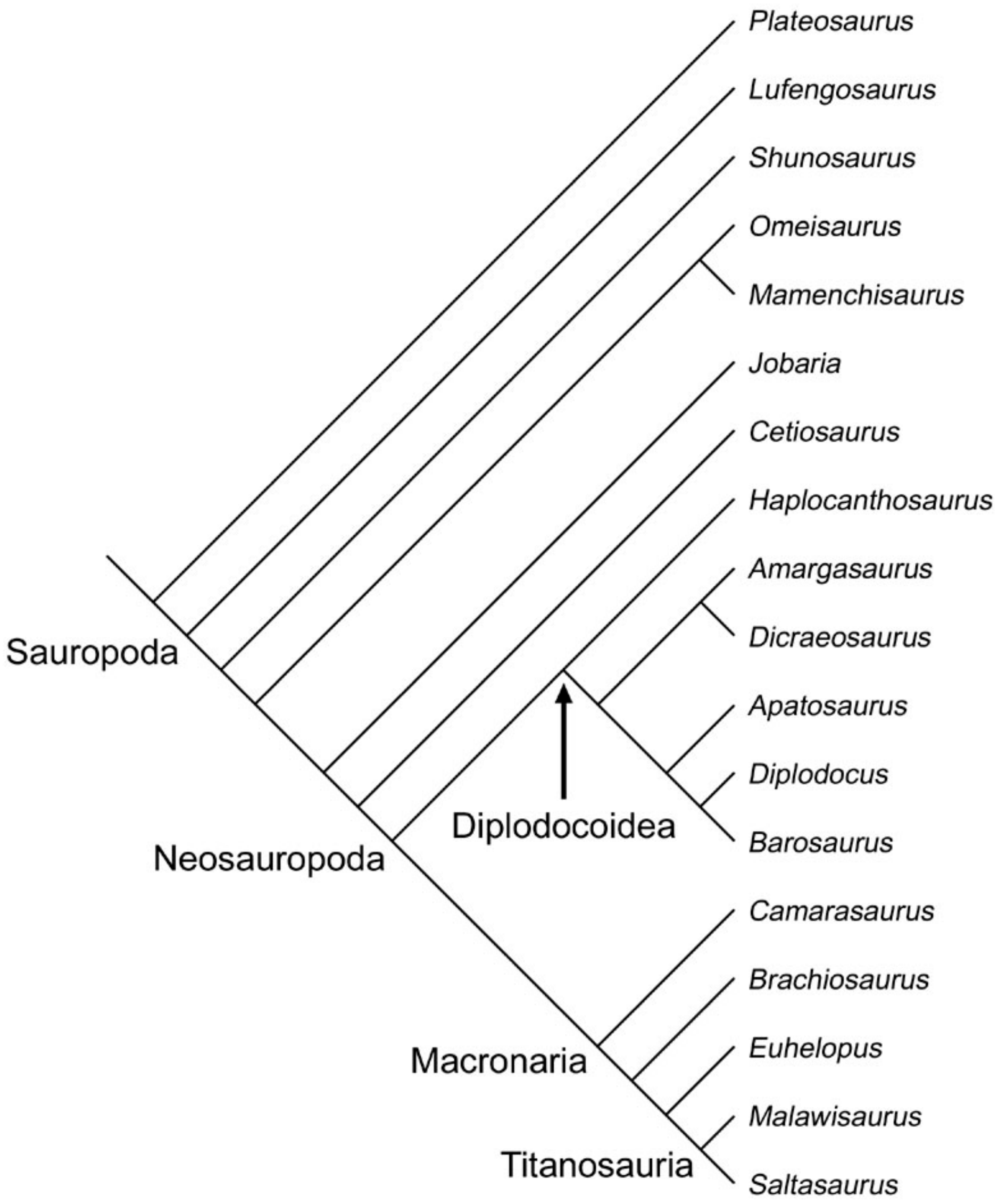

FIGURE 4-6. Taxa used in this study, arranged according to Upchurch et al. (2004). Plateosaurus and Lufengosaurus were added based on Upchurch et al. (2007).

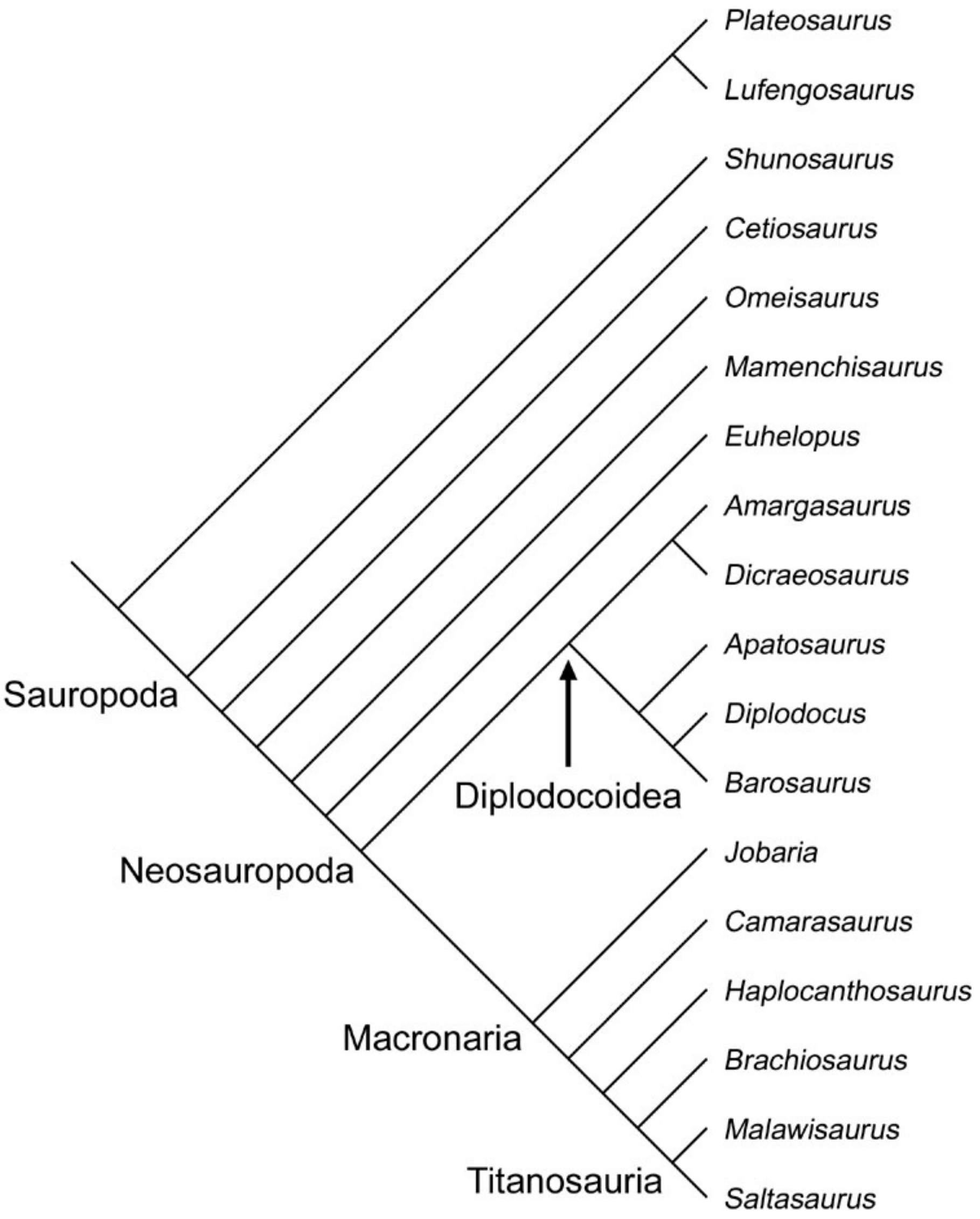

FIGURE 4-7. Graphical representation of the independent contrasts analyses. Lines between variables represent statistically significant correlations ($p \leq 0.05$). Heavy lines are strong correlations ($r^2 > 0.65$), and light dotted lines are weak correlations ($0.46 \leq r^2 \leq 065$). See Tables 4-4 and 4-5 for $r^2$ and p values. Abbreviations follow Table 4-2.

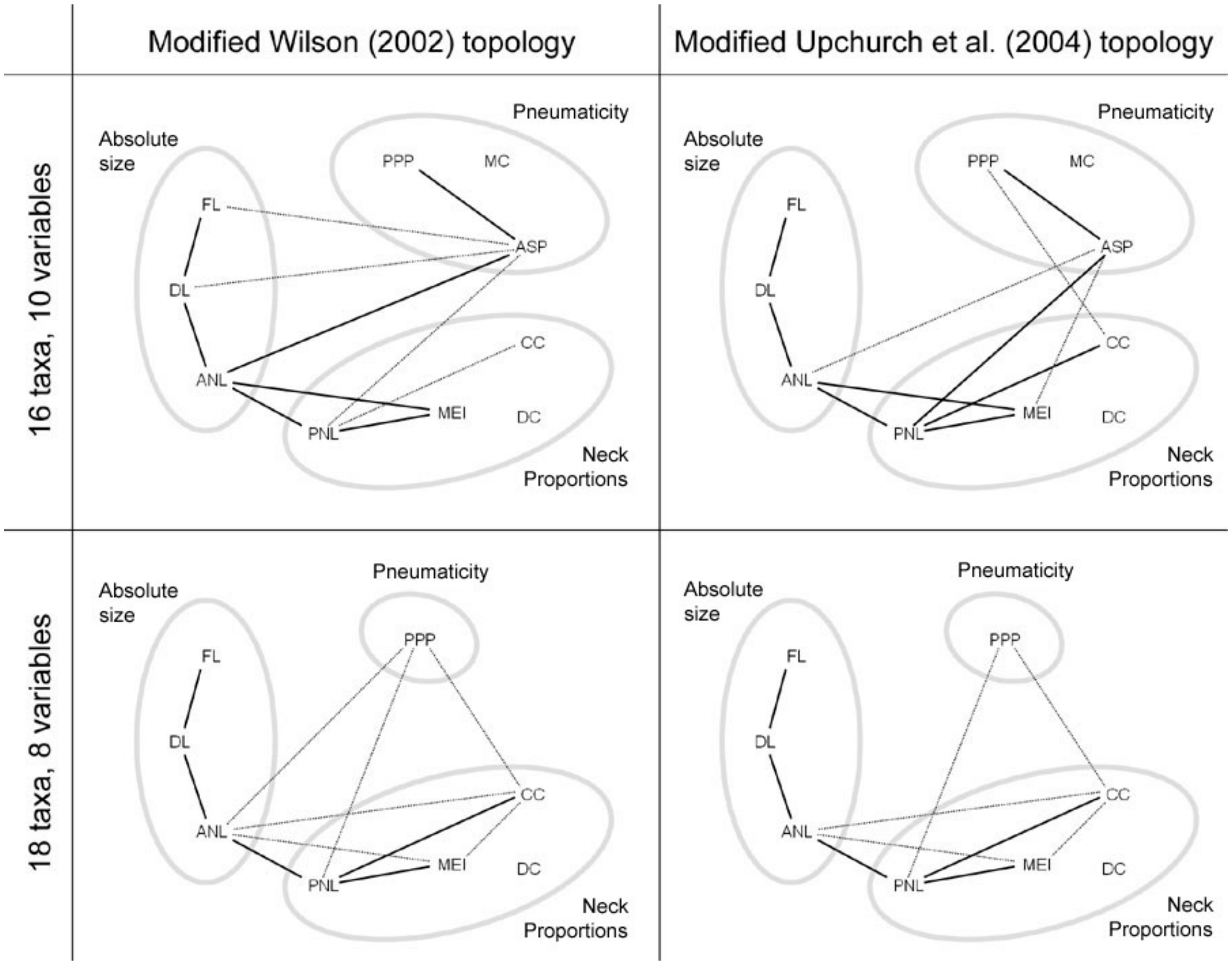
Modified Wilson (2002) topology
Modified Upchurch et al. (2004) topology
16 taxa, 10 variables
18 taxa, 8 variables
Absolute size
Pneumaticity
Neck Proportions
FL
DL
ANL
PPP
MC
ASP
CC
MEI
DC
PNL

APPENDIX I

The method of calculating the volumes of bone removed by pneumatization in Coelophysis and Pantydraco (see Palaeobiological Implications, above) is provided here. To estimate the whole body volumes of the dinosaurs I used graphic double integration (GDI: Jerison 1973; Hurlburt 1999; Murry and Vickers-Rich 2004). I traced over the skeletal reconstructions of Colbert (1989, fig. 103) and Benton et al. (2000, fig. 19) to make lateral view body outlines. Dorsal view body outlines were drawn by hand based on those of Paul (1997) and digitally manipulated to match the dimensions of the skeletal reconstructions. Using GDI, I obtained whole body volumes of 23.5 litres for Coelophysis and 3.3 litres for the holotypic individual of Pantydraco caducus; the latter animal is a small juvenile. Adjusted for scale, these results are consistent with previous mass estimates for both taxa (Peczkis 1994).

Pneumaticity is present throughout the cervical series of Coelophysis. The total length of the cervical series is c. 50 cm, and the vertebral centra have a mean diameter of 1 cm, based on measurements of uncatalogued CM specimens. The neural spines are roughly the same size as the centra. The combined cervical centra are treated as a simple cylinder 50 cm long with a diameter of 1 cm, which yields a volume of 40 $cm^3$. If the neural spines are assumed to be equal in volume to the centra, the combined volume of the cervical vertebrae is 80 $cm^3$. The cervical vertebrae of Coelophysis are probably not more than 50 per cent air by volume, based on observations of broken specimens, so the volume of bone removed during pneumatization of the cervical vertebrae was c. 40 $cm^3$, or 0.17 per cent of the volume of the body.

For Pantydraco caducus it is simpler to calculate the volumes of the individual fossae. The fossae on cervicals 6–8 are each c. 5 mm long, 2.5 mm tall, and 1.25 mm deep. We can think of the paired fossae on each vertebra as forming the two halves of an oblate spheroid with x, y and z diameters of 5, 2.5 and 2.5 mm. The volume of this spheroid, and thus the volume of the paired fossae, is 0.016 $cm^3$. The fossae on cervicals 6–8 are all roughly the same size, and the visible fossa on the ninth cervical is only about half as deep. The volume of bone removed during fossa formation is therefore 0.057 $cm^3$, or 0.0017 per cent of the volume of the body.

These calculations are all approximate, but they are sufficient to demonstrate that PSP did not have a noticeable effect on the skeletal mass of basal saurischians. In the case that I have underestimated the volume of the pneumatic chambers in Coelophysis relative to the body volume by a factor of six: the volume of these chambers would still only be one percent of the volume of the body. In contrast, the volume of air in the pneumatic vertebrae of Tyrannosaurus and Diplodocus accounted for 4–6 per cent of the volume of the animals. These air spaces replaced bone, a relatively dense tissue, and lightened the animals by 7–10 per cent (Wedel 2004, 2005).

APPENDIX II

The method of calculating the relative densities of pneumatic and apneumatic bones is given here. The thickness of the walls of tubular bones is typically expressed as the variable K, which is the inner diameter of the bone (r) divided by the outer diameter (R). For a large sample (>150) of avian long bones, Cubo and Casinos (2000) reported average K values of 0.65 for apneumatic bones and 0.77 for pneumatic bones. The amount of the cross-sectional area that is not occupied by bone can be found by taking the square of K ($\pi r^2/\pi R^2 = r^2/R^2 = K^2$). For the apneumatic bones, this is $0.65^2$ or 0.42, and for the pneumatic bones it is $0.77^2$ or 0.59. This means that on average, the cross-sectional area of pneumatic element is 59% air and 41% bone, and the cross-sectional area of an apneumatic element is 42% marrow and 58% bone. The specific gravity of marrow is 0.95 (Currey and Alexander 1985), and that of avian compact bone is 1.8 (Spector 1956). The mass of air is negligible. On average, the density of pneumatic avian long bones is 1.8 x 0.41 = 0.74 g/cm$^3$, and the density of apneumatic avian long bones is (1.8 x 0.58) + (0.95 x 0.42) = 1.4 g/cm$^3$. It may seem surprising that pneumatic bones that differ from apneumatic bones by 10% of K are only half as dense. However, the cross-sectional geometry of the bones is proportional to the square of K, and that the diaphyses of apneumatic bones are filled with marrow which contributes much of the mass of the elements. For more information on the impact of pneumaticity on body mass in birds and other dinosaurs, see Wedel (2005).

APPENDIX III

Specimens personally examined by the author during the research for Chapter Three.

| Taxon | Specimen Number |
|---|---|
| Non-dinosaurian archosauromorphs | |
| Erythrosuchus africanus | BMNH R533, R3592, R3762 |
| "Mandasuchus" sp. | BMNH U2/28, U22/2 |
| Prestosuchus chiniquensis | AMNH 3856 |
| Postosuchus kirkpatricki | CM 73372 |
| Arizonasaurus babbitti | MSM 4590 |
| Lotosaurus adentus | IVPP V4881 |
| Effigia okeeffeae | AMNH 30587-30590 |
| Eoraptor lunensis | PVSJ 512 |
| Sauropodomorphs | |
| Pantydraco caducus | BMNH P24 |
| Plateosaurus trossingensis | AMNH 6810, SMNS 13200 |
| Anchisaurus polyzeus | YPM 1883 |
| Mamenchisaurus hochuanensis | CCG V 20401 |
| Jobaria tiguidensis | MNN TIG3, MNN TIG4 |
| Cetiosaurus oxoniensis | LCM G468.1968, OUMNH lectotype series[1] |
| Haplocanthosaurus priscus | CM 572, CM 879 |

| Taxon | Specimen Number |
|---|---|
| Amphicoelias altus | AMNH 5764 |
| Apatosaurus louisae | CM 3018 |
| Apatosaurus excelsus | FMNH P25112 |
| Apatosaurus sp. | CM 555, OMNH 1331, 1436 |
| "Apatosaurus" minimus | AMNH 675 |
| Diplodocus carnegii | CM 84/94 |
| Diplodocus longus | AMNH 223 |
| Diplodocus sp. | AMNH 7532, AMNH 7533, CM 36039 |
| Barosaurus lentus | AMNH 6341, AMNH 7535 |
| Bellusaurus sui | IVPP V8300 |
| Zizhongosaurus chuanchensis | IVPP V9067 |
| Camarasaurus lewisi | BYU 9047 |
| Camarasaurus supremus | AMNH 5761 |
| Brachiosaurus altithorax | FMNH P 25107 |
| Brachiosaurus sp. | BYU 12866, BYU 12867 |
| Erketu ellisoni | IGM 100/1803 |
| Malawisaurus dixeyi | MAL holotype series[2] |
| Rapetosaurus krausei | FMNH |
| Neosauropoda indet. | AMNH 860 |
| Titanosauria indet. | YPM 5103, 5104, 5116, 5147, 5151, 5152, 5199, 5294, 5449 |

| Taxon | Specimen Number |
|---|---|
| | Theropods |
| Coelophysis bauri | AMNH FR 7223, 7224, CM 76864, UCMP 129618 |
| Dilophosaurus wetherilli | UCMP 37302, 37303 |
| Allosaurus fragilis | AMNH 666 |
| Acrocanthosaurus atokensis | OMNH 10146 |
| Sinraptor dongi | IVPP 10600 |
| Chilantaisaurus maortuensis | IVPP AS 2885 |
| Archaeornithomimus sp. | AMNH 6570, 21790 |
| Sinornithomimus dongi | IVPP V11797 |
| Caudipteryx dongi | IVPP V12344 |
| Deinonychus antirrhopus | AMNH 3015, YPM 5204, 5210 |
| Microraptor sp. | IVPP V12330, 13352, 13320 |
| Sinovenator changii | IVPP V 12583 |

[1]Includes numerous specimens; see Upchurch and Martin (2002) for complete list.

[2]Includes numerous specimens; see Gomani (2005) for complete list.